\pdfoutput=1
\documentclass[cernpreprint, atlasdraft=false, coverpage=false, texlive=2023, UKenglish, orcidlogo, texmf, orcidlogo]{atlasdoc}
 
\usepackage{atlaspackage}
\usepackage{atlasbiblatex}
 
\usepackage{atlasphysics}

\graphicspath{{logos/}{figures/}}
 
\usepackage{ANA-HIGG-2020-16-PAPER-defs}
 
\usepackage{adjustbox}
\usepackage{forest}
\usepackage{xifthen}
\usepackage{multirow,bigdelim,makecell}
\usepackage{cellspace}
\usepackage{lscape}
\usepackage{caption}
\usepackage{pdflscape}
\usepackage{siunitx}
\usepackage{xparse}
\usepackage{bm}
\usepackage{graphbox}

\AtlasTitle{Measurement of the properties of Higgs boson production at $\sqrt{s} = 13$~TeV in the $H\to\gamma\gamma$ channel using $139$~fb$^{-1}$ of $pp$ collision data with the ATLAS experiment}

\AtlasAbstract{
Measurements of Higgs boson production cross-sections are carried out in the diphoton decay channel using 139~fb$^{-1}$ of $pp$ collision data at $\sqrt{s} = 13$~TeV collected by the ATLAS experiment at the LHC. The analysis is based on the definition of 101 distinct signal regions using machine-learning techniques. The inclusive Higgs boson signal strength in the diphoton channel is measured to be $1.04^{+0.10}_{-0.09}$. Cross-sections for gluon-gluon fusion, vector-boson fusion, associated production with a $W$ or $Z$ boson, and top associated production processes are reported. An upper limit of 10 times the Standard Model prediction is set for the associated production process of a Higgs boson with a single top quark, which has a unique sensitivity to the sign of the top quark Yukawa coupling. Higgs boson production is further characterized through measurements of Simplified Template Cross-Sections (STXS). In total, cross-sections of 28 STXS regions are measured. The measured STXS cross-sections are compatible with their Standard Model predictions, with a $p$-value of $93\%$. The measurements are also used to set constraints on Higgs boson coupling strengths, as well as on new interactions beyond the Standard Model in an effective field theory approach. No significant deviations from the Standard Model predictions are observed in these measurements, which provide significant sensitivity improvements compared to the previous ATLAS results.
}

\AtlasRefCode{HIGG-2020-16}
 
\PreprintIdNumber{CERN-EP-2022-094}

\AtlasJournalRef{JHEP 07 (2023) 088}
\AtlasDOI{DOI:10.1007/JHEP07(2023)088}

\hypersetup{pdftitle={ATLAS document},pdfauthor={The ATLAS Collaboration}}
 
\begin{document}
 
\maketitle
 
\tableofcontents

\section{Introduction}
 
The experimental characterization of the Higgs boson discovered by the ATLAS and CMS experiments~\cite{HIGG-2012-27, CMS-HIG-12-028} is not only crucial for our understanding of the mechanism of electroweak symmetry breaking~\cite{Englert:1964et,Higgs:1964pj,Guralnik:1964eu} but also for providing insight into physics beyond the Standard Model (SM). Despite a small Higgs boson to diphoton (\Hyy) branching ratio of ($0.227 \pm 0.007$)\%~\cite{deFlorian:2016spz} in the SM, measurements in the diphoton final state have yielded some of the most precise determinations of Higgs boson properties~\cite{HIGG-2013-08,HIGG-2013-10,HIGG-2013-12,HIGG-2016-21,HIGG-2016-33}, thanks to the excellent performance of photon reconstruction and identification with the ATLAS detector.
 
The signature of the Higgs boson in the diphoton final state is a narrow peak in the diphoton invariant mass (\mgg) distribution with a width consistent with detector resolution, rising above a smoothly falling background. The diphoton mass resolution for such a resonance is typically between 1~\GeV\ and 2~\GeV, depending on the event kinematics. The mass and event yield of the Higgs boson signal can be extracted through fits of the \mgg\ distribution. Properties of the Higgs boson have been studied extensively in the diphoton final state by the ATLAS and CMS experiments~\cite{HIGG-2013-17,HIGG-2016-21,HIGG-2018-57,HIGG-2016-33,HIGG-2019-01,CMS-HIG-19-004,CMS-HIG-17-031,CMS-HIG-17-025,CMS-HIG-19-015,CMS-HIG-19-013}.
This paper reports measurements of Higgs boson production cross-sections in the diphoton decay channel, using a data set of proton--proton collisions at $\sqrt{s} = 13$~\TeV\ collected by the ATLAS experiment from 2015 to 2018, a period known as Run~2 of the Large Hadron Collider (LHC). Its integrated luminosity is 139~\ifb~\cite{ATLAS-CONF-2019-021,LUCID2}, a roughly fourfold increase compared to the previous ATLAS publication of such measurements in the diphoton channel~\cite{HIGG-2016-21}. Apart from the increased data set size, the most significant improvement in the sensitivity is due to redesigned and refined event selection and categorization techniques compared to Ref.~\cite{HIGG-2016-21}. Uncertainties on the modeling of continuum background have been reduced through the use of a smoothing procedure based on a Gaussian kernel~\cite{Frate:2017mai}. The performance of the reconstruction and selection of the physics objects used in these measurements has also been generally improved.
 
The analysis is optimized to measure production cross-sections in the Simplified Template Cross-Section (STXS) framework~\cite{Badger:2016bpw,deFlorian:2016spz,STXS,Amoroso:2020lgh}, in which the Higgs boson production phase space is partitioned by production process as well as by kinematic and event properties. Thanks to the increased integrated luminosity and an improved analysis method, a total of 28 STXS regions are measured in this analysis, compared to 10 in Ref.~\cite{HIGG-2016-21}. By combining several STXS regions, the analysis provides strong sensitivity to the cross-sections of the main Higgs boson production modes, gluon-gluon fusion (\ggF), vector-boson fusion (\VBF), and associated production with a vector boson (\VH\, where $V$ = $W$ or $Z$), or a top quark pair (\ttH). The analysis is furthermore specifically optimized for the detection of single-top associated production of the Higgs boson (\tH), which has a unique sensitivity to the sign of the top-quark Yukawa coupling. While the analysis does not reach sensitivity to the small \tH\ event yield predicted by the SM, it can set constraints on enhanced \tH\ rates due to potential effects from physics beyond the Standard Model (BSM)~\cite{Ellis:2013yxa}. A measurement of the inclusive Higgs boson production yield within $|y_H|<2.5$ in the diphoton channel is also reported. Uncertainties and correlations of the production mode cross-section measurements are reduced, and in particular, the uncertainties in the measurements of \VH\ and top-associated production modes are reduced by more than a factor of four.
 
Two sets of interpretations of these measurements are also performed to provide constraints on potential effects arising from BSM physics: one in terms of Higgs boson coupling strengths within the $\kappa$-framework~\cite{deFlorian:2016spz}, and the other in terms of  Wilson coefficients describing potential BSM interactions in the context of a Standard Model effective field theory (SMEFT) model~\cite{Buchmuller:1985jz,Grzadkowski:2010es,SMEFTsim3}.

This paper is organized as follows. Section~\ref{sec:detector} describes the ATLAS detector, Section~\ref{sec:dataMC} details the data and Monte Carlo simulation samples used in this analysis, Section~\ref{sec:reco} explains the object reconstruction and event selection. The design of the measurement is discussed in Section~\ref{sec:design}, and the modelling of the diphoton mass distribution is discussed in Section~\ref{sec:modeling}. Systematic uncertainties are described in Section~\ref{sec:systs}, and Section~\ref{sec:results} presents the measurement results. Sections~\ref{sec:results:kappas} and~\ref{sec:results:smeft}  respectively report the results of interpretations in the context of the $\kappa$-framework and the SMEFT model. Conclusions are presented in Section~\ref{sec:conclusion}.


\section{ATLAS detector}
\label{sec:detector}

The ATLAS detector~\cite{PERF-2007-01} at the LHC covers nearly the entire solid angle around the collision point.\footnote{\AtlasCoordFootnote}
It consists of an inner tracking detector surrounded by a thin superconducting solenoid, electromagnetic and hadronic calorimeters,
and a muon spectrometer incorporating three large superconducting toroidal magnets.
 
The inner-detector system (ID) is immersed in a \SI{2}{\tesla} axial magnetic field
and provides charged-particle tracking in the range \(|\eta| < 2.5\).
The high-granularity silicon pixel detector covers the vertex region and typically provides four measurements per track,
the first hit normally being in the insertable B-layer installed before Run~2~\cite{ATLASDET-IBL,PIX-2018-001}.
It is followed by the silicon microstrip tracker, which usually provides eight measurements per track.
These silicon detectors are complemented by the transition radiation tracker (TRT),
which enables radially extended track reconstruction up to \(|\eta| = 2.0\).
The TRT also provides electron identification information
based on the fraction of hits (typically 30 in total) above a higher energy-deposit threshold corresponding to transition radiation.
 
The calorimeter system covers the pseudorapidity range \(|\eta| < 4.9\).
Within the region \(|\eta|< 3.2\), electromagnetic calorimetry is provided by barrel and
endcap high-granularity lead/liquid-argon (LAr) calorimeters,
with an additional thin LAr presampler covering \(|\eta| < 1.8\)
to correct for energy loss in material upstream of the calorimeters.
Hadronic calorimetry is provided by the steel/scintillator-tile calorimeter,
segmented into three barrel structures within \(|\eta| < 1.7\), and two copper/LAr hadronic endcap calorimeters.
The solid angle coverage is completed with forward copper/LAr and tungsten/LAr calorimeter modules
optimized for electromagnetic and hadronic measurements respectively.
 
The muon spectrometer (MS) comprises separate trigger and
high-precision tracking chambers measuring the deflection of muons in a magnetic field generated by superconducting air-core toroids.
The field integral of the toroids ranges between \num{2.0} and \SI{6.0}{\tesla$\cdot$\metre}
across most of the detector.
A set of precision chambers covers the region \(|\eta| < 2.7\) with three layers of monitored drift tubes,
complemented by cathode-strip chambers in the forward region, where the background is highest.
The muon trigger system covers the range \(|\eta| < 2.4\) with resistive-plate chambers in the barrel, and thin-gap chambers in the endcap regions.
 
Interesting events are selected to be recorded by the first-level trigger system implemented in custom hardware,
followed by selections made by algorithms implemented in software in the high-level trigger~\cite{TRIG-2016-01}.
The first-level trigger accepts events from the \SI{40}{\MHz} bunch crossings at a rate below \SI{100}{\kHz},
which the high-level trigger reduces to about \SI{1}{\kHz} in order to record events to disk. An extensive software suite~\cite{ATL-SOFT-PUB-2021-001} is used in the reconstruction and analysis of real and simulated data, in detector operations, and in the trigger and data acquisition systems of the experiment.


\section{Data and simulation samples}
\label{sec:dataMC}
 
\subsection{Data}
 
This study uses a data set of $\sqrt{s}=13$~\TeV\ proton--proton collisions recorded by the ATLAS detector during a period ranging from 2015 to 2018, corresponding to Run 2 of the LHC. After data quality requirements~\cite{DAPR-2018-01} are applied to ensure that all detector components are in good working condition, the data set amounts to an integrated luminosity of $139.0 \pm 2.4$~fb$^{-1}$~\cite{ATLAS-CONF-2019-021,LUCID2}. The mean number of interactions per bunch crossing, averaged over all colliding bunch pairs, was $\langle \mu \rangle=33.7$ for this data set.
 
Events are selected if they pass either a diphoton or single-photon trigger. The diphoton trigger has transverse energy thresholds of 35~\GeV\ and 25~\GeV\ for the leading and subleading photon candidates, respectively~\cite{TRIG-2018-05}, with photon identification selections based on calorimeter shower shape variables. In 2015--2016, a \emph{loose} photon identification requirement was used in the trigger, while in 2017--2018, a tighter requirement was used to cope with higher instantaneous luminosity. The single-photon trigger requires the transverse energy of the leading photon be greater than 120~\GeV\ in data collected between 2015 and 2017, with the threshold rising to 140~\GeV\ for data collected in 2018. The photon candidate used in the trigger decision is required to pass the \emph{loose} photon identification requirement mentioned above. On average, the trigger efficiency is greater than 98\% for events that pass the diphoton event selection described in Section~\ref{sec:reco}, with no substantial variations over the data-taking period. The addition of the single-photon trigger improves the selection efficiency by 1\% overall, and by up to 2\% for high-\pt\ Higgs boson candidates.
 
\subsection{Simulation samples}
\label{sec:mc}
 
Major Higgs boson production processes, including \ggF, \VBF, \VH, \ttH, and associated production with a pair of bottom quarks (\bbH) were generated using \POWHEGBOX[v2]~\cite{Alioli:2010xd,Nason:2004rx,Frixione:2007vw,Alioli:2010xd,Hartanto:2015uka}. The \ggF\ simulation achieves next-to-next-to-leading-order (NNLO) accuracy for  inclusive \ggF\ observables by reweighting the Higgs boson rapidity spectrum in {\textsc{Hj-MiNLO}}~\cite{Hamilton:2012np,Campbell:2012am,Hamilton:2012rf} to that of HNNLO~\cite{Catani:2007vq}. The Higgs boson transverse momentum spectrum obtained with this sample is found to be compatible with the fixed-order HNNLO calculation and the  \HRES[2.3] calculation~\cite{Bozzi:2005wk,deFlorian:2011xf} performing resummation at next-to-next-to-leading-logarithm accuracy matched to a NNLO fixed-order calculation (NNLL+NNLO).  The \VBF\ process was simulated at next-to-leading-order (NLO) accuracy in QCD. The simulation of the \WH\ and \qqqgZH\ processes is accurate to NLO in QCD with up to one extra jet in the event, while the simulation for the \ggtoZH\ process was performed at leading order in QCD. The \ttH\ and \bbH\ processes were simulated at NLO in the strong coupling constant \alphas\ in the five-flavour scheme.
The \PDFforLHC[15]\ sets~\cite{Butterworth:2015oua} of parton distribution functions (PDFs) were used for all the processes listed above. The NNLO set was used for \ggF, and the NLO set for other processes.
 
The \tHqb\ (\tHW) samples were produced with \MGNLO[2.6]~\cite{Alwall:2014hca} in the four-flavour (five-flavour) scheme with the \NNPDF[3.0nnlo] PDF. The same flavour scheme was used in the matrix element calculation and the PDF. The top quark and $W$ boson decays were handled by \MADSPIN~\cite{Artoisenet:2012st} to account for spin correlations in the decay products. The overlap of the \tHW\ process with \ttH\ at NLO was removed by using a diagram removal technique~\cite{Frixione:2008yi,Demartin:2016axk}
The \pptotHb\ process has a small cross-section and was not considered in the modelling of \tH\ production.
 
All generated events for the processes listed above were interfaced to \PYTHIA[8.2]~\cite{pythia8,Sjostrand:2014zea} to model parton showering, hadronization and the underlying event using the AZNLO set of parameter values tuned to data~\cite{Aad:2014xaa}.
The decays of bottom and charm hadrons were simulated using the \EVTGEN[1.6.0]\ program~\cite{Lange:2001uf}. Systematic uncertainties related to the signal
modeling are estimated using a set of samples where \HERWIG{7}~\cite{Bellm:2017jjp,Bellm:2015jjp} is used for parton showering.
 
Major Higgs boson production processes were also simulated using alternative generator programs, in order to check the signal model and associated uncertainties (see Section~\ref{sec:theo_stxs}). The \ggF\ process was also generated with \MGNLO, using an NLO-accurate matrix element for up to two additional partons and applying the \textsc{FxFx} merging scheme to obtain an inclusive sample~\cite{Alwall:2014hca,Frederix:2012ps}. The generation used an effective vertex with a point-like coupling between the Higgs boson and gluons in the infinite top-mass limit. The events were showered using \PYTHIA[8.2] with the A14 set of tuned parameters~\cite{ATL-PHYS-PUB-2014-021}. The  \VBF\ alternative sample was generated with \MGNLO\ at NLO accuracy in the matrix element. It was then showered with \HERWIG[7.1.6].
The \VH\ alternative sample was simulated with \MGNLO, and the simulation is accurate to NLO in QCD for zero or one additional parton merged with the FxFx merging scheme. The \ggtoZH\ process was also simulated at LO with \MGNLO\ and showered with \PYTHIA[8.2]. The \ttH\ alternative sample was simulated with \MGNLO at NLO and the parton showering was performed with \PYTHIA[8.2].
 
All Higgs boson signal events were generated with a Higgs boson mass ($m_H$) of $125\,\GeV$ and an intrinsic width ($\Gamma_H$) of $4.07\,\MeV$~\cite{LHCHiggsCrossSectionWorkingGroup:2013rie}.
The cross-sections of Higgs production processes are reported for a centre-of-mass energy of
$\sqrt{s}=13$~\TeV\ and a Higgs boson with mass $m_H = 125.09$~\GeV~\cite{HIGG-2014-14}. These cross-sections
~\cite{deFlorian:2016spz,
Anastasiou:2016cez,Anastasiou:2015ema,Dulat:2018rbf,Harlander:2009mq,Harlander:2009bw,Harlander:2009my,Pak:2009dg,Actis:2008ug,Actis:2008ts,Bonetti:2018ukf,Aglietti:2004nj,Aaboud:2017vzb,
Bagnaschi:2011tu,Hamilton:2015nsa,
Ciccolini:2007jr,Ciccolini:2007ec,Bolzoni:2010xr,
Brein:2012ne,Harlander:2018yio,Brein:2003wg,Brein:2011vx,Altenkamp:2012sx,Harlander:2014wda,Denner:2014cla,Ciccolini:2003jy,
Beenakker:2002nc,Dawson:2003zu,Yu:2014cka,Frixione:2015zaa,
Dawson:2003kb,Dittmaier:2003ej,Harlander:2011aa,
Demartin:2015uha,Demartin:2016axk}, shown in Table~\ref{tab:mc_table}, are used together with the Higgs boson branching ratio to diphotons~\cite{deFlorian:2016spz,Djouadi:1997yw,Spira:1997dg,Djouadi:2006bz,Bredenstein:2006ha,Bredenstein:2006rh,Bredenstein:2006nk} to scale the simulated signal samples to their SM predictions.

Prompt diphoton production (\gamgam) was simulated with the
\Sherpa[2.2.4]~\cite{Bothmann:2019yzt} generator. In this set-up, NLO-accurate
matrix elements for up to one parton, and LO-accurate matrix elements for up
to three partons were calculated with the Comix~\cite{Gleisberg:2008fv} and
\OPENLOOPS~\cite{Buccioni:2019sur,Cascioli:2011va,Denner:2016kdg} libraries. They were matched
with the \Sherpa\ parton shower~\cite{Schumann:2007mg} using the MEPS@NLO
prescription~\cite{Hoeche:2011fd,Hoeche:2012yf,Catani:2001cc,Hoeche:2009rj}
with a dynamic merging cut~\cite{Siegert:2016bre} of 10~\GeV.
Photons were required to be isolated according to a smooth-cone isolation
criterion~\cite{Frixione:1998jh}. Samples were generated using the
\NNPDF[3.0nnlo] PDF set~\cite{Ball:2014uwa},
along with the dedicated set of tuned parton-shower parameters developed by the \Sherpa\ authors.
 
The production of $V\gamma\gamma$ events was simulated with the
\Sherpa[2.2.4]~\cite{Bothmann:2019yzt} generator. QCD LO-accurate matrix elements
for up to one additional parton emission were matched and merged with the \Sherpa\ parton shower based on the Catani--Seymour
dipole factorization~\cite{Gleisberg:2008fv,Schumann:2007mg} using the MEPS@LO
prescription~\cite{Hoeche:2011fd,Hoeche:2012yf,Catani:2001cc,Hoeche:2009rj}.
Samples were generated using the same PDF set and parton-shower parameters as the \gamgam\ sample.
The production of $t\bar{t}\gamma\gamma$\ events was modelled using the \MGNLO[2.3.3]
generator at LO with the \NNPDF[2.3lo]~\cite{Ball:2012cx} PDF.
The parton-showering and underlying-event simulation were performed using \PYTHIA[8.2].

The effect of multiple interactions in the same and neighbouring bunch
crossings (pile-up) was modelled by overlaying the original hard-scattering event with simulated inelastic
proton--proton ($pp$) events generated with 
\PYTHIA[8.1] using the \NNPDF[2.3lo] PDF set and the
A3 tune~\cite{ATL-PHYS-PUB-2016-017}.
The generated signal and background events were passed through a simulation of the ATLAS detector~\cite{SOFT-2010-01} using the~\GEANT~toolkit~\cite{Agostinelli:2002hh}. The only exception is the prompt diphoton sample: due to the large size of the sample, the generated events were instead processed using a fast simulation of the ATLAS detector~\cite{ATL-PHYS-PUB-2010-013} where the full simulation of the calorimeter is replaced with a parameterization of the calorimeter response.

A summary of the simulated signal and background samples is shown in Table~\ref{tab:mc_table}.
 
\begin{table}[ht]
\caption{Event generators and PDF sets used to model signal and background processes.
The cross-sections of Higgs boson production processes~\cite{deFlorian:2016spz,Anastasiou:2015ema, Anastasiou:2016cez, Actis:2008ug, Anastasiou:2008tj,Ciccolini:2007jr,Ciccolini:2007ec,Bolzoni:2010xr,Brein:2003wg,Altenkamp:2012sx,Denner:2011id,Beenakker:2002nc,Dawson:2003zu,Yu:2014cka,Frixione:2015zaa,Dawson:2003kb,Dittmaier:2003ej,Harlander:2011aa,Demartin:2015uha}
are reported for a centre-of-mass energy of
$\sqrt{s}=13$~\TeV\ and a Higgs boson mass of $m_H = 125.09\,\GeV$. The order of the calculated
cross-section is reported in each case.
The cross-sections for the background processes are omitted,
since the background normalization is determined in fits to the data.
}
\label{tab:mc_table}
\centering
\resizebox{0.95\textwidth}{!}{
\begin{tabular}{lccccc}
\toprule
\multirow{2}{*}{Process} &
\multirow{2}{*}{Generator} & \multirow{2}{*}{Showering} & \multirow{2}{*}{PDF set} & $\sigma~[\mathrm{pb}] $ & \multirow{2}{*}{Order of $\sigma$ calculation} \\
&            &            &             & $\sqrt{s}=13$~\TeV & \\
\midrule
\ggF    & \nnlops               & \PYTHIA[8.2] & \PDFforLHC[15]  & 48.5~~ & N$^3$LO(QCD)+NLO(EW)         \\
\VBF    & \POWHEGBOX            & \PYTHIA[8.2] & \PDFforLHC[15]  & 3.78  & approximate-NNLO(QCD)+NLO(EW) \\
\WH     & \POWHEGBOX            & \PYTHIA[8.2] & \PDFforLHC[15]  & 1.37  & NNLO(QCD)+NLO(EW)             \\
\qqqgZH & \POWHEGBOX            & \PYTHIA[8.2] & \PDFforLHC[15]  & 0.76  & NNLO(QCD)+NLO(EW)             \\
\ggtoZH & \POWHEGBOX            & \PYTHIA[8.2] & \PDFforLHC[15]  & 0.12  & NLO(QCD)                      \\
\ttH    & \POWHEGBOX            & \PYTHIA[8.2] & \PDFforLHC[15]  & 0.51  & NLO(QCD)+NLO(EW)              \\
\bbH    & \POWHEGBOX            & \PYTHIA[8.2] & \PDFforLHC[15]  & 0.49  & NNLO(QCD)                     \\
\tHqb   & \MGNLO & \PYTHIA[8.2] & \NNPDF[3.0nnlo] & 0.074  & NLO(QCD)                                    \\
\tHW    & \MGNLO & \PYTHIA[8.2] & \NNPDF[3.0nnlo] & 0.015  & NLO(QCD)                                    \\
\midrule
\gamgam                         & \Sherpa  & \Sherpa    & \NNPDF[3.0nnlo] \\
$V\gamma\gamma$                 & \Sherpa  & \Sherpa    & \NNPDF[3.0nnlo] \\
$t\bar{t}\gamma\gamma$          & \MGNLO   & \PYTHIA[8] & \NNPDF[2.3lo]   \\
\bottomrule
\end{tabular}
}
\end{table}


\section{Event reconstruction and selection}
\label{sec:reco}

Events in this analysis are selected using the following procedure.
Reconstructed photon candidates are first required to satisfy a set of {\itshape preselection}-level identification criteria. The two
highest-\pt preselected photons are then used to define the diphoton system, and an algorithm is used to
identify the event primary vertex.
Finally, the photons are required to satisfy isolation criteria and additional identification criteria. Jets (including $b$-tagged jets), muons, electrons, and missing transverse energy (\met) are used in the analysis in order to categorize diphoton events and measure Higgs boson properties.   
 
\subsection{Photon reconstruction and identification}
Photons are reconstructed from energy deposits in the calorimeter that are formed  using a dynamical, topological cell-clustering algorithm \cite{EGAM-2018-01}. The photon candidate is classified as \emph{converted} if it is matched to either two tracks forming a conversion vertex, or one track with the signature of an electron track without hits in the innermost pixel layer; otherwise, it is classified as \emph{unconverted}. The fraction of converted photons varies from about 25\% in the central region to about 50\% in the forward region.
The photon candidate's energy is calibrated using a procedure described in Ref.~\cite{EGAM-2018-01}.

Reconstructed photon candidates must satisfy $|\eta|<2.37$ in order to fall inside the region of the electromagnetic (EM)
calorimeter with a finely segmented first layer, and outside the range $1.37<|\eta|<1.52$ corresponding to the
transition region between the barrel and endcap EM calorimeters. Photon candidates are distinguished from jet backgrounds using identification
criteria based on calorimeter shower shape variables \cite{EGAM-2018-01}. A
{\itshape loose} working point is used for preselection, and the final selection of photon candidates
is made using a {\itshape tight} selection.
The efficiency of the {\itshape tight} identification for reconstructed photon candidates ranges from about 84\% (85\%) at $\pt = 25~\GeV$ to 94\% (98\%) for unconverted (converted) photons with $\pt > 100~\GeV$.
 
The final selection of photons includes both calorimeter- and track-based isolation requirements to further suppress jets misidentified
as photons. The calorimeter isolation variable is defined as the total energy of calorimeter clusters
in a cone of size $\Delta R=0.2$ around the photon candidate, excluding the energy in a fixed-size window containing the photon shower; a correction is applied for leakage of photon energy from this window into the surrounding cone~\cite{EGAM-2018-01}. Contributions from pile-up and the underlying event are subtracted~\cite{Cacciari:2008gn,Cacciari:2009dp,PERF-2013-04,STDM-2010-08,EGAM-2018-01}.
The calorimeter-based isolation must be less than 6.5\% of the photon transverse energy for each photon candidate.
The track-based isolation variable is defined as the scalar sum of the transverse momenta of tracks within
a $\Delta R = 0.2$ cone around the photon candidate. The tracks considered in the isolation variable
are restricted to those with $\pt>1$~\GeV\ that are matched to the selected diphoton primary vertex described below and
not associated with the photon conversion vertex, if present. Each photon must have a track isolation less than 5\% of the photon transverse energy.
 
\subsection{Event selection and selection of the diphoton primary vertex}
Events are selected by first requiring at least two photons satisfying the {\itshape loose}
identification preselection criteria. The two highest-\pt\ preselected photons
are designated as the candidates for the diphoton system.
The  {\itshape diphoton primary vertex} of the event is determined using a neural-network algorithm~\cite{HIGG-2013-08}.  Information about the reconstructed vertices in the event and the trajectories of the two photons, measured using the depth segmentation of the calorimeter and completed by photon conversion information if present, is used as input to the network.
\cite{HIGG-2013-08}.
The algorithm is trained on simulation and leads to an 8\% improvement in the mass resolution for inclusive Higgs boson production, relative to the default primary vertex selection~\cite{PERF-2015-01}, and results in better analysis sensitivity. The improvement is the largest for the \ggtoH\ production process, which has the lowest vertex selection efficiency among the main production modes. The algorithm
performance was validated using studies of $Z{\rightarrow}ee$ events in data and simulation, in which the electrons were treated as photon candidates and their track information ignored. 
This performance is weakly dependent on the event pile-up, and its residual dependence is well described by simulation.
 
The two preselected photon candidates are required to satisfy the {\itshape tight} identification criteria and the isolation selection described above.
Finally, the highest-\pt\ and second-highest-\pt\ photon candidates are required to satisfy $\pt / m_{\gamma\gamma} > 0.35$ and 0.25, respectively.
As discussed in Sections~\ref{sec:design} and~\ref{sec:modeling}, events that fail the tight identification or the isolation selection are used as a control sample for background estimation and modelling purposes.
 
The trigger, photon and
event selections described above are used to define the events that are selected for further
analysis for Higgs boson properties.
In total,
about 1.2~million events are selected in this data set with a diphoton invariant mass between
105 and 160~\GeV.
The total selection efficiency for a SM Higgs boson signal with $|y_H|<2.5$ obtained from simulation is $39\%$.

\subsection{Reconstruction and selection of hadronic jets, $b$-jets, leptons, top quarks and missing transverse momentum}

Jets are reconstructed using a particle-flow algorithm~\cite{PERF-2015-09} from noise-suppressed positive-energy topological clusters~\cite{PERF-2014-07} in the calorimeter using the anti-$k_t$ algorithm~\cite{Cacciari:2008gp,Fastjet} with a radius parameter $R$~=~0.4. Energy deposited in the calorimeter by charged particles is subtracted and replaced by the momenta of tracks that are matched to those topological clusters.
The jet four-momentum is corrected for the non-compensating calorimeter response, signal losses due to noise threshold effects, energy lost in non-instrumented regions, and contributions from pile-up~\cite{JETM-2018-05}. Jets are required to have $\pt > 25\,\GeV$ and an absolute value of rapidity $y$~less than 4.4.
A jet-vertex-tagger (JVT) multivariate discriminant~\cite{PERF-2014-03} is applied to jets with $\pt < 60\,\GeV$ and $|\eta|<2.4$, to suppress jets from pile-up; in the $|\eta|$ range beyond 2.5, a forward version of the JVT~\cite{ATL-PHYS-PUB-2019-026} is applied to jets with $\pt < 120\,\GeV$.
Jets with $|\eta|<2.5$ containing $b$-hadrons are identified using the DL1r $b$-tagging algorithm and its 60\%, 70\%, 77\% and 85\% efficiency working points, which are combined into a pseudo-continuous $b$-tagging score~\cite{FTAG-2018-01}.
 
Electrons are reconstructed by matching tracks in the ID to topological clusters
formed using the same dynamical, topological cell-clustering algorithm as
in the photon reconstruction~\cite{EGAM-2018-01}.
Electron candidates are required to have $\pt>10\,\GeV$ and $|\eta|<2.47$, excluding the EM calorimeter transition region of $1.37<|\eta|<1.52$, and  must satisfy the {\itshape medium} identification selection based on a likelihood discriminant using calorimeter shower shapes and track parameters~\cite{EGAM-2018-01}.
Isolation criteria are applied to electrons, using calorimeter- and track-based
information.
The reconstructed track matched to the electron candidate must be consistent with the diphoton
vertex, which is ensured by requiring its longitudinal impact parameter $z_0$ relative to the vertex to satisfy
$|z_0 \sin\theta|<0.5$~mm. In addition, the electron track's transverse impact parameter with respect to the
beam axis divided by its uncertainty, $|d_0|/\sigma_{d_0}$, must be less than 5.
 
Muons are reconstructed by matching tracks from the MS and ID
subsystems. In the pseudorapidity range of $2.5<|\eta|<2.7$, muons without an ID track but whose MS track is compatible with originating from the
interaction point are also considered. Muon candidates are required to have $\pt>10$~\GeV\ and $|\eta|<2.7$, and must satisfy the {\itshape medium}
identification requirements~\cite{MUON-2018-03}.
Muons are required to satisfy calorimeter- and track-based isolation requirements that are
95\%--97\% efficient for muons with $10 \le \pt \le 60$~\GeV\ and 99\% efficient for $\pt>60$~\GeV.
Muon tracks must satisfy $|z_0 \sin\theta|<0.5$~mm and $|d_0|/\sigma_{d_0}<3$.
 
Top quark candidates are reconstructed and identified using a boosted decision tree (BDT) discriminant, using the same procedure as in Ref.~\cite{HIGG-2019-01} applied to the particle-flow jets described above.
The BDT targets both leptonic top quark signatures, in which the  top quark decays to a $W$ boson that decays promptly to an electron or a muon, and hadronic signatures in which the $W$ boson decays to hadrons or to a $\tau$-lepton.
 
An overlap removal procedure is performed in order to avoid double-counting objects.
First, electrons overlapping with any photons ($\Delta R<0.4$) that
pass the isolation and identification requirements are removed.
Jets overlapping with the selected photons ($\Delta R<0.4$) and electrons ($\Delta R<0.2$) are removed. In the calculation of the $\Delta R$ between a jet and another object, the jet rapidity is used. Electrons overlapping with the remaining jets ($\Delta R<0.4$) are removed to match the requirements imposed when
measuring isolated electron efficiencies.
Finally, muons overlapping with photons or jets ($\Delta R<0.4$) are removed.
 
The missing transverse momentum is defined as the negative vector sum of the
transverse momenta of the selected photon, electron, muon and jet objects, plus the
transverse momenta of remaining low-\pt particles, estimated using tracks matched to the diphoton primary
vertex but not assigned to any of the selected objects~\cite{PERF-2016-07}.
Its magnitude is denoted by \met.
 
Finally, an event veto is applied to suppress the overlap between the selection described here and that of the search for Higgs boson pair production in the $b\bar{b}\gamma\gamma$ final state~\cite{HDBS-2018-34}, to facilitate the statistical combination of the two results at a later stage. Most of the vetoed events would enter the \ttH\ and \tH\ classes defined in Section~\ref{sec:design}. This veto has a negligible impact on the analysis results.


\section{Design of the measurement}
\label{sec:design}

\subsection{Overview}
\label{sec:design:overview}
 
The analysis is designed to measure the production cross-sections in the STXS framework~\cite{STXS}. The regions considered in this paper are based on the Stage 1.2 STXS binning.
They are defined in the Higgs boson rapidity range of $|y_H| < 2.5$, separately for mutually exclusive Higgs boson production processes: the \ggtoH\ process, which includes both \ggF\ production and \ggtoZH\ production followed by a hadronic decay of the $Z$ boson; the electroweak \qqtoHqq\ process, encompassing both \VBF\ production and \qqtoVH\ production followed by a hadronic decay of the vector boson; the \VlepH\ process, corresponding to \pptoVH\ production followed by a leptonic decay of the vector boson (in the case of \ZH, including both decays to charged leptons and to neutrinos); and top-associated \ttH\ and \tH\ production. The Higgs boson decay information is not used in the definition of STXS regions. For each process, non-overlapping fiducial regions are defined. These are based on the kinematics of the Higgs boson and of the associated jets and $W$ and $Z$ bosons, as well as the numbers of jets, leptons and top quarks. Jets are reconstructed at the particle level from all stable particles with a lifetime greater than $10\,\ps$, excluding the decay products of the Higgs boson and leptons from $W$ and $Z$ boson decays, using the anti-$k_t$ algorithm with a jet radius parameter $R = 0.4$, and must have a transverse momentum larger than $30\,\GeV$.
 
Compared to the Stage 1.2 STXS definition, two sets of modified STXS regions are defined at the particle level: a set of \emph{analysis regions} which is used in the design of the analysis strategy, and is defined below; and a set of \emph{measurement regions}, in which some analysis regions are merged, which are used to present the measurement results and are defined at the beginning of Section~\ref{sec:results:STXS}.
The 45 STXS analysis regions are listed in Figure~\ref{fig:design:stxs_forest}. They follow the Stage 1.2 definitions with the following modifications:
\begin{itemize}
\item The \bbH\ production mode is experimentally difficult to separate from \ggtoH, and these two production modes have similar selection efficiencies. The two modes are therefore measured as a single process, with each STXS region of the combined process corresponding to the sum of \ggtoH\ and \bbH\ contributions.
 
\item For \ggtoH\ and \qqtoHqq\ processes, STXS regions requiring two or more jets are not split by the transverse momentum of the system consisting of the Higgs boson and two highest-\pt\ jets,  \ptHjj, since the measurement does not provide sufficient sensitivity to this split. In addition, the STXS region defined by $\mjj \ge 700~\GeV$, where \mjj\ is the invariant mass of the two highest-\pT\ jets, is split into two bins corresponding to \mjj\ above or below 1~\TeV. An additional splitting at $\mjj = 700~\GeV$ is also introduced in the $\pTH \ge 200~\GeV$ region of the \qqtoHqq\ process.
\item The \ggtoZH\ and \qqtoZH\ production modes with a leptonic $Z$ boson decay similarly cannot be distinguished by the analysis selections, and are therefore considered as a single \pptoZH\ process. In addition, each region of this process is split into separate regions for charged  (\pptoHll) and neutral (\pptoHnn) dileptons.
\item Production of \tH\ is split into separate \pptotHW\ and \pptotHqb\ contributions, since the two processes have different acceptances for the analysis selections. The $s$-channel \pptotHb\ process is neglected due to its small cross-section.
\item The \VlepH\ regions are not separated according to the number of jets in the event.
\end{itemize}

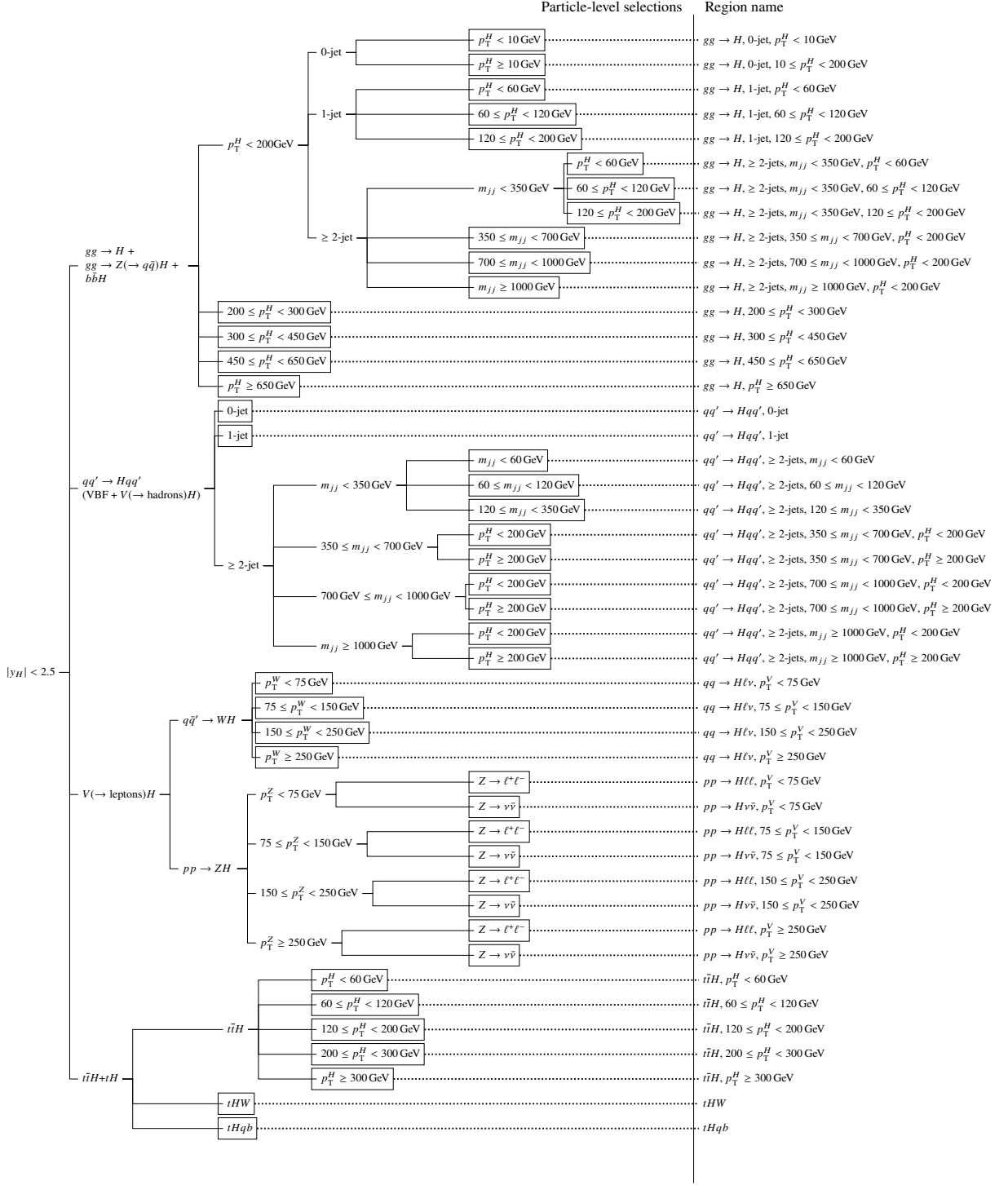
\begin{figure}[!tbp]
\adjustbox{max width=1.0\textwidth}{ 

\begin{forest} 
for tree={
edge path={
\noexpand\path[\forestoption{edge}](!u.parent anchor) -- +(10pt,0) |- (.child anchor) -- +(5pt,0) \forestoption{edge label};},
grow=0,s sep=1mm,reversed,parent anchor=east,child anchor=west,anchor=west,fit=band,align=left},
[$|y_H| < 2.5$,align=left, base=bottom
[~~~$\ggtoH \,\,+ $ \\ ~~~$gg\to Z(\to q\bar{q})H \,\, + $~~ \\ ~~~\bbH{},tier=0,
[~~$\ptH < 200\gev$,tier=one
[~~0-jet,tier=two
[~~$\ptH < 10\,\GeV$,tier=three,rectangle,draw [~~\ggHjPt{0}{}{10}{}   ,tier=reco,edge={dotted,line width=1.5pt}, name=topreco]]
[~~$\ptH \ge 10\,\GeV$,tier=three,rectangle,draw [~~\ggHjPt{0}{10}{200}{},tier=reco,edge={dotted,line width=1.5pt}]]
]
[~~1-jet,tier=two
[~~$\ptH < 60\,\GeV$         ,tier=three,rectangle,draw [~~\ggHjPt{1}{}{60}{}   ,tier=reco,edge={dotted,line width=1.5pt}]]
[~~$60 \le \ptH <120\,\GeV$ ,tier=three,rectangle,draw [~~\ggHjPt{1}{60}{120}{} ,tier=reco,edge={dotted,line width=1.5pt}]]
[~~$120 \le \ptH <200\,\GeV$,tier=three,rectangle,draw [~~\ggHjPt{1}{120}{200}{},tier=reco,edge={dotted,line width=1.5pt}]]
]
[~~$\geq 2$-jet,tier=two
[~~$\mjj < 350\,\GeV$, tier=three
[~~$\ptH < 60\,\GeV$         ,tier=four,rectangle,draw [~~\ggHmPt{}{350}{}{60}{}   ,tier=reco,edge={dotted,line width=1.5pt}]]
[~~$60 \le \ptH <120\,\GeV$ ,tier=four,rectangle,draw [~~\ggHmPt{}{350}{60}{120}{} ,tier=reco,edge={dotted,line width=1.5pt}]]
[~~$120 \le \ptH <200\,\GeV$,tier=four,rectangle,draw [~~\ggHmPt{}{350}{120}{200}{},tier=reco,edge={dotted,line width=1.5pt}]]
]
[\textcolor{black}{~~$350 \le \mjj < 700\,\GeV$}, tier=three,rectangle,draw
[~~\ggHmPt{350}{700}{}{200}{} ,tier=reco,edge={dotted,line width=1.5pt}]
]
[\textcolor{black}{~~$700 \le \mjj < 1000\,\GeV$}, tier=three,rectangle,draw
[~~\ggHmPt{700}{1000}{}{200}{},tier=reco,edge={dotted,line width=1.5pt}]
]
[\textcolor{black}{~~$\mjj \ge 1000\,\GeV$}, tier=three,rectangle,draw
[~~\ggHmPt{1000}{}{}{200}{},tier=reco,edge={dotted,line width=1.5pt}]
]
]
]
[~~$200 \le \ptH < 300\,\GeV$,tier=one,rectangle,draw [~~\ggHPt{200}{300}{},tier=reco,edge={dotted,line width=1.5pt}]]
[~~$300 \le \ptH < 450\,\GeV$,tier=one,rectangle,draw [~~\ggHPt{300}{450}{},tier=reco,edge={dotted,line width=1.5pt}]]
[~~$450 \le \ptH < 650\,\GeV$,tier=one,rectangle,draw [~~\ggHPt{450}{650}{},tier=reco,edge={dotted,line width=1.5pt}]]
[~~$\ptH \ge 650\,\GeV$      ,tier=one,rectangle,draw [~~\ggHPt{650}{}{}   ,tier=reco,edge={dotted,line width=1.5pt}]]
]
[~~\qqtoHqq \\ ~~(VBF + \VhadH){},tier=zero
[~~0-jet       ,tier=one,rectangle,draw [~~\Hqqj{0}{} ,tier=reco,edge={dotted,line width=1.5pt}]]
[~~1-jet       ,tier=one,rectangle,draw [~~\Hqqj{1}{} ,tier=reco,edge={dotted,line width=1.5pt}]]
[~~$\geq 2$-jet,tier=one
[~~$\mjj < 350\,\GeV$, tier=two
[~~$\mjj < 60\,\GeV$         ,tier=three,rectangle,draw [~~\Hqqm{}{60}{}   ,tier=reco,edge={dotted,line width=1.5pt}]]
[~~$60 \le \mjj <120\,\GeV$ ,tier=three,rectangle,draw [~~\Hqqm{60}{120}{} ,tier=reco,edge={dotted,line width=1.5pt}]]
[~~$120 \le \mjj <350\,\GeV$,tier=three,rectangle,draw [~~\Hqqm{120}{350}{},tier=reco,edge={dotted,line width=1.5pt}]]
]
[~~\textcolor{black}{$350 \le \mjj < 700\,\GeV$}, tier=two
[~~$\ptH < 200\,\GeV$,tier=three,rectangle,draw
[~~\HqqmPt{350}{700}{}{200}{} ,tier=reco,edge={dotted,line width=1.5pt}]]
[~~$\ptH \ge 200\,\GeV$,tier=three,rectangle,draw [~~\HqqmPt{350}{700}{200}{}{},tier=reco,edge={dotted,line width=1.5pt}]]
]
[~~\textcolor{black}{$700\,\GeV \le \mjj < 1000\,\GeV$}, tier=two
[~~$\ptH < 200\,\GeV$,tier=three,rectangle,draw
[~~\HqqmPt{700}{1000}{}{200}{} ,tier=reco,edge={dotted,line width=1.5pt}]
]
[~~$\ptH \ge 200\,\GeV$,rectangle,draw [~~\HqqmPt{700}{1000}{200}{}{},tier=reco,edge={dotted,line width=1.5pt}]]
]
[~~\textcolor{black}{$ \mjj \ge 1000\,\GeV$}, tier=two
[~~$\ptH < 200\,\GeV$,tier=three,rectangle,draw
[~~\HqqmPt{1000}{}{}{200}{} ,tier=reco,edge={dotted,line width=1.5pt}]
]
[~~$\ptH \ge 200\,\GeV$,tier=three,rectangle,draw [~~\HqqmPt{1000}{}{200}{}{},tier=reco,edge={dotted,line width=1.5pt}]]
]
]
]
[~~\VlepH,tier=zero
[~~\qqtoWH\,
[~~$\pT^W < 75\,\GeV$         ,rectangle,draw [~~\HlnPt{}{75}{}   ,tier=reco,edge={dotted,line width=1.5pt}]]
[~~$75  \le \pT^W < 150\,\GeV$,rectangle,draw [~~\HlnPt{75}{150}{},tier=reco,edge={dotted,line width=1.5pt}]]
[~~\textcolor{black}{$150 \le \pT^W < 250\,\GeV$},rectangle,draw
[~~\HlnPt{150}{250}{},tier=reco,edge={dotted,line width=1.5pt}]
]
[~~$\pT^W \ge 250\,\GeV$,rectangle,draw [~~\HlnPt{250}{}{},tier=reco,edge={dotted,line width=1.5pt}]]
]
[~~\pptoZH\,
[~~$\pT^Z < 75\,\GeV$,
[~~\textcolor{black}{$Z \rightarrow \ell^+\ell^-$}       ,tier=three,rectangle,draw [~~\HllPt{}{75}{}{},tier=reco,edge={dotted,line width=1.5pt}]]
[~~\textcolor{black}{$Z \rightarrow \nu\bar{\nu}$},tier=three,rectangle,draw [~~\HnunuPt{}{75}{}{},tier=reco,edge={dotted,line width=1.5pt}]]
]
[~~$75 \le \pT^Z < 150\,\GeV$,
[~~\textcolor{black}{$Z \rightarrow \ell^+\ell^-$}       ,tier=three,rectangle,draw [~~\HllPt{75}{150}{}{},tier=reco,edge={dotted,line width=1.5pt}]]
[~~\textcolor{black}{$Z \rightarrow \nu\bar{\nu}$},tier=three,rectangle,draw [~~\HnunuPt{75}{150}{}{},tier=reco,edge={dotted,line width=1.5pt}]]
]
[~~$150 \le \pT^Z < 250\,\GeV$
[~~\textcolor{black}{$Z \rightarrow \ell^+\ell^-$}       ,tier=three,rectangle,draw [~~\HllPt{150}{250}{}{},tier=reco,edge={dotted,line width=1.5pt}]]
[~~\textcolor{black}{$Z \rightarrow \nu\bar{\nu}$},tier=three,rectangle,draw [~~\HnunuPt{150}{250}{}{},tier=reco,edge={dotted,line width=1.5pt}]]
]
[~~$\pT^Z \ge 250\,\GeV$,
[~~\textcolor{black}{$Z \rightarrow \ell^+\ell^-$}       ,tier=three,rectangle,draw [~~\HllPt{250}{}{}{},tier=reco,edge={dotted,line width=1.5pt}]]
[~~\textcolor{black}{$Z \rightarrow \nu\bar{\nu}$},tier=three,rectangle,draw [~~\HnunuPt{250}{}{}{},tier=reco,edge={dotted,line width=1.5pt}]]
]
]
]
[~~\ttH+$tH$,tier=zero
[~~\ttH,tier=one,
[~~$\ptH < 60\,\GeV$         ,tier=two,rectangle,draw [~~\ttHPt{}{60}{}   ,tier=reco,edge={dotted,line width=1.5pt}]]
[~~$60 \le \ptH <120\,\GeV$ ,tier=two,rectangle,draw [~~\ttHPt{60}{120}{},tier=reco,edge={dotted,line width=1.5pt}]]
[~~$120 \le \ptH <200\,\GeV$,tier=two,rectangle,draw [~~\ttHPt{120}{200}{},tier=reco,edge={dotted,line width=1.5pt}]]
[~~$200 \le \ptH < 300\,\GeV$  ,tier=two,rectangle,draw [~~\ttHPt{200}{300}{},tier=reco,edge={dotted,line width=1.5pt}]]
[~~$\ptH \ge 300\,\GeV$        ,tier=two,rectangle,draw [~~\ttHPt{300}{}{}   ,tier=reco,edge={dotted,line width=1.5pt}]]
]
[~~$tHW$ ,tier=one,rectangle,draw [~~\tHW, tier=reco,edge={dotted,line width=1.5pt}]]
[~~$tHqb$,tier=one,rectangle,draw [~~\tHqb,tier=reco,edge={dotted,line width=1.5pt},name=botreco]]
]
]
{\node[anchor=south west,align=left] at (topreco.north west) {\Large ~~Region name\\};}
{\node[anchor=south east,align=left] at (topreco.north west) {\Large ~~Particle-level selections~~~\\};}
{\draw (botreco.south west) +(0,-1.5) -- (topreco.north west) -- +(0,0.5);}
\end{forest}


} 
\includegraphics[width=0.001\textwidth]{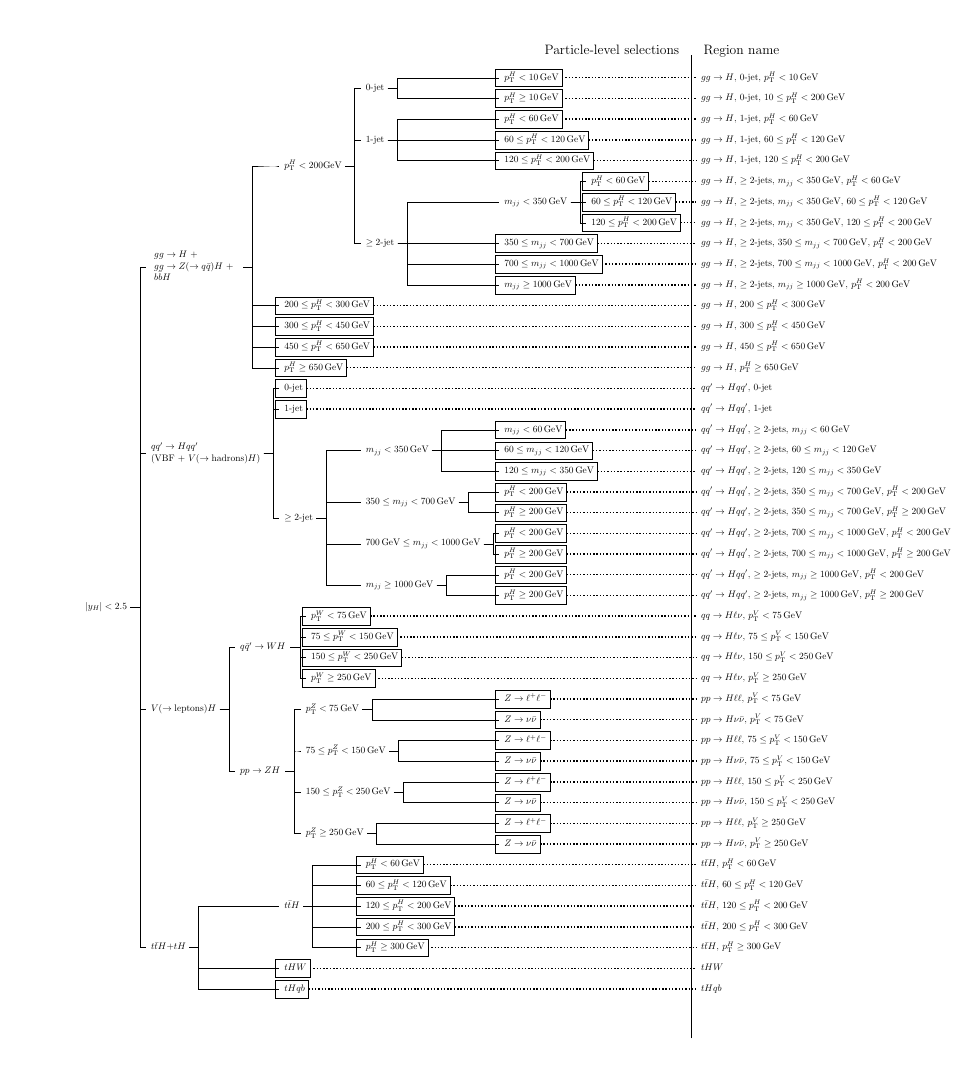}
\caption{
Summary of the STXS regions considered in the analysis design. The left part of the plot shows the selections applied to particle-level quantities in simulated signal events, with the selections applied sequentially along the branches of the graph. The final selection for each region is indicated by a box, and the name of each region, used in the rest of this paper, is shown on the right.
}
\label{fig:design:stxs_forest}
\end{figure}


\subsection{Categorization}
\label{sec:cats}
 
The events passing the selection described in Section~\ref{sec:reco} are classified into mutually exclusive event \emph{categories}, each targeted towards a particular STXS analysis region.\footnote{In this paper, \emph{categories} refers to event groupings defined from reconstructed quantities, while \emph{regions} refers to the particle-level selections defined in the STXS framework. \emph{Classes} refers to groups of categories targeting the same STXS region.}
This follows a technique similar to the one used in Ref.~\cite{HIGG-2016-21}, but the definition of the categories has been improved significantly.
The categorization in Ref.~\cite{HIGG-2016-21} was implemented sequentially over production modes, in order of increasing cross-section. In the present analysis, the categories are instead defined using a unified technique covering all processes simultaneously, and are designed to maximize a global criterion of sensitivity in the measurement of the cross-sections in all STXS regions.
 
The technique proceeds in several steps. First, simulated Higgs boson production event samples are used to train a multiclass BDT to separate signal events coming from different STXS analysis regions. This multiclass BDT classifier outputs one discriminant value for each of the 45 STXS analysis region.
The output discriminant values are then used to assign signal events to 45 STXS \emph{classes}. Each of these detector-level classes targets events from a particular STXS analysis region defined at the particle level. Finally, each class is further divided into multiple categories using a binary multivariate classifier. This classifier is trained to separate signal from continuum background and Higgs boson events from other STXS regions in each class.
 
The inputs to all the classifiers are variables describing the kinematic and identification properties of the reconstructed particles presented in Section~\ref{sec:reco}:
\begin{itemize}
\item the kinematics of the diphoton system;
\item the numbers of reconstructed jets, $b$-jets, electrons, muons and top quarks;
\item the kinematics of the system composed of the two photons and one or more jets, if jets are present, and of the system composed of the two highest-\pt\ jets in the event, if at least two jets are present;
\item the kinematics of the reconstructed leptons and top quarks;
\item the reconstruction score of the top quarks, computed from the kinematics of the top quark decay products as described in Ref.~\cite{HIGG-2019-01};
\item other event quantities such as the missing transverse momentum.
\end{itemize}
Among the top-associated production processes, the \tHqb\ mode can be separated from both \ttH\ and \tHW\ due to differences in kinematics and event topology, in particular the presence of a forward jet and the absence of a second well reconstructed top quark candidate in the event.
 
In order to avoid distorting the smoothly falling shapes of the background \mgg\ distributions, any variable found to have a linear correlation coefficient of 5\% or more with \mgg\ in the signal or background training samples is removed from the list of inputs to the binary classifiers. The training variables used in the analysis are summarized in Tables~\ref{tab:design:trainingvariables} and~\ref{tab:design:trainingvariables2}.

\begin{table}[ht]
\caption{Training variables used as input to the multiclass BDT. The dagger symbol $\dagger$ denotes variables that have two versions with different jet \pt\ requirements. One version of such a variable is defined using jets with $\pt > 25\,\GeV$, and the other version is defined using jets with $\pt > 30\,\GeV$. Both versions are used in the training of the multiclass BDT. The two highest-\pt\ photons are denoted as $\gamma_1$ and $\gamma_2$, the two highest-\pt\ jets as $j_1$ and $j_2$, the two highest-\pt\ top quarks as $t_1$ and $t_2$ and the most forward jet as $j_F$. $\Delta R(W,b)$ is the $\Delta R$ between the $W$ and $b$ components of a top-quark candidate.}
\label{tab:design:trainingvariables}
\begin{center}
\begin{tabular}{c}
\toprule
$\eta_{\gamma_1}$, $\eta_{\gamma_2}$, $\ptgg$, $y_{\gamgam}$,\\
$p_{\mathrm{T},jj}^{\dagger}$, $m_{jj}$, and $\Delta y$, $\Delta\phi$, $\Delta\eta$ between $j_1$ and $j_2$, \\
$p_{\mathrm{T},\gamgam j_1}$, $m_{\gamgam j_1}$, $p_{\mathrm{T},\gamgam jj} \dagger$, $m_{\gamgam jj}$ \\
$\Delta y$, $\Delta\phi$ between the \gamgam\ and $jj$ systems, \\
minimum $\Delta R$ between jets and photons,\\
invariant mass of the system comprising all jets in the event, \\
dilepton \pt, di-$e$ or di-$\mu$ invariant mass (leptons are required to be oppositely charged), \\
\met, \pt\ and transverse mass of the lepton + \met system, \\
\pt, $\eta$, $\phi$ of top-quark candidates, $m_{t_1 t_2}$ \\
Number of jets${\dagger}$, of central jets ($|\eta|<2.5$)${\dagger}$, of $b$-jets${\dagger}$ and of leptons, \\
\pt\ of the highest-\pt\ jet, scalar sum of the \pt\ of all jets, \\
scalar sum of the transverse energies of all particles ($\sum E_\mathrm{T}$), \met\ significance, \\
$\left|\met\ - \met (\text{primary vertex with the highest }\sum p_\mathrm{T,track}^2)\right| > 30\,\GeV$ \\
Top reconstruction BDT of the top-quark candidates, \\
$\Delta R(W, b)$ of $t_2$, \\
$\eta_{j_F}$, $m_{\gamgam j_F}$ \\
Average number of interactions per bunch crossing. \\
\bottomrule
\end{tabular}
\end{center}
\end{table}
 
\begin{table}[ht]
\caption{Training variables used for the binary classifiers. The sets of classes to which the classifiers are applied are specified in the first column, and the corresponding variables in each case are listed in the second column. The asterisk symbol~$^{*}$ denotes \tH\ training variables that are only used for the classifiers suppressing the continuum background. Other \tH\ training variables are used in all three \tH\ classifiers. The $\gamma\gamma$ and $jj$ notations refer to the systems composed of the two highest-\pt\ photons and jets, respectively. The two highest-\pt\ photons are denoted as $\gamma_1$ and $\gamma_2$, the two highest-\pt\ top quarks as $t_1$ and $t_2$, and the most forward jet as $j_F$. The differences in $\eta$ and $\phi$ between $\gamma_1$ and $\gamma_2$ are denoted respectively as $\Delta\phi_{\gamma\gamma}$ and $\Delta\eta_{\gamma\gamma}$. $\Delta R(W,b)$ is the $\Delta R$ between the $W$ and $b$ components of a top-quark candidate.}
\label{tab:design:trainingvariables2}
\begin{center}
\begin{tabular}{cc}
\toprule
STXS classes  & Variables  \\
\midrule
\begin{tabular}{c}
Individual \\
STXS classes from \\
\ggtoH \\
\qqtoHqq \\
\qqtoHln \\
\pptoHll \\
\pptoHnn
\end{tabular}
&
\begin{tabular}{c}
All multiclass BDT variables, \\
$\vec{p}_{\mathrm{T}}^{\gamgam}$ projected to the thrust axis of the \gamgam\ system ($p_\mathrm{Tt}^{\gamma\gamma}$), \\
$\Delta\eta_{\gamma\gamma}$, $\eta^{\text{Zepp}} = \frac{\eta_{\gamgam} - \eta_{jj}}{2}$, \\
$\phi^{*}_{\gamma\gamma} = \tan\left(\frac{\pi-|\Delta\phi_{\gamma\gamma}|}{2}\right) \sqrt{1-\tanh^{2}\left(\frac{\Delta\eta_{\gamma\gamma}}{2}\right)}$, \\
$\cos\theta^{*}_{\gamma\gamma}= \left|\frac{(E^{\gamma_1} + p_z^{\gamma_1}) \cdot (E^{\gamma_2} - p_z^{\gamma_2}) -  (E^{\gamma_1} - p_z^{\gamma_1}) \cdot (E^{\gamma_2} + p_z^{\gamma_2})}{ m_{\gamma\gamma} + \sqrt{m_{\gamma\gamma}^2 + (p_\mathrm{T}^{\gamma\gamma})^2} }\right|$ \\
Number of electrons and muons. \\
\end{tabular}
\\
\midrule
\begin{tabular}{c}
all \ttH\ and \tHW\ \\
STXS classes \\
combined
\end{tabular}
&
\begin{tabular}{c}
\pt, $\eta$, $\phi$ of $\gamma_1$ and $\gamma_2$, \\
\pt, $\eta$, $\phi$ and $b$-tagging scores of the six highest-\pt\ jets,\\
\met, \met\ significance, \met\ azimuthal angle, \\
Top reconstruction BDT scores of the top-quark candidates, \\
\pt, $\eta$, $\phi$ of the two highest-\pt\ leptons. \\
\end{tabular}
\\
\midrule
\tHqb &
\begin{tabular}{c}
$\pTgg/\mgg$, $\eta_{\gamma\gamma}$, \\
\pt, invariant mass, BDT score and $\Delta R(W, b)$ of $t_1$,\\
\pt, $\eta$ of $t_2$, \\
\pt, $\eta$ of $j_F$, \\
Angular variables: $\Delta\eta_{\gamma\gamma t_1}$, $\Delta\theta_{\gamma\gamma t_2}$, $\Delta\theta_{t_1 j_F}$ , $\Delta\theta_{t_2 j_F}$, $\Delta\theta_{\gamma\gamma j_F}$ \\
Invariant mass variables: $m_{\gamma\gamma j_F}$, $m_{t_1 j_F}$, $m_{t_2 j_F}$, $m_{\gamma\gamma t_1}$ \\
Number of jets with $\pt > 25\,\GeV$,  Number of $b$-jets with $\pt > 25\,\GeV^{*}$;\\
Number of leptons$^{*}$, \met\ significance$^{*}$
\end{tabular}
\\
\bottomrule
\end{tabular}
\end{center}
\end{table}


The multiclass BDT used in the initial step of the classification is trained on a data set obtained by merging the \ggF, \VBF, \VH, \ttH\ and \tH\ signal samples described in Section~\ref{sec:mc}. A weight is applied to the events in each STXS analysis region so that the regions have equal event yields in the training sample. This configuration improves the performance of the discrimination. For each event, the output of the BDT consists of a set of class discriminants $y_i$, where the index $i$ runs over the 45 STXS regions defined in Table~\ref{fig:design:stxs_forest}. This output is then normalized into the parameters $z_i = \exp(y_i)/\sum_j \exp(y_j)$, a procedure also known as a softmax layer. The training is performed by minimizing the cross-entropy of the $z_i$ with respect to the true STXS analysis region assignments\footnote{The cross-entropy loss function is computed as $-\sum\limits_{k=1}^n \omega_k \sum\limits_{i=1}^{45} \delta_{i,k} \ln(z_i)$, where $k$ runs over the $n$ events in the training sample, $\omega_k$ are event weights applied to balance the class yields as described in the text, $i$ runs over the classes, and $\delta_{k,i}$ has a value of $1$ if class $i$ is the correct assignment for event $k$, and $0$ otherwise.} using the LightGBM package~\cite{NIPS2017_6907}.
 
A second training phase is then performed to optimize the classification procedure in terms of the analysis sensitivity itself. The sensitivity is estimated as the inverse determinant $|C|^{-1}$ of the covariance matrix of the measurement of the signal event yields in each analysis region. This \emph{D-optimality} (determinant) criterion leads in particular to a reduction of the expected statistical uncertainty of the measurement, and is suggested by the fact that $|C|^{-1}$ is a known measure of the information provided by the measurement~\cite{Lindley1956}. The classification procedure is performed so that events are assigned to the STXS class $i$ corresponding to the maximum value of $w_i z_i$, where the $w_i$ are a set of per-class weights. These weights are initially set to $1$, and then iteratively updated so as to maximize $|C|^{-1}$: for each value of the $w_i$, a simulated data set is generated for each analysis region by mixing events from each signal sample in  proportion to their SM production cross-sections, together with a sample of simulated continuum background events normalized to data in the control region $95 \le \mgg < 105\,\GeV$. A simplified statistical model approximating the full model described in Section~\ref{sec:modeling} is then used to estimate $|C|^{-1}$, and the procedure is iterated until a maximum is found for $|C|^{-1}$.
 
Figure~\ref{fig:design:BDT_outputs_multiclass} shows distributions of the weighted multiclass discriminant output $w_i z_i$ for four representative STXS classes, illustrating the discrimination provided by the multiclass BDT. While events with high BDT output values for a given analysis region tend to be selected in the corresponding class, this does not manifest itself as a sharp cut, due to the interplay between the selections for the different classes.
Compared to the simple selection based only on the $z_i$, the selection based on the $w_i z_i$ provides both higher purity and higher selection efficiency for classes associated with rare processes such as \tH, \ttH, \VH\ and \VBF, as well as production at high values of \ptH\ or \mjj. This leads to measurements with generally smaller uncertainties and lower correlations.
 
This multiclass training allows the  selection of target process events that otherwise would fail a requirement based on detector-level quantities corresponding to the STXS region definition.  For example, in the STXS region \ggHjPt{1}{}{60}{}, detector-level events that originate from the target process but have no reconstructed jets would fail requirements defined by the number of jets and \ptH; however, those events could be selected by the multiclass discriminant. For this STXS region, 20\% of events from the target process have no reconstructed jets. The recovery of these events leads to a reduction of about $6\%$ in the measurement uncertainty. It is also robust against pile-up in the determination of jet multiplicity in \ggtoH.
 
\begin{figure}[!htbp]
\centering
\subfloat[\ggtoH, 1-jet, $120 \le~\ptH \le 200\,\GeV$]{\includegraphics[width=0.49\linewidth]{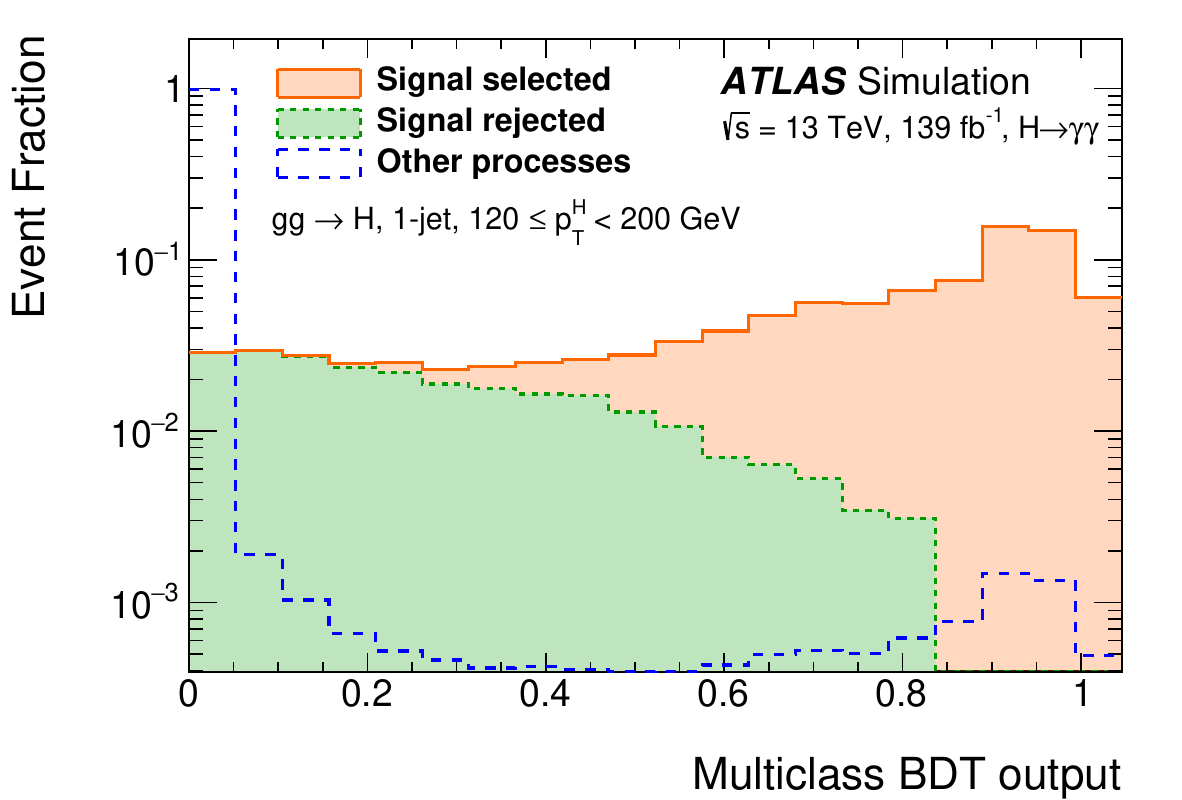}
}
\subfloat[\qqtoHqq, $\ge 2$-jets, $700 < \mjj < 1000\,\GeV$, $\ptH < 200\,\GeV$]{\includegraphics[width=0.49\linewidth]{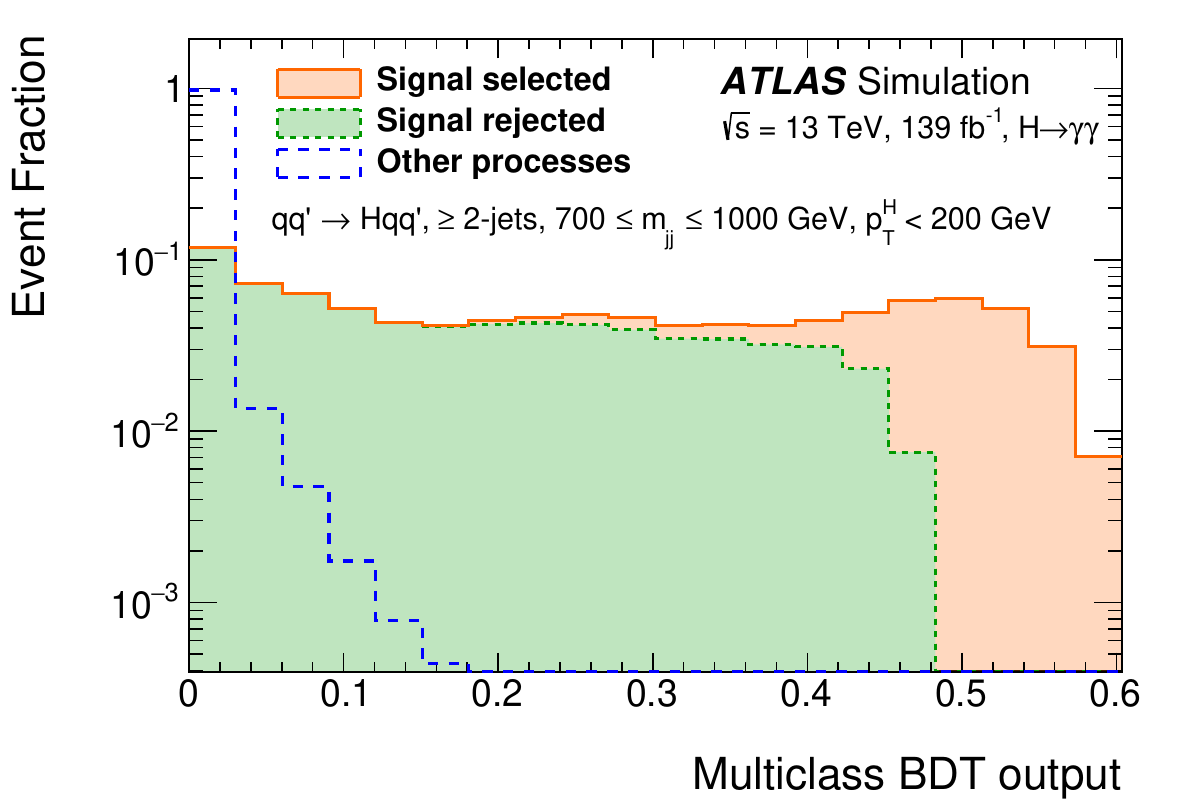}
}\\
\subfloat[\qqtoHln, $75 \le \ptV < 150\,\GeV$]{\includegraphics[width=0.49\linewidth]{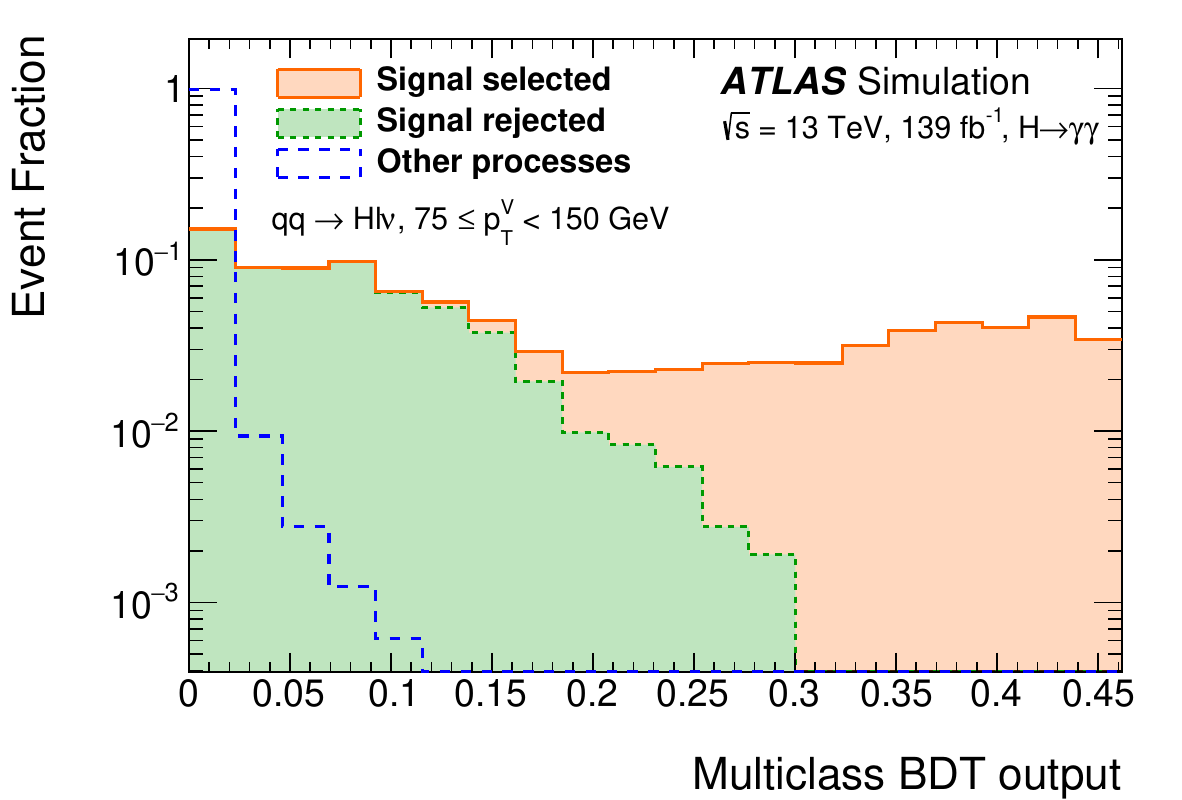} 
}
\subfloat[\ttH, $60 \le \ptH < 120\,\GeV$  ]{\includegraphics[width=0.49\linewidth]{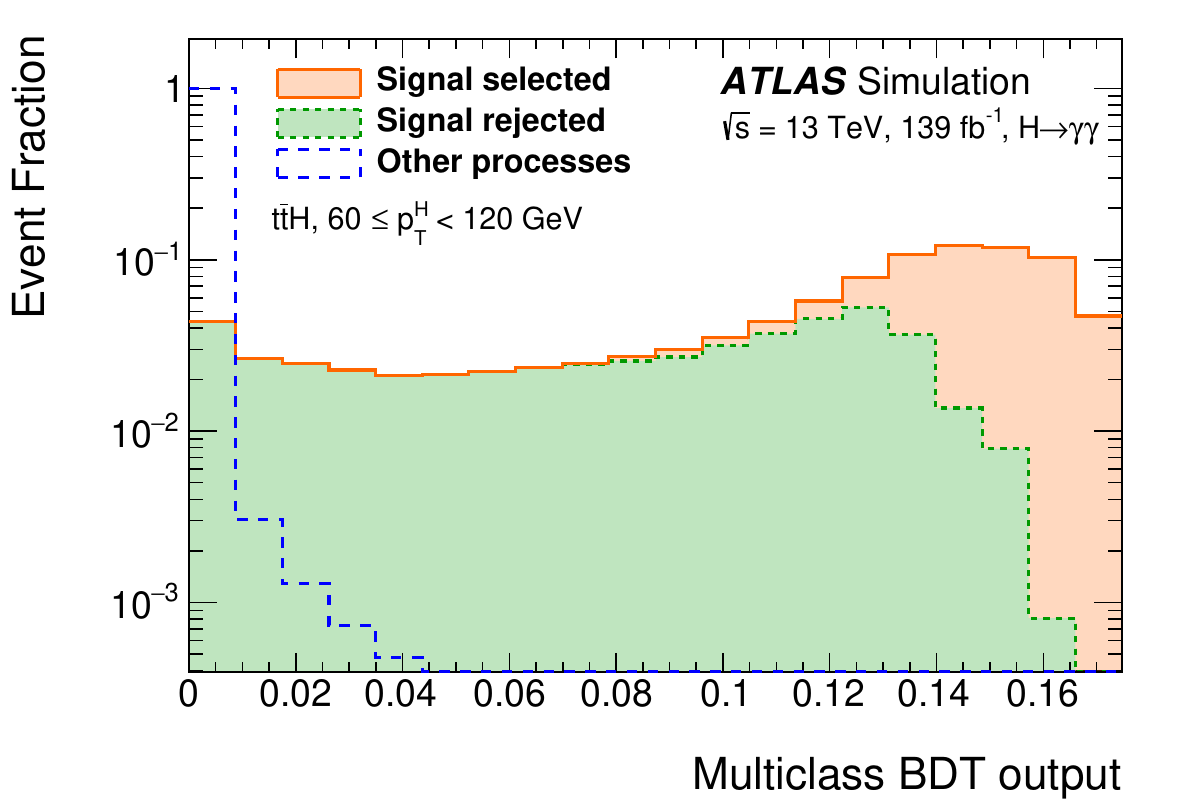}
}
\caption{Distributions of the weighted multiclass discriminant output $w_i z_i$, where $z_i$ is the raw discriminant output and $w_i$ the per-class weight defined in the text, for four representative STXS classes.
In each plot, the distribution is shown separately for events corresponding to the target STXS analysis region (solid) and events in other STXS analysis regions (long-dashed). The target STXS analysis region is further broken down into the subset of events assigned to the correct class at detector level (orange-solid), and the subset of events that are assigned to other classes (green-dashed). The orange-solid component is stacked on top of the dashed component. An event is assigned to the class with the largest $w_i z_i$ value.
}
\label{fig:design:BDT_outputs_multiclass}
\end{figure}

After the classes are defined, binary classifiers are then trained and used to further divide each class into multiple categories, to improve the measurement sensitivity. For each of the classes targeting \ggtoH,  \qqtoHqq\ and \VlepH\ processes, a binary BDT classifier is trained to distinguish between simulated signal events of the corresponding STXS analysis region and both simulated continuum background events and Higgs boson events from other STXS analysis regions.
 
For the \ttH\ and \tHW\ classes, a binary BDT classifier is trained to separate \ttH\ signal and the continuum background using all events assigned to various \ttH\ classes targeting different \ptH\ regions. Similarly, a binary BDT classifier is trained to separate \tHW\ signal and the continuum background using events assigned to the \tHW\ class.
 
To enhance the sensitivity to the sign of the top-Yukawa coupling modifier $\kappa_t$ (defined in more detail in Section~\ref{sec:results:kappas}), a specialization is introduced for the \tHqb\ class. First, the class is divided into two sub-classes based on a neural-network (NN) binary classifier that separates \tHqb\ production with $\kappa_t = 1$ from \tHqb\ production with $\kappa_t = -1$. In each sub-class, the events are then further divided into categories based on NN binary classifiers trained to separate the corresponding \tHqb\ signal events from continuum background events and Higgs boson events from other processes.
 
The binary classifiers used to suppress continuum background processes in the \ttH, \tHW, and \tHqb\ classes are trained on events from control regions in data, which provide larger event yields than the available simulated background samples. These regions are defined using the same selections as the classes, but reversing the photon identification requirement, the photon isolation requirement, or both.
 
\begin{figure}[t]
\centering
\subfloat[\ggtoH, 1-jet, $120 \le~\ptH < 200\,\GeV$]{\includegraphics[width=0.49\linewidth]{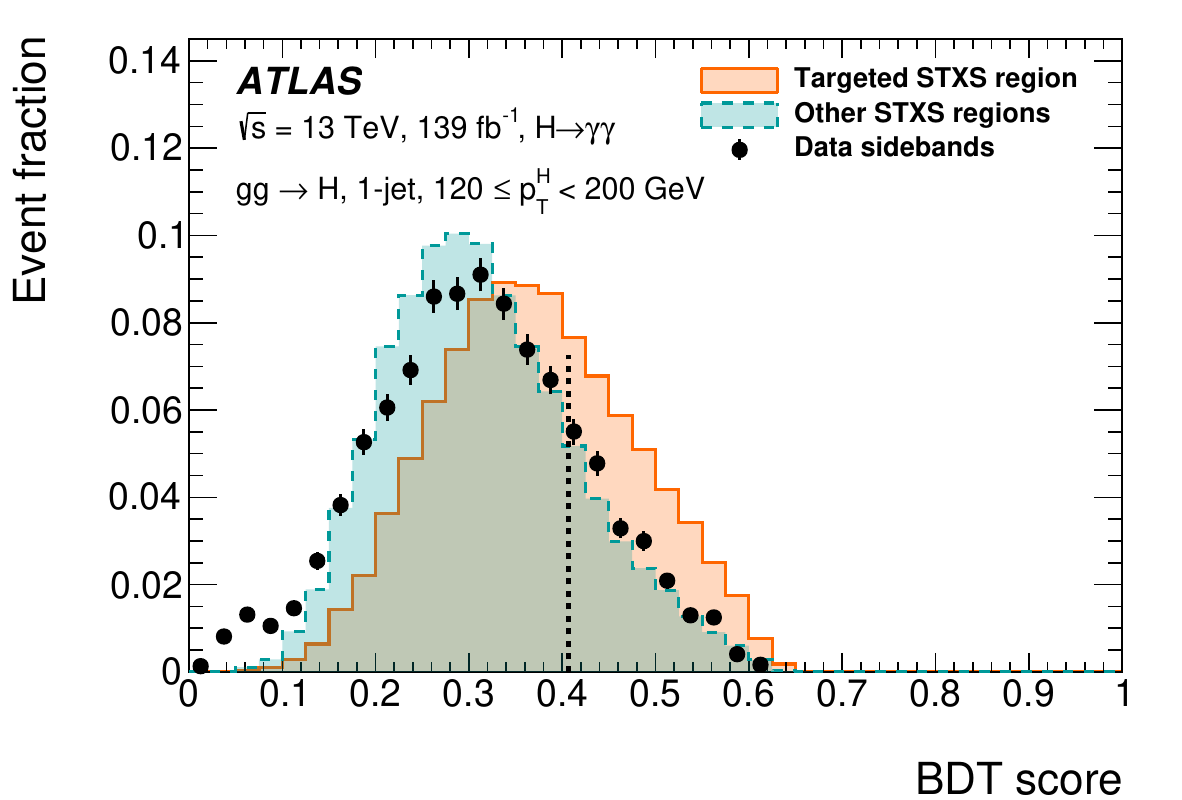}
}
\subfloat[\qqtoHqq, $\ge 2$-jets, $700 < \mjj < 1000\,\GeV$, $\ptH < 200\,\GeV$]{\includegraphics[width=0.49\linewidth]{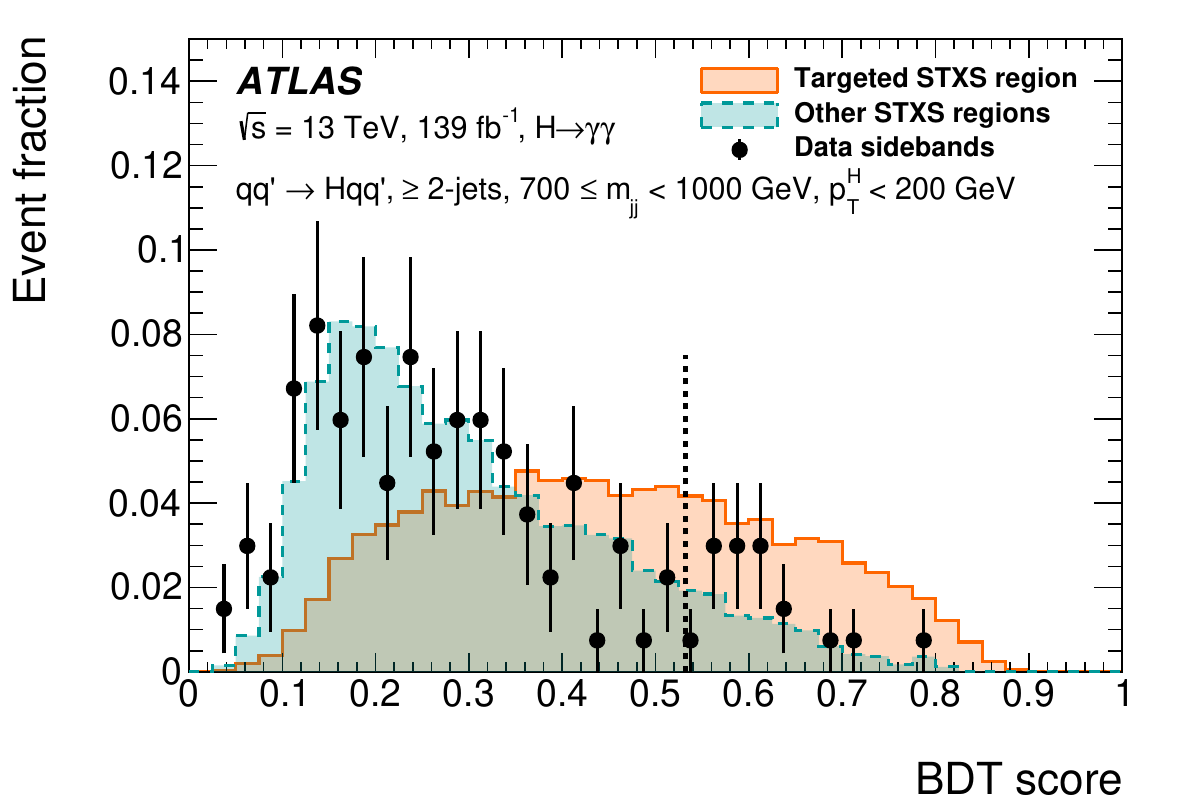}
}	\\
\subfloat[\qqtoHln, $75 \le \ptV < 150\,\GeV$]{\includegraphics[width=0.49\linewidth]{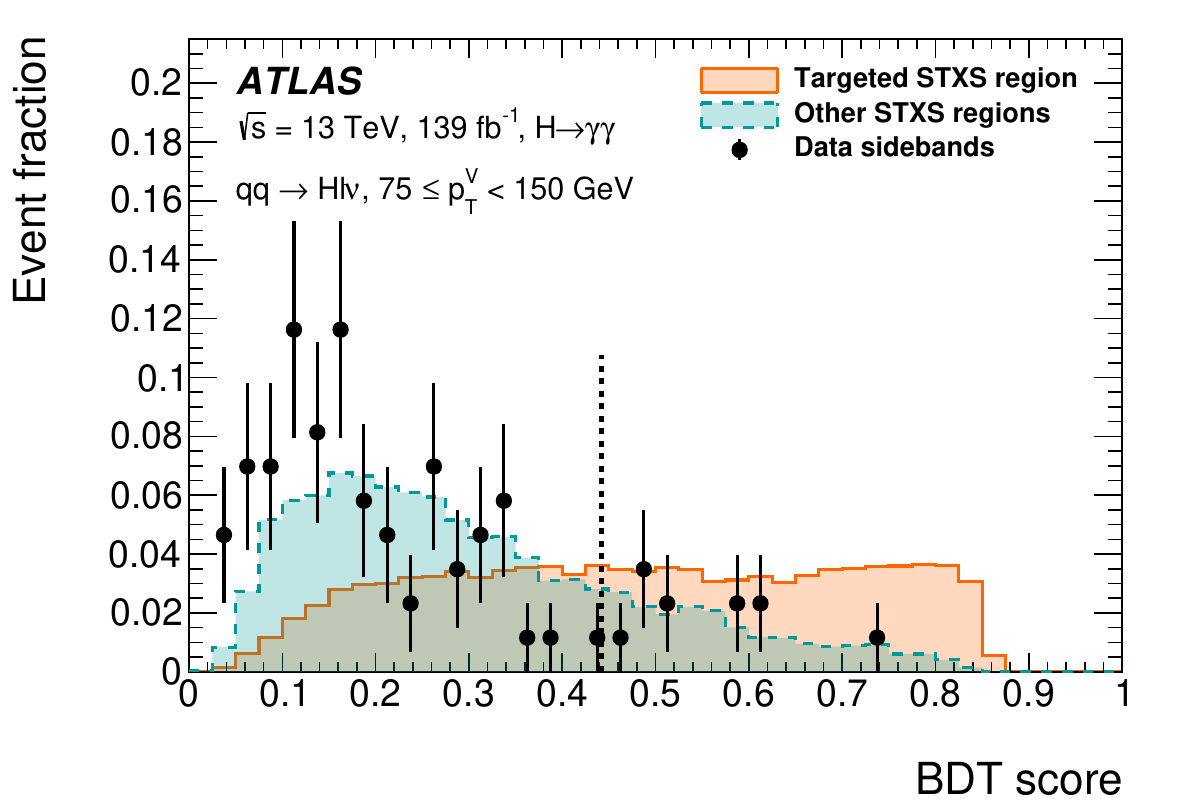}
}
\subfloat[\ttH, $60 \le \ptH < 120\,\GeV$]{\includegraphics[width=0.49\linewidth]{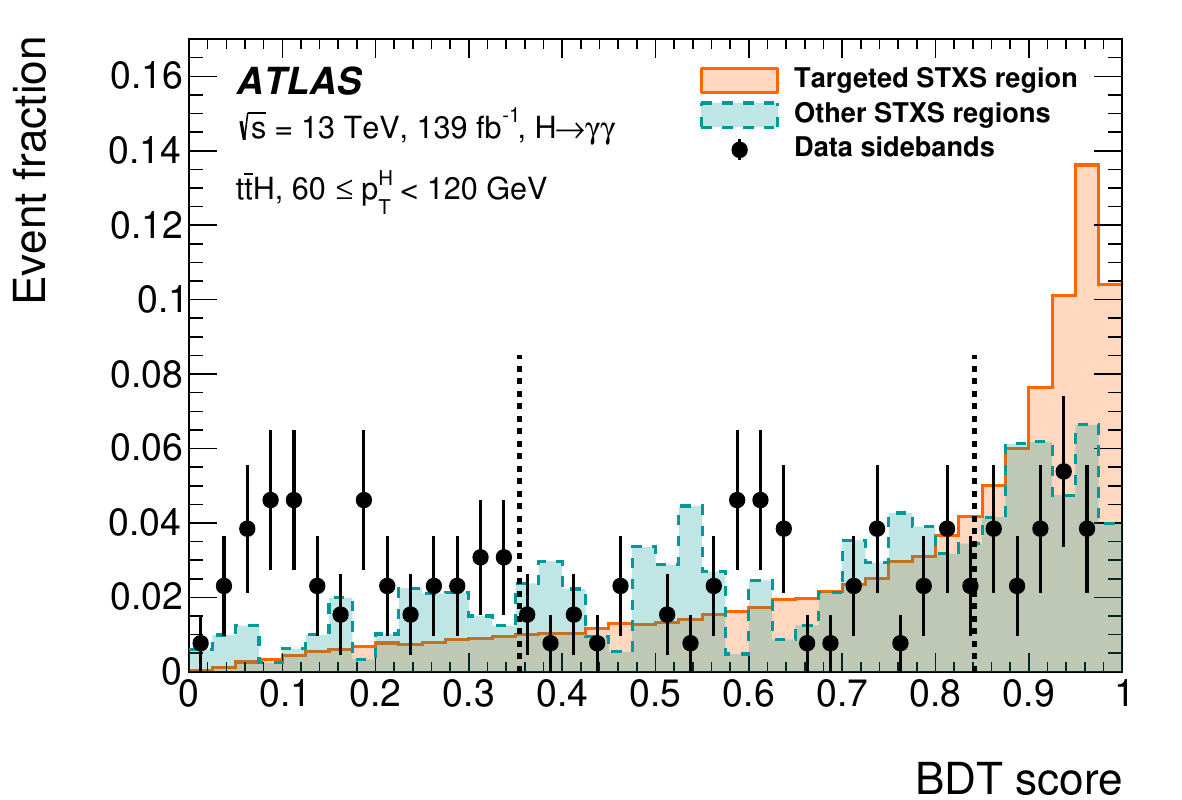}
}
\caption{Binary BDT discriminant distributions in four representative STXS classes. The binary BDT discriminant distribution is shown for simulated signal events in the target STXS analysis region (solid) and in other STXS analysis regions (dashed), and by the events in the diphoton mass sidebands ($105 \le \mgg < 120\,\GeV$ or $130 \le \mgg < 160\,\GeV$) representing background (dots). The vertical lines delimit the categories defined in the analysis within each class. }
\label{fig:design:BDT_outputs_binary}
\end{figure}
 
In each class, events are then assigned to categories corresponding to ranges of binary classifier output values. Up to three categories are defined in this way, depending on the targeted STXS region. The category boundaries in the BDT output are determined by scanning over all possible values and finding the set that maximizes the sum in quadrature of the expected significance values in these categories. The expected significance is computed as $Z = \sqrt{2( (S+B) \ln(1 + S/B) - S)}$~\cite{Cowan:2010st}, where $S$ and $B$ are the expected signal yield and background yield in the targeted STXS analysis region in the smallest range of \mgg\ around the signal peak position that contains 90\% of signal events. The background $B$ includes contributions from continuum background and Higgs boson events from other STXS analysis regions. The continuum background is computed from the \mgg\ distribution in simulation, normalized to the data control region $95 \le \mgg < 105\,\GeV$. A class is split into two categories if this leads to an improvement of more than 5\% in the expected significance, and into three categories if a further improvement of at least 5\% relative to the two-category configuration can be achieved. The categories are referred to as \emph{High-purity}, \emph{Med-purity} and, in the case of a 3-category split, \emph{Low-purity} in order of decreasing BDT output values. No events are removed at the categorization stage, since the lower-purity categories bring non-negligible contributions to the analysis sensitivity. Figure~\ref{fig:design:BDT_outputs_binary} shows binary BDT discriminant distributions as well as category boundaries for four representative STXS classes.
 
The categorization for the \tHqb\ class follows a different procedure, which aims to maximize both the sensitivity to a \tHqb\ signal and the sensitivity to the sign of $\kappa_t$. A boundary is placed in the NN classifier that separates the \tHqb\ signal with $\kappa_t = 1$ from the \tHqb\ signal with $\kappa_t = -1$. Different boundaries are also placed in the two binary NN classifiers that separate \tHqb\ signals from continuum background. These boundaries are determined simultaneously. Finally, a \emph{low-purity top} category is formed by grouping together the events with the lowest binary classifier output values in both the \ttH\ and \tH\ classes.
 
The entire categorization procedure results in the definition of 101 categories in total. The expected signal and background yields in these categories are summarized in Table~\ref{tab:design:yields}. The expected signal purity, defined as the expected signal yield divided by the expected yield from both the signal and background processes, in the smallest \mgg\ window containing 90\% of signal events, ranges from 0.03\% to 78\%. Figure~\ref{fig:design:yields} shows the contributions to the expected event yields from each of the 28 STXS measurement regions defined in Section~\ref{sec:results:STXS}. The contributions are shown as fractions of events originating from each STXS analysis region, in groups of analysis categories targeting the same region.
They are obtained as a weighted sum of the fractions for each category in the group, with weights given by the signal-over-background ratio $f$ in each category as defined in Table~\ref{tab:design:yields}.

\begin{table}[ht]
\caption{
Expected signal ($S$) and background ($B$) yields in each category within the smallest mass window containing 90\% of signal events. The half-width of this window is given by $\sigma$. The signal purity $f = S/(S + B)$ and expected significance $Z = \sqrt{2( (S+B) \ln(1 + S/B) - S)}$ are also shown. Only the signal process corresponding to the targeted STXS region is considered in the signal yield. }
\label{tab:design:yields}
\centering
\renewcommand{\arraystretch}{1.35}
\resizebox{0.49\textwidth}{!}{
\adjustbox{valign=t}{
\begin{tabular}{lrrrrr}
\toprule
\multirow{2}{*}{Category}  &  {\multirow{2}{*}{$S$}}  & {\multirow{2}{*}{$B$}}  &  {$\sigma$}  &    {$f$}  &  {\multirow{2}{*}{$Z$}} \\
&  &   & {[GeV]} & {[\%]}  & \\
\midrule
\multicolumn{6}{c}{\ggtoH}\\
\midrule
\tabggHjPt{0}{}{10}{}                 & \numRF{ 694.74}{3} & \numRF{26011.74}{3} & 3.43 & \numRF{ 2.60}{2} & \numRF{4.29}{2} \\
\tabggHjPt{0}{10}{}{}                 & \numRF{1435.38}{3} & \numRF{47039.25}{3} & 3.41 & \numRF{ 2.96}{2} & \numRF{6.58}{2} \\
\tabggHjPt{1}{}{60}{High}             & \numRF{ 168.27}{3} & \numRF{ 4252.94}{3} & 3.20 & \numRF{ 3.81}{2} & \numRF{2.56}{2} \\
\tabggHjPt{1}{}{60}{Med}              & \numRF{ 197.18}{3} & \numRF{11535.99}{3} & 3.38 & \numRF{ 1.68}{2} & \numRF{1.83}{2} \\
\tabggHjPt{1}{60}{120}{High}          & \numRF{ 185.78}{3} & \numRF{ 3313.89}{3} & 3.10 & \numRF{ 5.31}{2} & \numRF{3.20}{2} \\
\tabggHjPt{1}{60}{120}{Med}           & \numRF{ 180.29}{3} & \numRF{ 7779.75}{3} & 3.37 & \numRF{ 2.26}{2} & \numRF{2.04}{2} \\
\tabggHjPt{1}{120}{200}{High}         & \numRF{  23.01}{3} & \numRF{  182.46}{3} & 2.61 & \numRF{11.20}{2} & \numRF{1.67}{2} \\
\tabggHjPt{1}{120}{200}{Med}          & \numRF{  40.65}{3} & \numRF{  717.02}{3} & 3.00 & \numRF{ 5.37}{2} & \numRF{1.50}{2} \\
\tabggHmPt{}{350}{}{60}{High}         & \numRF{  23.46}{3} & \numRF{ 1053.36}{3} & 3.08 & \numRF{ 2.18}{2} & \numRF{0.72}{2} \\
\tabggHmPt{}{350}{}{60}{Med}          & \numRF{  43.13}{3} & \numRF{ 4363.41}{3} & 3.39 & \numRF{ 0.98}{2} & \numRF{0.65}{2} \\
\tabggHmPt{}{350}{}{60}{Low}          & \numRF{  47.50}{3} & \numRF{16788.57}{3} & 3.51 & \numRF{ 0.28}{2} & \numRF{0.37}{2} \\
\tabggHmPt{}{350}{60}{120}{High}      & \numRF{  49.08}{3} & \numRF{  900.70}{3} & 3.03 & \numRF{ 5.17}{2} & \numRF{1.62}{2} \\
\tabggHmPt{}{350}{60}{120}{Med}       & \numRF{  93.92}{3} & \numRF{ 6436.75}{3} & 3.30 & \numRF{ 1.44}{2} & \numRF{1.17}{2} \\
\tabggHmPt{}{350}{120}{200}{High}     & \numRF{  15.50}{3} & \numRF{   74.80}{3} & 2.64 & \numRF{17.16}{2} & \numRF{1.73}{2} \\
\tabggHmPt{}{350}{120}{200}{Med}      & \numRF{  22.70}{3} & \numRF{  343.13}{3} & 2.97 & \numRF{ 6.20}{2} & \numRF{1.21}{2} \\
\tabggHmPt{350}{700}{}{200}{High}    & \numRF{   4.31}{3} & \numRF{   47.47}{3} & 2.72 & \numRF{ 8.31}{2} & \numRF{0.62}{2} \\
\tabggHmPt{350}{700}{}{200}{Med}     & \numRF{  15.35}{3} & \numRF{  380.28}{3} & 3.02 & \numRF{ 3.88}{2} & \numRF{0.78}{2} \\
\tabggHmPt{350}{700}{}{200}{Low}     & \numRF{  10.53}{3} & \numRF{ 1079.22}{3} & 3.31 & \numRF{ 0.97}{2} & \numRF{0.32}{2} \\
\tabggHmPt{700}{1000}{}{200}{High}   & \numRF{   2.34}{3} & \numRF{   33.29}{3} & 2.84 & \numRF{ 6.56}{2} & \numRF{0.40}{2} \\
\tabggHmPt{700}{1000}{}{200}{Med}    & \numRF{   4.23}{3} & \numRF{  135.51}{3} & 3.07 & \numRF{ 3.02}{2} & \numRF{0.36}{2} \\
\tabggHmPt{700}{1000}{}{200}{Low}    & \numRF{   3.34}{3} & \numRF{  428.59}{3} & 3.26 & \numRF{ 0.77}{2} & \numRF{0.16}{2} \\
\tabggHmPt{1000}{}{}{200}{High}      & \numRF{   1.14}{3} & \numRF{   14.49}{3} & 2.97 & \numRF{ 7.29}{2} & \numRF{0.30}{2} \\
\tabggHmPt{1000}{}{}{200}{Med}       & \numRF{   2.52}{3} & \numRF{   47.45}{3} & 3.10 & \numRF{ 5.04}{2} & \numRF{0.36}{2} \\
\tabggHmPt{1000}{}{}{200}{Low}       & \numRF{   2.49}{3} & \numRF{  142.39}{3} & 3.37 & \numRF{ 1.72}{2} & \numRF{0.21}{2} \\
\tabggHPt{200}{300}{High}             & \numRF{  15.31}{3} & \numRF{   38.01}{3} & 2.28 & \numRF{28.72}{2} & \numRF{2.34}{2} \\
\tabggHPt{200}{300}{Med}              & \numRF{  29.40}{3} & \numRF{  235.59}{3} & 2.64 & \numRF{11.09}{2} & \numRF{1.88}{2} \\
\tabggHPt{300}{450}{High}             & \numRF{   1.52}{3} & \numRF{    2.13}{3} & 2.02 & \numRF{41.68}{2} & \numRF{0.95}{2} \\
\tabggHPt{300}{450}{Med}              & \numRF{   6.75}{3} & \numRF{   17.65}{3} & 2.16 & \numRF{27.67}{2} & \numRF{1.52}{2} \\
\tabggHPt{300}{450}{Low}              & \numRF{   4.66}{3} & \numRF{   43.05}{3} & 2.46 & \numRF{ 9.77}{2} & \numRF{0.70}{2} \\
\tabggHPt{450}{650}{High}             & \numRF{   1.00}{3} & \numRF{    1.25}{3} & 1.85 & \numRF{44.57}{2} & \numRF{0.81}{2} \\
\tabggHPt{450}{650}{Med}              & \numRF{   0.80}{3} & \numRF{    2.00}{3} & 1.98 & \numRF{28.62}{2} & \numRF{0.53}{2} \\
\tabggHPt{450}{650}{Low}              & \numRF{   0.83}{3} & \numRF{   10.67}{3} & 2.19 & \numRF{ 7.22}{2} & \numRF{0.25}{2} \\
\tabggHPt{650}{}{}                    & \numRF{   0.22}{3} & \numRF{    1.08}{3} & 1.73 & \numRF{16.78}{2} & \numRF{0.20}{2} \\
\midrule
\multicolumn{6}{c}{\qqtoHqq}\\
\midrule
\tabHqqj{0}{High}                     & \numRF{ 0.33}{3}  & \numRF{   25.00}{3} & 3.33 & \numRF{ 1.29}{2}  & \numRF{0.07}{1} \\
\tabHqqj{0}{Med}                      & \numRF{ 1.27}{3}  & \numRF{  470.99}{3} & 3.35 & \numRF{ 0.27}{2}  & \numRF{0.06}{1} \\
\tabHqqj{0}{Low}                      & \numRF{10.68}{3}  & \numRF{18822.68}{3} & 3.48 & \numRF{ 0.06}{1}  & \numRF{0.08}{1} \\
\tabHqqj{1}{High}                     & \numRF{ 1.08}{3}  & \numRF{    2.78}{3} & 2.99 & \numRF{28.00}{2}  & \numRF{0.61}{2} \\
\tabHqqj{1}{Med}                      & \numRF{ 3.50}{3}  & \numRF{   26.07}{3} & 3.11 & \numRF{11.83}{2}  & \numRF{0.67}{2} \\
\tabHqqj{1}{Low}                      & \numRF{ 2.88}{3}  & \numRF{  144.55}{3} & 3.24 & \numRF{ 1.95}{2}  & \numRF{0.24}{2} \\
\tabHqqm{}{60}{High}                  & \numRF{ 0.35}{3}  & \numRF{    2.10}{3} & 2.71 & \numRF{14.37}{2}  & \numRF{0.24}{2} \\
\tabHqqm{}{60}{Med}                   & \numRF{ 0.67}{3}  & \numRF{   19.02}{3} & 2.79 & \numRF{ 3.38}{2}  & \numRF{0.15}{2} \\
\tabHqqm{}{60}{Low}                   & \numRF{ 1.92}{3}  & \numRF{  243.11}{3} & 2.93 & \numRF{ 0.78}{2}  & \numRF{0.12}{2} \\
\tabHqqm{60}{120}{High}               & \numRF{ 3.45}{3}  & \numRF{    6.34}{3} & 2.65 & \numRF{35.23}{2}  & \numRF{1.27}{2} \\
\tabHqqm{60}{120}{Med}                & \numRF{ 4.99}{3}  & \numRF{   42.99}{3} & 2.85 & \numRF{10.39}{2}  & \numRF{0.75}{2} \\
\tabHqqm{60}{120}{Low}                & \numRF{ 2.99}{3}  & \numRF{   87.32}{3} & 3.01 & \numRF{ 3.31}{2}  & \numRF{0.32}{2} \\
\tabHqqm{120}{350}{High}              & \numRF{ 2.98}{3}  & \numRF{   24.44}{3} & 2.93 & \numRF{10.87}{2}  & \numRF{0.59}{2} \\
\tabHqqm{120}{350}{Med}               & \numRF{ 6.73}{3}  & \numRF{  203.77}{3} & 2.94 & \numRF{ 3.20}{2}  & \numRF{0.47}{2} \\
\tabHqqm{120}{350}{Low}               & \numRF{ 8.78}{3}  & \numRF{ 1361.89}{3} & 2.99 & \numRF{ 0.64}{2}  & \numRF{0.24}{2} \\
\tabHqqmPt{350}{700}{}{200}{High}    & \numRF{ 2.52}{3}  & \numRF{    2.75}{3} & 2.96 & \numRF{47.86}{2}  & \numRF{1.35}{2} \\
\tabHqqmPt{350}{700}{}{200}{Med}     & \numRF{ 9.15}{3}  & \numRF{   34.71}{3} & 3.06 & \numRF{20.86}{2}  & \numRF{1.49}{2} \\
\tabHqqmPt{350}{700}{}{200}{Low}     & \numRF{ 5.97}{3}  & \numRF{  106.22}{3} & 3.27 & \numRF{ 5.32}{2}  & \numRF{0.57}{2} \\
\tabHqqmPt{700}{1000}{}{200}{High}   & \numRF{ 2.91}{3}  & \numRF{    3.00}{3} & 2.90 & \numRF{49.24}{2}  & \numRF{1.48}{2} \\
\tabHqqmPt{700}{1000}{}{200}{Med}    & \numRF{ 5.60}{3}  & \numRF{   22.70}{3} & 3.11 & \numRF{19.80}{2}  & \numRF{1.13}{2} \\
\tabHqqmPt{1000}{}{}{200}{High}      & \numRF{10.79}{3}  & \numRF{    3.89}{3} & 3.01 & \numRF{73.50}{2}  & \numRF{4.17}{2} \\
\tabHqqmPt{1000}{}{}{200}{Med}       & \numRF{10.67}{3}  & \numRF{   18.95}{3} & 3.23 & \numRF{36.03}{2}  & \numRF{2.26}{2} \\
\bottomrule
\end{tabular}}}
\resizebox{0.49\textwidth}{!}{
\adjustbox{valign=t}{
\begin{tabular}{lrrrrr}
\toprule
\multirow{2}{*}{Category}  &  {\multirow{2}{*}{$S$}} &  {\multirow{2}{*}{$B$}}  &  {$\sigma$}  &  {$f$}  &  {\multirow{2}{*}{$Z$}}  \\
&  &  & {[GeV]} & {[\%]}  & \\
\midrule
\tabHqqmPt{350}{700}{200}{}{High}    & 1.31 & \numRF{ 2.19}{3} & 2.48 & \numRF{37.35}{2} & \numRF{0.81}{2} \\
\tabHqqmPt{350}{700}{200}{}{Med}     & 1.40 & \numRF{ 9.22}{3} & 2.49 & \numRF{13.18}{2} & \numRF{0.45}{2} \\
\tabHqqmPt{350}{700}{200}{}{Low}     & 1.16 & \numRF{65.50}{3} & 2.54 & \numRF{ 1.74}{2} & \numRF{0.14}{2} \\
\tabHqqmPt{700}{1000}{200}{}{High}   & 2.51 & \numRF{ 3.02}{3} & 2.43 & \numRF{45.33}{2} & \numRF{1.29}{2} \\
\tabHqqmPt{700}{1000}{200}{}{Med}    & 1.49 & \numRF{47.42}{3} & 2.54 & \numRF{ 3.04}{2} & \numRF{0.22}{2} \\
\tabHqqmPt{1000}{}{200}{}{High}      & 5.65 & \numRF{ 1.57}{3} & 2.39 & \numRF{78.24}{2} & \numRF{3.27}{2} \\
\tabHqqmPt{1000}{}{200}{}{Med}       & 2.96 & \numRF{ 6.31}{3} & 2.55 & \numRF{31.89}{2} & \numRF{1.10}{2} \\
\midrule
\multicolumn{2}{c}{\qqtoHln}\\
\midrule
\tabHlnPt{}{75}{High}       & 1.91 & \numRF{ 4.91}{3} & 3.17 & \numRF{27.98}{2} & \numRF{0.81}{2} \\
\tabHlnPt{}{75}{Med}        & 2.59 & \numRF{20.15}{3} & 3.28 & \numRF{11.41}{2} & \numRF{0.57}{2} \\
\tabHlnPt{75}{150}{High}     & 2.62 & \numRF{ 2.05}{3} & 3.02 & \numRF{56.10}{2} & \numRF{1.57}{2} \\
\tabHlnPt{75}{150}{Med}      & 2.08 & \numRF{12.37}{3} & 3.23 & \numRF{14.42}{2} & \numRF{0.58}{2} \\
\tabHlnPt{150}{250}{High}    & 1.74 & \numRF{ 2.06}{3} & 2.78 & \numRF{45.82}{2} & \numRF{1.08}{2} \\
\tabHlnPt{150}{250}{Med}     & 0.16 & \numRF{ 2.90}{3} & 3.17 & \numRF{ 5.24}{2} & \numRF{0.09}{1} \\
\tabHlnPt{250}{}{High}       & 1.36 & \numRF{ 1.79}{3} & 2.41 & \numRF{43.12}{2} & \numRF{0.91}{2} \\
\tabHlnPt{250}{}{Med}        & 0.02 & \numRF{ 3.12}{3} & 3.15 & \numRF{ 0.78}{2} & \numRF{0.01}{1} \\
\midrule
\multicolumn{2}{c}{\pptoHll}\\
\midrule
\tabHllPt{}{75}{High}       & 1.14 & \numRF{  1.82}{3}  & 3.25  & \numRF{38.60}{2} & \numRF{0.78}{2} \\
\tabHllPt{}{75}{Med}        & 1.06 & \numRF{214.99}{3}  & 3.29  & \numRF{ 0.49}{2} & \numRF{0.07}{1} \\
\tabHllPt{75}{150}{High}     & 1.07 & \numRF{  1.58}{3}  & 3.08  & \numRF{40.28}{2} & \numRF{0.77}{2} \\
\tabHllPt{75}{150}{Med}      & 0.02 & \numRF{  1.81}{3}  & 3.06  & \numRF{ 1.15}{2} & \numRF{0.02}{1} \\
\tabHllPt{150}{250}{High}    & 0.71 & \numRF{  1.79}{3}  & 2.78  & \numRF{28.36}{2} & \numRF{0.50}{2} \\
\tabHllPt{150}{250}{Med}     & 0.10 & \numRF{ 16.52}{3}  & 2.88  & \numRF{ 0.62}{2} & \numRF{0.03}{1} \\
\tabHllPt{250}{}{}           & 0.27 & \numRF{  2.06}{3}  & 2.48  & \numRF{11.53}{2} & \numRF{0.18}{2} \\
\midrule
\multicolumn{2}{c}{\pptoHnn}\\
\midrule
\tabHnnPt{}{75}{High}     & 0.60 & \numRF{ 170.16}{3} & 3.50 & \numRF{ 0.35}{2} & \numRF{0.05}{1} \\
\tabHnnPt{}{75}{Med}      & 1.15 & \numRF{1021.05}{3} & 3.57 & \numRF{ 0.11}{2} & \numRF{0.04}{1} \\
\tabHnnPt{}{75}{Low}      & 0.87 & \numRF{2629.27}{3} & 3.67 & \numRF{ 0.03}{1} & \numRF{0.02}{1} \\
\tabHnnPt{75}{150}{High}   & 0.58 & \numRF{   2.30}{3} & 2.97 & \numRF{20.13}{2} & \numRF{0.37}{2} \\
\tabHnnPt{75}{150}{Med}    & 1.83 & \numRF{  17.77}{3} & 3.26 & \numRF{ 9.32}{2} & \numRF{0.43}{2} \\
\tabHnnPt{75}{150}{Low}    & 2.18 & \numRF{ 287.86}{3} & 3.44 & \numRF{ 0.75}{2} & \numRF{0.13}{2} \\
\tabHnnPt{150}{250}{High}  & 0.92 & \numRF{   2.00}{3} & 2.75 & \numRF{31.58}{2} & \numRF{0.61}{2} \\
\tabHnnPt{150}{250}{Med}   & 0.75 & \numRF{   2.54}{3} & 2.94 & \numRF{22.73}{2} & \numRF{0.45}{2} \\
\tabHnnPt{150}{250}{Low}   & 0.26 & \numRF{  11.68}{3} & 3.28 & \numRF{ 2.16}{2} & \numRF{0.08}{1} \\
\tabHnnPt{250}{}{High}     & 0.67 & \numRF{   1.55}{3} & 2.46 & \numRF{30.15}{2} & \numRF{0.50}{2} \\
\tabHnnPt{250}{}{Med}      & 0.05 & \numRF{   1.97}{3} & 3.05 & \numRF{ 2.59}{2} & \numRF{0.04}{1} \\
\midrule
\multicolumn{2}{c}{\ttH}\\
\midrule
\tabttHPt{}{60}{High}      & 3.04 & \numRF{ 4.01}{3} & 3.18 & \numRF{43.15}{2} & \numRF{1.37}{2} \\
\tabttHPt{}{60}{Med}       & 2.78 & \numRF{13.27}{3} & 3.37 & \numRF{17.32}{2} & \numRF{0.74}{2} \\
\tabttHPt{60}{120}{High}   & 4.30 & \numRF{ 4.09}{3} & 3.06 & \numRF{51.23}{2} & \numRF{1.86}{2} \\
\tabttHPt{60}{120}{Med}    & 2.99 & \numRF{ 8.61}{3} & 3.31 & \numRF{25.77}{2} & \numRF{0.97}{2} \\
\tabttHPt{120}{200}{High}  & 4.65 & \numRF{ 3.52}{3} & 2.73 & \numRF{56.86}{2} & \numRF{2.11}{2} \\
\tabttHPt{120}{200}{Med}   & 1.66 & \numRF{ 4.16}{3} & 2.93 & \numRF{28.57}{2} & \numRF{0.77}{2} \\
\tabttHPt{200}{300}{}      & 3.39 & \numRF{ 2.26}{3} & 2.46 & \numRF{59.99}{2} & \numRF{1.89}{2} \\
\tabttHPt{300}{}{}         & 2.73 & \numRF{ 1.66}{3} & 2.12 & \numRF{62.18}{2} & \numRF{1.76}{2} \\
\midrule
\multicolumn{2}{c}{\tH}\\
\midrule
\tHqb, High-purity             & 0.55 & \numRF{ 2.16}{3} & 3.04 & \numRF{20.18}{2} & \numRF{0.36}{2} \\
\tHqb, Med-purity              & 0.14 & \numRF{ 2.78}{3} & 3.45 & \numRF{ 4.92}{2} & \numRF{0.09}{1} \\
\tHqb, BSM   ($\kappa_t = -1$) & 0.12 & \numRF{ 1.86}{3} & 3.25 & \numRF{ 6.03}{2} & \numRF{0.09}{1} \\
\tHW                           & 0.16 & \numRF{ 6.91}{3} & 2.74 & \numRF{ 2.25}{2} & \numRF{0.06}{1} \\
\midrule
Low-purity top            & 5.18 & \numRF{65.77}{3} &  3.32   & \numRF{ 7.30}{2} & \numRF{0.63}{2} \\
\bottomrule
\end{tabular}}}
\end{table}


\begin{landscape}
\begin{figure}[ht]
\centering
\includegraphics[width=1.3\textwidth]{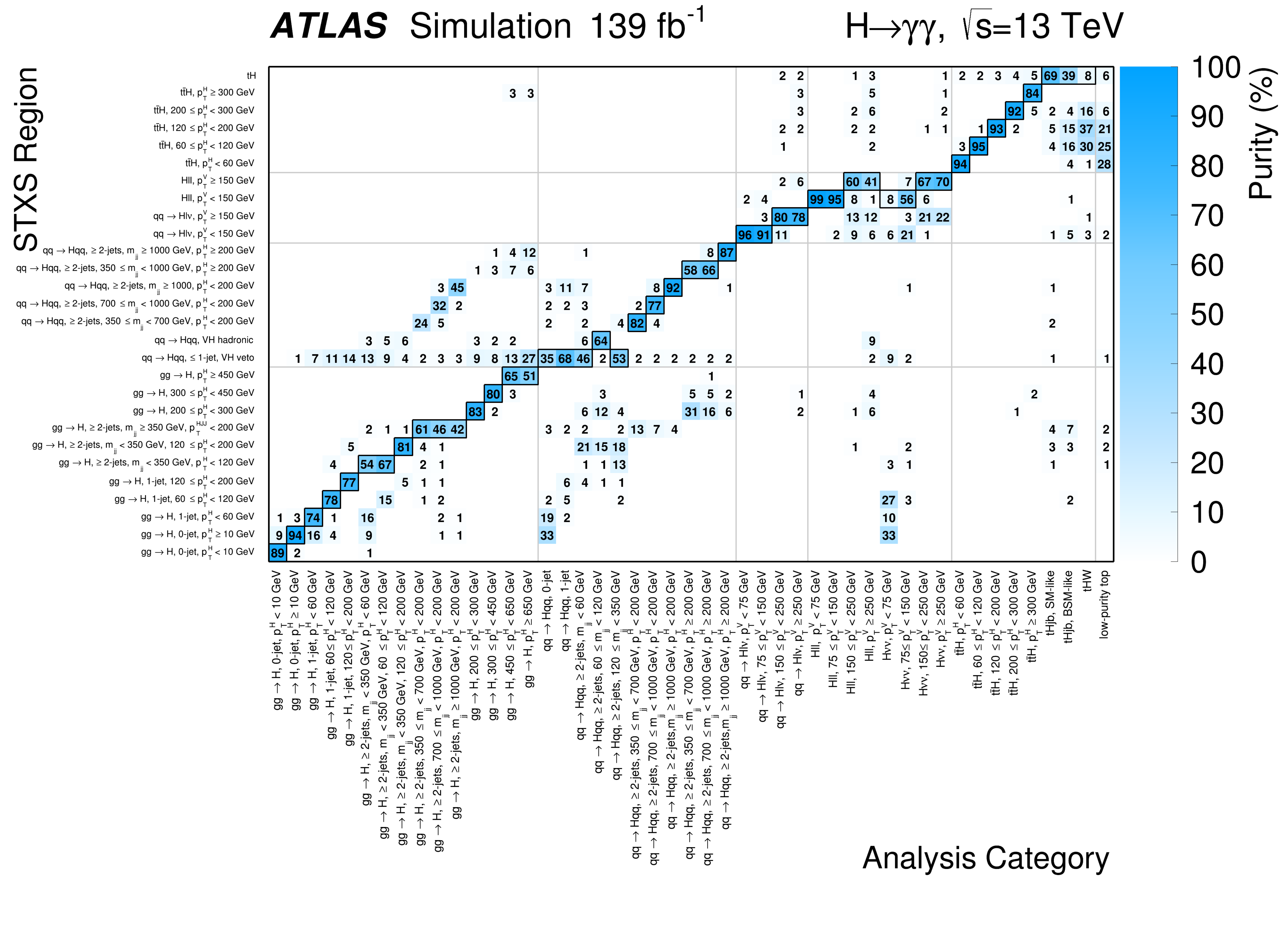}
\caption{Contributions of STXS measurement regions to the expected event yields in groups of analysis categories. The vertical axis lists the 28 STXS measurement regions defined in Section~\ref{sec:results:STXS}, while the horizontal axis lists groups of analysis categories that target the same STXS measurement region, weighted by their $f$ value as given in Table~\ref{tab:design:yields}.
Entries correspond to the percentage of the signal yield in each group of analysis categories (on the $x$-axis) that is contributed by a given STXS measurement region (on the $y$-axis).
Entries with a value below $1\%$ are not shown. The entries in each column, corresponding to the same group of analysis categories, add up to 100 (\%), except for rounding effects and values below $1\%$.
}
\label{fig:design:yields}
\end{figure}
\end{landscape}


\section{Modelling of diphoton mass distributions}
\label{sec:modeling}

The \mgg\ distribution in each category is described by an extended probability density function (pdf) in which the signal and background shapes are analytic functions of \mgg. As in the previous measurement~\cite{HIGG-2016-21}, the analytic functions are defined over the range of $105 \le \mgg < 160\,\GeV$. The analysis results are obtained by a simultaneous fit of these pdfs to the \mgg\ distributions in the categories defined in Section~\ref{sec:cats}. Systematic uncertainties related to signal yield, signal shape and background modelling are incorporated into the likelihood model as nuisance parameters. For each of these nuisance parameters, a Gaussian or log-normal constraint pdf is included in the likelihood function. Gaussian constraints are used for uncertainties related to the background modelling, the peak position of the signal \mgg\ distribution, and the Higgs boson mass. Log-normal constraints are used for other uncertainties, including multiplicative uncertainties in expected signal yields and in the \mgg\ mass resolution. Asymmetric log-normal forms are used when the corresponding uncertainties are themselves asymmetric. The Higgs boson mass $m_H$ is assumed to be $125.09 \pm 0.24\,\GeV$, as measured in Ref.~\cite{HIGG-2014-14}.

The effects of interference between the \Hyy\ signal process and continuum \gamgam\ production lead to a small change in the expected Higgs boson production rate (a 2\% reduction in the inclusive rate~\cite{Campbell:2017rke}) as well as a shift in the signal peak position that is small compared to the uncertainty in $m_H$~\cite{Dixon:2013haa}. Both effects are neglected.
 
In each category $i$, the normalization of the background pdf is a free parameter in the fit, as well as the parameters describing the shapes of the background pdfs, as discussed in Section~\ref{sec:modeling:bkg} below. The normalization of the signal pdf is expressed as
\begin{equation}
N_i = \sum\limits_t (\sigma_t \times \Byy) \, \epsilon_{it}  \,\mathcal{L} \, K_i(\bm{\theta}_{\text{yield}}) + N_{\text{spur},i}  \,\theta_{\text{spur},i}
\label{eq:yield}
\end{equation}
where the sum runs over all regions defined in the Stage 1.2 STXS scheme, $(\sigma_t \times \Byy)$ is the measurement parameter for region $t$, $\epsilon_{it}$ describes the  efficiency for events from region $t$ to be reconstructed in category $i$, and $\mathcal{L}$ is the integrated luminosity of the fitted sample. The factor $K_i(\bm{\theta}_{\text{yield}})$ corresponds to multiplicative corrections to the signal yields from systematic uncertainty effects detailed in Section~\ref{sec:systs}, as a function of nuisance parameters collectively denoted by $\bm{\theta}_{\text{yield}}$; $N_{\text{spur},i}$ is the value of the background modelling uncertainty described in Section~\ref{sec:modeling:bkg}, implemented as an additive correction to the signal yield proportional to the nuisance parameter $\theta_{\text{spur},i}$.
The values of the measurement parameters are obtained from a maximum-likelihood fit to the data.
 
\subsection{Modelling of the signal shape}
 
The signal component in each category corresponds to the sum of the contributions from each STXS analysis region, which are all assumed to follow the same \mgg\ distribution in this category. The shape is described using a double-sided Crystal Ball (DSCB) function~\cite{Oreglia:1980cs,HIGG-2014-04}, consisting of a Gaussian distribution in the region around the peak position, continued by power-law tails at lower and higher \mgg\ values. An intrinsic shape difference between the DSCB function and signal \mgg\ distribution is found to cause only a negligible bias in the estimated signal yield~\cite{HIGG-2016-21}.
 
The parameters of the Crystal Ball function in each category are obtained by a fit to a mixture of the \ggF, \VBF, \VH, \ttH\ and \tH\ samples, described in Section~\ref{sec:mc}, in proportion to their SM cross-sections. A shift of $0.09\,\GeV$ is applied to the position of the signal peak to account for the difference between the reference Higgs boson mass used in this analysis ($m_H = 125.09\,\GeV$) and the mass for which the samples were generated ($m_H = 125\,\GeV$).
Simulated signal \mgg\ distributions and their corresponding DSCB functions are shown for two groups of categories in Figure~\ref{fig:modeling:signal}.
\begin{figure}[ht]
\centering
 
\subfloat[\ggtoH, 1-jet, $120 \le p_\mathrm{T}^{H} < 200\,\GeV$]{\includegraphics[width=0.49\textwidth]{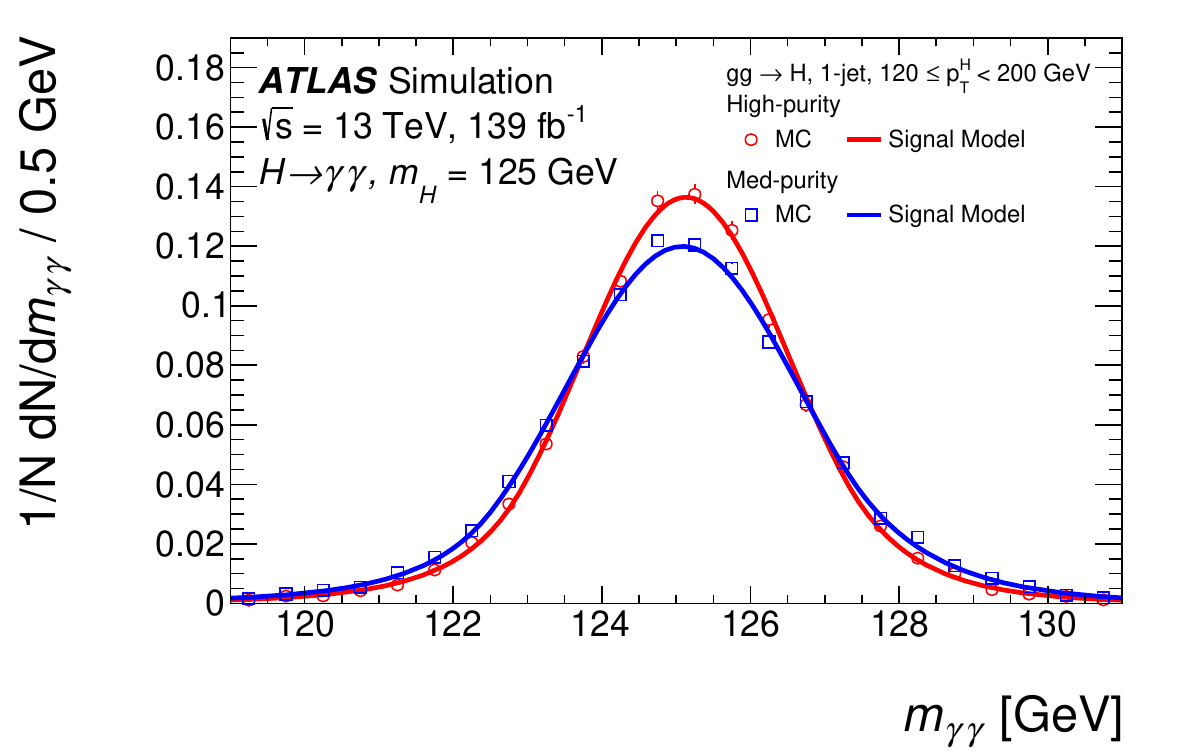}\label{fig:modeling:signal:gg2H}}
\subfloat[\ttH]{\includegraphics[width=0.49\textwidth]{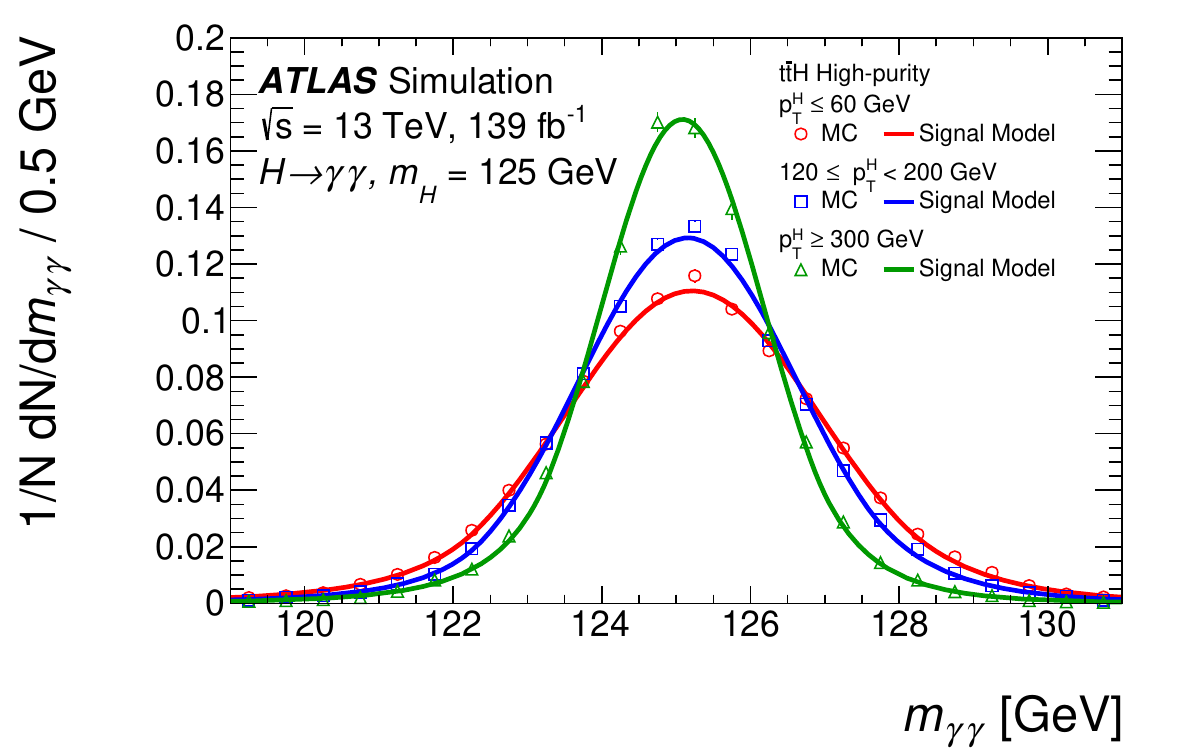}\label{fig:modeling:signal:ttH}}
\caption{
Shape of the signal \mgg\ distribution for two groups of categories. Panel \ref{fig:modeling:signal:gg2H} compares the signal \mgg\ shapes for the two categories targeting the \ggtoH, 1-jet, $120 \le p_\mathrm{T}^{H} < 200\,\GeV$ region. Panel \ref{fig:modeling:signal:ttH} compares the signal \mgg\ shapes for three High-purity categories targeting different \ptH\ regions of the \ttH\ process.
The markers represent distributions in MC samples with $m_H = 125\,\GeV$, while the solid lines represent the corresponding fitted DSCB functions. }
\label{fig:modeling:signal}
\end{figure}
 
\subsection{Modelling of the continuum background shape}
\label{sec:modeling:bkg}
 
The background in the selected diphoton sample mainly consists of continuum \gamgam\ production, \gamjet\ and \jetjet\ production where one or more jets in the event are misidentified as photons. In the categories targeting \VlepH\ production, the main contribution is from the $V\gamma\gamma$ processs, while in categories targeting \ttH\ and \tH\ production the main contributions are from $t\bar{t}\gamma\gamma$ and other processes involving top quarks. The modelling of this continuum background follows the same procedure as in previous analyses~\cite{HIGG-2016-21}. This procedure involves two main steps: first, a background \mgg\ template is constructed from a combination of simulation samples and data control samples; secondly, a background function is selected from a number of candidate functions, using the \emph{spurious-signal test}, with the goal of identifying an analytic function that is flexible enough to fit the \mgg\ distribution in data and results in a sufficiently small potential bias compared to the statistical uncertainty.
 
In categories targeting the \ggtoH\ and \qqtoHqq\ processes, the template is defined as a combination of \gamgam, \gamjet, and \jetjet\ processes, each of which is weighted according to its fraction in the selected analysis category. The fractions of these processes are determined by a data-driven method, known as the double two-dimensional sideband method~\cite{STDM-2017-30}, which uses three control regions in data in which one (for the \gamjet\ process) or both (for the \jetjet\ process) photons fail to satisfy the identification and/or isolation criteria, respectively. The fraction of the total background that is from the \gamgam\ process ranges between 75\% and 95\%, the fraction from the \gamjet\ process is between 2\% and 25\%, and the \jetjet\ process contributes less than 6\%.
 
While a simulation sample is used to model the \gamgam\ process in this study, it is computationally prohibitive to generate sufficiently large samples of \gamjet\ and \jetjet\ production events passing analysis selections due to their large cross-sections and the high jet-rejection performance of the ATLAS photon identification algorithms. To avoid this issue, the \mgg\ shapes of the \gamjet\ or \jetjet\ components are obtained from data control samples defined by inverting the identification requirement of one  or both  photons as described above. The ratio of the \mgg\ distribution of the \gamjet\ and \jetjet\ components to that of the simulated \gamgam\ sample is
well described by a linear function of \mgg. A linear fit to the ratio of these distributions is therefore used to  provide an \mgg-dependent weight that is applied to the \gamgam\ sample to obtain a final template that also accounts for the \gamjet\ and \jetjet\ components. Changing the fraction of the \gamjet\ and \jetjet\ components within the uncertainties of their determination is found to have a negligible impact on the spurious-signal test described below.
 
For categories targeting the \VlepH\ STXS regions, the background template is built using simulated events of $V\gamma\gamma$ and prompt $\gamma\gamma$ production. Since the available yields for the latter are not sufficient to build the template directly, the following procedure is followed: first a linear function of \mgg is fitted to the ratio of the \mgg\ distribution of both samples to that of the $V\gamma\gamma$ sample alone; the  resulting linear function from the fit is then applied to the \mgg\ distribution of the $V\gamma\gamma$ sample as an \mgg-dependent weight to obtain the final template describing both contributions.
For categories targeting the \ttH\ and \tH\ processes, the diphoton events are primarily from $t\bar{t}\gamma\gamma$ production. As such, a sample of simulated $t\bar{t}\gamma\gamma$ events is used to construct the background template for those categories.
Contributions from processes with jets misidentified as photons are not considered in categories targeting \VH, \ttH\ and \tH\ STXS regions as they do not significantly alter the background shape.
The background templates constructed for four categories targeting the \ggtoH, \qqtoHqq, \VH\ and \ttH\ processes are shown as examples in Figure~\ref{fig:design:bkg}. While the background template and data \mgg\ distribution have slightly different shapes in some categories, the selected background analytic functions are flexible enough to absorb these small differences.
 
\begin{figure}[ht]
\centering
-\subfloat[\ggtoH, 1-jet, $\ptH < 60\,\GeV$, High-purity]{\includegraphics[width=0.49\textwidth]{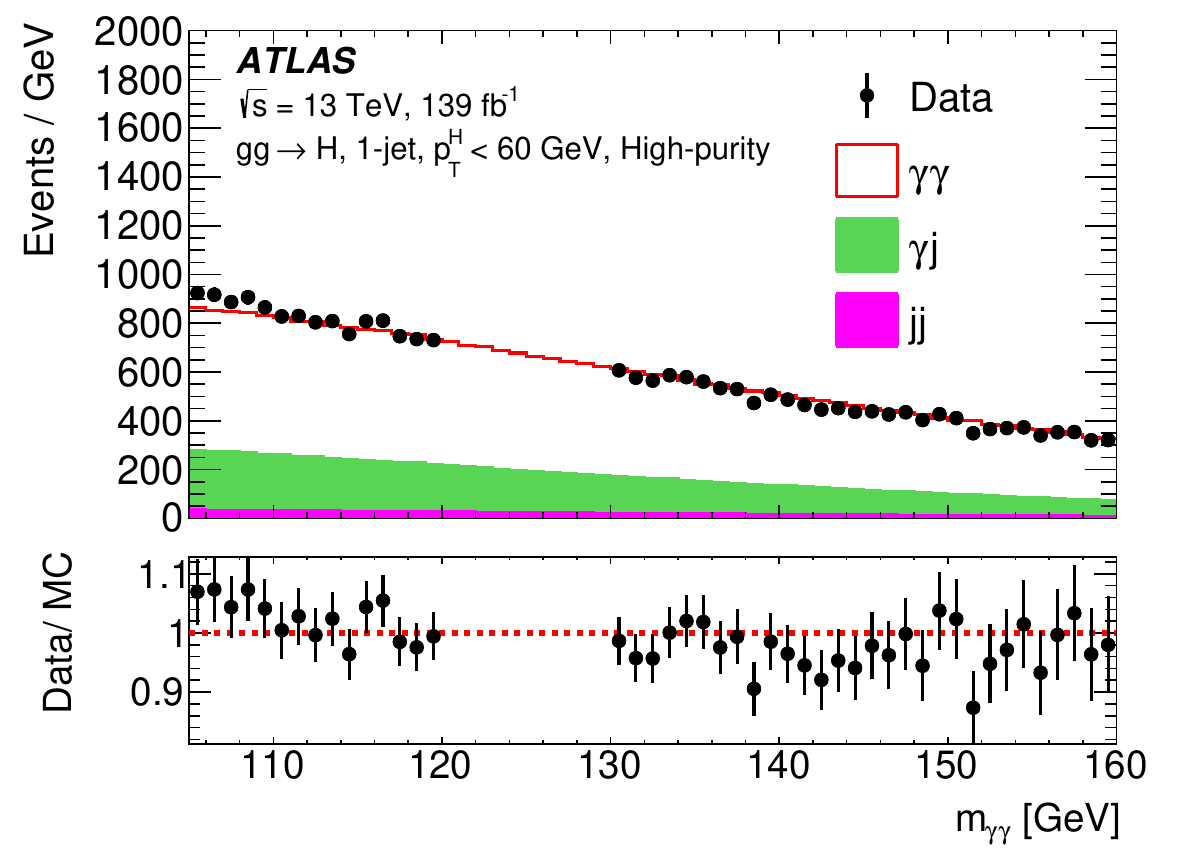}\label{fig:design:bkg_ggH}}
\subfloat[\qqtoHqq, $\ge 2$-jets, $350 \le \mjj < 700\,\GeV$, $\ptH < 200\,\GeV$, Med-purity]{\includegraphics[width=0.49\textwidth]{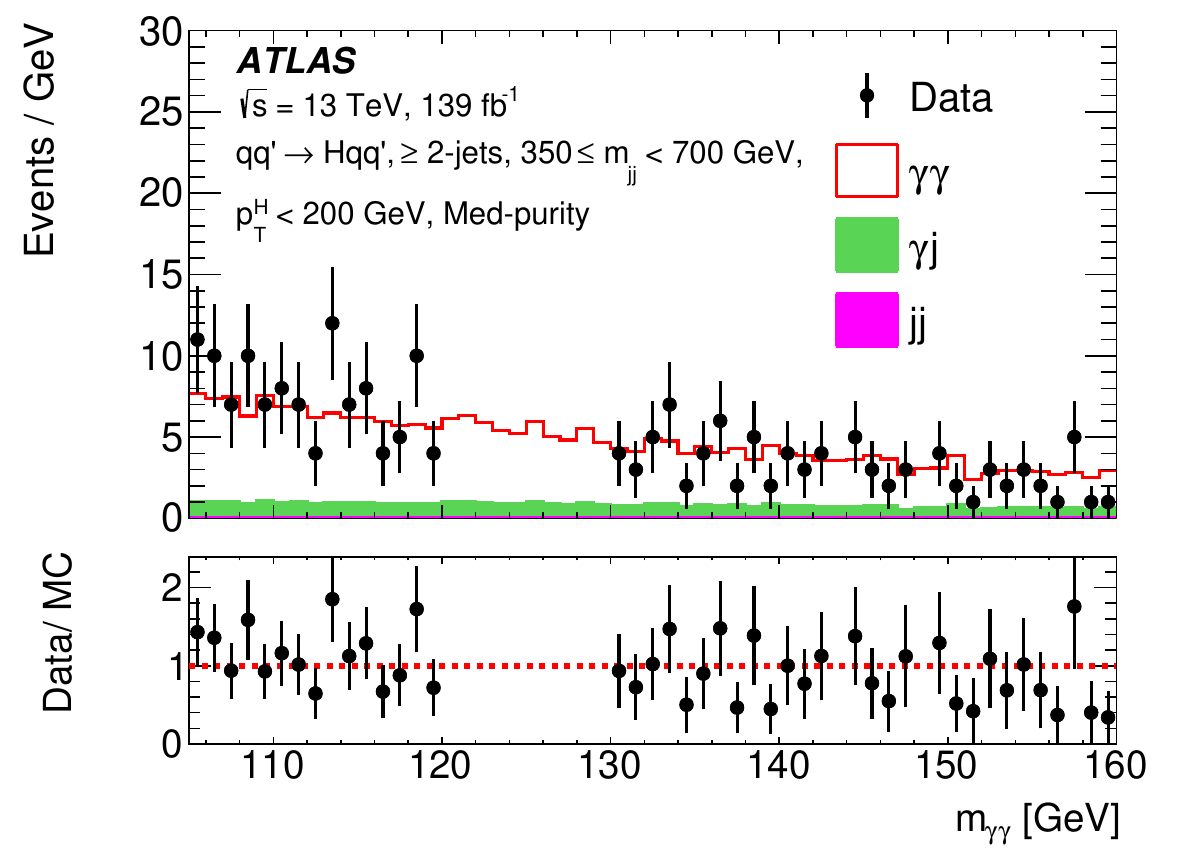}\label{fig:design:bkg_VBF}}	\\
\subfloat[\qqtoHln, $75 \le \ptV < 150\,\GeV$, Med-purity]{\includegraphics[width=0.49\textwidth]{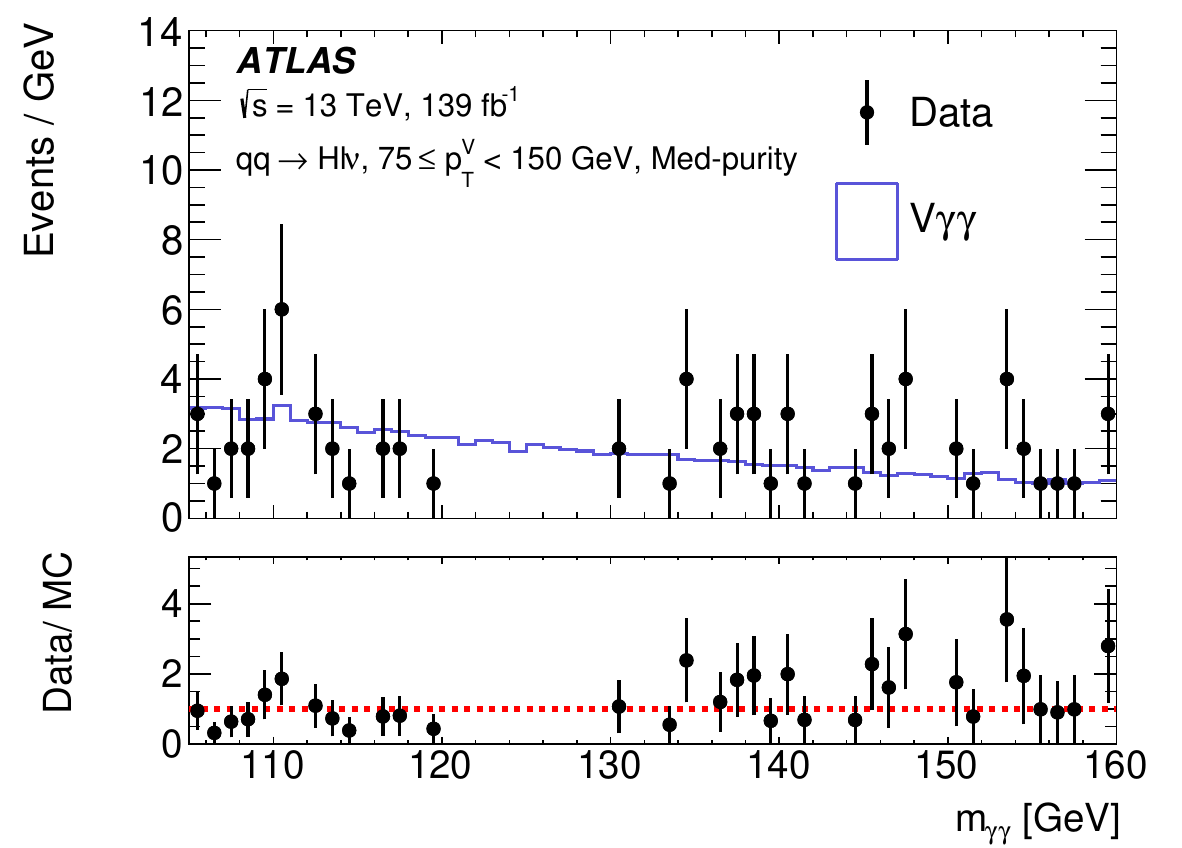} \label{fig:design:bkg_VH}}
\subfloat[\ttH, $60 \le \ptH < 120\,\GeV$, Med-purity]{\includegraphics[width=0.49\textwidth]{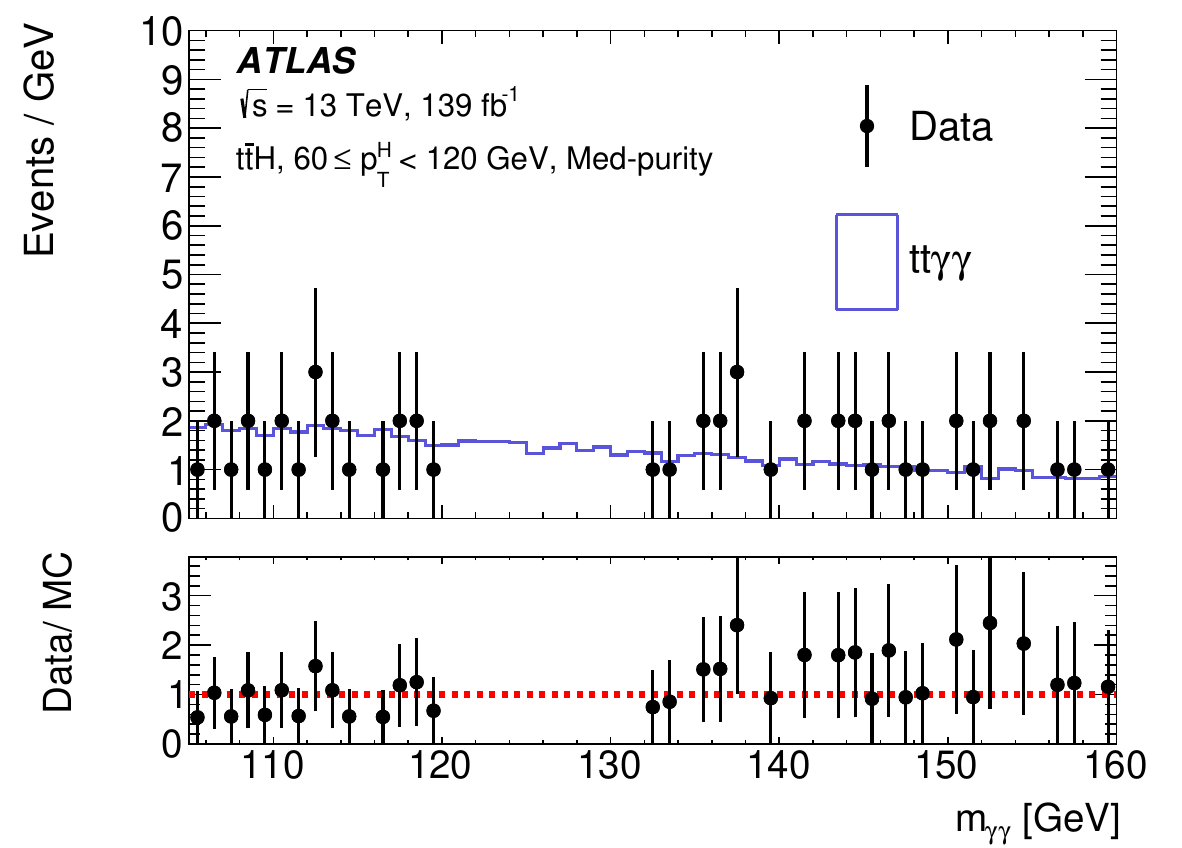}\label{fig:design:bkg_ttH}}
\caption{The diphoton invariant mass \mgg\ distribution in data (black points) and continuum background templates (histograms) in four representative STXS categories. The data are shown excluding the region $120 \le \mgg < 130\,\GeV$ containing the signal. In panels~\ref{fig:design:bkg_ggH} and~\ref{fig:design:bkg_VBF}, stacked histograms corresponding to the \gamgam\ (white), \gamjet\ (green) and \jetjet\ (magenta) background contributions are shown. In panel~\ref{fig:design:bkg_VH}, the histogram represents contributions from $V\gamgam$ and other sources of prompt \gamgam\ production. In~\ref{fig:design:bkg_ttH}, the histogram corresponds to simulated $\ttbar\gamma\gamma$ events. The templates do not represent the background shapes used in the analysis, but are used to identify flexible functions used to model the background in each category as described in the text.
}
\label{fig:design:bkg}
\end{figure}
 
The background templates are defined over the range $105 \le \mgg < 160\,\GeV$ with 220 uniform-width bins. A template smoothing procedure based on a Gaussian kernel~\cite{Frate:2017mai} is applied to analysis categories where the average bin occupancy in the background template is at least 20 entries. This procedure suppresses statistical fluctuations in the background templates, decreasing the systematic uncertainty on the modeling of the background. A study using pseudo-experiments generated with known template shapes was performed to verify that the smoothing procedure does not introduce a significant bias in the estimate of the spurious signal.
 
Three families of analytic functions are tested as  candidates to model the \mgg\ distribution for a given analysis category. They include power-law functions, Bernstein polynomials~\cite{Bernstein}, and exponential functions of a polynomial. These functions and the number of degrees of freedom tested are summarized in Table~\ref{tab:modeling:functions}.
The parameters of these functions are considered to be independent across categories and always left free to vary.
\begin{table}[!htp]
\footnotesize
\centering
\caption{Summary of the functions used for the modelling of the continuum background component. The free parameters used to define the function shape are denoted as $a$ or $a_i$, and their total number by $N_{\text{pars}}$. For the definition of the Bernstein polynomials, $x = (\mgg - m_{\text{min}})/(m_{\text{max}} - m_{\text{min}})$, where $m_{\text{min}} = 105\,\GeV$ and $m_{\text{max}} = 160\,\GeV$ are respectively the lower and upper bounds of the fitted \mgg\ range.}
\label{tab:modeling:functions}
\begin{tabular}{lccc}
\toprule
Type & Function & $N_{\text{pars}}$ & Acronym \\
\midrule
Power law                              & $\mgg^a$                                   & 1                 & \texttt{PowerLaw} \\
Bernstein polynomial & $(1 - x)^n + a_1 nx(1 - x)^{n-1} + \cdots + a_n x^n$         & $n = \text{1--5}$ & \texttt{Bern1}--\texttt{Bern5} \\
Exponential                            & $\exp(a \mgg)$                             & 1                 & \texttt{Exp} \\
Exponential of second-order polynomial & $\exp(a_1 \mgg + a_2 \mgg^2)$              & 2                 & \epoly{2} \\
Exponential of third-order polynomial  & $\exp(a_1 \mgg + a_2 \mgg^2 + a_3 \mgg^3)$ & 3                 & \epoly{3} \\
\bottomrule
\end{tabular}
\end{table}
The main criterion used to select the functional form in each category is a bias test performed by fitting the background template using a model with free parameters for both the signal and the background event yields. In this fit, the background template is normalized to the number of events observed in data in this category.
The potential bias due to the mis-modelling of background \mgg\ distribution is estimated from the fitted signal yield (the \emph{spurious} signal). This test is performed for $m_H$ values ranging from 123\,\GeV\ to 127\,\GeV, in steps of 0.5\,\GeV. In order to avoid accidentally small bias values at the nominal Higgs boson mass of $m_H~=~125.09\,\GeV$, the maximum absolute value of fitted signal yield $|S_{\text{spur}}|$ over the range $123 < m_H < 127\,\GeV$ is considered as the potential bias.
For categories where the original background \mgg\ templates (before normalization to the data yields) have at least 20 entries per bin on average,
the background functions are required to yield a value of $|S_{\text{spur}}|$ that is smaller than either 10\% of the total expected Higgs boson signal yield or 20\% of the statistical uncertainty of the fitted signal yield. If multiple functions pass these requirements, the one with the smallest number of degrees of freedom is chosen.
 
An additional check is performed for the nine categories in which a fit of the analytic function to the background template is found to yield a $\chi^2$ $p$-value that is below $1\%$.\footnote{The $\chi^2$ is computed in a background template uniformly binned over $105 \le \mgg < 160\,\GeV$. It has 55 bins, and the number of degrees of freedom used in the computation is $54 - N_{\text{pars}}$, where the $N_{\text{pars}}$ is the number of free function parameters. The normalization of the template removes one degree of freedom. The background \mgg\ templates before the smoothing procedure are used.}
For each of these categories, a set of samples is randomly drawn from the background template, each with a number of events equal to the observed yield in the data sidebands. The fit of the analytic function and the computation of the $\chi^2$ are then repeated for each sample. In all nine categories, more than 90\% of the samples yield a $\chi^2$ $p$-value above 5\%. This shows that the chosen functions provide a sufficiently good fit to the data in these categories, in spite of the low $p$-values observed in the fit to the nominal background template.
For categories where the average number of entries per bin is less than 20, candidate background functions are limited to \expfunc, \epoly{2}\ and \epoly{3}\ (as defined in Table~\ref{tab:modeling:functions}) in order to avoid unphysical fits due to large statistical fluctuations in the sidebands.  The function is chosen using a Wald test~\cite{Wald:1945}: first the quantity $q_{12} = -2\ln L_1/L_2$ is computed from the maximum likelihood values $L_1$ and $L_2$ of background-only fits to the data sideband regions using respectively the \expfunc\ and \epoly{2}\ descriptions of the backgrounds. The \epoly{2} model is chosen if the $p$-value computed from $q_{12}$ is less than $0.05$, assuming that $q_{12}$ follows a $\chi^2$ distribution with one degree of freedom. Similarly, the \epoly{3}\ form is chosen over \epoly{2}\ if the $p$-value for the corresponding Wald test is $0.05$ or less. For 32 out of the 101 categories, the Wald-test-based condition was used and the \expfunc\ function was selected in each case.
 
In all cases, the $|S_{\text{spur}}|$ value of the selected background function provides an estimate of the possible bias in the fitted signal yield introduced by the intrinsic difference between the background \mgg\ shape and the selected function. It is used to define the systematic uncertainty for the background modelling in category $i$, denoted as $N_{\text{spur},i}$ in Eq.~(\ref{eq:yield}). The selected functional form for each category is shown in Table~\ref{tab:design:cats}.

\begin{table}[ht]
\caption{Selected background functional form, number of observed data events in the range $105 \le \mgg < 160\,\GeV$ ($N_{\text{data}}$), and modelling uncertainty ($N_{\text{spur}}$) for each of the 101 analysis categories. The last column indicates whether the Wald test is used to determine the functional form, as described in the text.}
\label{tab:design:cats}
\centering
\renewcommand{\arraystretch}{1.1}
\resizebox{0.49\textwidth}{!}{
\adjustbox{valign=t}{
\begin{tabular}{lrrrc}
\toprule
Category   & Function & $N_{\text{data}}$ & $N_{\text{spur}}$  & Wald \\
\midrule
\multicolumn{2}{c}{\ggtoH}\\
\midrule
\tabggHjPt{0}{}{10}{}               & \epoly{2}         & 191623 & 64.8  & \\
\tabggHjPt{0}{10}{}{}               & \epoly{2}         & 349266 & 50.4  & \\
\tabggHjPt{1}{}{60}{High}           & \epoly{2}         & 32644  & 20.7  & \\
\tabggHjPt{1}{}{60}{Med}            & \epoly{2}         & 85229  & 24.9  & \\
\tabggHjPt{1}{60}{120}{High}        & \epoly{1}         & 26236  & 23.7  & \\
\tabggHjPt{1}{60}{120}{Med}         & \epoly{2}         & 56669  & 21.3  & \\
\tabggHjPt{1}{120}{200}{High}       & \epoly{2}         & 1570   & 1.48  & \\
\tabggHjPt{1}{120}{200}{Med}        & \epoly{2}         & 6163   & 5.33  & \\
\tabggHmPt{}{350}{}{60}{High}       & \epoly{2}         & 8513   & 1.51  & \\
\tabggHmPt{}{350}{}{60}{Med}        & \epoly{2}         & 31163  & 13.6  & \\
\tabggHmPt{}{350}{}{60}{Low}        & \epoly{2}         & 120357 & 15.7  & \\
\tabggHmPt{}{350}{60}{120}{High}    & \epoly{2}         & 7582   & 2.26  & \\
\tabggHmPt{}{350}{60}{120}{Med}     & \epoly{2}         & 48362  & 6.21  & \\
\tabggHmPt{}{350}{120}{200}{High}   & \epoly{2}         & 728    & 0.004 & \\
\tabggHmPt{}{350}{120}{200}{Med}    & \texttt{PowerLaw} & 3007   & 0.983 & \\
\tabggHmPt{350}{700}{}{200}{High}  & \epoly{1}         & 432    & 0.487 & \\
\tabggHmPt{350}{700}{}{200}{Med}   & \epoly{2}         & 3084   & 1.33  & \\
\tabggHmPt{350}{700}{}{200}{Low}   & \epoly{1}         & 7999   & 5.78  & \\
\tabggHmPt{700}{1000}{}{200}{High} & \epoly{1}         & 302    & 0.560 & \\
\tabggHmPt{700}{1000}{}{200}{Med}  & \epoly{1}         & 1033   & 1.44  & \\
\tabggHmPt{700}{1000}{}{200}{Low}  & \epoly{1}         & 3187   & 4.32  & \\
\tabggHmPt{1000}{}{}{200}{High}    & \epoly{1}         & 113    & 0.192 & \\
\tabggHmPt{1000}{}{}{200}{Med}     & \epoly{1}         & 332    & 0.804 & \\
\tabggHmPt{1000}{}{}{200}{Low}     & \texttt{PowerLaw} & 1020   & 1.09  & \\
\tabggHPt{200}{300}{High}           & \epoly{1}         & 420    & 1.68  & \\
\tabggHPt{200}{300}{Med}            & \epoly{1}         & 2296   & 0.714 & \\
\tabggHPt{300}{450}{High}           & \epoly{1}         & 25     & 0.407 & $\checkmark$ \\
\tabggHPt{300}{450}{Med}            & \epoly{1}         & 186    & 0.259 & \\
\tabggHPt{300}{450}{Low}            & \epoly{1}         & 422    & 0.121 & \\
\tabggHPt{450}{650}{High}           & \epoly{1}         & 15     & 0.138 & $\checkmark$ \\
\tabggHPt{450}{650}{Med}            & \epoly{1}         & 25     & 0.391 & $\checkmark$ \\
\tabggHPt{450}{650}{Low}            & \epoly{1}         & 109    & 0.031 &  \\
\tabggHPt{650}{}{}                  & \epoly{1}         & 14     & 0.448 & $\checkmark$\\
\midrule
\multicolumn{2}{c}{\qqtoHqq}\\
\midrule
\tabHqqj{0}{High}                   & \epoly{1}         & 176    & 0.180 & \\
\tabHqqj{0}{Med}                    & \epoly{2}         & 3238   & 4.73  & \\
\tabHqqj{0}{Low}                    & \epoly{2}         & 133314 & 49.7  & \\
\tabHqqj{1}{High}                   & \epoly{1}         & 19     & 0.125 & $\checkmark$\\
\tabHqqj{1}{Med}                    & \epoly{1}         & 187    & 0.361 & \\
\tabHqqj{1}{Low}                    & \texttt{PowerLaw} & 1040   & 1.97  & \\
\tabHqqm{}{60}{High}                & \epoly{1}         & 17     & 0.499 & $\checkmark$\\
\tabHqqm{}{60}{Med}                 & \epoly{1}         & 157    & 0.489 & \\
\tabHqqm{}{60}{Low}                 & \texttt{PowerLaw} & 1978   & 1.29  & \\
\tabHqqm{60}{120}{High}             & \epoly{1}         & 53     & 0.165 & $\checkmark$\\
\tabHqqm{60}{120}{Med}              & \epoly{1}         & 329    & 0.520 & \\
\tabHqqm{60}{120}{Low}              & \texttt{PowerLaw} & 709    & 1.15  & \\
\tabHqqm{120}{350}{High}            & \epoly{1}         & 214    & 1.08  & \\
\tabHqqm{120}{350}{Med}             & \epoly{2}         & 1671   & 1.07  & \\
\tabHqqm{120}{350}{Low}             & \texttt{PowerLaw} & 11195  & 6.34  & \\
\tabHqqmPt{350}{700}{}{200}{High}  & \epoly{1}         & 25     & 0.162 & $\checkmark$\\
\tabHqqmPt{350}{700}{}{200}{Med}   & \epoly{1}         & 260    & 0.443 & \\
\tabHqqmPt{350}{700}{}{200}{Low}   & \epoly{1}         & 753    & 1.17  & \\
\tabHqqmPt{700}{1000}{}{200}{High} & \epoly{1}         & 25     & 0.670 & $\checkmark$ \\
\tabHqqmPt{700}{1000}{}{200}{Med}  & \epoly{1}         & 166    & 0.713 &  \\
\tabHqqmPt{1000}{}{}{200}{High}    & \epoly{1}         & 48     & 1.47  & $\checkmark$\\
\tabHqqmPt{1000}{}{}{200}{Med}     & \epoly{1}         & 142    & 0.270 & \\
\bottomrule
\end{tabular}
}
}
\resizebox{0.49\textwidth}{!}{
\adjustbox{valign=t}{
\begin{tabular}{lrrrc}
\toprule
Category   & Function & $N_{\text{data}}$ & $N_{\text{spur}}$  & Wald \\
\midrule
\tabHqqmPt{350}{700}{200}{}{High}  & \epoly{1} & 18     & 0.189 & $\checkmark$\\
\tabHqqmPt{350}{700}{200}{}{Med}   & \epoly{1} & 84     & 0.513 & $\checkmark$\\
\tabHqqmPt{350}{700}{200}{}{Low}   & \epoly{1} & 595    & 0.721 & \\
\tabHqqmPt{700}{1000}{200}{}{High} & \epoly{1} & 19     & 0.110 & $\checkmark$\\
\tabHqqmPt{700}{1000}{200}{}{Med}  & \epoly{1} & 411    & 0.193 & \\
\tabHqqmPt{1000}{}{200}{}{High}    & \epoly{1} & 23     & 1.30  & $\checkmark$ \\
\tabHqqmPt{1000}{}{200}{}{Med}     & \epoly{1} & 56     & 0.329 & $\checkmark$\\
\midrule
\multicolumn{2}{c}{\qqtoHln}\\
\midrule
\tabHlnPt{}{75}{High}              & \epoly{1} & 40     & 0.277 & \\
\tabHlnPt{}{75}{Med}               & \epoly{1} & 158    & 0.609 & \\
\tabHlnPt{75}{150}{High}           & \epoly{1} & 15     & 0.069 & \\
\tabHlnPt{75}{150}{Med}            & \epoly{1} & 104    & 0.255 & \\
\tabHlnPt{150}{250}{High}          & \epoly{1} & 17     & 0.128 & $\checkmark$\\
\tabHlnPt{150}{250}{Med}           & \epoly{1} & 21     & 0.150 & \\
\tabHlnPt{250}{}{High}             & \epoly{1} & 16     & 0.237 & $\checkmark$\\
\tabHlnPt{250}{}{Med}              & \epoly{1} & 27     & 0.054 & $\checkmark$\\
\midrule
\multicolumn{2}{c}{\pptoHll}\\
\midrule
\tabHllPt{}{75}{High}              & \epoly{1}         & 12     & 0.027 & \\
\tabHllPt{}{75}{Med}               & \texttt{PowerLaw} & 1620   & 2.28  & \\
\tabHllPt{75}{150}{High}           & \epoly{1}         & 13     & 0.015 & \\
\tabHllPt{75}{150}{Med}            & \epoly{1}         & 18     & 0.016 & \\
\tabHllPt{150}{250}{High}          & \epoly{1}         & 14     & 0.059 & $\checkmark$\\
\tabHllPt{150}{250}{Med}           & \epoly{1}         & 136    & 0.194 & \\
\tabHllPt{250}{}{}                 & \epoly{1}         & 14     & 0.311 & $\checkmark$\\
\midrule
\multicolumn{2}{c}{\pptoHnn}\\
\midrule
\tabHnnPt{}{75}{High}              & \epoly{1} & 1174   & 12.3  & $\checkmark$\\
\tabHnnPt{}{75}{Med}               & \epoly{1} & 6897   & 4.13  & \\
\tabHnnPt{}{75}{Low}               & \epoly{3} & 18084  & 9.95  & \\
\tabHnnPt{75}{150}{High}           & \epoly{1} & 16     & 0.407 & $\checkmark$\\
\tabHnnPt{75}{150}{Med}            & \epoly{1} & 124    & 1.30  & $\checkmark$\\
\tabHnnPt{75}{150}{Low}            & \epoly{1} & 2019   & 1.96  & \\
\tabHnnPt{150}{250}{High}          & \epoly{1} & 16     & 0.121 & $\checkmark$\\
\tabHnnPt{150}{250}{Med}           & \epoly{1} & 17     & 0.184 & $\checkmark$\\
\tabHnnPt{150}{250}{Low}           & \epoly{1} & 87     & 0.644 & $\checkmark$\\
\tabHnnPt{250}{}{High}             & \epoly{1} & 15     & 0.237 & $\checkmark$\\
\tabHnnPt{250}{}{Med}              & \epoly{1} & 18     & 0.201 & $\checkmark$\\
\midrule
\multicolumn{2}{c}{\ttH}\\
\midrule
\tabttHPt{}{60}{High}              & \epoly{1} & 35     & 0.040  & \\
\tabttHPt{}{60}{Med}               & \epoly{1} & 96     & 0.192  & \\
\tabttHPt{60}{120}{High}           & \epoly{1} & 34     & 0.038 & \\
\tabttHPt{60}{120}{Med}            & \epoly{1} & 74     & 0.274  & \\
\tabttHPt{120}{200}{High}          & \epoly{1} & 39     & 0.018  & \\
\tabttHPt{120}{200}{Med}           & \epoly{1} & 37     & 0.057  & \\
\tabttHPt{200}{300}{}              & \epoly{1} & 23     & 0.261  & \\
\tabttHPt{300}{}{}                 & \epoly{1} & 19     & 0.180  & $\checkmark$\\
\midrule
\multicolumn{2}{c}{\tH}\\
\midrule
\tHqb, High-purity                 & \epoly{1} & 17     & 0.371 & $\checkmark$\\
\tHqb, Med-purity                  & \epoly{1} & 19     & 0.320 & $\checkmark$\\
\tHqb, BSM   ($\kappa_t = -1$)     & \epoly{1} & 14     & 0.496 & $\checkmark$\\
\tHW                               & \epoly{1} & 38     & 0.070 & \\
\midrule
Low-purity top                & \epoly{1} & 500    & 0.870 & \\
\bottomrule
\end{tabular}}}
\end{table}



\section{Systematic uncertainties}
\label{sec:systs}
 
Systematic uncertainties considered in this analysis can be grouped into two categories: uncertainties in the modelling of the \mgg\ distribution for the signal and background processes, and uncertainties in the expected signal yields in each category arising from either experimental or theory sources. These systematic uncertainties are incorporated into the likelihood model of the measurement as nuisance parameters, as explained in Section~\ref{sec:modeling}. More details about the uncertainties are provided in this section.
 
\subsection{Experimental systematic uncertainties}
 
Experimental systematic uncertainties relevant to the modelling of the signal \mgg\ distribution include the uncertainties in the energy scale and energy resolution of photon candidates, as well as in the Higgs boson mass. The photon energy scale uncertainties are propagated to the peak position of the signal DSCB shape, with an impact that is usually less than $0.3\%$ relative to the peak position value, depending on the category. The photon energy resolution uncertainties are propagated to the Gaussian width of the signal DSCB shape, with a relative impact between 1\% and 15\%, depending on the category. The estimation and implementation of the photon energy scale and resolution uncertainties follow the procedure outlined in Ref.~\cite{EGAM-2018-01}.
The total uncertainty in the measured Higgs boson mass, $0.24\,\GeV$, is considered as an additional uncertainty of the peak position of the signal DSCB shape.
 
The modelling of the background \mgg\ distribution with an analytic function can produce a potential bias in the fitted signal yield. An uncertainty in the modelling of the background, computed using the spurious-signal method described in Section~\ref{sec:modeling:bkg}, is included as an additive contribution to the signal yield in each category as shown in Eq.~(\ref{eq:yield}). This uncertainty is considered to be uncorrelated between different categories. Out of the 101 analysis categories, 46 categories have a background modelling uncertainty that is no more than 10\% of the background statistical uncertainty, and only two categories (\HqqmPt{1000}{}{200}{}{High} and \HnnPt{}{75}{High}) have a background modelling uncertainty that is at least 50\% of the background statistical uncertainty.

Experimental uncertainties affecting the expected signal yields include: the efficiency of the diphoton trigger~\cite{TRIG-2018-05}, the photon identification efficiencies~\cite{EGAM-2018-01},  the photon isolation efficiencies, the impact of the photon energy scale and resolution uncertainties on the selection efficiency~\cite{EGAM-2018-01}, the modelling of pile-up in the simulation, which is evaluated by varying by $\pm 9\%$ the value of the visible inelastic cross-section used to reweight the pile-up distribution in the simulation to that in the data~\cite{STDM-2015-05}, the jet energy scale and resolution~\cite{JETM-2018-05}, the efficiency of the jet vertex tagger, the efficiency of the $b$-tagging algorithm~\cite{FTAG-2018-01}, the electron~\cite{EGAM-2018-01} and muon~\cite{MUON-2018-03} reconstruction, identification and isolation efficiencies,
the electron~\cite{EGAM-2018-01} and muon~\cite{MUON-2018-03} energy and momentum scale and resolution, as well as the contribution to $\ET^\mathrm{miss}$ from charged-particle tracks that are not associated with high-$\pT$ electrons, muons, jets, or photons~\cite{PERF-2016-07}. The uncertainty in the combined 2015--2018 integrated luminosity is 1.7\%~\cite{ATLAS-CONF-2019-021}, obtained using the LUCID-2 detector~\cite{LUCID2} for the primary luminosity measurements.

\subsection{Theory modelling uncertainties} \label{sec:theo_stxs}

The main theory uncertainties arise from missing higher-order terms in the perturbative QCD calculations, the modelling of parton showers, the PDFs and the value of \alphas. For measurements of the \ttH\ and \tH\ processes, the modelling of heavy-flavour jets in the \ggF, \VBF, and \VH\ processes is also important.
 
QCD uncertainties for each production mode are estimated by varying the renormalization and factorization scales used in the event generation, and the resulting variations in the predicted event yield in each STXS regions are considered as uncertainties.
 
For the \ggtoH\ processes, the QCD uncertainty model is implemented using four components~\cite{deFlorian:2016spz,Liu:2013hba, Stewart:2013faa, Boughezal:2013oha} accounting for modelling uncertainties in the jet multiplicity, three describing uncertainties in the modelling the \ptH\ distribution, one~\cite{Stewart:2011cf,Gangal:2013nxa} accounting for the uncertainty in the distribution of the \ptHjj\ variable, four accounting for the uncertainty in the distribution of the \mjj\ variable, and six covering modelling uncertainties in STXS regions with high \ptH\ ($> 300\,\GeV$). Following the principles of the Stewart--Tackmann procedure~\cite{Stewart:2011cf}, two components account for uncertainties in the inclusive \ggtoH\ event yields, while the others describe \emph{migration} uncertainties in the fraction of events passing the selections defining the STXS regions. The model provides a full description of the uncertainty in each STXS region, in which the uncertainty components are each assigned to one nuisance parameter that is treated as statistically independent from the others. These uncertainties are typically less than 22\% of the expected signal yield in analysis categories targeting the \ggtoH\ process.
 
For each of the \WH, \qqqgZH, and \ggtoZH\ processes, the QCD uncertainty model includes four independent components to account for uncertainties in the distribution of $p_\mathrm{T}^{V}$, and two independent components for uncertainties in the jet multiplicity distribution. For \qqtoHqq\ (\VBF\ and \VhadH) processes, the QCD uncertainty model includes a similar set of independent components: two for the modelling of the jet multiplicity and the \ptHjj\ distribution, one for migration between the $\ptH <200$\,\GeV\ and $\ptH \ge 200$\,\GeV\ regions, and six for the modelling of the \mjj\ distribution. These uncertainties are less than 10\% of the expected signal yield in analysis categories targeting these processes, with the exception of the \ggtoZH\ process where the uncertainty can be as large as 26\%.  
 
For the \ttH\ process, QCD uncertainties include five components covering migrations between \ttH\ STXS regions with different \ptH. These uncertainties are less than 10\% of the expected signal yield in the \ttH\ STXS regions in their targeted analysis categories.
In the case of \tHW, \tHqb\ and \bbH, an overall QCD uncertainty is used, taking the value from Ref.~\cite{deFlorian:2016spz}.

To check the robustness of the uncertainty model, a comparison between the efficiency factors of the nominal Higgs signal sample and the alternative sample generated by \MGNLO, as described in Section~\ref{sec:mc}, is performed for the \VBF, \VH, and \ttH\ processes. The differences between the efficiency factors of the nominal and alternative samples are significantly larger than the uncertainties from QCD scale variations and can reach values of up to $20\%$ in some phase-space regions of the \VBF\ process. The differences between the efficiency factors of the nominal and alternative samples are therefore considered as an additional systematic uncertainty. A similar comparison was not performed for the \ggtoH\ process since the corresponding alternative samples are already used in the derivation of the QCD uncertainty model described above.
 
The modelling of the parton shower, underlying event, and hadronization is assessed  separately for each Higgs boson production mode by comparing the efficiency factors of simulated signal samples showered by \PYTHIA[8] with those of samples showered by \HERWIG[7]. The uncertainties estimated from the differences between these factors typically do not exceed 20\%, and increase with jet multiplicity.
 
Uncertainties on the PDFs and the value of \alphas\ are taken from Ref.~\cite{deFlorian:2016spz} for the \tHW, \tHqb\ and \bbH\ processes. For other modes, the uncertainties are estimated using the PDF4LHC15 recommendations~\cite{Butterworth:2015oua}. Their effects are usually small compared to the those of the two other main sources of uncertainty mentioned at the start of this subsection.

In categories targeting the \ttH\ and \tH\ processes, the predicted \ggF, \VBF\ and \VH\ yields are each assigned a conservative 100\% uncertainty (correlated between categories), which is due to the theoretical uncertainty associated with the radiation of additional heavy-flavour jets in these Higgs boson production modes.  This is supported by measurements using $H {\rightarrow} ZZ^* {\rightarrow} 4\ell$~\cite{HIGG-2018-29}, $t\bar{t}b\bar{b}$~\cite{TOPQ-2014-10}, and $Vb$~\cite{STDM-2017-38,STDM-2012-11} events. The impact of this uncertainty on the results is generally negligible compared to the statistical uncertainties, since the contributions from non-\ttH\ processes are generally low.
 
A total uncertainty of $2.9\%$ is assigned to the \Hyy\ decay branching ratio, based on calculations from the HDECAY~\cite{Djouadi:1997yw,Spira:1997dg,Djouadi:2006bz} and \PROPHECY~\cite{Bredenstein:2006ha,Bredenstein:2006rh,Bredenstein:2006nk} programs, which also includes the uncertainty arising from its dependence on quark masses and \alphas.
 
Theory uncertainties, such as missing higher-order QCD corrections and PDF-induced uncertainties, affect both the expected signal yields from each production process and the signal efficiency factors ($\epsilon_{it}$ in Eq.~(\ref{eq:yield})) in each category. Uncertainties in signal efficiency factors are included in all the measurements presented in this paper. Signal yield uncertainties, including the uncertainty in the \Hyy\ branching ratio, are included only for the measurement of the Higgs boson signal strength and interpretations within the $\kappa$-framework and SMEFT models, which rely on comparisons between the observed event yields and their SM predictions.
Uncertainties on the parton shower, underlying event, and hadronization effects are included in all the measurements, without a separation into yield and acceptance components.
In addition, cross-section measurements spanning multiple STXS regions require assumptions about the expected event yields in each region, as explained in Section~\ref{sec:results:STXS}, which introduces a weak dependence on the signal yield uncertainties.
 
Table~\ref{tab:results:systematics} shows the expected experimental and theoretical systematic uncertainties of the cross-section measurements in the SM hypothesis, computed as described in Section~\ref{sec:stat}.

\begin{table}[!htp]
\begin{center}
\caption{Expected contributions from the main sources of systematic uncertainty to the total uncertainty in the measurement of the cross-section times \Hyy\ branching ratio for each of of the main Higgs boson production processes. The uncertainty from each source ($\Delta \sigma$) is shown as a fraction of the total expected cross-section ($\sigma$).}
\label{tab:results:systematics}
\resizebox{0.995\textwidth}{!}{
\begin{tabular}{lllllllllllll}
\toprule
& \multicolumn{1}{c}{\ggF\ + $b\bar{b}$H} & \multicolumn{1}{c}{\VBF\ } & \multicolumn{1}{c}{ \WH } & \multicolumn{1}{c}{ \ZH } & \multicolumn{1}{c}{\ttH} & \multicolumn{1}{c}{\tH}\\
\midrule
\multirow{1}{*}{Uncertainty source}& $\Delta \sigma$[\%] & $\Delta \sigma$[\%] & $\Delta \sigma$[\%] & $\Delta \sigma$[\%] & $\Delta \sigma$[\%] & $\Delta \sigma$[\%] \\

\midrule
\multicolumn{2}{l}{Theory uncertainties} \\
\midrule
\hspace{3mm}Higher-order QCD terms				                & ~~~\,$\pm 1.4$	& ~~~\,$\pm 4.1$	& ~~~\,$\pm 4.1$	& ~~~\,$\pm 12$	& ~~~\,$\pm 2.8$	& ~~~\,$\pm 16$	\\
\hspace{3mm}Underlying event and parton shower				    & ~~~\,$\pm 2.5$	& ~~~\,$\pm 16$	& ~~~\,$\pm 2.5$	& ~~~\,$\pm 4.0$	& ~~~\,$\pm 3.6$	& ~~~\,$\pm 48$	\\
\hspace{3mm}PDF and $\alphas$                                   &$< \pm 1$	& ~~~\,$\pm 2.0$	& ~~~\,$\pm 1.4$	& ~~~\,$\pm 2.3$	&$< \pm 1$	& ~~~\,$\pm 5.8$	\\
\hspace{3mm}Matrix element				                                &$< \pm 1$	& ~~~\,$\pm 3.2$	&$< \pm 1$	& ~~~\,$\pm 1.2$	& ~~~\,$\pm 2.5$	& ~~~\,$\pm 8.2$	\\
\hspace{3mm}Heavy-flavour jet modelling in non-$t\bar{t}H$ processes	&$< \pm 1$	&$< \pm 1$	&$< \pm 1$	&$< \pm 1$	&$< \pm 1$	& ~~~\,$\pm 13$	\\
\midrule
\multicolumn{2}{l}{Experimental uncertainties} \\
\midrule
\hspace{3mm}Photon energy resolution				                & ~~~\,$\pm 3.0$	& ~~~\,$\pm 3.0$	& ~~~\,$\pm 3.8$	& ~~~\,$\pm 4.8$	& ~~~\,$\pm 3.0$	& ~~~\,$\pm 12$	\\
\hspace{3mm}Photon efficiency				                            & ~~~\,$\pm 2.7$	& ~~~\,$\pm 2.7$	& ~~~\,$\pm 3.3$	& ~~~\,$\pm 3.6$	& ~~~\,$\pm 2.9$	& ~~~\,$\pm 9.3$	\\
\hspace{3mm}Luminosity				                                    & ~~~\,$\pm 1.8$	& ~~~\,$\pm 2.0$	& ~~~\,$\pm 2.4$	& ~~~\,$\pm 2.7$	& ~~~\,$\pm 2.2$	& ~~~\,$\pm 6.6$ \\
\hspace{3mm}Pile-up				                                        & ~~~\,$\pm 1.4$	& ~~~\,$\pm 2.2$	& ~~~\,$\pm 2.0$	& ~~~\,$\pm 2.3$	& ~~~\,$\pm 1.4$	& ~~~\,$\pm 7.3$	\\
\hspace{3mm}Background modelling				                            & ~~~\,$\pm 2.0$	& ~~~\,$\pm 4.6$	& ~~~\,$\pm 3.6$	& ~~~\,$\pm 7.2$	& ~~~\,$\pm 2.5$	& ~~~\,$\pm 63$	\\
\hspace{3mm}Photon energy scale				                    &$< \pm 1$	&$< \pm 1$	&$< \pm 1$	& ~~~\,$\pm 1.3$	&$< \pm 1$	& ~~~\,$\pm 5.6$	\\
\hspace{3mm}Jet/\met				                                    &$< \pm 1$	& ~~~\,$\pm 6.8$	&$< \pm 1$	& ~~~\,$\pm 2.2$	& ~~~\,$\pm 3.5$	& ~~~\,$\pm 22$ \\
 
\hspace{3mm}Flavour tagging				                                &$< \pm 1$	&$< \pm 1$	&$< \pm 1$	&$< \pm 1$	& ~~~\,$\pm 1.5$	& ~~~\,$\pm 3.4$	\\
\hspace{3mm}Leptons				                                        &$< \pm 1$	&$< \pm 1$	&$< \pm 1$	&$< \pm 1$	&$< \pm 1$	& ~~~\,$\pm 1.8$	\\
\hspace{3mm}Higgs boson mass				                            &$< \pm 1$	&$< \pm 1$	&$< \pm 1$	&$< \pm 1$	&$< \pm 1$	&$< \pm 1$	\\

\hline
\hline
\end{tabular}
}
\end{center}
\end{table}




\section{Results}
\label{sec:results}
 
Results are presented in terms of several descriptions of Higgs boson production: the overall signal strength of Higgs boson production measured in the diphoton decay channel (Section~\ref{sec:results:mu}), separate cross-sections for the main Higgs boson production modes (Section~\ref{sec:results:prodXS}), and cross-sections in a set of merged STXS regions defined for each production process (Section~\ref{sec:results:STXS}).

\subsection{Statistical procedure}
\label{sec:stat}
 
The results for each measurement reported in this paper are obtained by expressing the signal event yields in each analysis category in terms of the measurement parameters, and fitting the model to the data in all categories simultaneously. Both positive and negative values are allowed for all parameters, unless otherwise indicated. Best-fit values are reported along with uncertainties corresponding to $68\%$ confidence level (CL) intervals obtained from a profile likelihood technique~\cite{Cowan:2010st}. The endpoints of the intervals are defined by the condition $-2 \ln \Lambda(\mu) = 1$, where $\Lambda(\mu)$ is the ratio of the profile likelihood at a value $\mu$ of the parameters of interest to the profile likelihood at the best-fit point. Similarly, $95\%$ CL intervals are defined by the condition $-2 \ln \Lambda(\mu) = 3.84$. In some cases, uncertainties are presented as a decomposition into separate components: the \emph{statistical} component is obtained from a fit in which the nuisance parameters associated with systematic uncertainties are fixed to their best-fit values; the \emph{systematic} component, corresponding to the combined effect of systematic uncertainties, is computed as the square root of the difference between the squares of the total uncertainty and the statistical uncertainty. Uncertainty components corresponding to smaller groups of nuisance parameters are obtained by iteratively fixing the parameters in each group, subtracting the square of the uncertainty obtained in this configuration from that obtained when the parameters are profiled, and taking the square root.
 
Expected results for the SM are obtained from a fit to an Asimov data set~\cite{ATL-PHYS-PUB-2011-11,Cowan:2010st} built from the likelihood model with signal and background components. The nuisance parameters of the likelihood model are determined in a fit to the observed data where the STXS parameters defining the signal normalization in each category are left free. The STXS parameters are set to their SM expectations to generate the Asimov data set.
Compatibility with the Standard Model is assessed from the value of the profile likelihood ratio of the model in data under the SM hypothesis; a $p$-value for compatibility with the SM is computed assuming that the profile likelihood follows a $\chi^2$ distribution with a number of degrees of freedom equal to the number of parameters of interest~\cite{Cowan:2010st}. In the case of cross-section measurements, uncertainties in the SM predictions are not accounted for in the $p$-value computation.
 
\subsection{Overall Higgs boson signal strength}
\label{sec:results:mu}
 
To quantify the overall size of the Higgs boson signal in the diphoton channel, the inclusive signal strength, $\mu$, defined as the ratio of the observed value of the product of the Higgs boson production cross-section and the \Hyy\ branching ratio $(\sigma \times \Byy)$ in $|y_H|<2.5$ to that of its SM prediction, is measured by simultaneously fitting the \mgg\ distributions of the 101 analysis categories. The signal strength $\mu$ is treated in the likelihood function as a single parameter of interest which scales the expected yields in all STXS regions and is found to be
 
\begin{equation*}
\mu = 1.04^{+0.10}_{-0.09} =
\HyynumRF{1.044}{3}
\pm 0.06 \text{ (stat.)}
\Hyynumpmerr{+0.06}{-0.05}{1} \text{ (theory syst.) }
\Hyynumpmerr{+0.05}{-0.04}{1} \text{ (exp. syst.). }
\end{equation*}
 
The overall \mgg\ distribution of the selected diphoton sample is shown in Figure~\ref{fig:result:inclusivemgg}. The events are weighted by the value of $\ln(1+S/B)$ of their category, where $S$ and $B$ are the expected signal and background yields within the smallest \mgg\ window containing 90\% of the signal events, shown in Table~\ref{tab:design:yields}. This choice of event weight is designed to enhance the contribution of events from categories with higher signal-to-background ratio in a way that approximately matches the impact of these events in the categorized analysis of the data.
\begin{figure}
\centering
\includegraphics[width=0.8\textwidth]{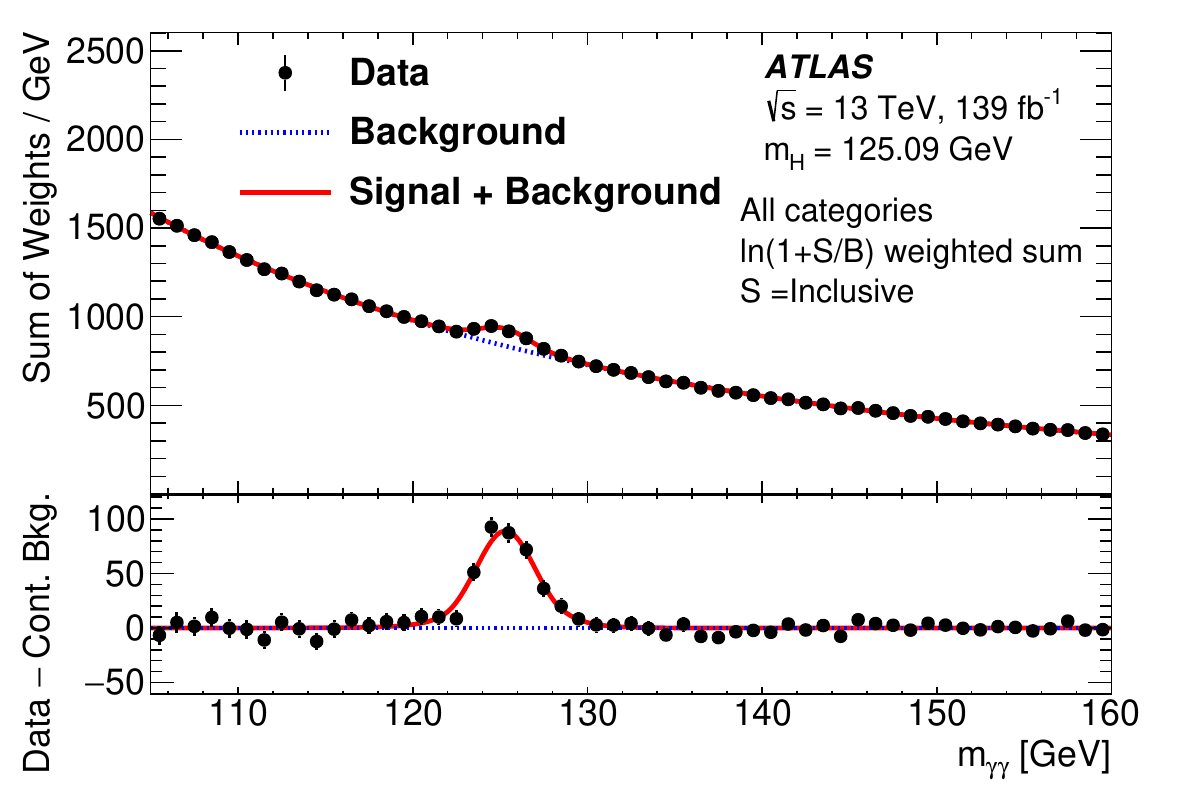}
\caption{The inclusive diphoton invariant mass distribution of events from all analysis categories. The data events (dots) in each category are weighted by the value of $\ln(1+S/B)$, where $S$ and $B$ are the expected signal and background yields in this category within the smallest \mgg\ window containing 90\% of the signal events. The expected signal is considered inclusively over all STXS regions. The fitted signal-plus-background pdfs from all categories are also weighted and summed, shown as the solid line. The blue dotted line represents the weighted sum of the fitted background functions from all categories. The error bars on the data points are computed following Ref.~\cite{bohm2010introduction}.}
\label{fig:result:inclusivemgg}
\end{figure}
 
Table~\ref{tab:inclusive_mu_syst} further breaks down the impact of systematic uncertainties on the signal-strength measurement. The leading sources of experimental systematic uncertainty are the photon energy resolution uncertainty (2.8\%) and photon efficiency uncertainty (2.6\%), while the leading sources of theoretical uncertainty are the QCD scale uncertainty (3.8\%) and \Hyy\ branching ratio uncertainty (3.0\%).
 
\begin{table}[ht]
\caption{Summary of the leading sources of systematic uncertainty in the measurement of the Higgs boson signal strength.}
\label{tab:inclusive_mu_syst}
\centering
\begin{tabular}{ll}
\toprule
\multirow{1}{*}{Uncertainty source} & $\Delta \mu$ [\%] \\
\midrule
\multicolumn{2}{l}{Theory uncertainties} \\
\midrule
Higher-Order QCD Terms                                              & $\pm 3.8$ \\
Branching Ratio                                                     & $\pm 3.0$ \\
Underlying Event and Parton Shower                                  & $\pm 2.5$ \\
PDF and $\alpha_s$                                                  & $\pm 2.1$ \\
Matrix Element                                                      & $\pm 1.0$ \\
Modeling of Heavy Flavor Jets in non-$t\bar{t}H$ Processes          & $< \pm 1$ \\
\midrule
\multicolumn{2}{l}{Experimental uncertainties} \\
\midrule
Photon energy resolution                                      & ~~~\,$\pm 2.8$ \\
Photon efficiency                                                   & ~~~\,$\pm 2.6$ \\
Luminosity                                                          & ~~~\,$\pm 1.8$ \\
Pile-up                                                              & ~~~\,$\pm 1.5$ \\
Background modelling                                                 & ~~~\,$\pm 1.3$ \\
Photon energy scale                                           & $< \pm 1$ \\
Jet/\met            	                                			& $< \pm 1$	\\
Flavour tagging			                                        	& $< \pm 1$	\\
Leptons		                                                 		& $< \pm 1$	\\
Higgs boson mass                                     				& $< \pm 1$	\\
\bottomrule
\end{tabular}
\end{table}

\subsection{Production cross-sections}
\label{sec:results:prodXS}

The mechanism of Higgs boson production is probed by considering the \ggF, \VBF, \WH, \ZH, \ttH, and \tH\  production processes separately. The measurement is reported in terms of the $(\sigma \times \Byy)$ value in each case, with the cross-sections defined in $|y_H|<2.5$. As in the STXS region definition, the contribution from the \bbH\ process is included in the \ggF\ component.
Figure~\ref{fig:results:prodXS_spectra} shows the  \mgg\ distributions for analysis categories targeting different production modes separately.
The same weighting procedure as in Figure~\ref{fig:result:inclusivemgg} is used, except that the signal yield only includes the contributions from the targeted production process, while other signal production processes are included in the background yield.
 
\begin{figure}[tpb!]
\centering
\subfloat[\ggF~+~\bbH]{\includegraphics[width=.475\textwidth]{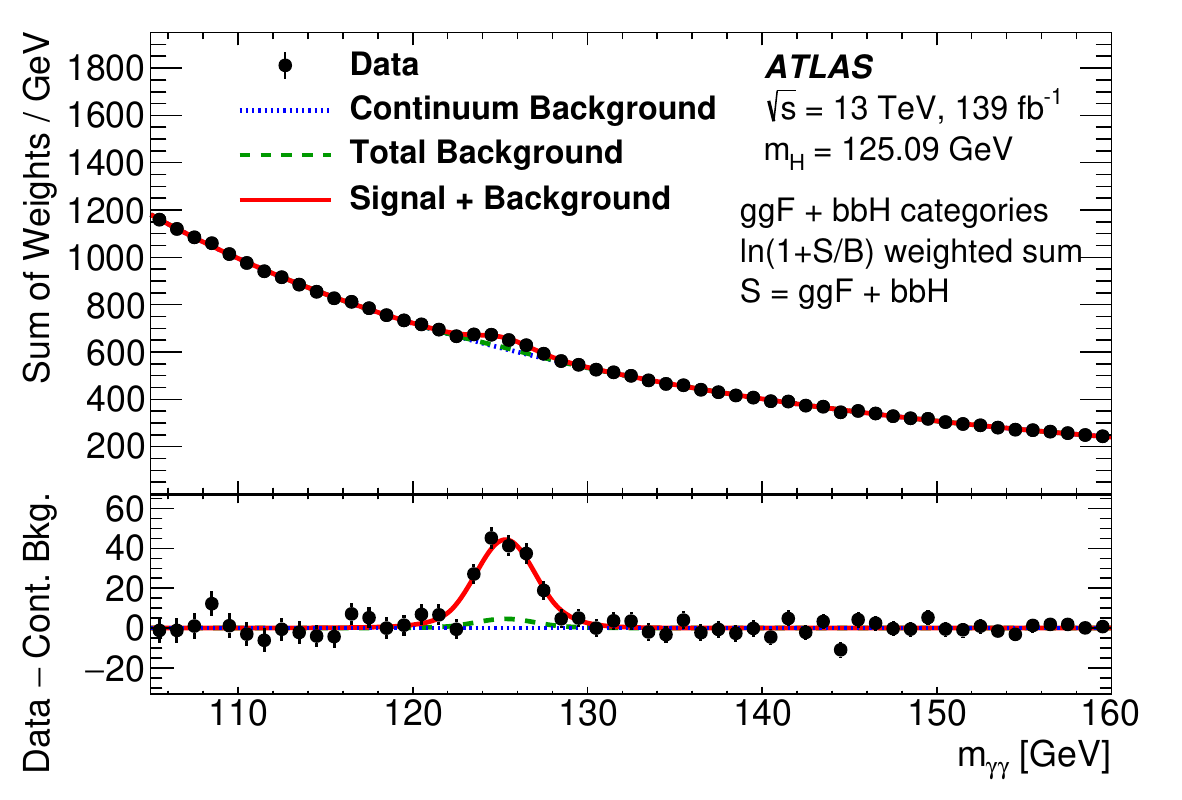}}
\subfloat[\VBF]{\includegraphics[width=.475\textwidth]{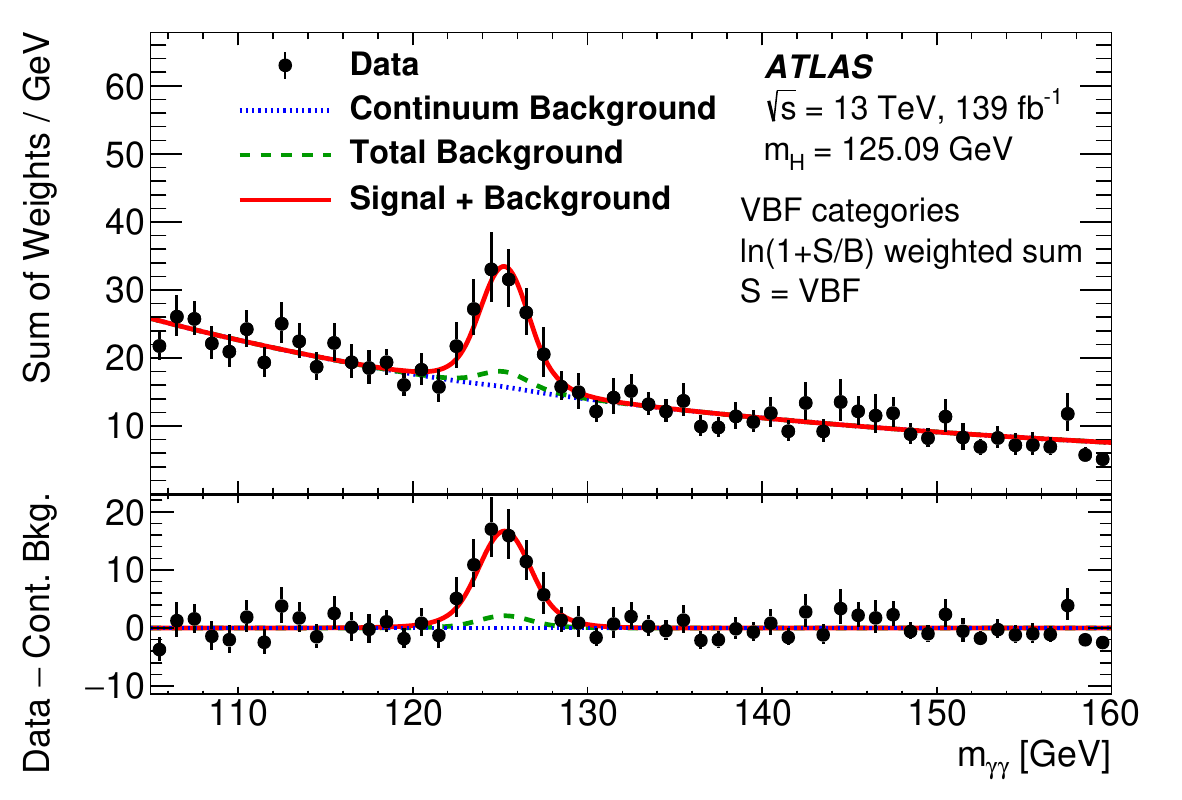} }\\
\subfloat[\WH]{\includegraphics[width=.475\textwidth]{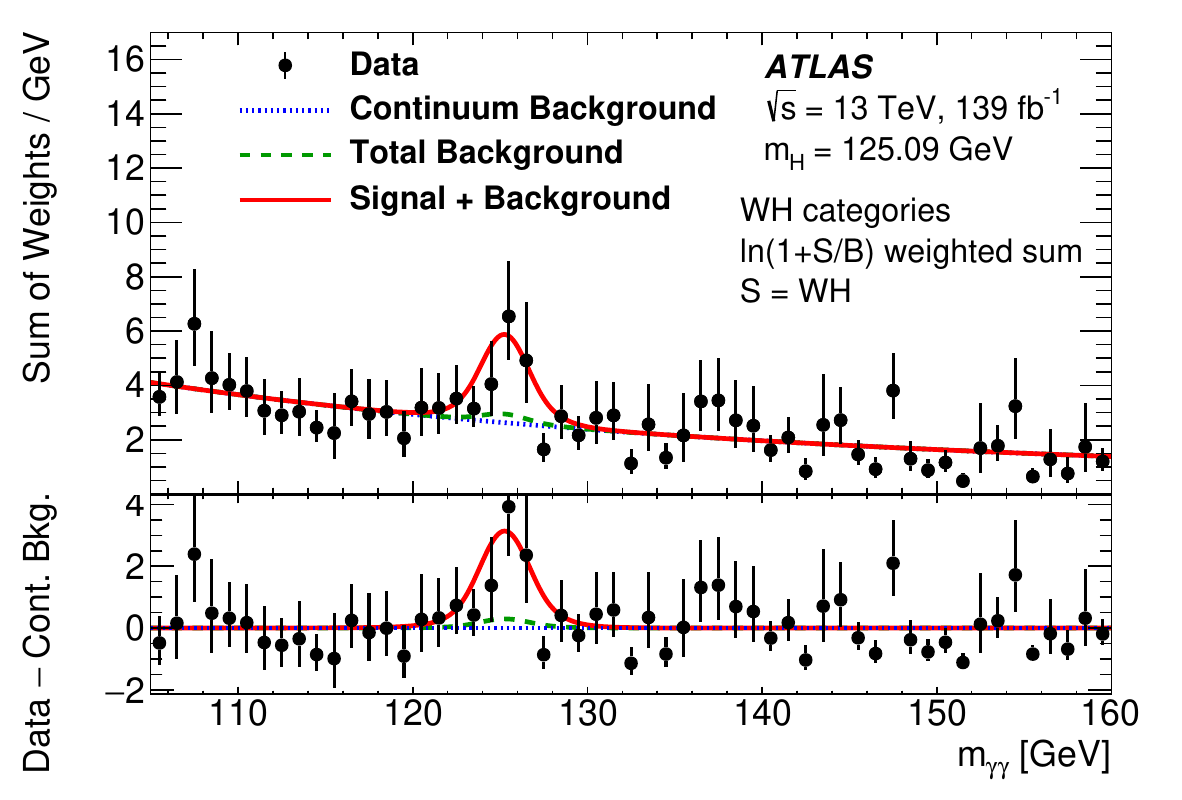}}
\subfloat[\ZH]{\includegraphics[width=.475\textwidth]{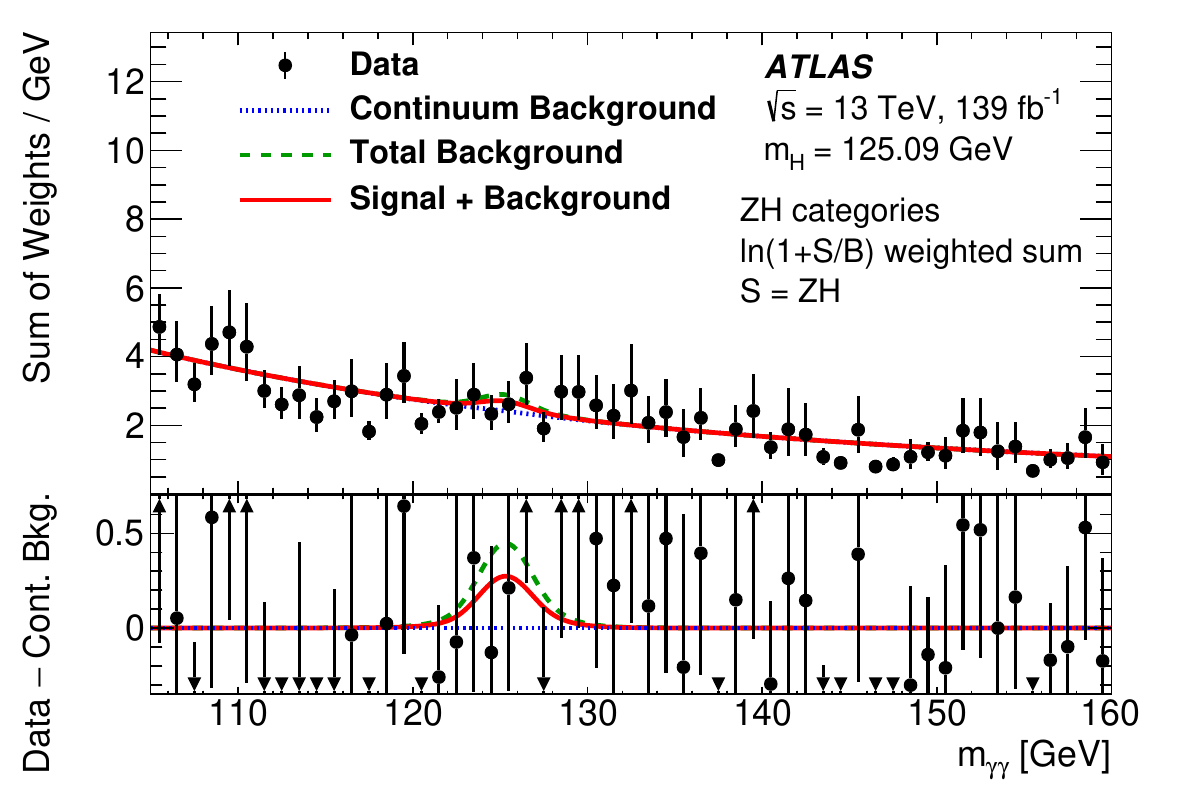}}\\
\subfloat[\ttH]{\includegraphics[width=.475\textwidth]{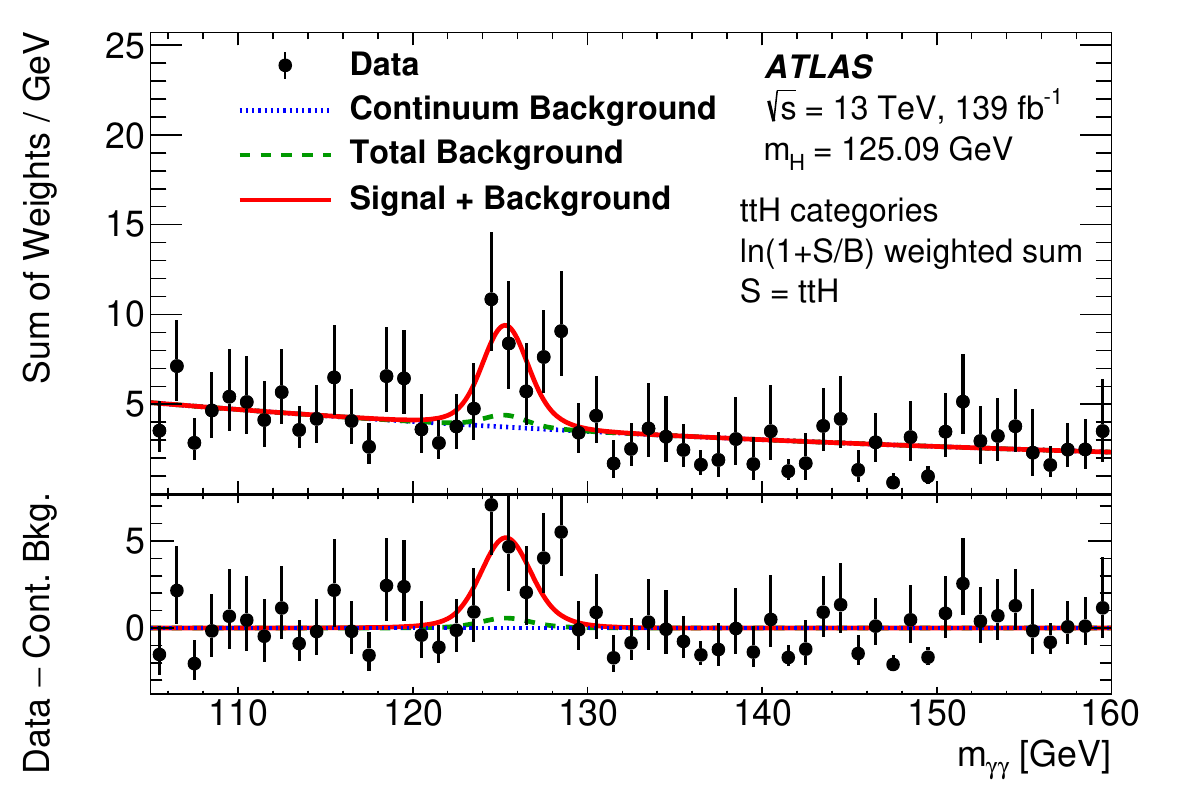}}
\subfloat[\tH]{\includegraphics[width=.475\textwidth]{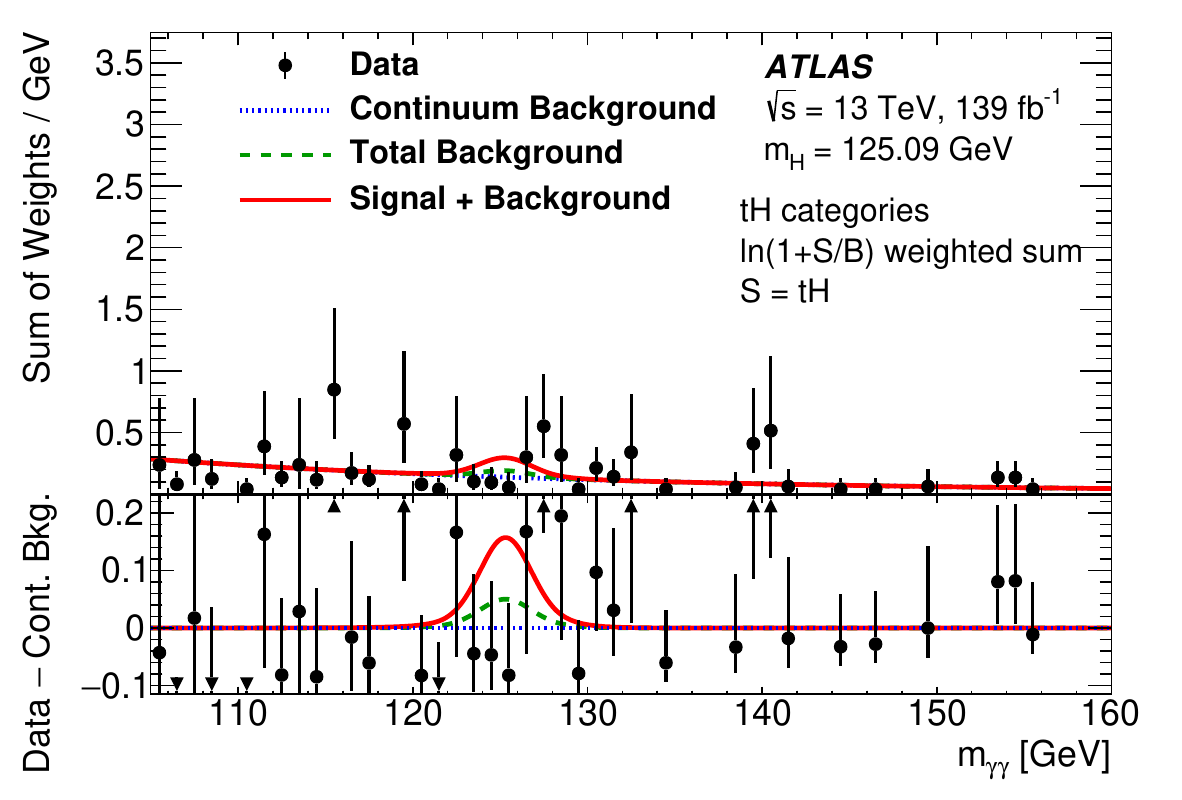}}\\
\caption{Combined diphoton invariant mass distributions for categories targeting the same production processes.
The data (black dots) are weighted by $\mathrm{ln}(1+S/B)$ where $S$ and $B$ are respectively the expected signal and background yields in the smallest \mgg\ window containing 90\% of the signal events. In this calculation, only Higgs boson events from the targeted production processes are considered as signal events. Higgs boson events from other processes as well as the continuum background events are considered as background.
The fitted signal-plus-background pdfs from the relevant categories are summed, and represented by a solid line. The blue dotted line represents the weighted sum of the fitted continuum background pdfs, while the dashed line combines the contributions of continuum background and other Higgs boson events. The error bars on the data points are computed following Ref.~\cite{bohm2010introduction}. The weighted combination of categories with low event counts leads in some cases to data errors that are highly asymmetric and change by large amounts from point to point.}
\label{fig:results:prodXS_spectra}
\end{figure}

The best-fit values of the production cross-sections and their uncertainties are summarized in Figure~\ref{fig:results:prodXS} and Table~\ref{tab:results:prodXS}.
A negative best-fit value is observed for the cross-section of the \ZH\ process, which corresponds to a total observed event yield that is below the background expectation. The $p$-value for compatibility of the cross-section measurement and the SM prediction is $55\%$.
The correlations between these measurements are shown in Figure~\ref{fig:results:prodXS_corr}. Compared to Ref.~\cite{HIGG-2016-21}, correlations between measurements are reduced, and in particular, the anti-correlation between the \ggF\ and \VBF\ measurements is now $-13\%$, corresponding to a $30\%$ reduction. This is driven by a reduction in the \ggF\ contamination in categories targeting  the \VBF\ process, mainly resulting from the use of the D-optimality criterion in the categorization.
An anti-correlation of $-37\%$ is observed between the \WH\ and \ZH\ measurements, mainly due to contamination by \qqtoHln\ events in the categories targeting the \pptoHnn\ process. This correlation is mitigated by the separation of the \pptoHll\ and \pptoHnn\ processes that is introduced in the analysis categorization. Similarly, \ttH\ contamination in the categories targeting \tH\ lead to an anti-correlation of $-44\%$ between these two processes.
 
\begin{figure}[!ht]
\centering
\includegraphics[width=0.9\textwidth]{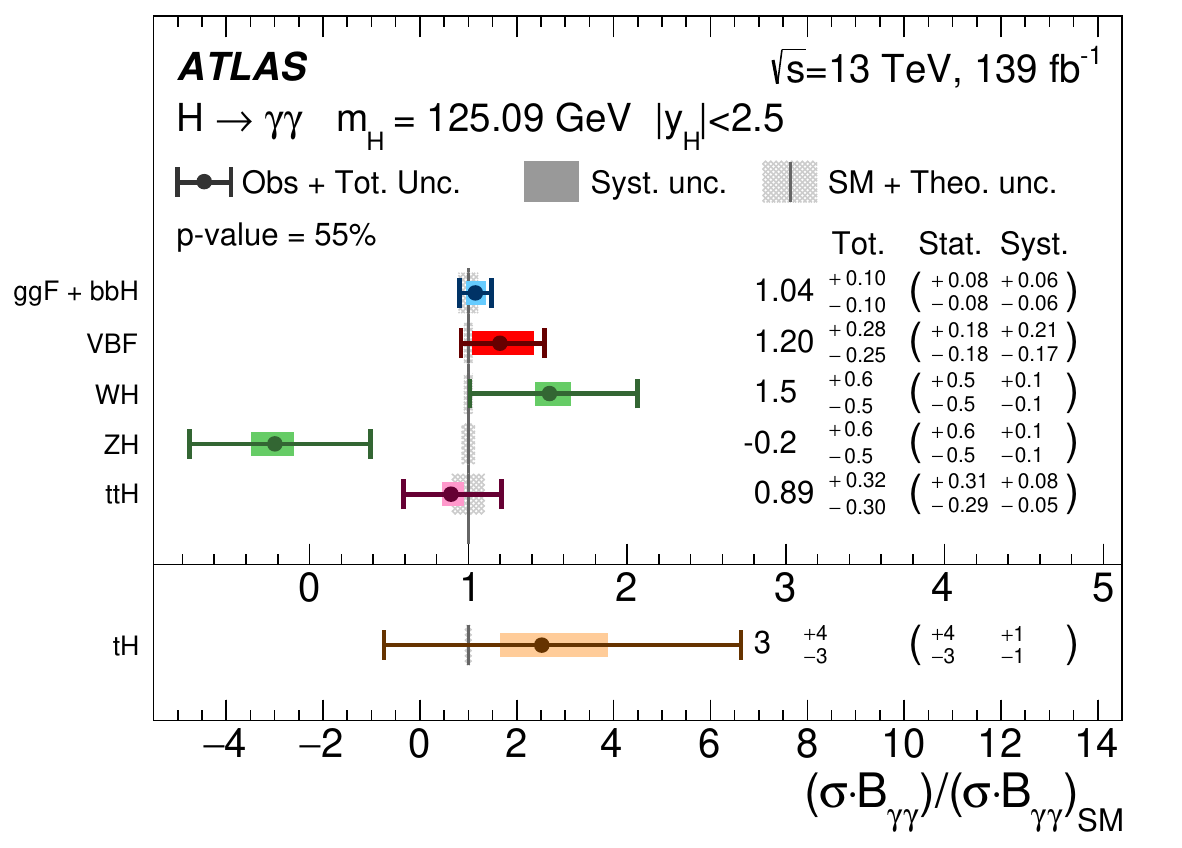}
\caption{Cross-sections times \Hyy\ branching ratio for \ggF~+~\bbH, \VBF, \VH, \ttH, and \tH\ production, normalized to their SM predictions. The values are obtained from a simultaneous fit to all categories. The error bars and shaded areas show respectively the total and systematic uncertainties in the measurements. The grey bands show the theory uncertainties in the predictions, including uncertainties due to missing higher-order terms in the perturbative QCD calculations, the PDFs and the value of \alphas, as well as the \Hyy\ branching ratio uncertainty.}
\label{fig:results:prodXS}
\end{figure}

\begin{table}
\caption{Best-fit values and uncertainties for the production cross-sections of the Higgs boson times the \Hyy\ branching ratio. The total uncertainties are decomposed into statistical (Stat.) and systematic (Syst.) uncertainties. SM predictions are shown for the cross-section of each production process. These are obtained from the total cross-sections and associated uncertainties reported in Ref.~\cite{deFlorian:2016spz}, multiplied by an acceptance factor for the region $|y_H| < 2.5$ computed using the Higgs boson simulation samples described in Section~\ref{sec:mc}.}
\label{tab:results:prodXS}
\begin{center}
\renewcommand{\arraystretch}{1.4}
\begin{tabular}{c|rlll|cl}
\hline \hline
Process       & \multicolumn{1}{c}{Value} & \multicolumn{3}{c|}{Uncertainty [fb] } & \multicolumn{1}{c}{SM pred.} \\
($|y_H|<2.5$) & \multicolumn{1}{c}{[fb]}  & Total  & \small{Stat.}      & \small{Syst.}        & \multicolumn{1}{c}{[fb]} \\
\hline
 
\ggF~+~\bbH\ &  \HyynumRF{106.1056}{3}~~~ &  \Hyynumpmerr{+10.4290}{-10.0523}{2} &  \Hyynumpmerr{+8.2804}{-8.2586}{1} & \Hyynumpmerr{+6.3403}{-5.7310}{1} & ${\HyynumRF{101.5990}{3}}^{+\HyynumRF{6.2497}{1}}_{\HyynumRF{-6.3200}{1}}$~~~ \\
\VBF\      &  \HyynumRF{9.5267}{2}   &  \Hyynumpmerr{+2.2388}{-1.9488}{2}   &  \Hyynumpmerr{+1.4694}{-1.4044}{2} & \Hyynumpmerr{+1.6892}{-1.3511}{2} & ${\HyynumRF{7.9450}{2}}^{+\HyynumRF{0.2126}{1}}_{\HyynumRF{-0.2154}{1}}$  \\
\WH\       &  \HyynumRF{4.1715}{2}   &  \Hyynumpmerr{+1.5309}{-1.3847}{2}   &  \Hyynumpmerr{+1.4884}{-1.3629}{2} & \Hyynumpmerr{+0.3582}{-0.2450}{1} & 
${\HyynumRF{2.7600}{2}}^{+\HyynumRF{0.1}{1}}_{\HyynumRF{-0.1}{1}}$  \\
\ZH\       &  \HyynumRF{-0.3946}{1}  &  \Hyynumpmerr{+1.0883}{-1.0}{2}   &  \Hyynumpmerr{+1.0679}{-1.0}{2} & \Hyynumpmerr{+0.2102}{-0.2668}{1} &
${\HyynumRF{1.8070}{2}}^{+\HyynumRF{0.1}{1}}_{\HyynumRF{-0.1}{1}}$  \\
\ttH\      &  \HyynumRF{1.0111}{2}   &  \Hyynumpmerr{+0.3615}{-0.3382}{1}   &
\Hyynumpmerr{+0.3499}{-0.3331}{1} & \Hyynumpmerr{+0.1}{-0.1}{1} & ${\HyynumRF{1.1350}{2}}^{+\HyynumRF{0.1164}{1}}_{\HyynumRF{-0.1160}{1}}$  \\
\tH\       &  \HyynumRF{0.4833}{1}   &  \Hyynumpmerr{+0.7904}{-0.6255}{1}   &  \Hyynumpmerr{+0.7459}{-0.6032}{1} & \Hyynumpmerr{+0.2616}{-0.1657}{1} & ${\HyynumRF{0.1920}{2}}^{+\HyynumRF{0.0126}{1}}_{\HyynumRF{-0.0246}{1}}$  \\
\bottomrule
 
\hline \hline
\end{tabular}
\end{center}
\end{table}
 
The largest theoretical systematic uncertainty in these measurements arises from the modelling of the parton showering and underlying event, and its impact on the measured cross-sections ranges from 38\% for the \tH\ process to 14\% for the \VBF\ process and to 3\%--4\% for the \ggtoH\ and \WH\ processes. For the \ggtoH\ process, the leading experimental systematic uncertainty is the photon energy resolution uncertainty (3\%). For the \VBF\ and \ttH\ processes, the leading experimental uncertainty is related to the properties of jets and missing transverse momentum, reaching 5.4\% for \VBF. For other processes, the leading experimental systematic uncertainty is the background modelling uncertainty, ranging from 3.7\% for \WH\ to 24\% for \tH.
 
\begin{figure}[tpb!]
\centering
\includegraphics[width=.7\textwidth]{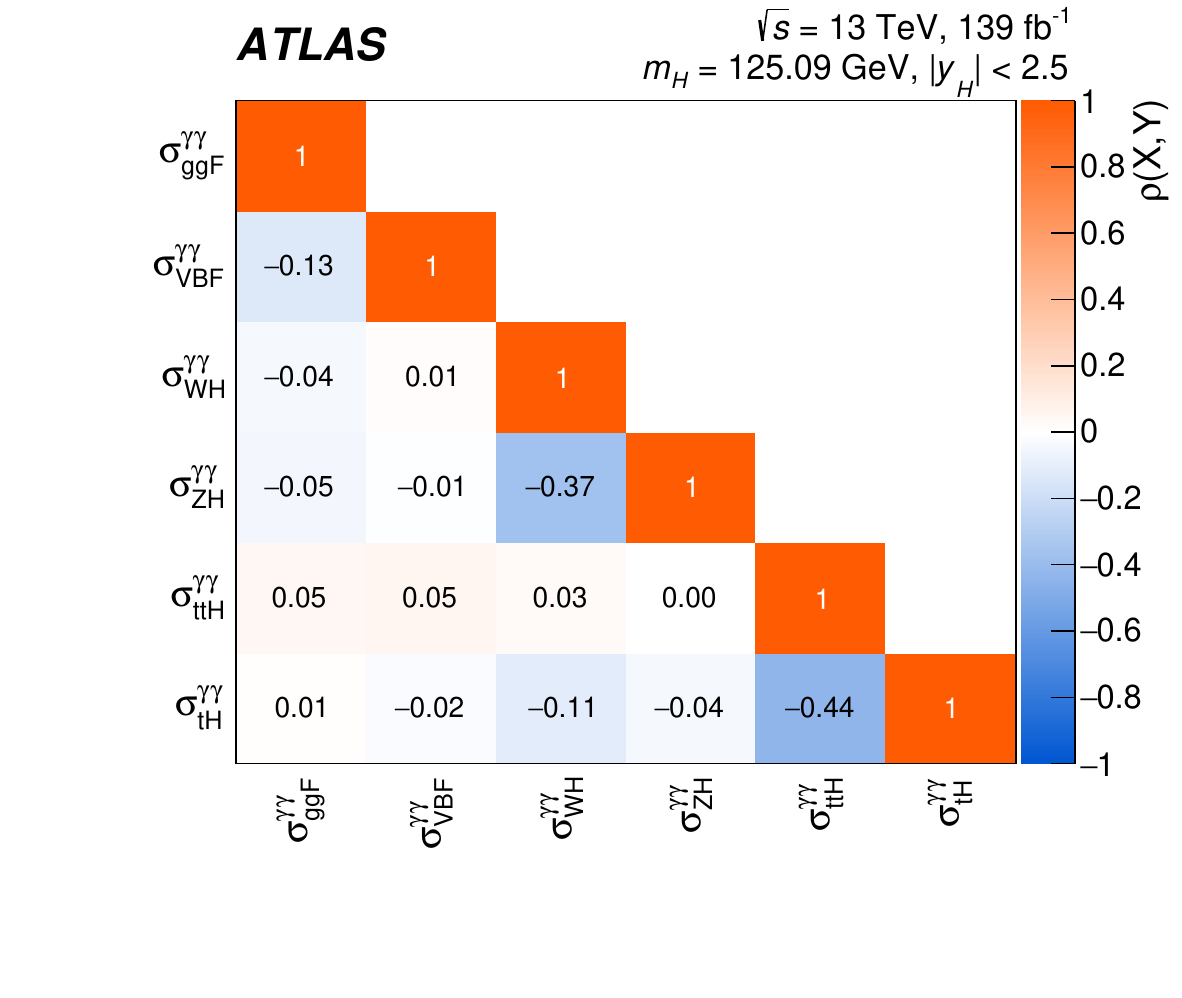}
\caption{Correlation matrix for the measurement of  production cross-sections of the Higgs boson times the \Hyy\ branching ratio.
}
\label{fig:results:prodXS_corr}
\end{figure}

An upper limit on the rate of \tH\ production is obtained by treating the normalization of other Higgs boson production processes as nuisance parameters. Using the CL$_\mathrm{s}$ method~\cite{Read:2002hq}, this excludes a \tH\ production rate of 10 times its SM prediction or greater at 95\% CL
while the expected 95\% CL limit is 6.8 times the SM \tH\ production cross-section.


\subsection{Cross-sections in STXS regions}
\label{sec:results:STXS}
 
A measurement of the cross-sections defined in the STXS scheme is performed using the 45 STXS analysis regions described in Section~\ref{sec:design}. In order to avoid large uncertainties and large absolute correlations between the measurements, a set of 28 measurement regions is obtained by merging some of the analysis regions as follows:
\begin{itemize}
\item For the \ggtoH\ process, within the phase space of $\ge~$2-jets, $\mjj < 350\,\GeV$, the two regions with $\ptH < 60\,\GeV$ and $60 \le \ptH < 120\,\GeV$ are merged into one region corresponding to $\ptH < 120\,\GeV$. Within $\ge~$2-jets,  $\ptH<200\,\GeV$, the three bins defined in the \mjj\ variable are merged into a single $\mjj> 350\,\GeV$ region. Finally, the $\ptH > 650\,\GeV$ bin is merged with the neighbouring $450 \le \ptH < 650\,\GeV$ bin to form a single region corresponding to  $\ptH \ge 450\,\GeV$.
 
\item For the \qqtoHqq\ process, the 0-jet and 1-jet regions, as well as the regions corresponding to $\mjj < 60\,\GeV$ and $120 < \mjj < 350\,\GeV$, are combined into a new region, referred to as \qqtoHqq, $\le 1$-jet and \VH-veto. The regions corresponding to $\ge~$2-jets, $\ptH \ge 200\,\GeV$, $350 <\mjj< 1000 \,\GeV$, are merged into a single region.
 
\item For both the \qqtoWH\ and \pptoZH\ processes, only the two regions $\ptV < 150\,\GeV$ and $\ptV \ge 150\,\GeV$ are retained, removing the intermediate splits at $\ptV = 75\,\GeV$ and $\ptV = 250\,\GeV$. For \pptoZH\ processes, no distinction is made between regions with charged leptons and regions with neutrinos.
 
\item The \tHqb\ and \tHW\ regions are merged into a single \tH\ region.
\end{itemize}
This scheme is based on the expected analysis sensitivity under the SM hypothesis, independently of the observed data, and is illustrated in Figure~\ref{fig:results:STXS_scheme}. The merging reduces the number of regions for which a measurement is reported to 28 in the scheme described above. The 101 categories in which the measurement is performed, described in Section~\ref{sec:cats}, remain unchanged.
\begin{figure}[tpb!]
\centering
\includegraphics[align=t, width=0.69\textwidth]{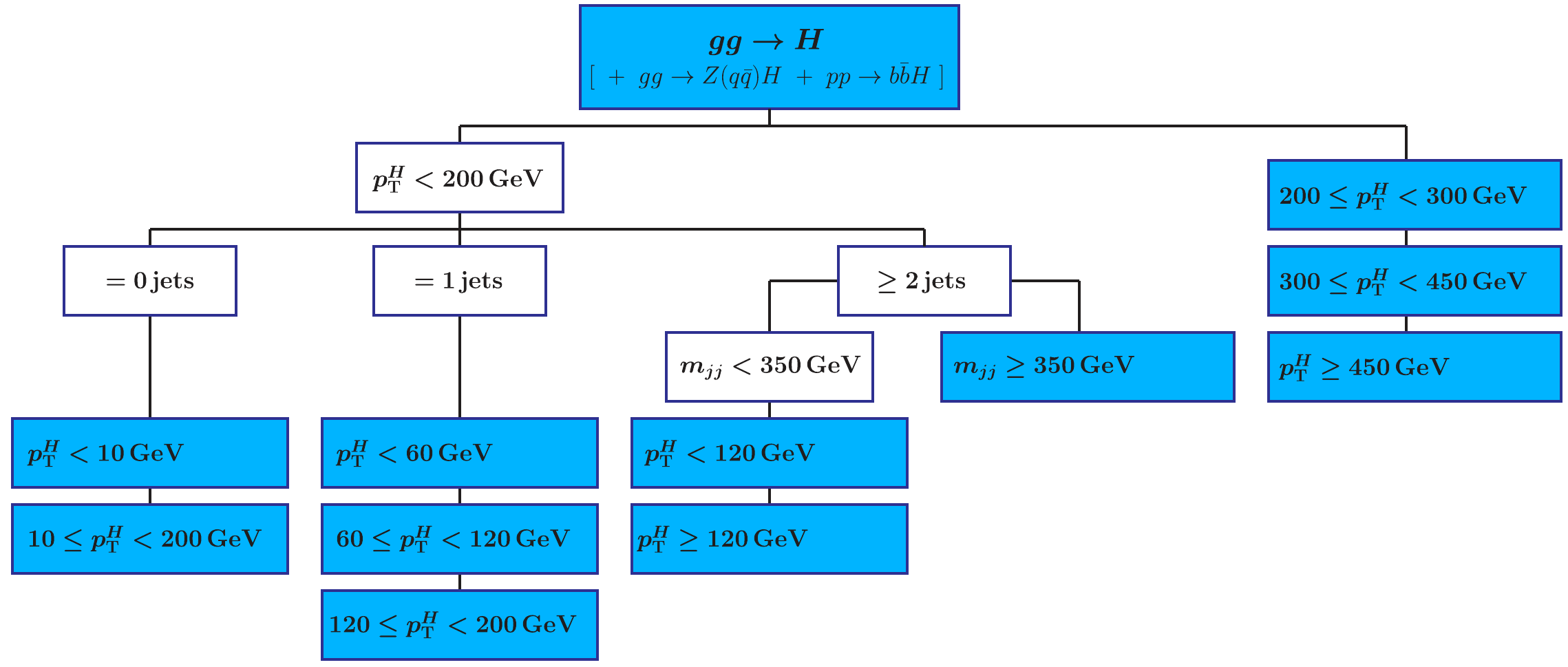}
\hfill
\includegraphics[align=t, width=0.29\textwidth]{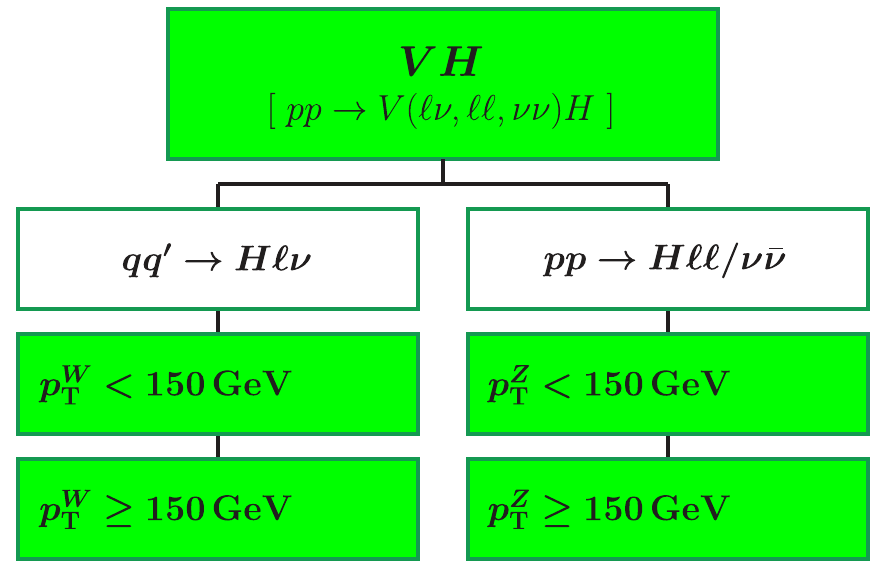}\\
\vspace{1cm}
\includegraphics[align=t, width=0.59\textwidth]{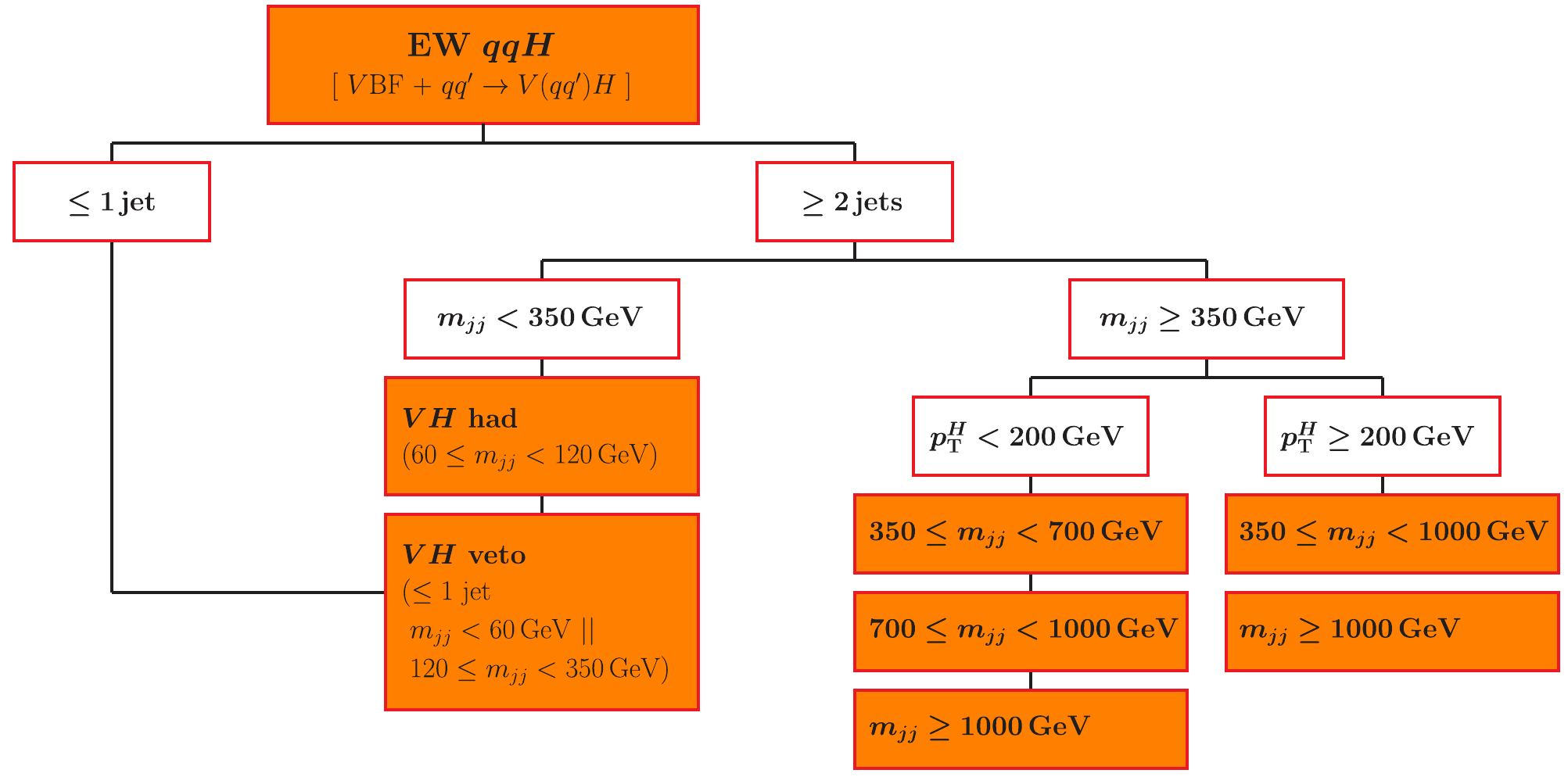}
\hfill
\includegraphics[align=t, width=0.39\textwidth]{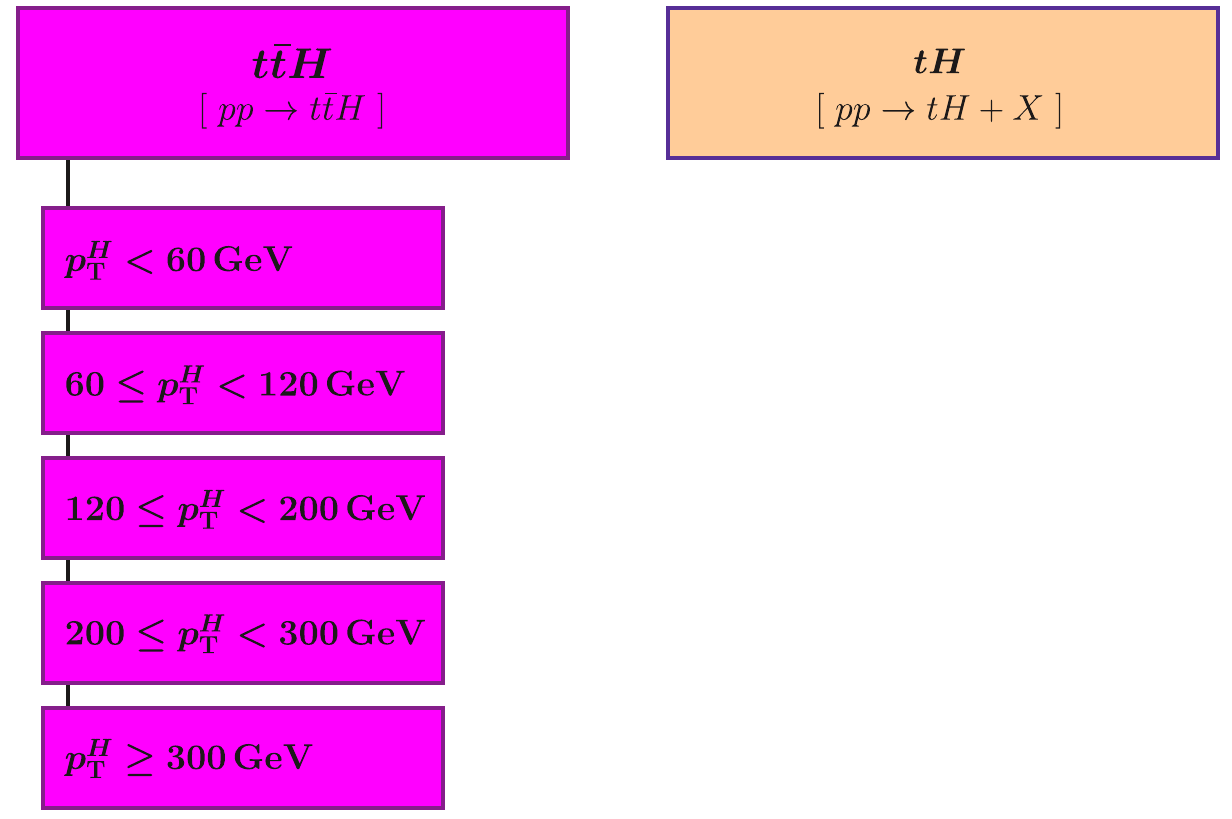}\\
\caption{Summary of the 28 regions for which STXS measurements are reported.}
\label{fig:results:STXS_scheme}
\end{figure}
The efficiency factors for merged STXS regions are computed as weighted averages of those for the original STXS regions, with the weights corresponding to the expected cross-sections in the SM. The uncertainty on the efficiency factor of the merged STXS region is then calculated from the uncertainties in the efficiency factors and expected cross-sections of the original STXS regions.
 
Results are shown in Table~\ref{tab:results:STXS} and Figure~\ref{fig:results:STXS}.
The correlation matrix of the measurements is shown in Figure~\ref{fig:results:STXS_corr}. The correlation between most STXS region measurements is small, and the largest correlation is $-51\%$, observed for the measurements of STXS regions \ggHmPt{350}{}{}{200}{} and \HqqmPt{350}{700}{}{200}{}. The Higgs boson production processes in these STXS regions have similar event topologies and are intrinsically difficult to separate.
The relative uncertainties in the measurements range from $20\%$ to more than $100\%$. Smaller uncertainties are associated with the 0-jet and 1-jet regions of \ggtoH, as well as the $200 \le \ptH < 300\,\GeV$ region of \ggtoH\ and the $\mjj \ge 700\,\GeV$ region of \qqtoHqq. Larger uncertainties occur especially in regions of high \ptH\ and \ptV, as well as the low-\mjj\ regions of \qqtoHqq. The systematic component of the uncertainties is everywhere smaller than the statistical component, but reaches similar values for the 0-jet regions of \ggtoH.
No significant deviations from the SM expectation are observed and the $p$-value for compatibility of the measurements and the SM predictions is $93\%$. 
Results in a finer set of 33 STXS measurements regions are also presented in Table~\ref{tab:results:STXS33} of Appendix~\ref{app:stxs}. Results are not reported using the full granularity of the 45 STXS analysis regions, since the statistical power of the \Hyy\ analysis alone is currently insufficient to perform this measurement. This configuration is however used in the \Hyy\ inputs to combinations with other Higgs boson processes, which allow more granular measurements. This includes for example combinations~\cite{HIGG-2021-23} with the ATLAS \VH, \Hbb\ analysis~\cite{HIGG-2018-51,HIGG-2018-52}, which reports STXS results in the \qqtoWH\ and \pptoZH\ processes using a finer granularity than in the present paper.

\begin{table}[ht]
\caption{
Best-fit values and uncertainties for the production cross-section times \Hyy\ branching ratio $(\sigma_i \times \Byy)$ in each STXS region. The values for the \ggtoH\ process also include the contributions from \bbH\ production. The total uncertainties are decomposed into statistical (Stat.) and systematic (Syst.) uncertainties. The uncertainties for the $\pptoHllnn, \ptV < 150\,\GeV$ region are truncated at the value for which the model pdf becomes negative. SM predictions~\cite{deFlorian:2016spz} are also shown for each quantity with their total uncertainties.
}
\label{tab:results:STXS}
\centering
\renewcommand{\arraystretch}{1.4}
\resizebox{0.9\textwidth}{!}{
\begin{tabular}{lrlll@{ }l@{ }}
\toprule
\multirow{2}{*}{STXS region $(\sigma_i \times \Byy)$} & \multicolumn{1}{c}{Value} & \multicolumn{3}{c}{ Uncertainty [fb]} & \multicolumn{1}{c}{SM prediction} \\
&  \multicolumn{1}{c}{[fb]}                & Total   & Stat.                & Syst.   & \multicolumn{1}{c}{[fb]}   \\
\midrule
\ggHjPt{0}{}{10}{}          & \numRF{10.1603}{2}~~~~~ &  \Hyynumpmerr{+4.2680}{-4.0739}{1} &  \Hyynumpmerr{+3.8254}{-3.8106}{1} &  \Hyynumpmerr{+1.8928}{-1.4407}{1} & \quad~~~~~${\numRF{15.0660}{2}}^{+\numRF{1.9682}{1}}_{\numRF{-1.9652}{1}}$ \\
\ggHjPt{0}{10}{}{}          & \numRF{57.5597}{2}~~~~~ &  \Hyynumpmerr{+8.5770}{-8.1189}{1} &  \Hyynumpmerr{+7.0256}{-7.0140}{1} &  \Hyynumpmerr{+4.9199}{-4.0892}{1} & \quad~~~~~${\numRF{46.8570}{2}}^{+\numRF{3.5802}{1}}_{\numRF{-3.6136}{1}}$ \\
\ggHjPt{1}{}{60}{}          & \numRF{15.7703}{2}~~~~~ &  \Hyynumpmerr{+5.3772}{-5.2168}{1} &  \Hyynumpmerr{+4.9838}{-4.9690}{1} &  \Hyynumpmerr{+2.0189}{-1.5888}{1} & \quad~~~~~${\numRF{14.7580}{2}}^{+\numRF{2.0413}{1}}_{\numRF{-2.0459}{1}}$ \\
\ggHjPt{1}{60}{120}{}       & \numRF{11.3879}{2}~~~~~ &  \Hyynumpmerr{+3.8390}{-3.4704}{1} &  \Hyynumpmerr{+3.0914}{-3.0861}{1} &  \Hyynumpmerr{+2.2762}{-1.5873}{1} & \quad~~~~~${\numRF{10.2220}{2}}^{+\numRF{1.4311}{1}}_{\numRF{-1.4320}{1}}$ \\
\ggHjPt{1}{120}{200}{}      & \numRF{1.6173}{2}~~  &  \Hyynumpmerr{+0.9325}{-0.8648}{1} &  \Hyynumpmerr{+0.8600}{-0.8446}{1} &  \Hyynumpmerr{+0.3606}{-0.1860}{1} & \quad~~~~${\numRF{ 1.6960}{2}}^{+\numRF{0.2960}{1}}_{\numRF{-0.2958}{1}}$ \\
\ggHmPt{}{350}{}{120}{}     & \numRF{3.9262}{1}~~~~~  &  \Hyynumpmerr{+3.6340}{-3.4900}{1} &  \Hyynumpmerr{+3.3815}{-3.3577}{1} &  \Hyynumpmerr{+1.3311}{-1.0}{1} & \quad~~~~~~~${\numRF{ 6.7270}{1}}^{+\numRF{1.4181}{1}}_{\numRF{-1.4188}{1}}$ \\
\ggHmPt{}{350}{120}{200}{}  & \numRF{2.7735}{2}~~  &
$\ensuremath{^{+1.0}_{-1.0}}$ &
$\ensuremath{^{+1.0}_{-1.0}}$ &
\Hyynumpmerr{+0.3191}{-0.2120}{1} & \quad~~~~${\numRF{ 2.1410}{2}}^{+\numRF{0.4921}{1}}_{\numRF{-0.4919}{1}}$ \\
\ggHmPt{350}{}{}{200}{}     & \numRF{2.0205}{1}~~~~~  &  \Hyynumpmerr{+1.7620}{-1.7510}{1} &  \Hyynumpmerr{+1.6380}{-1.6048}{1} &
$\ensuremath{^{+1}_{-1}}$ &
\quad~~~~${\numRF{ 1.9920}{2}}^{+\numRF{0.4737}{1}}_{\numRF{-0.4736}{1}}$ \\
\ggHPt{200}{300}{}          & \numRF{1.6151}{2}~~  &  \Hyynumpmerr{+0.4498}{-0.4172}{1} &  \Hyynumpmerr{+0.4084}{-0.3941}{1} &  \Hyynumpmerr{+0.1885}{-0.1369}{1} & \quad~~~~${\numRF{ 1.0400}{2}}^{+\numRF{0.2346}{1}}_{\numRF{-0.2343}{1}}$ \\
\ggHPt{300}{450}{}          & \numRF{0.0438}{1}  &  \Hyynumpmerr{+0.1283}{-0.1131}{2} &  \Hyynumpmerr{+0.1239}{-0.1078}{2} &  \Hyynumpmerr{+0.0333}{-0.0345}{1} & \quad~~${\numRF{ 0.2410}{2}}^{+\numRF{0.0612}{1}}_{\numRF{-0.0611}{1}}$ \\
\ggHPt{450}{}{}             & \numRF{0.0859}{1}  &  \Hyynumpmerr{+0.0589}{-0.0471}{1} &  \Hyynumpmerr{+0.0568}{-0.0462}{1} &
$\ensuremath{^{+0.02}_{-0.01}}$ &
\quad~~${\numRF{ 0.0410}{1}}^{+\numRF{0.0120}{1}}_{\numRF{-0.0119}{1}}$ \\
\qqtoHqq, $\le 1$-jet and \VH-veto  & \numRF{5.8236}{1}~~~~~ &  \Hyynumpmerr{+5.9351}{-5.3315}{1} &  \Hyynumpmerr{+5.5447}{-5.1634}{1} &  \Hyynumpmerr{+2.1172}{-1.3281}{1} &   \quad~~~~${\numRF{6.5780}{2}}^{+\numRF{0.1893}{1}}_{\numRF{-0.1917}{1}}$ \\
\qqtoHqq, \VH-had         & \numRF{0.1892}{2} &  \Hyynumpmerr{+0.8478}{-0.7279}{2} &  \Hyynumpmerr{+0.8302}{-0.7080}{2} &  \Hyynumpmerr{+0.1721}{-0.1688}{2} &   \quad~~${\numRF{1.1580}{3}}^{+\numRF{0.0428}{1}}_{\numRF{-0.0433}{1}}$ \\
\HqqmPt{350}{700}{}{200}{}  & \numRF{1.4580}{2}~~  &  \Hyynumpmerr{+0.9344}{-0.7034}{1} &  \Hyynumpmerr{+0.7068}{-0.6432}{1} &  \Hyynumpmerr{+0.6112}{-0.2848}{1} & \quad~~${\numRF{ 1.2150}{3}}^{+\numRF{0.0373}{1}}_{\numRF{-0.0377}{1}}$ \\
\HqqmPt{700}{1000}{}{200}{} & \numRF{0.7958}{1}~~  &  \Hyynumpmerr{+0.4601}{-0.3703}{1} &  \Hyynumpmerr{+0.3887}{-0.3478}{1} &  \Hyynumpmerr{+0.2461}{-0.1273}{1} & \quad~~${\numRF{ 0.5810}{2}}^{+\numRF{0.0188}{1}}_{\numRF{-0.0190}{1}}$ \\
\HqqmPt{1000}{}{}{200}{}    & \numRF{1.1883}{2}~~  &  \Hyynumpmerr{+0.4324}{-0.3530}{1} &  \Hyynumpmerr{+0.3276}{-0.2994}{1} &  \Hyynumpmerr{+0.2823}{-0.1869}{1} & \quad~~${\numRF{ 1.00}{3}}^{+\numRF{0.0315}{1}}_{\numRF{-0.0319}{1}}$ \\
\HqqmPt{350}{1000}{200}{}{} & \numRF{0.0396}{1}  &
$\ensuremath{^{+0.12}_{-0.10}}$ &
$\ensuremath{^{+0.12}_{-0.10}}$ &
\Hyynumpmerr{+0.0241}{-0.0229}{1} & \quad${\numRF{ 0.1670}{3}}^{+\numRF{0.0049}{1}}_{\numRF{-0.0049}{1}}$ \\
\HqqmPt{1000}{}{200}{}{}    & \numRF{0.2661}{2}  &
$\ensuremath{^{+0.11}_{-0.09}}$ &
$\ensuremath{^{+0.10}_{-0.08}}$ &
\Hyynumpmerr{+0.0452}{-0.0368}{1} & \quad${\numRF{ 0.1660}{3}}^{+\numRF{0.0052}{1}}_{\numRF{-0.0052}{1}}$ \\
\HlnPt{}{150}{}             & \numRF{1.4043}{2}~~  &  \Hyynumpmerr{+0.6476}{-0.5782}{1} &  \Hyynumpmerr{+0.6332}{-0.5709}{1} &
$\ensuremath{^{+0.1}_{-0.1}}$ &
\quad~~${\numRF{ 0.7930}{2}}^{+\numRF{0.0224}{1}}_{\numRF{-0.0229}{1}}$ \\
\HlnPt{150}{}{}             & \numRF{0.1968}{2}  &  \Hyynumpmerr{+0.1334}{-0.1067}{2} &  \Hyynumpmerr{+0.1324}{-0.1061}{2} &  \Hyynumpmerr{+0.0169}{-0.0117}{1} & \quad${\numRF{ 0.1210}{3}}^{+\numRF{0.0054}{1}}_{\numRF{-0.0054}{1}}$ \\
\HllnnPt{}{150}{}           & \numRF{-0.2918}{2} &
$\ensuremath{^{+0.40}_{-0.08}}$ &
$\ensuremath{^{+0.39}_{-0.08}}$ &
$\ensuremath{^{+0.07}_{-0.00}}$ &
\quad~~${\numRF{ 0.4510}{2}}^{+\numRF{0.0187}{1}}_{\numRF{-0.0189}{1}}$ \\
\HllnnPt{150}{}{}           & \numRF{0.0372}{1}  &  
$\ensuremath{^{+0.10}_{-0.08}}$ &
$\ensuremath{^{+0.10}_{-0.08}}$ &
\Hyynumpmerr{+0.0180}{-0.0162}{1} & \quad~~${\numRF{ 0.0920}{1}}^{+\numRF{0.0109}{1}}_{\numRF{-0.0110}{1}}$ \\
\ttHPt{}{60}{}             & \numRF{0.2159}{2}  &  \Hyynumpmerr{+0.2127}{-0.1775}{2} &  \Hyynumpmerr{+0.2102}{-0.1770}{2} &  \Hyynumpmerr{+0.0325}{-0.0143}{1} & \quad~~${\numRF{ 0.2680}{2}}^{+\numRF{0.0365}{1}}_{\numRF{-0.0365}{1}}$ \\
\ttHPt{60}{120}{}           & \numRF{0.3164}{2}  &  \Hyynumpmerr{+0.2339}{-0.1962}{2} &  \Hyynumpmerr{+0.2312}{-0.1954}{2} &  \Hyynumpmerr{+0.0354}{-0.0186}{1} & \quad~~${\numRF{ 0.4040}{2}}^{+\numRF{0.0455}{1}}_{\numRF{-0.0454}{1}}$ \\
\ttHPt{120}{200}{}          & \numRF{0.1832}{2}  &  \Hyynumpmerr{+0.1787}{-0.1477}{2} &  \Hyynumpmerr{+0.1748}{-0.1459}{2} &  \Hyynumpmerr{+0.0372}{-0.0231}{1} & \quad~~${\numRF{ 0.2870}{2}}^{+\numRF{0.0357}{1}}_{\numRF{-0.0357}{1}}$ \\
\ttHPt{200}{300}{}          & \numRF{0.1420}{2}  &  \Hyynumpmerr{+0.0925}{-0.0737}{1} &  \Hyynumpmerr{+0.0916}{-0.0734}{1} &  
$\ensuremath{^{+0.01}_{-0.01}}$ &
\quad~~${\numRF{ 0.1190}{2}}^{+\numRF{0.0171}{1}}_{\numRF{-0.0171}{1}}$ \\
\ttHPt{300}{}{}             & \numRF{0.0628}{1}  &  \Hyynumpmerr{+0.0511}{-0.0397}{1} &  \Hyynumpmerr{+0.0503}{-0.0392}{1} &
$\ensuremath{^{+0.01}_{-0.01}}$ &
\quad~~${\numRF{ 0.0550}{1}}^{+\numRF{0.01}{1}}_{\numRF{-0.01}{1}}$ \\
\tH                         & \numRF{0.3675}{1}~~  &  \Hyynumpmerr{+0.7536}{-0.5937}{1} &  \Hyynumpmerr{+0.7171}{-0.5701}{1} &  \Hyynumpmerr{+0.2319}{-0.1656}{1} & \quad~~${\numRF{ 0.1920}{2}}^{+\numRF{0.0126}{1}}_{\numRF{-0.0246}{1}}$ \\
\bottomrule
\end{tabular}}
\end{table}


\FloatBarrier
\begin{figure}[tp!]
\centering
\includegraphics[width=.875\textwidth]{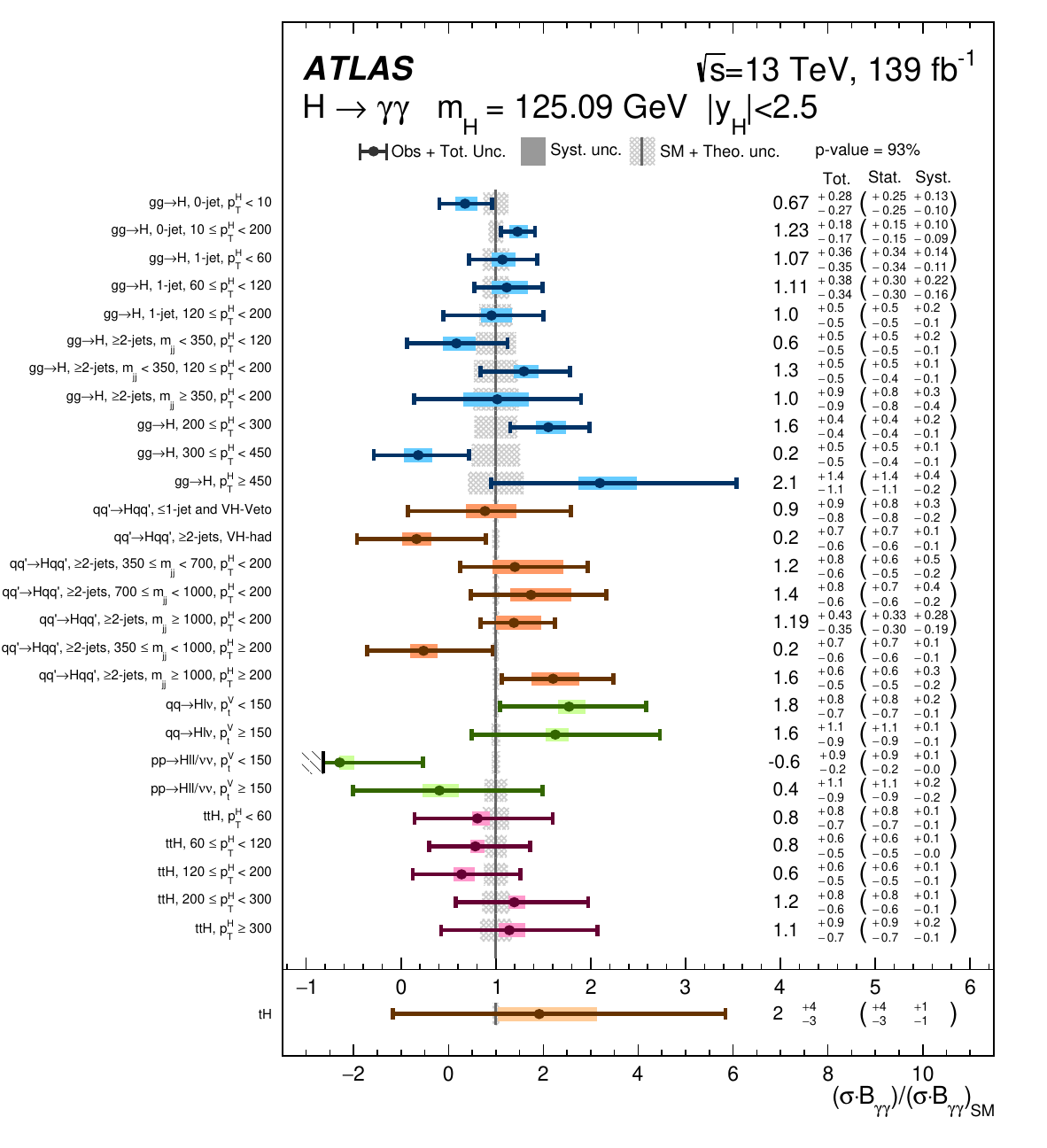}
\vspace{-10pt}
\caption{Best-fit values and uncertainties for STXS parameters in each of the \nSTXS\ regions considered, normalized to their SM predictions.
The values for the \ggtoH\ process also include the contributions from \bbH\ production. The error bars and shaded areas show the total and systematic uncertainties in the measurements, respectively. The uncertainties for the $\pptoHllnn, \ptV < 150\,\GeV$ region are truncated at the value for which the model pdf becomes negative. The grey bands around the vertical line at $\sigma^{\gamma\gamma}/\sigma^{\gamma\gamma}_{SM} = $1 show the theory uncertainties in the predictions, including uncertainties due to missing higher-order terms in the perturbative QCD calculations, the PDFs and the value of \alphas, as well as the \Hyy\ branching ratio uncertainty. The \pt\ and \mjj\ values in the region definitions are indicated in \GeV.}
\label{fig:results:STXS}
\end{figure}
\begin{figure}[tpb!]
\centering
\includegraphics[width=.995\textwidth]{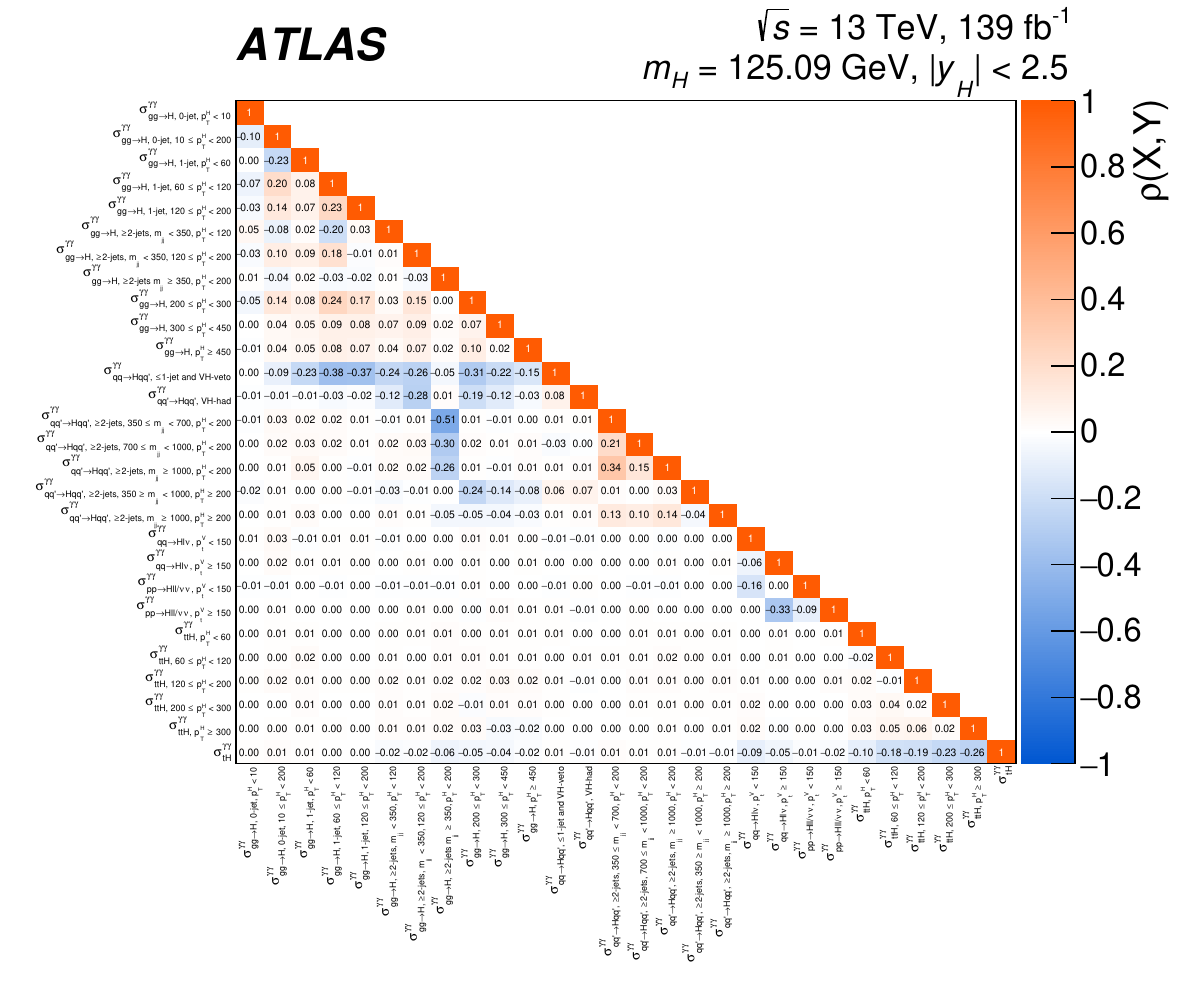}
\caption{Correlation matrix for the measurement of STXS parameters in each of the \nSTXS\ regions considered.
}
\label{fig:results:STXS_corr}
\end{figure}
\FloatBarrier


\section{Interpretation of the results in the $\kappa$-framework}
\label{sec:results:kappas}
 
Event rates for processes involving Higgs bosons can be expressed in terms of modifiers applied to the SM Higgs boson couplings, based on the leading-order contributions to each process~\cite{deFlorian:2016spz}. These coupling modifiers affect Higgs boson production cross-sections and decay partial widths and therefore provide a consistent framework for Higgs boson coupling measurements in both production and decay.
 
The Higgs boson production cross-section in STXS region $i$ followed by a \hgg\ decay is written in the narrow-width approximation as
\begin{equation*}
\sigma_i \cdot B_{\gamma\gamma} = \frac{\sigma_i(\bm{\kappa}) \cdot \Gamma_{\gamma\gamma}(\bm{\kappa})}{\Gamma_H(\bm{\kappa})}
\end{equation*}
where the coupling modifiers are collectively denoted as $\bm{\kappa}$, $\sigma_i$ is the production cross-section in region $i$, $B_{\gamma\gamma}$ and $\Gamma_{\gamma\gamma}$ are respectively the Higgs boson branching ratio and partial width into the $\gamma\gamma$ final state, and $\Gamma_H$ is the total width of the Higgs boson. The parameterizations $\sigma_i(\bm{\kappa})$,  $\Gamma_{\gamma\gamma}(\bm{\kappa})$ and $\Gamma_H(\bm{\kappa})$ are shown in Table~\ref{tab:kappas:parameterization} in Appendix~\ref{app:kappas}. They are similar to the ones used in Ref.~\cite{HIGG-2018-57}, except for the parameterization of the \tHW\ and \tHqb\ processes. These have been updated to reflect the fact that the efficiency factors in each analysis category are $\bm{\kappa}$-dependent due to changes in the process kinematics caused by interference effects between the different parton-level processes contributing to \tHW\ and \tHqb.
Separate parameterizations are therefore used in each analysis category for \tHW\ and \tHqb.
 
Two parameterizations are considered for the \ggtoH\ and \Hyy\ processes: a \emph{resolved} parameterization in which they are assumed to proceed through the same loop amplitudes as in the SM at leading order, and an \emph{effective} parameterization that makes no assumption about the internal structure of the interactions. For the latter, event rates are expressed using modifiers to the effective couplings of the Higgs boson to the gluon and the photon, respectively denoted by $\kappa_g$ and $\kappa_\gamma$.
The \ggtoZH\ loop process is always described in the resolved parameterization.
Sensitivity to the sign of coupling modifiers is obtained through interference between processes involving different combinations of modifiers. These include, in particular, the \Hyy\ and \ggtoH\ loops and the \ggtoZH\ and \tH\ processes.

Two specific models of coupling modifications are considered in this section, and two additional models are described in Appendix~\ref{app:kappas}. The first model focuses on the $\kappa_t$ modifier to the Higgs boson coupling with the top quark. Two configurations are used for the \ggtoH\ and \Hyy\ loop processes: in the first case, both are described using their resolved parameterization as a function of $\kappa_t$; in the second case, both are described using the effective couplings $\kappa_g$ and $\kappa_\gamma$. All other Higgs boson couplings are fixed to their SM values, in particular the coupling with the $W$ boson which enters the resolved parameterization of the \Hyy\ loop.
These two models also allow the sign of $\kappa_t$ to be probed, with sensitivity coming from interference effects in certain amplitudes. These occur in the \tH\ and \ggtoZH\ processes, as well as in the \Hyy\ process when its parameterization is resolved in terms of $\kappa_t$ and other coupling modifiers.
 
The negative log-likelihood scans for both configurations are shown in Figure~\ref{fig:kappas:kappa_t}.
\begin{figure}[h!]
\centering
\includegraphics[width=.8\textwidth]{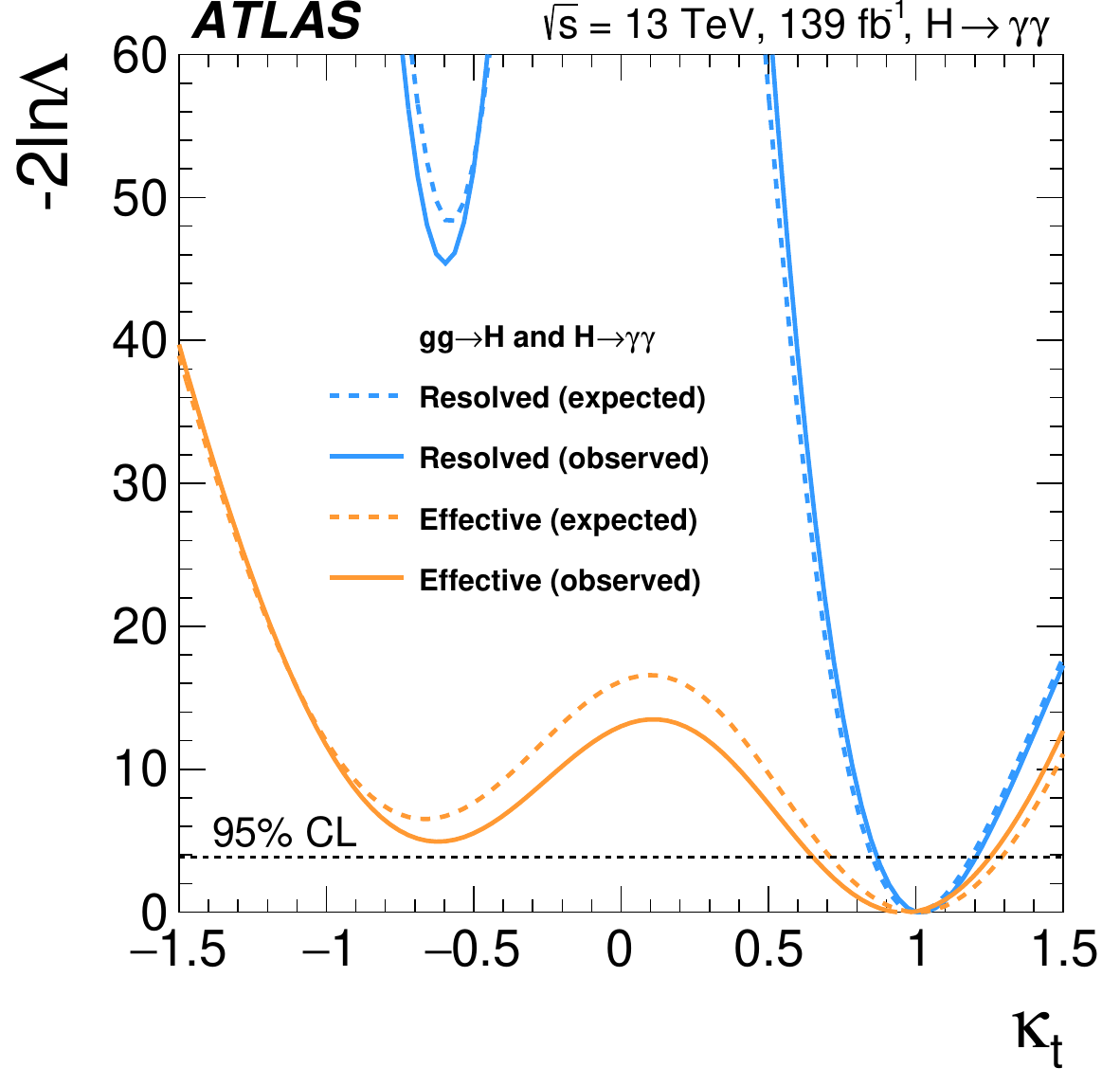}
\caption{Negative log-likelihood scans as a function of $\kappa_t$ in a model where other coupling modifiers are fixed to their SM values. The \Hyy\ and \ggtoH\ loops are either parameterized as a function of $\kappa_t$ (blue) or fixed to their SM expectation (orange). In the latter case, sensitivity to the sign of $\kappa_t$ is provided by the \tH\ process, and to a lesser degree by the \ggtoZH\ process. The solid curves correspond to observed data, and the dotted curves to an Asimov data set generated under the SM hypothesis.}
\label{fig:kappas:kappa_t}
\end{figure}
In both cases, good agreement with the SM expectation of $\kappa_t=1$ is seen.
When the \Hyy\ and \ggtoH\ loops are resolved, negative values of $\kappa_t$ are excluded with a significance of $6.7\sigma$ or above\footnote{Significances are computed as $\sqrt{-2\ln\Lambda(\kappa_t)}$, where $\Lambda(\kappa_t)$ is the profile likelihood ratio defined in Section~\ref{sec:stat} in terms of the parameter of interest $\kappa_t$.}. When effective loop couplings are used, an exclusion of $2.2\sigma$ or above is observed through the sensitivity provided by the \tH\ process, with a smaller contribution from the \ggtoZH\ process.
Values of $\kappa_t$ outside of the range $0.87 < \kappa_t < 1.20$ are excluded at 95\% CL in the first case ($0.85 < \kappa_t < 1.19$ expected under the SM hypothesis), as are values outside $0.65 < \kappa_t < 1.25$ in the second case ($0.71 < \kappa_t < 1.29$ expected).
 
In the second model, the \ggtoH\ and \hgg\ loop processes are described using the effective modifiers $\kappa_g$ and $\kappa_\gamma$. Both modifiers are assumed to be positive, since the measurement provides no sensitivity to their signs. Other modifiers are fixed to their SM values.
The measurement in the plane of $(\kappa_g, \kappa_\gamma)$ is shown in Figure~\ref{fig:kappas:kgky}. The best-fit values are
\begin{align*}
\kappa_g      &= 1.01^{+\;0.11}_{-\;0.09}\\
\kappa_\gamma &= 1.02^{+\;0.08}_{-\;0.07}.
\end{align*}
A linear correlation coefficient of $-79\%$ between the parameters is observed. 
\begin{figure}[h!]
\centering
\includegraphics[width=.8\textwidth]{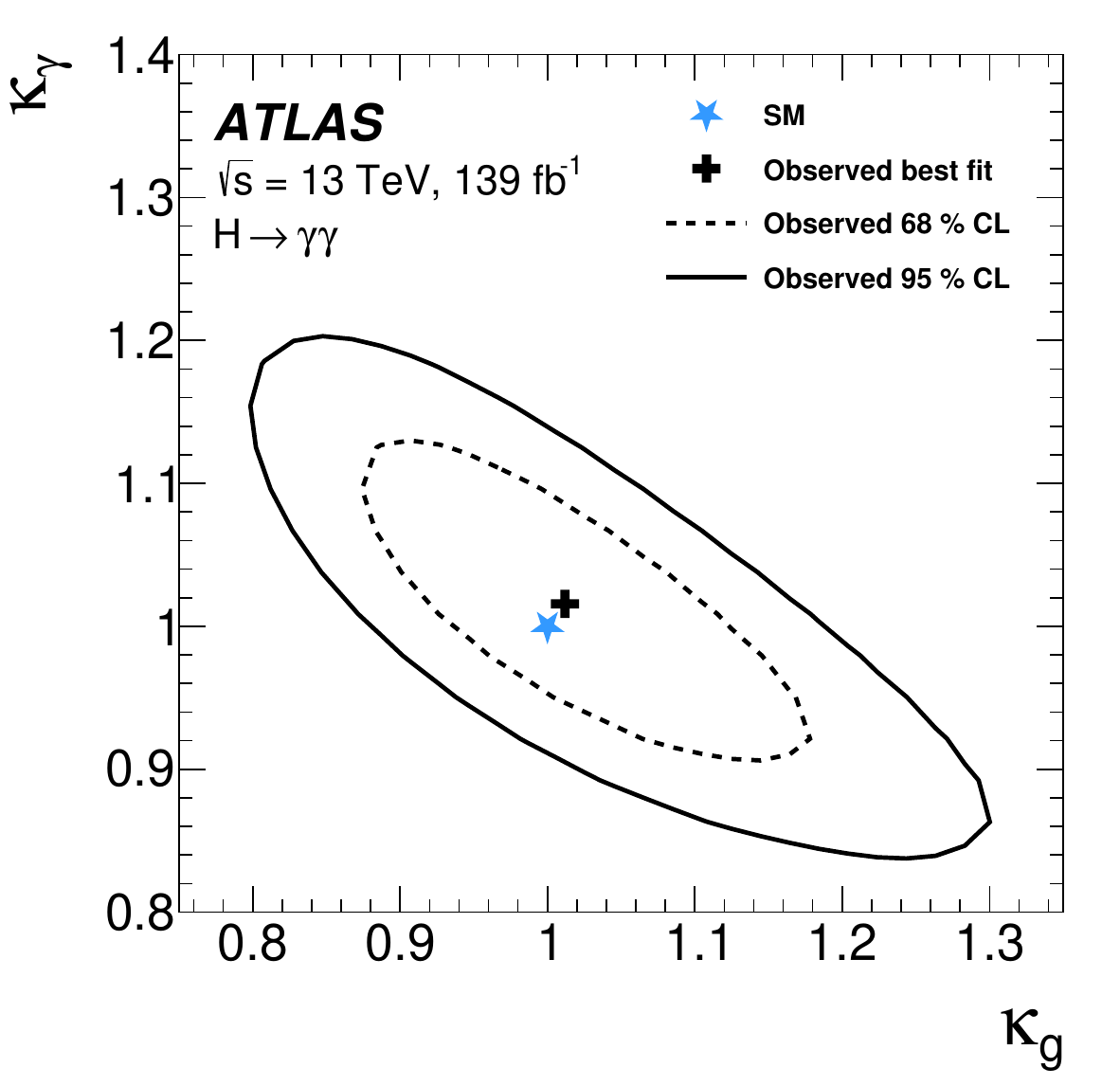}
\caption{Negative log-likelihood contours at 68\% (dashed line) and 95\% CL (solid line) in the ($\kappa_g$,~$\kappa_\gamma$)~plane, assuming that all other coupling-strength modifiers take their SM values. All other $\kappa$ modifiers are fixed to their SM values. The best-fit point is indicated by a cross while the SM prediction is indicated by a star.}
\label{fig:kappas:kgky}
\end{figure}

\FloatBarrier


\section{Interpretation of the results in the Standard Model effective field theory framework}
\label{sec:results:smeft}
 
\subsection{Interpretation framework}
 
The Standard Model effective field theory (SMEFT) framework provides a model-independent setting to describe deviations from SM predictions. New effective interactions involving Standard Model particles are introduced in the Lagrangian to describe the effect of physics beyond the SM occurring above a high scale $\Lambda$. These interactions are considered order by order in the mass dimension $d$ of the relevant operators, with leading-order effects occurring at $d=6$ when assuming that the lepton number $L$ and baryon number $B$ are conserved.
 
The effective Lagrangian up to dimension 6 is written as
\begin{equation*}
\mathcal{L} = \mathcal{L}_{\text{SM}} + \sum\limits_k \frac{c_k}{\Lambda^2} O_k
\end{equation*}
where the sum runs over the dimension-6 operators $O_k$ describing effective interactions in the SMEFT. The $c_k$ are the corresponding Wilson coefficients, which are considered as the measurement parameters of the model. Subleading contributions with dimension 8 and above are neglected, and only operators with even CP quantum numbers that conserve $B$ and $L$ are considered.
The selected operators are expressed in the Warsaw basis~\cite{Buchmuller:1985jz,Grzadkowski:2010es}.
The $U(3)^5$-symmetric model of fermion flavour~\cite{SMEFTsim3} is considered, assuming separate global flavour symmetries for each fermion type over the three fermion generations. In the cases where Wilson coefficients can have complex values, only their real parts are considered. The SM corresponds to all $c_k$ set to 0. The $c_k $ are defined for a scale $\Lambda = 1\,\TeV$.
 
The $c_k$ are determined through an interpretation of the STXS results presented in Appendix~\ref{app:stxs}. This is achieved by expressing as functions of the $c_k$ the signal-strength parameters
\begin{equation*}
\mu_i^{\gamma\gamma} = \frac{\sigma_i \cdot B_{\gamma\gamma}}{\sigma_i^{\text{SM}} \cdot B_{\gamma\gamma}^{\text{SM}}},
\end{equation*}
where $i$ runs over the 33 STXS regions listed in Table~\ref{tab:results:STXS33}, and $\sigma_i^{\text{SM}}$ and $B_{\gamma\gamma}^{\text{SM}}$ are the SM predictions for the production cross-section in STXS region $i$ and the \Hyy\ branching ratio, respectively. The 33 STXS regions used here correspond to a finer binning than the 28 regions for which results were reported in Section~\ref{sec:results:STXS}, and provide better granularity especially at high \mjj\ values in the \ggtoH\ and \qqtoHqq\ processes.
 
SMEFT effects are modelled as single insertions of the operators $O_k$ in each Higgs boson production and decay amplitude. These amplitudes are therefore linear in the $c_k$, so that the production cross-sections and decay widths are at most quadratic functions of the $c_k$. The signal strengths are thus written as
\begin{equation}
\mu_i^{\gamma\gamma} =
\left(1 + \sum\limits_k A_k^{i \to H} c_k  + \sum\limits_{kl} B_{kl}^{i \to H} c_k c_l \right)
\frac{\left(1 + \sum\limits_k A_k^{\hgg} c_k  + \sum\limits_{kl} B_{kl}^{\hgg} c_k c_l \right)}{\left(1 + \sum\limits_k A_k^{\Gamma} c_k  + \sum\limits_{kl} B_{kl}^{\Gamma} c_k c_l \right)}
\label{eq:eft:param}
\end{equation}
where $A_k^{i \to H}$, $A_k^{\hgg}$ and $A_k^{\Gamma}$ are respectively the coefficients describing the linear $c_k$-dependence of the production cross-section $\sigma_i$, the partial decay width $\Gamma_{\gamma\gamma}$ and the total width $\Gamma_H$. Similarly, the $B_{kl}^{i \to H}$, $B_{kl}^{\hgg}$ and $B_{kl}^{\Gamma}$ coefficients describe the quadratic dependence of the same quantities on the $c_k$.
 
Two SMEFT parameterizations are considered in the following: a \emph{linear} parameterization including only the effect of the $A$ coefficients and a \emph{linear+quadratic} parameterization including both the $A$ and $B$ terms. In the linear case, Eq.~(\ref{eq:eft:param}) is linearized to first order in the $c_k$ so that
\begin{equation}
\mu_i^{\gamma\gamma} =
1 + \sum\limits_k \left[A_k^{i \to H}  +  A_k^{\hgg} - A_k^{\Gamma} \right] c_k.
\label{eq:eft:param_lin}
\end{equation}
Results are derived using both parameterizations, and their difference is considered to be indicative of the impact of the neglected higher-order terms in the SMEFT expansion.
 
The values of the $A$ and $B$ coefficients are generally obtained using the \SMEFTsim~\cite{SMEFTsim,SMEFTsim3} and \SMEFTatNLO~\cite{SMEFTatNLO} programs. The coefficients are obtained by setting SMEFT parameters to non-zero values (one parameter at a time to compute $A$ coefficients, and in pairs to compute $B$ coefficients), and comparing the cross-sections obtained in this case with the ones for all coefficients set to 0. Events corresponding to each STXS region are selected using a RIVET routine~\cite{RIVET_STXS}. \SMEFTatNLO\ is used to obtain predictions for \ggtoH\ and \ggtoZH\ loop processes, while \SMEFTsim\ is used for all other processes. The dependence of the Higgs boson total decay width on the SMEFT parameters is computed by considering all decays of the Higgs boson with up to four particles in the final state.
The $A$ coefficients for the \hgg\ decay are taken from an analytic calculation~\cite{Dawson:2018liq}, which includes NLO electroweak corrections that are not implemented in the programs mentioned above. The coefficients are obtained for the full phase space of the decay, relying on the fact that they are only weakly dependent on acceptance, as discussed below. The $B$ coefficients for the \hgg\ decay are computed using \SMEFTsim.

The SMEFT operators considered in the analysis are shown in Table~\ref{tab:eft:ops}.

\begin{table}[!htbp]
\centering
\caption{Wilson coefficients $c_i$ and corresponding dimension-6 SMEFT operators $\mathcal{O}_i$ used in this analysis. The notations follow that of Ref.~\cite{SMEFTsim3}. Hermitian conjugates of non-Hermitian operators are implicitly considered in addition to the expression shown in the table. The operators indicated by a checkmark are the ones included in the measurement, due to having a significant impact on STXS cross-sections or on the \Hyy\ branching ratio.}
\label{tab:eft:ops}
\vspace{5pt}
\begin{tabular}{c c c | c c c}
\hline
\hline
Coeff. & Operator & Incl. & Coeff. & Operator & Incl. \\
\hline
\cG         & \opG         & $\checkmark$ & \cfn{qq}{3}     & \opqqTriplet        & $\checkmark$  \\
\cW         & \opW         & $\checkmark$ & \cfn{qq}{3}[\prime] & \opqqTripletPrime & $\checkmark$ \\
\cH         & \opH         &              & \cfn{qq}{1}     & \opqqSinglet        & $\checkmark$ \\
\cHbox      & \opHbox      & $\checkmark$ & \cfn{qq}{1}[\prime] & \opqqSingletPrime & $\checkmark$ \\
\cHD        & \opHD        & $\checkmark$ & \cfn{lq}{3}     & \oplqTriplet        &              \\
\cHG        & \opHG        & $\checkmark$ & \cfn{lq}{1}     & \oplqSinglet        &              \\
\cHW        & \opHW        & $\checkmark$ & \cf{ee}         & \opee               &              \\
\cHB        & \opHB        & $\checkmark$ & \cf{eu}         & \opeu               &              \\
\cHWB       & \opHWB       & $\checkmark$ & \cf{ed}         & \oped               &              \\
\ceH        & \opeH        & $\checkmark$ & \cf{uu}         & \opuu               & $\checkmark$ \\
\cuH        & \opuH        & $\checkmark$ & \cf{uu}[\prime] & \opuuPrime          & $\checkmark$ \\
\cdH        & \opdH        & $\checkmark$ & \cf{dd}         & \opdd               &              \\
\ceW        & \opeW        &              & \cf{dd}[\prime] & \opddPrime          &              \\
\ceB        & \opeB        &              & \cfn{ud}{1}     & \opudSinglet        & $\checkmark$ \\
\cuG        & \opuG        & $\checkmark$ & \cfn{ud}{8}     & \opudOctet          & $\checkmark$ \\
\cuW        & \opuW        & $\checkmark$ & \cf{le}         & \ople               &              \\
\cuB        & \opuB        & $\checkmark$ & \cf{lu}         & \oplu               &              \\
\cdG        & \opdG        &              & \cf{ld}         & \opld               &              \\
\cdW        & \opdW        &              & \cf{qe}         & \opqe               &              \\
\cdB        & \opdB        &              & \cfn{qu}{1}     & \opquSinglet        & $\checkmark$ \\
\cfn{Hl}{3} & \opHlTriplet & $\checkmark$ & \cfn{qu}{8}     & \opquOctet          & $\checkmark$ \\
\cfn{Hl}{1} & \opHlSinglet & $\checkmark$ & \cfn{qd}{1}     & \opqdSinglet        & $\checkmark$ \\
\cf{He}     & \opHe        & $\checkmark$ & \cfn{qd}{8}     & \opqdOctet          & $\checkmark$ \\
\cfn{Hq}{3} & \opHqTriplet & $\checkmark$ & \cf{ledq}       & \opledq             &              \\
\cfn{Hq}{1} & \opHqSinglet & $\checkmark$ & \cfn{quqd}{1}   & \opquqdSinglet      &              \\
\cf{Hu}     & \opHu        & $\checkmark$ & \cfn{quqd}{1}[\prime] & \opquqdSingletPrime &        \\
\cf{Hd}     & \opHd        & $\checkmark$ & \cfn{quqd}{8}   & \opquqdOctet        &              \\
\cf{Hud}    & \opHud       &              & \cfn{quqd}{8}[\prime]  & \opquqdOctetPrime   &       \\
\cf{ll}     & \opll        &              & \cfn{lequ}{1}   & \oplequSinglet      &             \\
\cf{ll}[\prime]  & \opllPrime   & $\checkmark$ & \cfn{lequ}{3}     & \oplequOctet        &              \\
\hline
\hline
\end{tabular}
\end{table}


Initially, 60 operators are considered but only 34 are found to have significant impact in at least one STXS region or on the \Hyy\ branching ratio, defined by a value above 0.01 for the corresponding $A$ coefficient. Only these 34 operators are considered in the measurements presented in this paper. The impact of the most relevant SMEFT parameter in the measured STXS regions is summarized in Figure~\ref{fig:eft:Warsaw_impact}.
The efficiency factors which are applied to the observed event yields to obtain the $\mu_i^{\gamma\gamma}$ parameters, as shown in Eq.~(\ref{eq:yield}), can depend on the SMEFT parameters due to modifications of the Higgs signal characteristics within each STXS bin. The effect was studied for the main Warsaw basis operators affecting the measurement, and setting the corresponding Wilson coefficients to 1 individually is generally found to have an impact below 10\% on the efficiency factors. These changes are therefore neglected in the analysis.
\begin{figure}[!htbp]
\centering
\includegraphics[width=1.1\textwidth]{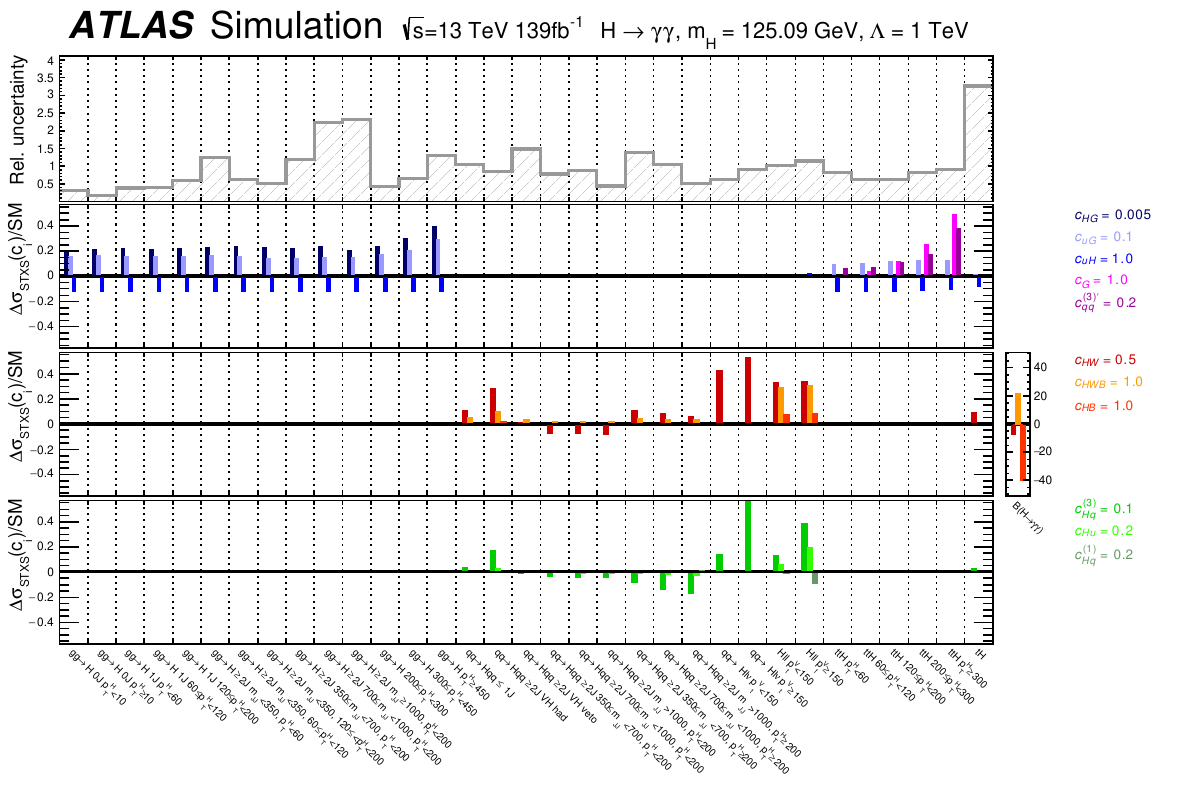}
\caption{Relative impact of the most relevant SMEFT operators on the STXS regions and \Hyy\ decay mode in the linear SMEFT model. Coloured bars indicate the relative impact of SMEFT parameters on the expected cross-section in the corresponding region. The impacts are computed for the parameter values shown on the right, relative to the SM prediction. The parameters are defined for a scale $\Lambda = 1\,\TeV$.
Three sets of operators with similar impacts on the measurement are shown in separate panels: those with impact mainly on the \ggtoH\ and \ttH\ processes (second from top), the \Hyy\ decay (third from top), and \VBF\ and \VH\ processes (bottom).
The expected total relative uncertainty in the measurement of the signal strength in each STXS region is shown in the top panel, as an indication of the experimental sensitivity of each region.
The \pt\ and \mjj\ values in the region definitions are indicated in \GeV, and the 0J, 1J and 2J shorthands refer respectively to the 0-jet, 1-jet and $\ge 2$-jets selections.
}
\label{fig:eft:Warsaw_impact}
\end{figure}
 
\subsection{Measurements of single SMEFT parameters}
\label{sec:results:smeft:singleWC}
 
In the measurements presented in this section, one SMEFT parameter at a time is left free to vary, while the others are fixed to $0$ as in the SM. This provides a measure of the sensitivity of the analysis for individual Wilson coefficients in the Warsaw basis, but the restrictive nature of this model limits the applicability of the measurements in probing effects beyond the SM.
 
The measurement results are summarized in Figure~\ref{fig:eft:singleWC}, and full results are provided in Table~\ref{tab:eft:singleWC_full} in Appendix~\ref{app:eft}. The SMEFT framework used in this measurement is considered valid only for parameter values of $O(10)$ or less, and confidence intervals that extend outside $|c_k| \le 20$ are therefore not shown.
\begin{figure}[!htbp]
\centering
\includegraphics[width=1.0\textwidth]{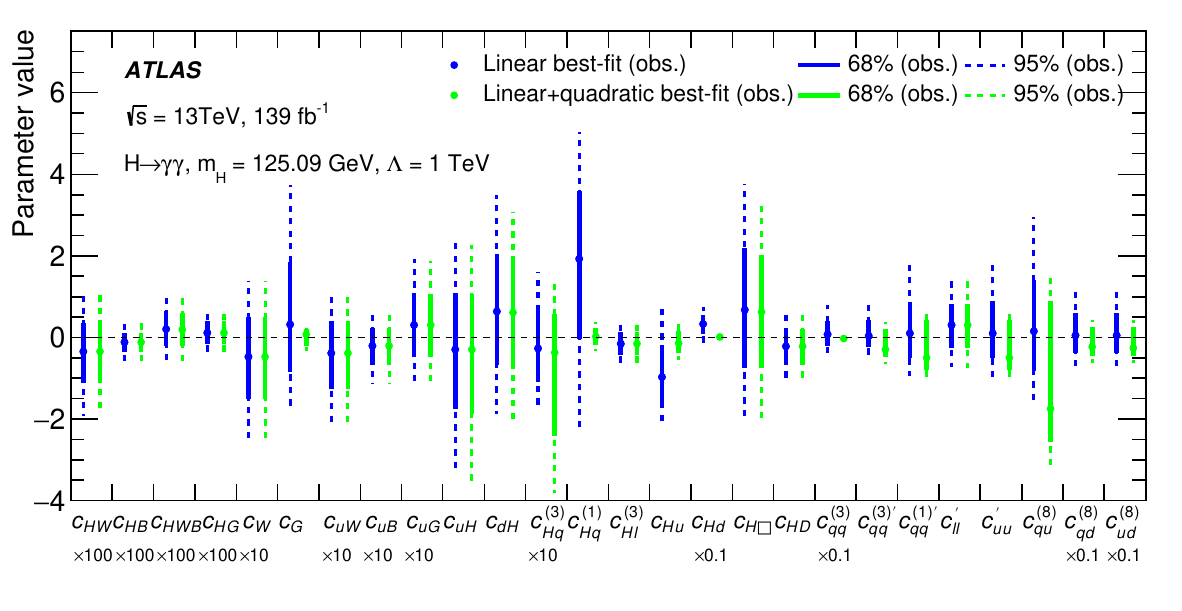}
\caption{Summary of the 68\% CL (solid lines) and 95\% CL (dashed lines) intervals for individual measurements of SMEFT parameters observed in data. In each case, SMEFT parameters other than the one measured are fixed to $0$. Blue and green curves correspond respectively to the linear and linear+quadratic SMEFT parameterizations. For presentation purposes, some parameters are scaled by a factor indicated below the parameter name. Results are not shown for coefficients $c_k$ where one or more of the intervals extend beyond the $|c_k| \le 20$ region, which is considered to be the region of validity of the SMEFT framework. }
\label{fig:eft:singleWC}
\end{figure}
Uncertainties in the parameter values range from below $\pm0.01$ to above the $|c_k| = 20$ threshold used to define the region of SMEFT validity. All values are compatible with the SM within measurement uncertainties. For parameters where the sensitivity mainly derives from inclusive event yields, such as $c_{HG}$, the linear and linear+quadratic parameterizations provide similar results. Conversely, operators with sensitivity to the high-\ptH\ bins of the \ttH\ or \ggtoH\ processes and the high-$\ptV$ regions of \qqtoVH\ show markedly smaller confidence intervals for the linear+quadratic case than for the linear case. The significant impact of the quadratic terms of the SMEFT parameterization in these cases may be indicative of significant
effects from missing higher-order terms in the SMEFT expansion.

\subsection{Simultaneous measurement of SMEFT parameters}
\label{sec:results:smeft:evn}
 
In this section, multiple SMEFT parameters are left free to vary simultaneously. The information present in the STXS measurement does not, however, allow constraints to be placed simultaneously on all the SMEFT parameters listed in Table~\ref{tab:eft:ops}. In addition, both the constrained and the unconstrained degrees of freedom generally consist in combinations of parameters, since predictions in each STXS region are affected by multiple SMEFT operators.
 
Unconstrained directions can be removed from consideration without loss of generality, since the corresponding measurement information is in any case negligible. This allows the number of measurement parameters to be reduced without incurring model-dependence, and also
avoids the probing of regions of parameter space beyond the bounds of SMEFT validity that occurs when confidence intervals along unconstrained directions extend beyond these bounds. Finally, this also avoids numerical issues in maximum-likelihood fits, since non-linear minimization algorithms can fail in cases where the local curvature of the likelihood function is too low.
 
The flat directions are identified by performing a principal component analysis of the information matrix $C^{-1}_{\text{SMEFT}}$ of the SMEFT parameter measurement. The information matrix is computed using the linear model with the assumption that the probability distribution function of the STXS measurement is approximately Gaussian. It is obtained as
\begin{equation*}
C^{-1}_{\text{SMEFT}} = P^T C^{-1}_{\text{STXS}} P
\end{equation*}
where $C^{-1}_{\text{STXS}}$ is the information matrix of the STXS measurement, computed in an Asimov data set generated under the SM hypothesis, and $P$ is the matrix representing the linear relation between the $\mu_i^{\gamma\gamma}$ and the $c_k$ in the linear SMEFT parameterization, with components given by $P_{ik} = A_k^{i \to H}  +  A_k^{\hgg} - A_k^{\Gamma}$ in the notation of Eq.~(\ref{eq:eft:param_lin}).
 
A rotation is then performed to align the measurement parameters with the eigenvectors \EVn\ of $C^{-1}_{\text{SMEFT}}$.
The unconstrained degrees of freedom of the measurement are identified with the eigenvectors corresponding to eigenvalues $\lambda_n$ of $C^{-1}_{\text{SMEFT}}$ with magnitude $\lambda_n < 0.005$. In the limit of a Gaussian measurement, each $\lambda_n$ is the inverse square of the measurement uncertainty along the direction of the corresponding \EVn, so the threshold for $\lambda_n$ corresponds to an uncertainty of about 14, which approximately corresponds to the region of SMEFT validity defined previously.
The unconstrained \EVn\ parameters are fixed to $0$ in the model, while the remaining 12 \EVn\ are considered as the measurement parameters. Their components along the $c_k$ SMEFT parameters are shown in Figure~\ref{fig:eft:rotation}. The full decomposition is shown in Table~\ref{tab:eft:EVs} of Appendix~\ref{app:eft}.
 
The EV1 parameter is mainly sensitive to the total event rates; EV2 and EV8 to the difference between the rate of \ggtoH\ and of the other production modes; EV3 and EV7 to the high-\ptV\ regions of the \pptoVH\ processes; EV4 and EV5 to the high-\pTH\ regions of the \ttH\ process; and EV6 to the rate of the \qqtoHqq\ process.
\begin{figure}[!htbp]
\centering
\includegraphics[width=1.0\textwidth]{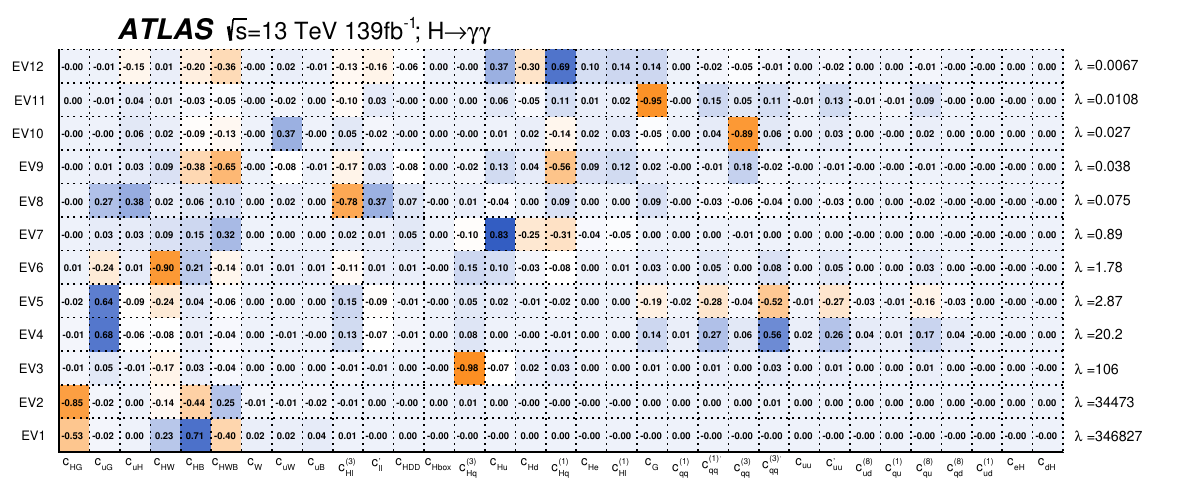}
\caption{Components of the \EVn\ parameters ($y$-axis) along each of the Warsaw-basis Wilson coefficients ($x$-axis). The \EVn\ are normalized to unit Euclidean norm. Coefficients below $0.01$ are not shown. The Warsaw-basis Wilson coefficients are defined for a scale $\Lambda = 1\,\TeV$.
The information-matrix eigenvalues ($\lambda_n$) corresponding to each eigenvector are shown on the right side of the plot.}
\label{fig:eft:rotation}
\end{figure}
Best-fit values and confidence intervals for each \EVn\ parameter are shown in Table~\ref{tab:eft:results} and illustrated in Figure~\ref{fig:eft:results}. No significant deviation from the SM is observed. In the linear parameterization, expected signal yields can become negative for some values of the SMEFT parameters, leading in some cases to a negative value of the model pdf, which invalidates the profile-likelihood computation. In these cases, the bounds of the confidence intervals are truncated at the point at which the pdf reaches 0. Results in the linear+quadratic parameterization are not affected since the expected signal yields are always positive by construction.
 
Profile likelihood scans for selected \EVn\ parameters are shown for illustration in Figure~\ref{fig:eft:ev_scans}. For some parameters, such as EV1, a broad shape is seen in the expected scan in the SM hypothesis for the linear+quadratic parameterization. This is caused in part by the presence of two degenerate minima, due to the quadratic dependence of the expected yields on the SMEFT parameters. The degeneracy is partially lifted by the fact that the observed data do not exactly correspond to the SM expectation, which leads to narrower profiles in the observed scans.
Similar scans are performed for eigenvectors corresponding to unconstrained directions with eigenvalues below $0.01$, with the measured \EVn\ also free to vary in the fits. The scans show that the measurement sensitivity in each of these directions is negligible.
 
The correlation matrix of the measurement is shown in Figure~\ref{fig:eft:correlations}. Non-zero values outside the diagonal are due to differences between the observed results and their expectations, and to the fact that the information matrix used in the principal component analysis is not an exact representation of the measurement, due to non-Gaussian effects. In the linear parameterization, these include in particular the effect of low expected event counts in some categories and the non-linear impact of some systematic uncertainties. In the linear+quadratic case, larger correlations are observed due to the effect of the quadratic terms, which are not considered in the principal component analysis. These correlations also contribute to the larger uncertainties reported in Table~\ref{tab:eft:results} for some \EVn\ parameters in the linear+quadratic parameterization, compared to the linear parameterization, and to the wider contours visible in Figure~\ref{fig:eft:ev_scans} for the linear+quadratic case. Linear parameterization results including corrections to the propagators of off-shell $W$ and $Z$ bosons, the Higgs bosons and the top quarks, as implemented in the \SMEFTsim\ generator~\cite{SMEFTsim3} are shown in Appendix~\ref{app:results:LP}.
 
\begin{table}[!htbp]
\caption{Summary of the \EVn\ parameter measurements in the linear and linear+quadratic parameterizations. The ranges correspond to 68\% CL intervals. All the \EVn\ parameters are free to vary in the fits. The upper bound of the intervals reported for EV9 and EV12 in the linear parameterization (shown in bold text) are truncated at the value for which the model pdf becomes negative.}
\label{tab:eft:results}
\centering
\renewcommand{\arraystretch}{1.5}
\adjustbox{max width=\textwidth}{
\begin{tabular}{lllll}
\toprule
Parameter &  \multicolumn{2}{c}{Linear} & \multicolumn{2}{c}{Linear+quadratic} \\
& Value & Uncertainty &  Value & Uncertainty \\
\midrule
EV1 &   $-0.0008$ & $^{       +0.0017}_{       -0.0018}$ &     ~~\,$0.0043$ & $^{       +0.0067}_{       -0.0095}$ \\
EV2 &    ~~\,$0.0004$ & $^{       +0.0059}_{       -0.0055}$ &    $-0.0061$ & $^{       +0.0084}_{       -0.0086}$ \\
EV3 &      ~~\,$0.039$ & $^{        +0.095}_{         -0.10}$ &      ~~\,$0.035$ & $^{         +0.11}_{        -0.081}$ \\
EV4 &     $-0.035$ & $^{         +0.25}_{         -0.22}$ &     $-0.079$ & $^{         +0.29}_{         -0.35}$ \\
EV5 &      $-0.22$ & $^{         +0.59}_{         -0.62}$ &       ~~\,$0.29$ & $^{         +0.30}_{         -0.69}$ \\
EV6 &       ~~\,$0.19$ & $^{         +0.81}_{         -0.80}$ &      ~~\,$0.011$ & $^{         +0.79}_{         -0.47}$ \\
EV7 &       $-1.7$ & $^{          +1.0}_{         -0.96}$ &      $-0.91$ & $^{          +1.2}_{         -0.53}$ \\
EV8 &      $-0.65$ & $^{          +3.5}_{          -3.2}$ &       $-1.2$ & $^{          +2.5}_{          -1.0}$ \\
EV9 &        ~~\,$7.5$ & $^{ \mathbf{+2.5}}_{          -5.2}$ &        ~~\,$1.7$ & $^{          +1.4}_{          -1.6}$ \\
EV10 &       ~~\,$0.48$ & $^{          +6.7}_{          -8.5}$ &       ~~\,$0.42$ & $^{         +0.46}_{         -0.60}$ \\
EV11 &       $-5.6$ & $^{          +9.4}_{          -9.6}$ &      ~~\,$0.045$ & $^{         +0.47}_{         -0.21}$ \\
EV12 &        ~~\,$2.6$ & $^{  \mathbf{+12}}_{           -13}$ &        ~~\,$1.2$ & $^{         +0.81}_{          -1.0}$ \\
\bottomrule
\end{tabular}
}
\end{table}
\begin{figure}[!htbp]
\centering
\includegraphics[width=0.8\textwidth]{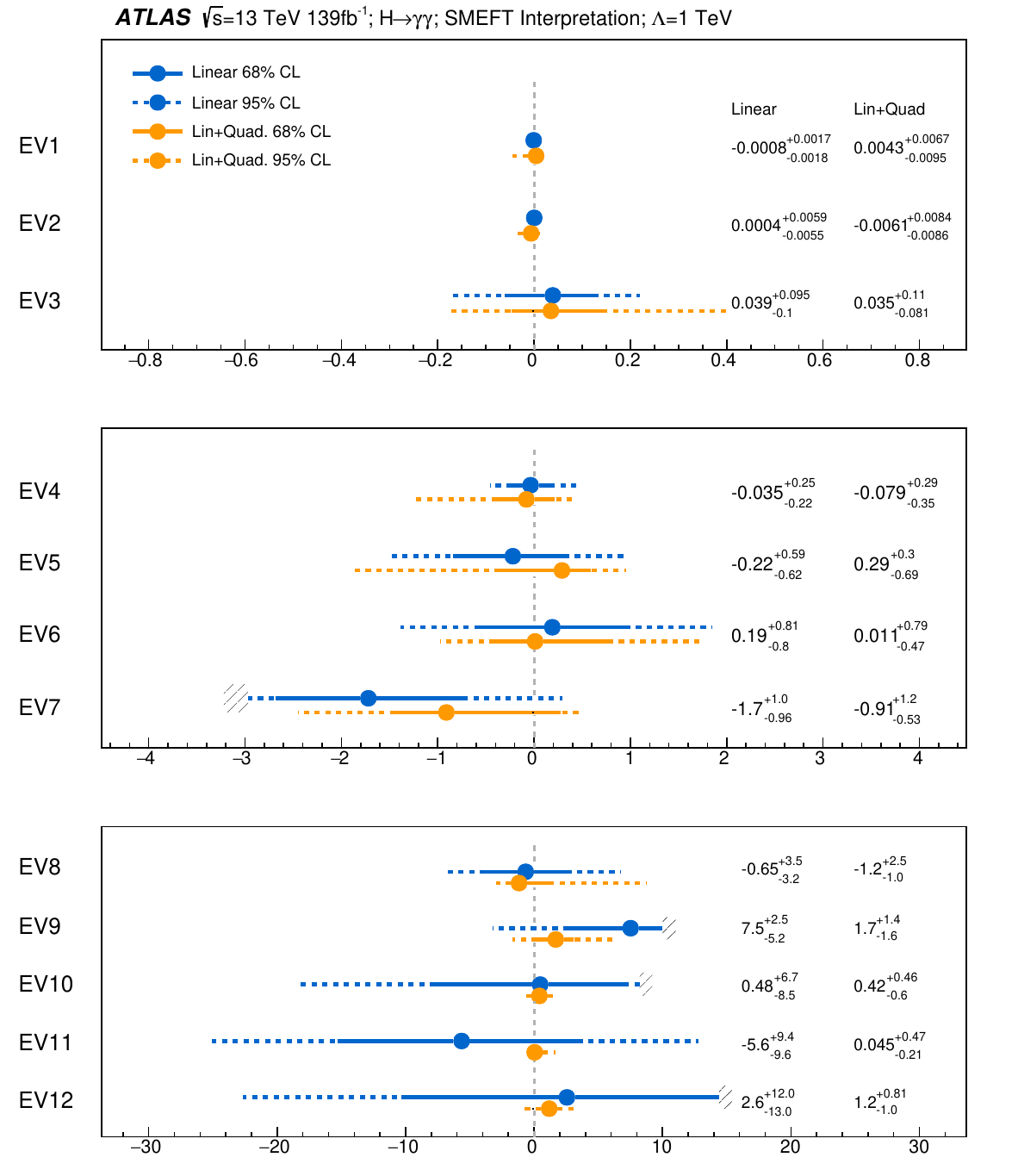}
\caption{Results of the \EVn\ parameter measurement in data, in the linear (blue) and linear+quadratic (orange) parameterizations of the SMEFT. All the \EVn\ parameters are free to vary in the fits. The ranges shown correspond to 68\% CL (solid) and 95\% CL (dashed) intervals. Cases where the intervals are truncated due to a negative model pdf are indicated by hashes.
}
\label{fig:eft:results}
\end{figure}
\begin{figure}[!htbp]
\centering
\includegraphics[width=0.49\textwidth]{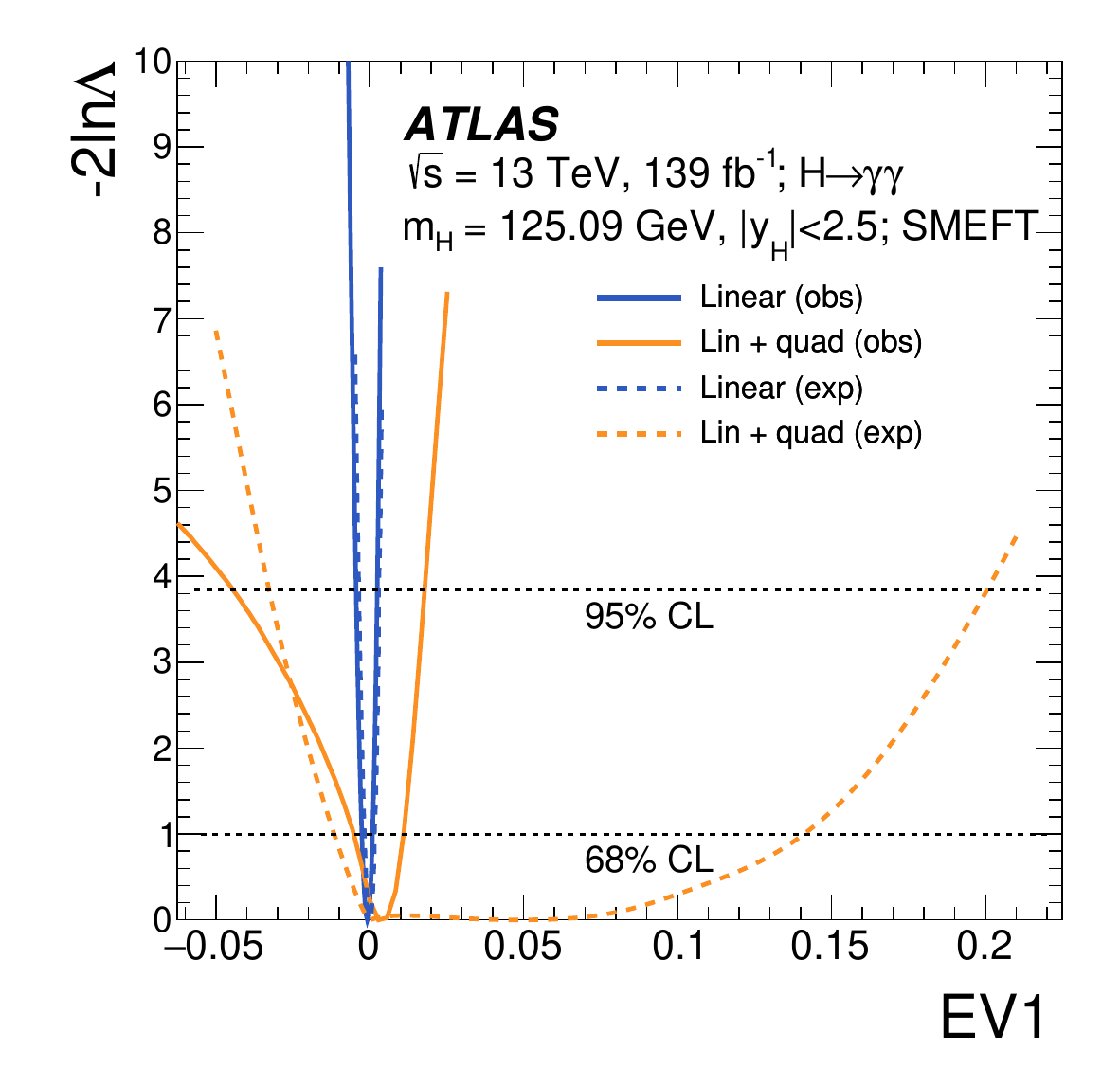}
\includegraphics[width=0.49\textwidth]{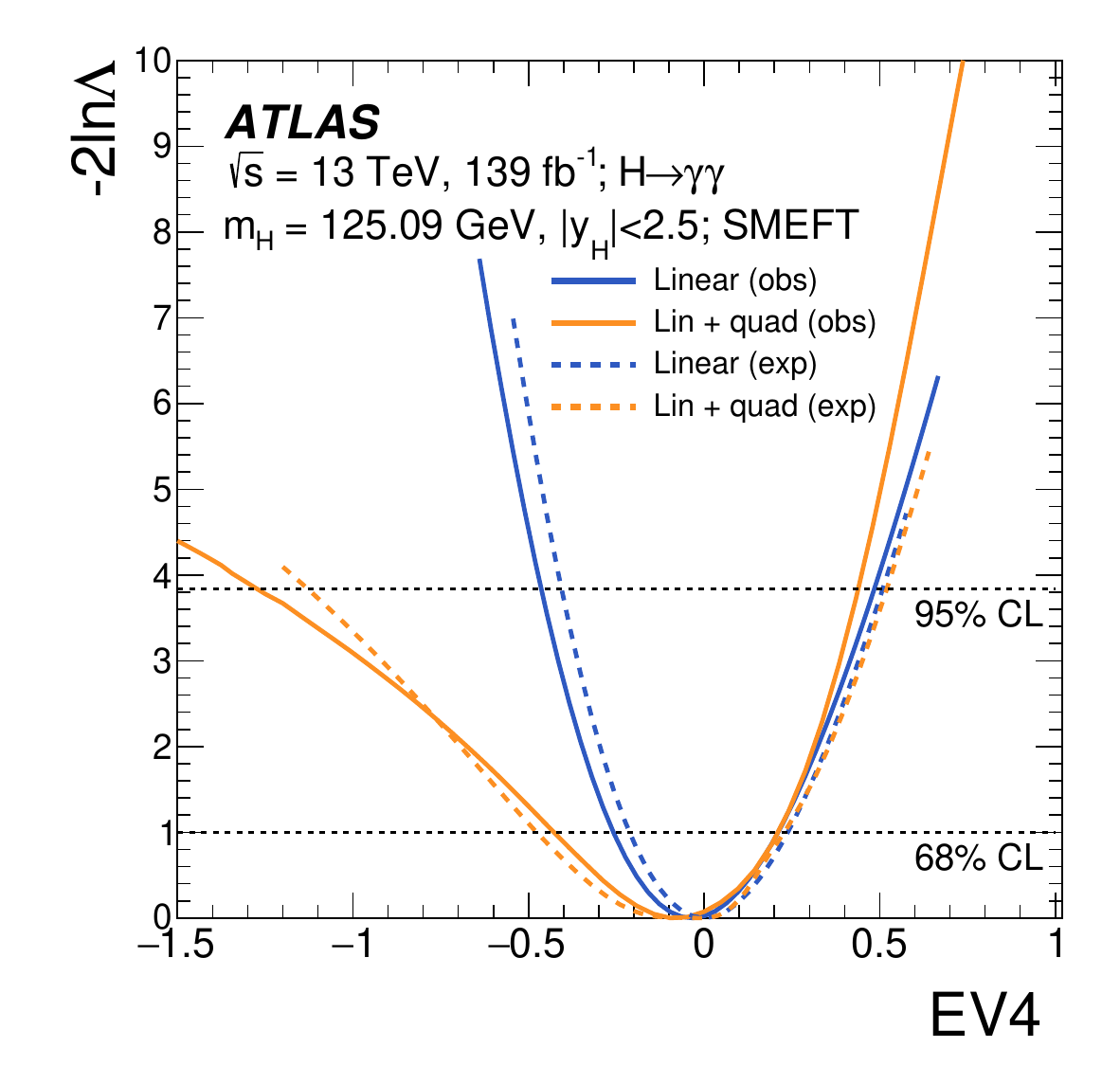}
\caption{Profile likelihood scans for the measurements of the EV1 (left) and EV4 (right) parameters in data. All measured \EVn\ parameters are free to vary in the fit. The blue and orange curves correspond respectively to the linear and linear+quadratic parameterizations. The dotted horizontal lines show the $-2\ln\Lambda=1$ and $-2\ln\Lambda=3.84$ levels that are used to define respectively the 68\% CL and 95\% CL intervals in the parameters.}
\label{fig:eft:ev_scans}
\end{figure}
\begin{figure}[!htbp]
\centering
\includegraphics[width=0.49\textwidth,page=1]{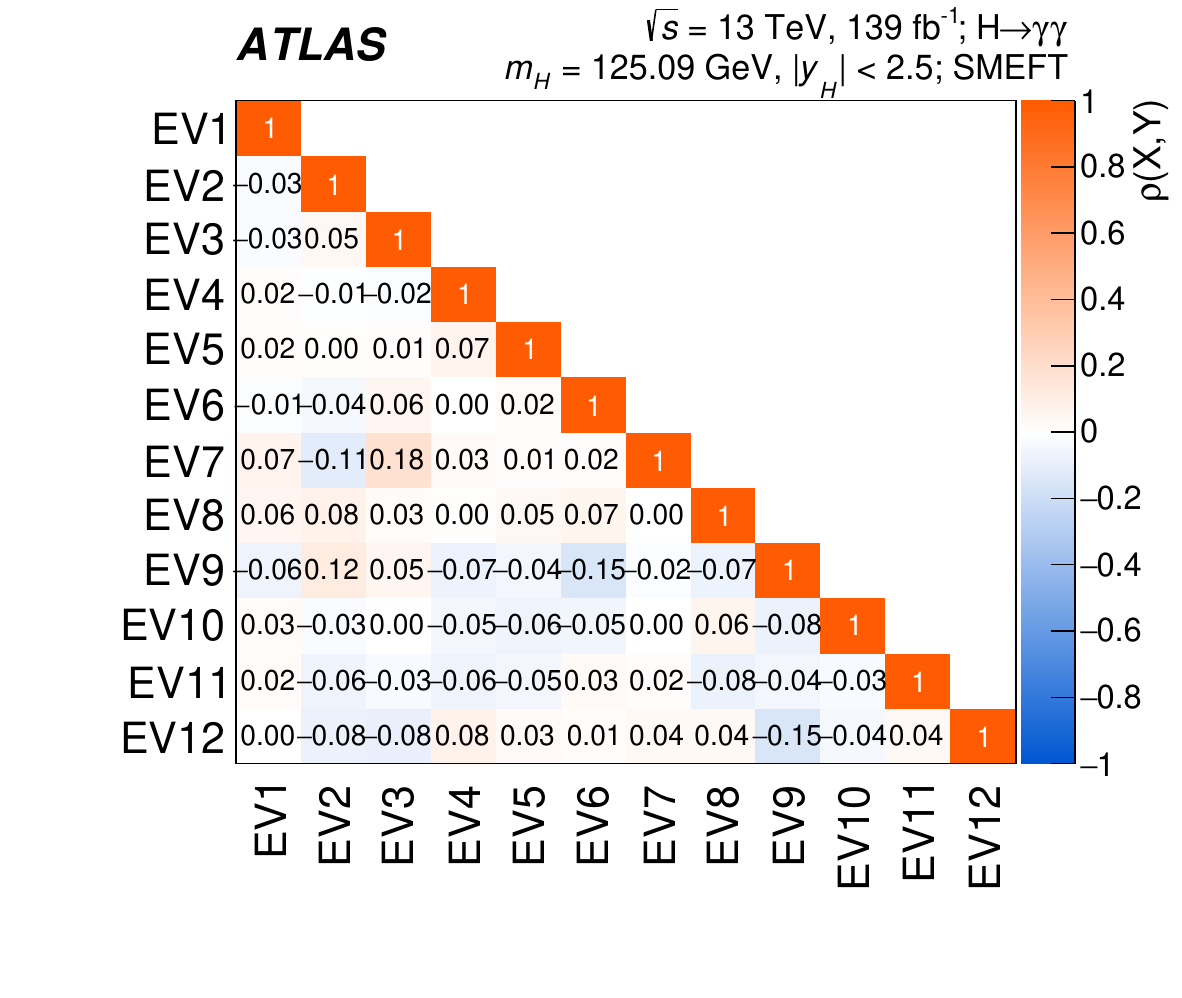}
\includegraphics[width=0.49\textwidth,page=1]{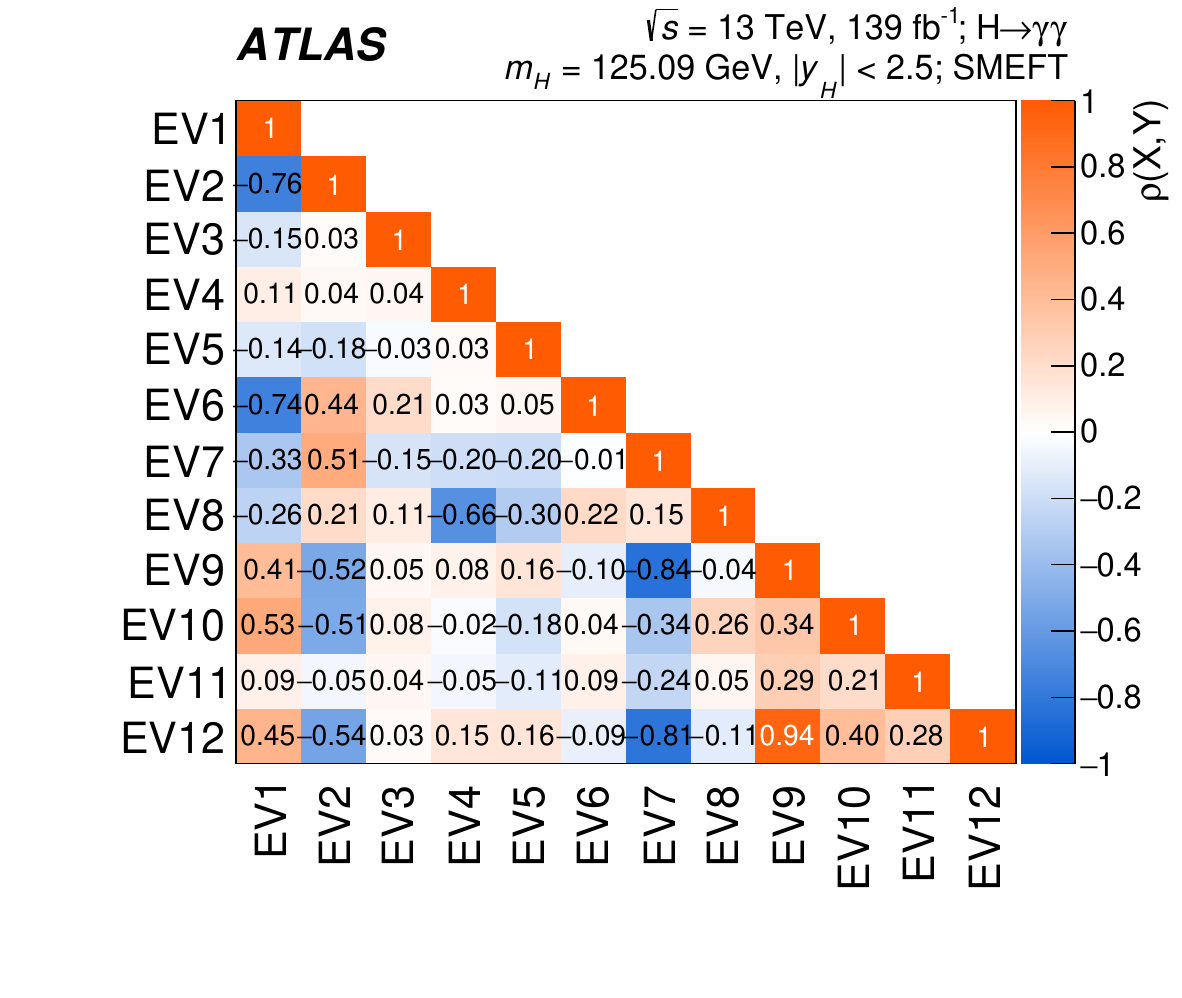}
\caption{Observed linear correlation coefficients of the \EVn\ parameters in the linear (left) and linear+quadratic (right) parameterization.}
\label{fig:eft:correlations}
\end{figure}
 
\FloatBarrier



\section{Conclusion}
\label{sec:conclusion}
 
Higgs boson production is measured in the diphoton decay channel using $139\,\ifb$ of 13~\TeV\ proton--proton collision data, corresponding to the full data set collected by ATLAS during Run~2 of the LHC.
 
The overall Higgs boson signal strength relative to its SM prediction is measured to be
\begin{equation*}
\mu = 1.04^{+0.10}_{-0.09} =
\HyynumRF{1.044}{3}
\pm 0.06 \text{ (stat.)}
\Hyynumpmerr{+0.06}{-0.05}{1} \text{ (theory syst.) }
\Hyynumpmerr{+0.05}{-0.04}{1} \text{ (exp. syst.). }
\end{equation*}
in good agreement with the SM.
 
Cross-sections for $\ggF+\bbH$, \VBF, \WH, \ZH, \ttH\ and \tH\ production are reported, with relative uncertainties of $10\%$ for $\ggF+\bbH$, $22\%$ for \VBF, and $35\%$ for \WH\ and \ttH.
An upper limit of ten times the SM prediction is set for the \tH\ process. This represents the most stringent experimental constraint on \tH\ production, superseding the previous ATLAS result from Run~2.
A fine-grained description of Higgs boson production is provided by cross-section measurements in 28 phase-space regions defined within the STXS framework, including additional measurements at high values of \ptH\ and \mjj\ compared to previous analyses. These measurements benefit from significant analysis improvements compared to previous ATLAS results~\cite{HIGG-2016-21}. A detailed classification of selected events into 101 separate categories based on multi-class machine learning techniques is used, and the uncertainties relative to the modeling of the continuum have been reduced through the use of Gaussian kernel smoothing.
 
Results are interpreted in models of Higgs boson coupling modifiers. All couplings are found to be compatible with their SM values. Sensitivity to the sign of the $\kappa_t$ modifier to the top quark coupling in the \tH\ process leads to an exclusion of the $\kappa_t < 0$ region with a significance of $2.2\sigma$.
An interpretation in the framework of SM effective field theory is used to set constraints on physics effects beyond the SM. Individual Wilson coefficients are measured while fixing the others to $0$. A simultaneous measurement of the linear combinations of Wilson coefficients that the STXS measurements are sensitive to is also performed. All results are in agreement with SM expectations.


\section*{Acknowledgements}


We thank CERN for the very successful operation of the LHC, as well as the
support staff from our institutions without whom ATLAS could not be
operated efficiently.
 
We acknowledge the support of
ANPCyT, Argentina;
YerPhI, Armenia;
ARC, Australia;
BMWFW and FWF, Austria;
ANAS, Azerbaijan;
CNPq and FAPESP, Brazil;
NSERC, NRC and CFI, Canada;
CERN;
ANID, Chile;
CAS, MOST and NSFC, China;
Minciencias, Colombia;
MEYS CR, Czech Republic;
DNRF and DNSRC, Denmark;
IN2P3-CNRS and CEA-DRF/IRFU, France;
SRNSFG, Georgia;
BMBF, HGF and MPG, Germany;
GSRI, Greece;
RGC and Hong Kong SAR, China;
ISF and Benoziyo Center, Israel;
INFN, Italy;
MEXT and JSPS, Japan;
CNRST, Morocco;
NWO, Netherlands;
RCN, Norway;
MEiN, Poland;
FCT, Portugal;
MNE/IFA, Romania;
MESTD, Serbia;
MSSR, Slovakia;
ARRS and MIZ\v{S}, Slovenia;
DSI/NRF, South Africa;
MICINN, Spain;
SRC and Wallenberg Foundation, Sweden;
SERI, SNSF and Cantons of Bern and Geneva, Switzerland;
MOST, Taiwan;
TENMAK, T\"urkiye;
STFC, United Kingdom;
DOE and NSF, United States of America.
In addition, individual groups and members have received support from
BCKDF, CANARIE, Compute Canada and CRC, Canada;
PRIMUS 21/SCI/017 and UNCE SCI/013, Czech Republic;
COST, ERC, ERDF, Horizon 2020 and Marie Sk{\l}odowska-Curie Actions, European Union;
Investissements d'Avenir Labex, Investissements d'Avenir Idex and ANR, France;
DFG and AvH Foundation, Germany;
Herakleitos, Thales and Aristeia programmes co-financed by EU-ESF and the Greek NSRF, Greece;
BSF-NSF and MINERVA, Israel;
Norwegian Financial Mechanism 2014-2021, Norway;
NCN and NAWA, Poland;
La Caixa Banking Foundation, CERCA Programme Generalitat de Catalunya and PROMETEO and GenT Programmes Generalitat Valenciana, Spain;
G\"{o}ran Gustafssons Stiftelse, Sweden;
The Royal Society and Leverhulme Trust, United Kingdom.
 
The crucial computing support from all WLCG partners is acknowledged gratefully, in particular from CERN, the ATLAS Tier-1 facilities at TRIUMF (Canada), NDGF (Denmark, Norway, Sweden), CC-IN2P3 (France), KIT/GridKA (Germany), INFN-CNAF (Italy), NL-T1 (Netherlands), PIC (Spain), ASGC (Taiwan), RAL (UK) and BNL (USA), the Tier-2 facilities worldwide and large non-WLCG resource providers. Major contributors of computing resources are listed in Ref.~\cite{ATL-SOFT-PUB-2023-001}.


\FloatBarrier
\clearpage
\appendix
\part*{Appendix}
\addcontentsline{toc}{part}{Appendix}

\clearpage
 
\section{Additional production mode cross-section and STXS measurement results}
\label{app:stxs}
 
Table~\ref{tab:results:STXS33} shows STXS results with a higher granularity than the baseline results presented in Section~\ref{sec:results:STXS}. A total of 33 regions are measured, with the following changes compared to the 28 regions in the baseline measurement:
\begin{itemize}
\item For the \ggtoH\ process, within the phase space of $\ge~$2-jets, $\mjj < 350\,\GeV$, the two regions with $\ptH < 60\,\GeV$ and $60 \le \ptH < 120\,\GeV$ are kept separate. The same also applies to the three bins in the \mjj\ variable within the $\ge~$2-jets,  $\ptH<200\,\GeV$ region, which are not  merged. Compared to the STXS analysis regions defined in Section~\ref{sec:design:overview}, the only merging that is performed is that of the $\ptH > 650\,\GeV$ bin with the neighbouring $450 \le \ptH < 650\,\GeV$ bin.
\item For the \qqtoHqq\ process, the 0-jet and 1-jet regions are merged into a single $\le 1$-jet bin, and the $\mjj < 60\,\GeV$ and $120 < \mjj < 350\,\GeV$ regions are also combined into a new \VH-veto region, but the two sets are not merged together as in the baseline results. The three regions in the \mjj\ variables within $\ge~$2-jets, $\ptH \ge 200\,\GeV$, are also not merged.
\end{itemize}
The correlation matrix of the measurement is shown in Fig.~\ref{fig:results:STXS_corr_full}.

\begin{table}[ht]
\caption{
Best-fit values and uncertainties for the production cross-section times \Hyy\ branching ratio $(\sigma_i \times \Byy)$ in each STXS region. The values for the \ggtoH\ process also include the contributions from \bbH\ production. The total uncertainties are decomposed into components for data statistics (Stat.) and systematic uncertainties (Syst.). SM predictions~\cite{deFlorian:2016spz} are also shown for each quantity with their total uncertainties.
}
\centering
\renewcommand{\arraystretch}{1.4}
\resizebox{0.9\textwidth}{!}{
\begin{tabular}{lrlll@{ }l@{ }}
\toprule
\multirow{2}{*}{STXS region $(\sigma_i \times \Byy)$} & \multicolumn{1}{c}{Value} & \multicolumn{3}{c}{ Uncertainty [fb]} & \multicolumn{1}{c}{SM prediction} \\
&  \multicolumn{1}{c}{[fb]}                & Total   & Stat.                & Syst.   & \multicolumn{1}{c}{[fb]}   \\
\midrule
\ggHjPt{0}{}{10}{}                 &  \numRF{10.0869}{2}~~~~~~~ &  \Hyynumpmerr{+4.2385}{-4.0395}{1} &  \Hyynumpmerr{+3.7905}{-3.7767}{1} &  \Hyynumpmerr{+1.8967}{-1.4333}{1} &  \quad~~~~~${\numRF{15.0660}{2}}^{+\numRF{1.9682}{1}}_{\numRF{-1.9652}{1}}$ \\
\ggHjPt{0}{10}{}{}                 &  \numRF{58.5821}{2}~~~~~~~ &  \Hyynumpmerr{+8.7205}{-8.2474}{1} &  \Hyynumpmerr{+7.0984}{-7.0870}{1} &  \Hyynumpmerr{+5.0655}{-4.2182}{1} &  \quad~~~~~${\numRF{46.8570}{2}}^{+\numRF{3.5802}{1}}_{\numRF{-3.6136}{1}}$ \\
\ggHjPt{1}{}{60}{}                 &  \numRF{17.2551}{2}~~~~~~~ &  \Hyynumpmerr{+5.6323}{-5.4273}{1} &  \Hyynumpmerr{+5.1618}{-5.1532}{1} &  \Hyynumpmerr{+2.2535}{-1.7028}{1} &  \quad~~~~~${\numRF{14.7580}{2}}^{+\numRF{2.0413}{1}}_{\numRF{-2.0459}{1}}$ \\
\ggHjPt{1}{60}{120}{}              &  \numRF{12.7946}{2}~~~~~~~ &  \Hyynumpmerr{+4.0979}{-3.6810}{1} &  \Hyynumpmerr{+3.2648}{-3.2857}{1} &  \Hyynumpmerr{+2.4768}{-1.6594}{1} &  \quad~~~~~${\numRF{10.2220}{2}}^{+\numRF{1.4311}{1}}_{\numRF{-1.4320}{1}}$ \\
\ggHjPt{1}{120}{200}{}             &   \numRF{1.9433}{2}~~~~ &  \Hyynumpmerr{+0.9909}{-0.9113}{1} &  \Hyynumpmerr{+0.9005}{-0.8889}{1} &  \Hyynumpmerr{+0.4136}{-0.2007}{1} &  \quad~~~~${\numRF{ 1.6960}{2}}^{+\numRF{0.2960}{1}}_{\numRF{-0.2958}{1}}$ \\
\ggHmPt{}{350}{}{60}{}             &   \numRF{0.2506}{1}~~~~ &  \Hyynumpmerr{+3.0618}{-2.8842}{2} &  \Hyynumpmerr{+2.8610}{-2.8402}{2} & $\ensuremath{^{+1.1}_{-0.5}}$ &
\quad~~~~${\numRF{ 2.6550}{2}}^{+\numRF{0.5770}{1}}_{\numRF{-0.5774}{1}}$ \\
\ggHmPt{}{350}{60}{120}{}          &   \numRF{1.3145}{2}~~~~ &  \Hyynumpmerr{+2.5061}{-2.4585}{2} &  \Hyynumpmerr{+2.4083}{-2.3852}{2} &  \Hyynumpmerr{+0.6934}{-0.5959}{1} &  \quad~~~~${\numRF{ 4.0730}{2}}^{+\numRF{0.8507}{1}}_{\numRF{-0.8508}{1}}$ \\
\ggHmPt{}{350}{120}{200}{}         &   \numRF{2.1999}{2}~~~~ &  \Hyynumpmerr{+1.0950}{-1.0517}{2} &  \Hyynumpmerr{+1.0606}{-1.0364}{2} &  \Hyynumpmerr{+0.2722}{-0.1785}{1} &  \quad~~~~${\numRF{ 2.1410}{2}}^{+\numRF{0.4921}{1}}_{\numRF{-0.4919}{1}}$ \\
\ggHmPt{350}{700}{}{200}{}         &   \numRF{2.6850}{2}~~~~ &  \Hyynumpmerr{+1.6109}{-1.5295}{2} &  \Hyynumpmerr{+1.5355}{-1.4974}{2} &  \Hyynumpmerr{+0.4872}{-0.3115}{1} &  \quad~~~~${\numRF{ 1.3890}{2}}^{+\numRF{0.3385}{1}}_{\numRF{-0.3384}{1}}$ \\
\ggHmPt{700}{1000}{}{200}{}        &  \numRF{-0.2025}{1}~~~~ &  \Hyynumpmerr{+0.7377}{-0.7537}{1} &  \Hyynumpmerr{+0.7077}{-0.6777}{1} &  \Hyynumpmerr{+0.2083}{-0.3299}{1} &  \quad~~~~${\numRF{ 0.3270}{1}}^{+\numRF{0.0950}{1}}_{\numRF{-0.0950}{1}}$ \\
\ggHmPt{1000}{}{}{200}{}           &  \numRF{-0.2562}{1}~~~~ &  \Hyynumpmerr{+0.5844}{-0.5699}{1} &  \Hyynumpmerr{+0.5641}{-0.5252}{1} &  \Hyynumpmerr{+0.1524}{-0.2213}{1} &  \quad~~${\numRF{ 0.2760}{2}}^{+\numRF{0.0821}{1}}_{\numRF{-0.0820}{1}}$ \\
\ggHPt{200}{300}{}                 &   \numRF{1.5106}{2}~~~~ &  \Hyynumpmerr{+0.4560}{-0.4241}{1} &  \Hyynumpmerr{+0.4192}{-0.4045}{1} &  \Hyynumpmerr{+0.1795}{-0.1272}{1} &  \quad~~~~${\numRF{ 1.0400}{2}}^{+\numRF{0.2346}{1}}_{\numRF{-0.2343}{1}}$ \\
\ggHPt{300}{450}{}                 &   \numRF{0.0110}{1}~~ &  \Hyynumpmerr{+0.1322}{-0.1186}{2} &  \Hyynumpmerr{+0.1282}{-0.1128}{2} &  \Hyynumpmerr{+0.0323}{-0.0365}{1} &  \quad~~${\numRF{ 0.2410}{2}}^{+\numRF{0.0612}{1}}_{\numRF{-0.0611}{1}}$ \\
\ggHPt{450}{}{}                    &   \numRF{0.0799}{1}~~ &  \Hyynumpmerr{+0.0598}{-0.0482}{1} &  \Hyynumpmerr{+0.0578}{-0.0473}{1} & 
$\ensuremath{^{+0.02}_{-0.01}}$ &
\quad~~${\numRF{ 0.0410}{1}}^{+\numRF{0.0120}{1}}_{\numRF{-0.0119}{1}}$ \\
\qqtoHqq, $\le 1$-jet            &   \numRF{0.6800}{1}~~~~ &  \Hyynumpmerr{+4.9487}{-4.4098}{2} &  \Hyynumpmerr{+4.8098}{-4.1819}{2} &  \Hyynumpmerr{+1.1645}{-1.3995}{2} &  \quad~~~~${\numRF{ 4.9090}{2}}^{+\numRF{0.1605}{1}}_{\numRF{-0.1620}{1}}$ \\
\qqtoHqq, \VH-veto                &   \numRF{4.3369}{1}~~~~~~~ &  \Hyynumpmerr{+2.7688}{-2.5692}{1} &  \Hyynumpmerr{+2.6243}{-2.4861}{1} & 
$\ensuremath{^{+1}_{-1}}$ &
\quad~~${\numRF{ 1.6690}{3}}^{+\numRF{0.0530}{1}}_{\numRF{-0.0536}{1}}$ \\
\qqtoHqq, \VH-had               &   \numRF{0.3623}{1}~~~~ &  \Hyynumpmerr{+0.8801}{-0.7526}{1} &  \Hyynumpmerr{+0.8616}{-0.7371}{1} &  \Hyynumpmerr{+0.1797}{-0.1516}{1} &  \quad~~${\numRF{ 1.1580}{3}}^{+\numRF{0.0428}{1}}_{\numRF{-0.0433}{1}}$ \\
\HqqmPt{350}{700}{}{200}{}         &   \numRF{1.2097}{2}~~~~ &  \Hyynumpmerr{+0.8951}{-0.7313}{1} &  \Hyynumpmerr{+0.7581}{-0.6874}{1} &  \Hyynumpmerr{+0.4759}{-0.2497}{1} &  \quad~~${\numRF{ 1.2150}{3}}^{+\numRF{0.0373}{1}}_{\numRF{-0.0377}{1}}$ \\
\HqqmPt{700}{1000}{}{200}{}        &   \numRF{1.0749}{2}~~~~ &  \Hyynumpmerr{+0.6325}{-0.4794}{1} &  \Hyynumpmerr{+0.4923}{-0.4404}{1} &  \Hyynumpmerr{+0.3971}{-0.1894}{1} &  \quad~~${\numRF{ 0.5810}{2}}^{+\numRF{0.0188}{1}}_{\numRF{-0.0190}{1}}$ \\
\HqqmPt{1000}{}{}{200}{}           &   \numRF{1.4211}{2}~~~~ &  \Hyynumpmerr{+0.5407}{-0.4286}{1} &  \Hyynumpmerr{+0.3911}{-0.3571}{1} &  \Hyynumpmerr{+0.3733}{-0.2369}{1} &  \quad~~${\numRF{ 1.00}{3}}^{+\numRF{0.0315}{1}}_{\numRF{-0.0319}{1}}$ \\
\HqqmPt{350}{700}{200}{}{}         &   \numRF{0.1176}{2}~~ &  \Hyynumpmerr{+0.1510}{-0.1234}{2} &  \Hyynumpmerr{+0.1471}{-0.1206}{2} &  \Hyynumpmerr{+0.0341}{-0.0259}{1} &  \quad${\numRF{ 0.1000}{3}}^{+\numRF{0.0030}{1}}_{\numRF{-0.0030}{1}}$ \\
\HqqmPt{700}{1000}{200}{}{}        &  \numRF{-0.0089}{1} &  \Hyynumpmerr{+0.0568}{-0.0414}{2} &  \Hyynumpmerr{+0.0565}{-0.0406}{2} &  \Hyynumpmerr{+0.0056}{-0.0080}{1} &  \quad${\numRF{ 0.0670}{2}}^{+\numRF{0.0021}{1}}_{\numRF{-0.0021}{1}}$ \\
\HqqmPt{1000}{}{200}{}{}           &   \numRF{0.2814}{2}~~ &  
$\ensuremath{^{+0.11}_{-0.09}}$ &
\Hyynumpmerr{+0.0993}{-0.0851}{1} &
\Hyynumpmerr{+0.0469}{-0.0381}{1} &  \quad${\numRF{ 0.1660}{3}}^{+\numRF{0.0052}{1}}_{\numRF{-0.0052}{1}}$ \\
\HlnPt{}{150}{}                    &   \numRF{1.4053}{2}~~~~ &  \Hyynumpmerr{+0.6477}{-0.5792}{1} &  \Hyynumpmerr{+0.6345}{-0.5719}{1} & 
$\ensuremath{^{+0.1}_{-0.1}}$ &
\quad~~${\numRF{ 0.7930}{2}}^{+\numRF{0.0224}{1}}_{\numRF{-0.0229}{1}}$ \\
\HlnPt{150}{}{}                    &   \numRF{0.1978}{2}~~ &  \Hyynumpmerr{+0.1336}{-0.1069}{2} &  \Hyynumpmerr{+0.1325}{-0.1062}{2} &  \Hyynumpmerr{+0.0170}{-0.0117}{1} &  \quad${\numRF{ 0.1210}{3}}^{+\numRF{0.0054}{1}}_{\numRF{-0.0054}{1}}$ \\
\HllnnPt{}{150}{}                  &  \numRF{-0.2881}{2}~~ &  
$\ensuremath{^{+0.40}_{-0.07}}$ &
$\ensuremath{^{+0.39}_{-0.07}}$ &
$\ensuremath{^{+0.07}_{-0.00}}$ &
\quad~~${\numRF{ 0.4510}{2}}^{+\numRF{0.0187}{1}}_{\numRF{-0.0189}{1}}$ \\
\HllnnPt{150}{}{}                  &   \numRF{0.0366}{1}~~ &  
$\ensuremath{^{+0.10}_{-0.08}}$ &
$\ensuremath{^{+0.10}_{-0.08}}$ &
\Hyynumpmerr{+0.0181}{-0.0162}{1} &  \quad~~${\numRF{ 0.0920}{1}}^{+\numRF{0.0109}{1}}_{\numRF{-0.0110}{1}}$ \\
\ttHPt{}{60}{}                    &   \numRF{0.2141}{2}~~ &  \Hyynumpmerr{+0.2128}{-0.1775}{2} &  \Hyynumpmerr{+0.2103}{-0.1769}{2} &  \Hyynumpmerr{+0.0328}{-0.0143}{1} &  \quad~~${\numRF{ 0.2680}{2}}^{+\numRF{0.0365}{1}}_{\numRF{-0.0365}{1}}$ \\
\ttHPt{60}{120}{}                  &   \numRF{0.3153}{2}~~ &  \Hyynumpmerr{+0.2338}{-0.1961}{2} &  \Hyynumpmerr{+0.2311}{-0.1952}{2} &  \Hyynumpmerr{+0.0355}{-0.0182}{1} &  \quad~~${\numRF{ 0.4040}{2}}^{+\numRF{0.0455}{1}}_{\numRF{-0.0454}{1}}$ \\
\ttHPt{120}{200}{}                 &   \numRF{0.1796}{2}~~ &  \Hyynumpmerr{+0.1786}{-0.1475}{2} &  \Hyynumpmerr{+0.1746}{-0.1457}{2} &  \Hyynumpmerr{+0.0378}{-0.0234}{1} &  \quad~~${\numRF{ 0.2870}{2}}^{+\numRF{0.0357}{1}}_{\numRF{-0.0357}{1}}$ \\
\ttHPt{200}{300}{}                 &   \numRF{0.1411}{2}~~ &  \Hyynumpmerr{+0.0927}{-0.0738}{1} &  \Hyynumpmerr{+0.0917}{-0.0735}{1} &  \Hyynumpmerr{+0.0137}{-0.01}{1} &  \quad~~${\numRF{ 0.1190}{2}}^{+\numRF{0.0171}{1}}_{\numRF{-0.0171}{1}}$ \\
\ttHPt{300}{}{}                    &   \numRF{0.0626}{1}~~ &  \Hyynumpmerr{+0.0510}{-0.0396}{1} &  \Hyynumpmerr{+0.0503}{-0.0392}{1} &  \Hyynumpmerr{+0.01}{-0.01}{1} &  \quad~~${\numRF{ 0.0550}{1}}^{+\numRF{0.01}{1}}_{\numRF{-0.01}{1}}$ \\
\tH                                &   \numRF{0.3624}{2}~~ &  \Hyynumpmerr{+0.7568}{-0.5960}{2} &  \Hyynumpmerr{+0.7188}{-0.5723}{2} &  \Hyynumpmerr{+0.2368}{-0.1664}{2} &  \quad~~${\numRF{ 0.1920}{2}}^{+\numRF{0.0126}{1}}_{\numRF{-0.0246}{1}}$ \\
\bottomrule
\end{tabular}}
\label{tab:results:STXS33}
\end{table}


\begin{figure}[tpb!]
\centering
\includegraphics[width=.995\textwidth]{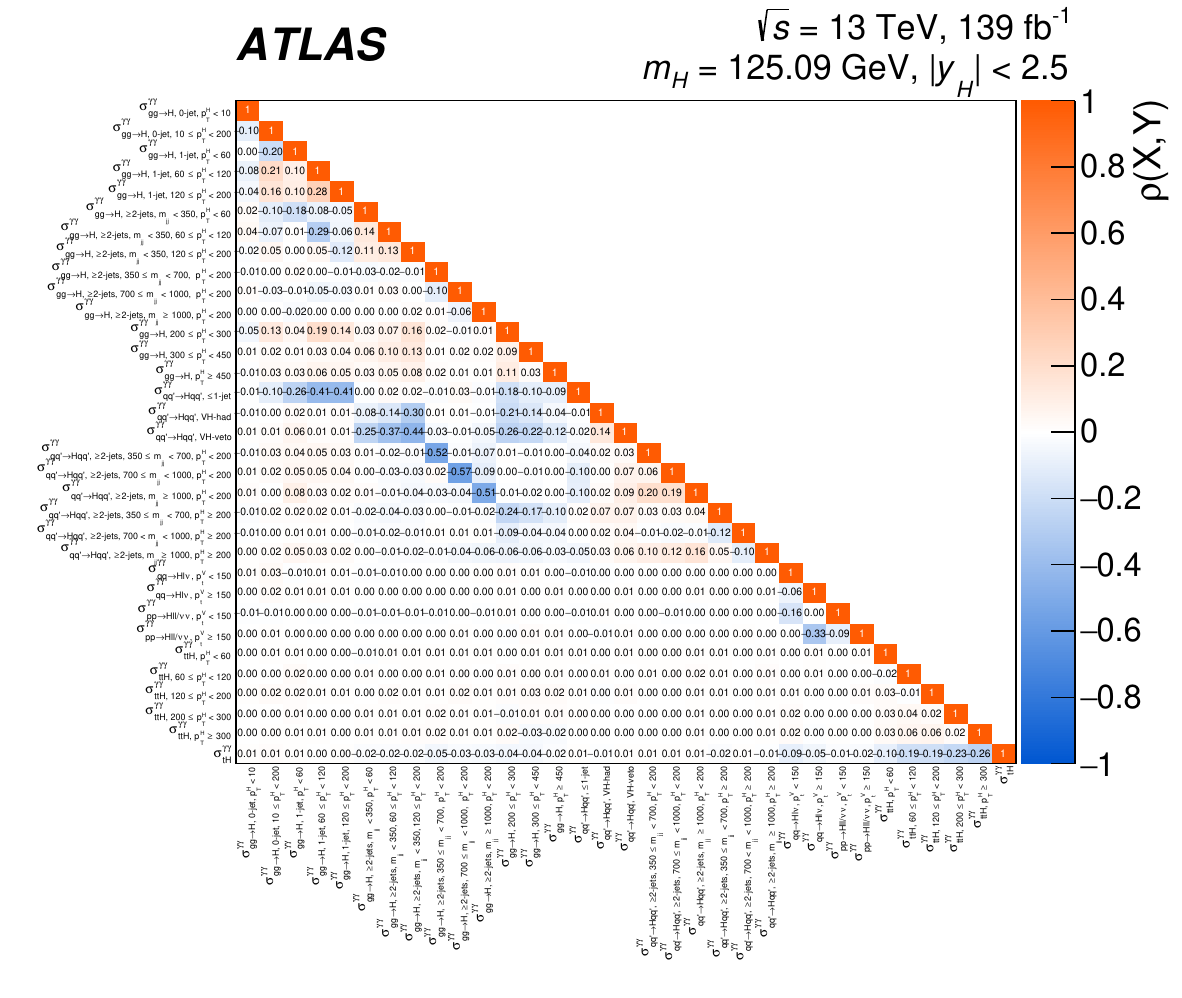}
\caption{Correlation matrix for the measurement of STXS parameters in each of the 33 regions considered.}
\label{fig:results:STXS_corr_full}
\end{figure}

\FloatBarrier

\section{Additional $\kappa$-framework interpretations}
\label{app:kappas}
 
\subsection{Parameterization of STXS cross-section parameters and the \hgg\ branching ratio}
 
Multiplicative modifiers are considered for Higgs boson couplings to the $W$ and $Z$ bosons (respectively $\kappa_W$ and $\kappa_Z$), and for couplings to the charm ($\kappa_c$), bottom ($\kappa_b$) and top ($\kappa_t$) quarks and the muon ($\kappa_\mu$) and $\tau$ ($\kappa_\tau$) leptons. Couplings to other SM particles are assumed to be equal to their SM predictions.
 
Table~\ref{tab:kappas:parameterization} presents the multiplicative corrections that are applied to the STXS cross-section parameters $\sigma_i$, the partial decay widths $\Gamma_{\yy}$, $\Gamma_{gg}$ and $\Gamma_{Z\gamma}$ of the \hgg, \Hgg\ and \Hzg\ decays, respectively, and the total width $\Gamma_H$. The symbols $\Gamma_{gg}^{\text{SM}}$ and $\Gamma_{Z\gamma}^{\text{SM}}$ denote the SM predictions for $\Gamma_{gg}$ and $\Gamma_{Z\gamma}$ respectively.
The total width $\Gamma_H$ is expressed as a function of the $\kappa$ modifiers, assuming no contributions from Higgs boson decays other than the ones present in the SM, except in the model in Appendix~\ref{sec:kappa:ratio} in which an effective description in terms of the $\kappa_H$ modifier is used instead.
 
The SM corresponds to the case $\kappa_W = \kappa_Z = \kappa_t = \kappa_b = \kappa_c = \kappa_\tau = \kappa_\mu = 1$, and in addition $\kappa_g = \kappa_\gamma = 1$ when effective parameterizations are used. The $\kappa_Z$ modifier is assumed to be positive, without loss of generality, since all predictions are invariant under a simultaneous flip of the sign of each $\kappa$ modifier. Sensitivity to $\kappa_b$ and $\kappa_\tau$ is achieved through the contributions of bottom quarks and $\tau$-leptons to these loop processes in the resolved description.

The SM predictions of $B_{\gamma\gamma}$ and the $\sigma_i$ are taken from Ref~\cite{deFlorian:2016spz}. These include the highest-order available computations in both the QCD and electroweak couplings.
 
\begin{table}[ht]
\caption{Parameterization of Higgs boson production cross-sections $\sigma_i$, the partial decay widths $\Gamma_{\yy}$, $\Gamma_{gg}$ and $\Gamma_{Z\gamma}$ of the \hgg, \Hgg\ and \Hzg\ decays, respectively, and the total width $\Gamma_H$, normalized to their SM values, as functions of the coupling-strength modifiers $\kappa$. The coefficients for $\sigma(\tHW)$ and $\sigma(\tHqb)$ include acceptance effects that differ between analysis categories as described in the text. Other coefficients are derived following the methodology in Refs.~\cite{LHCHiggsCrossSectionWorkingGroup:2013rie,deFlorian:2016spz}.}
\centering
\renewcommand{\arraystretch}{1.3}
\begin{tabular}{lccl}
\hline \hline
Production                          & Main                      & Effective          & \multirow{2}{*}{Resolved modifier}                                 \\
cross-section                       & interference              & modifier           &                                                                    \\
\hline
$\sigma({\ggF})$                    & $t$--$b$                  & $\kappa_g^2$       & $1.040\,\kappa_t^2 + 0.002 \kappa_b^2 - 0.038 \, \kappa_t \kappa_b - 0.005 \, \kappa_t \kappa_c$ \\
$\sigma({\VBF})$                    & -                         & -                  & $0.733\,\kappa_W^2 + 0.267\,\kappa_Z^2$                                       \\
$\sigma({\qqtoZH})$                   & -                         & -                  & $\kappa_Z^2$                                                                  \\
\multirow{2}{*}{$\sigma({\ggtoZH})$}  & \multirow{2}{*}{$t$--$Z$} & \multirow{2}{*}{-} & $2.456\,\kappa_Z^2  +  0.456\,\kappa_t^2 - 1.903\,\kappa_Z \kappa_t$          \\
&                           &                    & $\; -\,0.011\,\kappa_Z \kappa_b + 0.003\, \kappa_t \kappa_b$                  \\
$\sigma({\WH})$                     & -                         & -                  & $\kappa_W^2$                                                                  \\
$\sigma({\ttH})$                    & -                         & -                  & $\kappa_t^2$                                                                  \\
$\sigma({\tHW})$                    & $t$--$W$                  & -                  & $A\,\kappa_t^2 + B\,\kappa_W^2 + C\,\kappa_t \kappa_W$, category-dependent    \\
$\sigma({\tHqb})$                    & $t$--$W$                  & -                  & $A\,\kappa_t^2 + B\,\kappa_W^2 + C\,\kappa_t \kappa_W$, category-dependent     \\
$\sigma({\bbH})$                    & -                         & -                  & $\kappa_b^2$                                                                  \\
\hline\hline
\multicolumn{4}{l}{Partial and total decay widths}  \\
\hline
\multirow{2}{*}{$\Gamma_{\yy}$}     & \multirow{2}{*}{$t$--$W$} & \multirow{2}{*}{$\kappa_\gamma^2$}  & $1.589\,\kappa_W^2 + 0.072\,\kappa_t^2 - 0.674\,\kappa_W \kappa_t + 0.009\,\kappa_W \kappa_\tau$ \\
&                           &                                     & $\;{} + 0.008 \, \kappa_W \kappa_b - 0.002\,\kappa_t \kappa_b - 0.002 \, \kappa_t \kappa_\tau$   \\
$\Gamma_{gg}$      & $t$--$b$ & $\kappa_g^2$  & $1.111\,\kappa_t^2 + 0.012\,\kappa_b^2 - 0.123 \, \kappa_t \kappa_b$ \\
$\Gamma_{Z\gamma}$ & $t$--$W$ &               & $1.118\,\kappa_W^2 + 0.004\,\kappa_t^2 - 0.125 \, \kappa_W\kappa_t + 0.003 \, \kappa_W \kappa_b$ \\
\hline
\multirow{5}{*}{$\Gamma_H$}         & \multirow{5}{*}{-}        & \multirow{5}{*}{$\kappa_H^2$} & $0.581\,\kappa_b^2 + 0.215\,\kappa_W^2 + 0.063\,\kappa_\tau^2$  \\
&                           &                               & $\;{} + 0.026\,\kappa_Z^2 + 0.029\,\kappa_c^2 + 0.0023\,\kappa_\gamma^2$ \\
&                           &                               & $\;{} + 0.0004\,\kappa_s^2 + 0.00022\,\kappa_\mu^2$   \\
&                           &                               & $\;{} + 0.082\, (\Gamma_{gg}/\Gamma_{gg}^{\text{SM}})$  \\
&                           &                               & $\;{} + 0.0015\, (\Gamma_{Z\gamma}/\Gamma_{Z\gamma}^{\text{SM}})$  \\
\hline \hline
\end{tabular}
\label{tab:kappas:parameterization}
\end{table}

\FloatBarrier
 
\subsection{Parameterization with universal coupling modifiers to weak gauge bosons and fermions}
 
In this model, two universal coupling modifiers are considered: $\kappa_V = \kappa_W = \kappa_Z$ which modifies Higgs boson couplings to gauge bosons, and $\kappa_F = \kappa_t = \kappa_b = \kappa_c = \kappa_\tau = \kappa_\mu$, modifying couplings to fermions.
The \ggtoH, \Hyy\ and \ggtoZH\ loops are described using their resolved parameterizations as a function of $\kappa_V$ and $\kappa_F$. The measurement in the plane of $(\kappa_V, \kappa_F)$ is shown in Figure~\ref{fig:kappas:kVkF}. Only the region $\kappa_F > 0$ is considered, since $\kappa_F < 0$ was excluded with a significance larger than $4\sigma$ in analyses of the Run~1 data set~\cite{HIGG-2015-07}.
The best-fit values in data are
\begin{align*}
\kappa_V &= 1.02 ^{+\;0.06}_{-\;0.05}\\
\kappa_F &= 1.00 ^{+\;0.16}_{-\;0.13}.
\end{align*}
A linear correlation coefficient of $77\%$ between the parameters is observed.
\begin{figure}[h!]
\centering
\includegraphics[width=.8\textwidth]{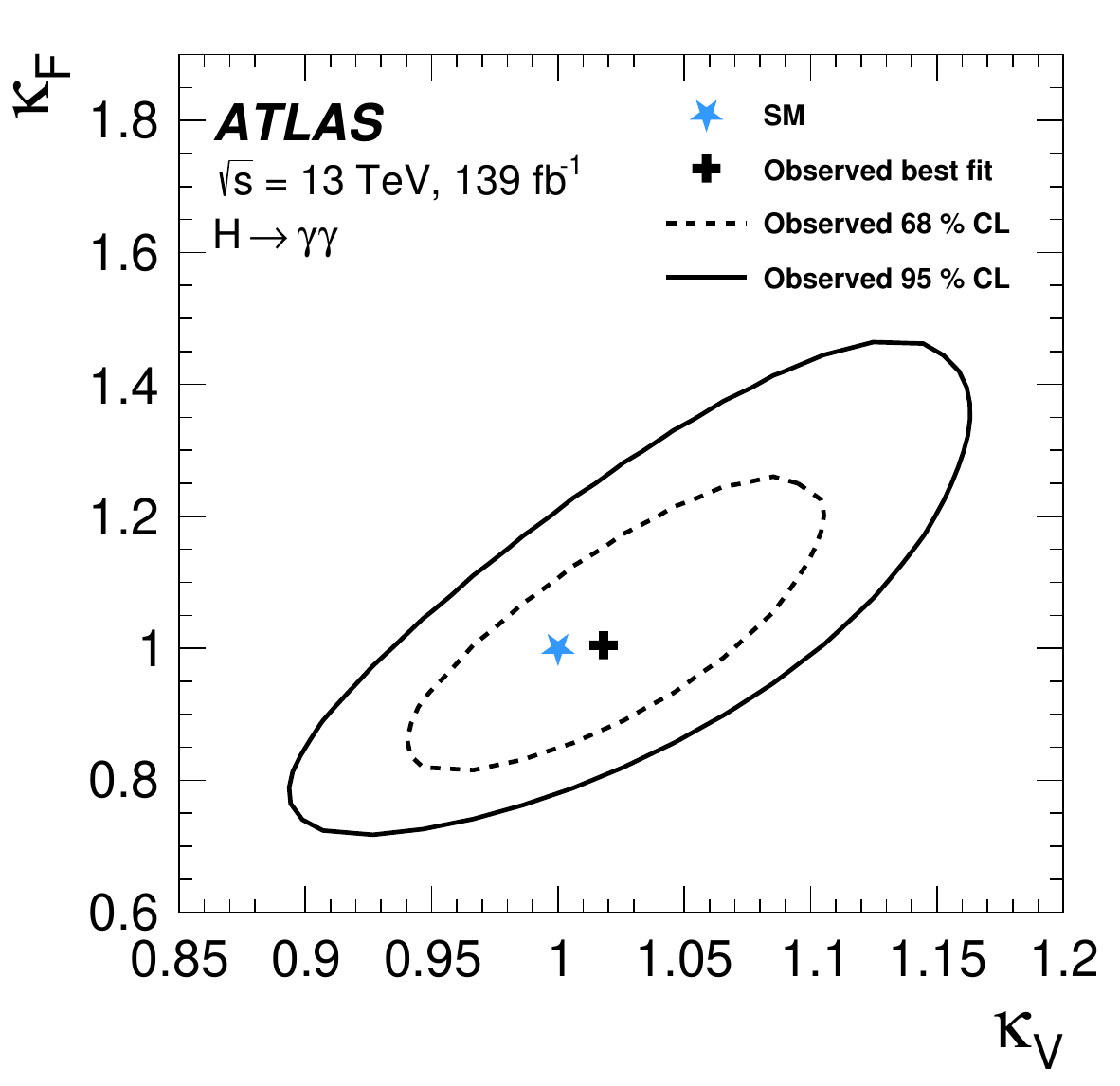}
\caption{Negative log-likelihood contours at 68\% CL (dashed line) and 95\% CL (solid line) in the ($\kappa_V$,~$\kappa_F$)~plane of modifiers applied to Higgs boson couplings to gauge bosons ($\kappa_V$) and fermions ($\kappa_F$). Loop processes and the Higgs boson total width $\Gamma_H$ are parameterized as a function of $\kappa_V$ and $\kappa_F$. The best-fit point is indicated by a cross, and the SM prediction by a star.}
\label{fig:kappas:kVkF}
\end{figure}
 
\FloatBarrier
 
\subsection{Generic parameterization using ratios of coupling modifiers}
\label{sec:kappa:ratio}
 
In this model, the effective parameterization of the \ggtoH\ and \hgg\ processes is used, and a common coupling modifier $\kappa_V = \kappa_W = \kappa_Z$ is introduced for couplings to both $W$ and $Z$ bosons. The $\kappa_\tau$ parameter is fixed to $1$ and $\kappa_b = \kappa_t$ is assumed. The total width of the Higgs boson is expressed using the effective parameterization $\Gamma_H = \kappa_H^2 \Gamma_H^{\text{SM}}$, where $\Gamma_H^{\text{SM}}$ is the SM value of the width and $\kappa_H$ is a coupling modifier.
The measurement parameters are
\begin{align*}
\kappa_{g\gamma} &= \kappa_g \kappa_\gamma/\kappa_H \\
\lambda_{Vg} &= \kappa_V/\kappa_g \\
\lambda_{tg} &= \kappa_t/\kappa_g,
\end{align*}
the first corresponding to the modifier for the $gg \rightarrow \hgg$ process, which is taken as a reference, and the others two to ratios of coupling modifiers that can be measured without assumptions about the total width of the Higgs boson. The $\lambda_{tg}$ parameter is allowed to take positive or negative values, while the other two parameters are positive by construction. Results are shown in Table~\ref{tab:kappas:results}. The negative log-likelihood scan of the $\lambda_{tg}$ parameter is shown in Figure~\ref{fig:kappas:lambda_tg}. Sensitivity to the sign of $\lambda_{tg}$ is provided by the \tH\ and \ggtoZH\ processes, and leads to exclusion of the region $\lambda_{tg} < 0$ with a significance of $2.1\sigma$.
\begin{table}[ht]
\caption{
Best-fit values and uncertainties in the coupling-modifier ratio model. The second column expresses the measured parameters in terms of the coupling modifiers. The SM corresponds to $\kappa_{g\gamma} = \lambda_{tg} = \lambda_{Vg} = 1$.}
\begin{center}
\renewcommand{\arraystretch}{1.3}
\begin{tabular}{lll}
\hline \hline
Parameter &  \begin{tabular}{@{}c@{}} Definition in terms \\ of $\kappa$ modifiers \end{tabular} & Result \\
\hline
$\kappa_{g\gamma}$        & $  \kappa_g\kappa_\gamma/\kappa_H$ & $1.02 \pm 0.06$ \\
$\lambda_{Vg}$       & $  \kappa_V/\kappa_g$         & $1.01 \pm 0.11$ \\
$\lambda_{tg}$       & $  \kappa_t/\kappa_g$         & $0.95\,\,^{+\;\,0.15}_{-\;\,0.16}$ \\
\hline \hline
\end{tabular}
\end{center}
\label{tab:kappas:results}
\end{table}
\begin{figure}[h!]
\centering
\includegraphics[width=0.6\textwidth]{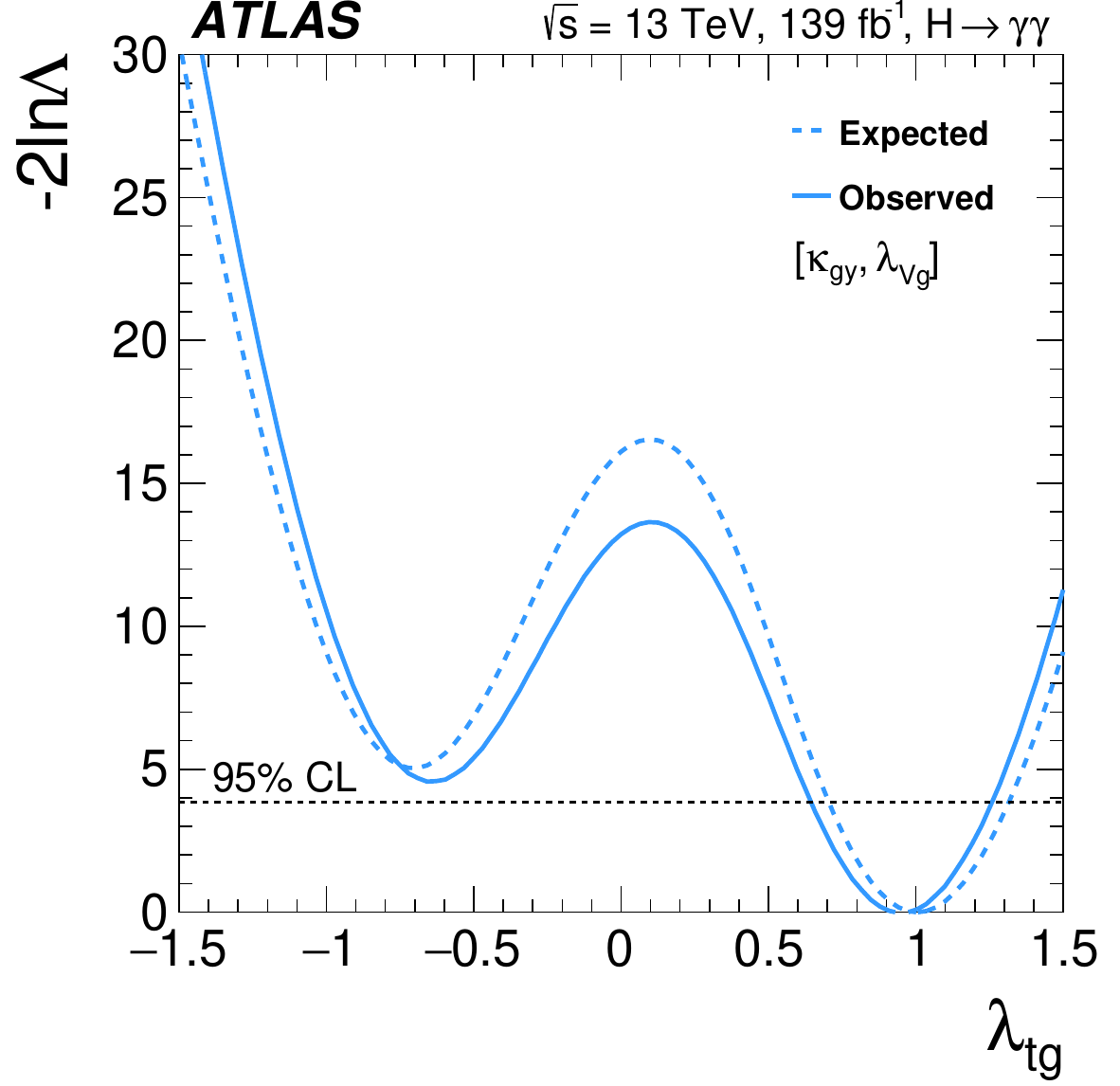}
\caption{Negative log-likelihood scan as a function of $\lambda_{tg} = \kappa_t/\kappa_g$ in the coupling-modifier ratio model described in the text. The solid curve corresponds to observed data, and the dotted curve to an Asimov data set generated under the SM hypothesis.}
\label{fig:kappas:lambda_tg}
\end{figure}

\FloatBarrier
 
\section{Effective field theory interpretation}
\label{app:eft}
 
\FloatBarrier
 
\subsection{Measurement of single SMEFT parameters}
This appendix presents the complete results of the single-parameter SMEFT measurements described in Section~\ref{sec:results:smeft:singleWC}, and illustrated in Figure~\ref{fig:eft:singleWC}. The SMEFT parameters corresponding to each operator in Table~\ref{tab:eft:ops} are individually measured, in each case while fixing the other SMEFT parameters to 0 as in the SM. Confidence intervals at 68\% and 95\% CL are computed both in observed data and in an Asimov data set generated under the SM hypothesis. Results for the parameters $c_k$ are reported in Table~\ref{tab:eft:singleWC_full}, except those where the confidence intervals extend beyond the region $|c_k|\le 20$ where the SMEFT framework is considered valid.

\begin{table}[ht]
\caption{Measurement results for each SMEFT parameter individually, obtained from profile-likelihood scans in which other SMEFT parameters are fixed to $0$. Confidence intervals at 68\% and 95\% CL are reported in data (observed) and in an Asimov dataset generated under the SM hypothesis (expected). Results in the linear and linear+quadratic SMEFT parameterizations are shown, for a scale $\Lambda=  1\,\TeV$. Confidence intervals for each parameter $c_k$ are reported, except if they extend beyond the $|c_k| \le 20$ region  where the SMEFT framework is considered valid.}
\label{tab:eft:singleWC_full}
\centering
\renewcommand{\arraystretch}{1.2}
\resizebox*{!}{0.80\textheight}{
\begin{tabular}{lllllllllll}
\toprule
\multirow{4}{*}{Parameter} & \multicolumn{6}{c}{Observed} & \multicolumn{4}{c}{Expected} \\
\cmidrule{2-11}
& \multicolumn{3}{c}{linear} & \multicolumn{3}{c}{linear+quadratic}
& \multicolumn{2}{c}{linear} & \multicolumn{2}{c}{linear+quadratic} \\
\cmidrule{2-11}
& \multirow{2}{*}{Value} & \multicolumn{2}{c}{Uncertainty}
& \multirow{2}{*}{Value} & \multicolumn{2}{c}{Uncertainty}
& \multicolumn{2}{c}{Uncertainty} & \multicolumn{2}{c}{Uncertainty} \\
\cmidrule{3-4}\cmidrule{6-11}
& & 68\% CL & 95\% CL  & & 68\% CL & 95\% CL  & 68\% CL & 95\% CL  & 68\% CL & 95\% CL \\
\midrule
\cHW &    $-0.0035$ & $^{ +0.0071}_{ -0.0077}$ & $^{  +0.014}_{  -0.016}$ &    $-0.0034$ & $^{ +0.0071}_{ -0.0073}$ & $^{  +0.014}_{  -0.014}$ & $^{ +0.0070}_{ -0.0075}$ & $^{  +0.013}_{  -0.015}$ & $^{ +0.0072}_{ -0.0072}$ & $^{  +0.014}_{  -0.014}$  \\
\cHB &    $-0.0011$ & $^{ +0.0023}_{ -0.0025}$ & $^{ +0.0044}_{ -0.0050}$ &    $-0.0011$ & $^{ +0.0023}_{ -0.0023}$ & $^{ +0.0046}_{ -0.0046}$ & $^{ +0.0022}_{ -0.0024}$ & $^{ +0.0043}_{ -0.0049}$ & $^{ +0.0023}_{ -0.0023}$ & $^{ +0.0046}_{ -0.0046}$  \\
\cHWB &     ~~\,$0.0020$ & $^{ +0.0044}_{ -0.0042}$ & $^{ +0.0090}_{ -0.0079}$ &     ~~\,$0.0019$ & $^{ +0.0042}_{ -0.0041}$ & $^{ +0.0083}_{ -0.0081}$ & $^{ +0.0043}_{ -0.0041}$ & $^{ +0.0088}_{ -0.0077}$ & $^{ +0.0042}_{ -0.0042}$ & $^{ +0.0083}_{ -0.0082}$  \\
\cHG &     ~~\,$0.0011$ & $^{ +0.0030}_{ -0.0028}$ & $^{ +0.0062}_{ -0.0053}$ &     ~~\,$0.0011$ & $^{ +0.0029}_{ -0.0028}$ & $^{ +0.0059}_{ -0.0055}$ & $^{ +0.0030}_{ -0.0027}$ & $^{ +0.0061}_{ -0.0052}$ & $^{ +0.0029}_{ -0.0028}$ & $^{ +0.0059}_{ -0.0054}$  \\
\cW &     $-0.047$ & $^{  +0.098}_{   -0.11}$ & $^{   +0.19}_{   -0.21}$ &     $-0.047$ & $^{  +0.098}_{   -0.11}$ & $^{   +0.19}_{   -0.21}$ & $^{  +0.096}_{   -0.10}$ & $^{   +0.18}_{   -0.21}$ & $^{  +0.096}_{   -0.10}$ & $^{   +0.18}_{   -0.21}$  \\
\cG &       ~~\,$0.32$ & $^{    +1.5}_{    -1.2}$ & $^{    +3.4}_{    -2.0}$ &      ~~\,$0.077$ & $^{   +0.13}_{   -0.30}$ & $^{   +0.22}_{   -0.40}$ & $^{    +1.5}_{    -1.1}$ & $^{    +3.4}_{    -1.9}$ & $^{   +0.18}_{   -0.20}$ & $^{   +0.28}_{   -0.30}$  \\
\midrule
\cuW &     $-0.039$ & $^{  +0.080}_{  -0.087}$ & $^{   +0.15}_{   -0.18}$ &     $-0.039$ & $^{  +0.080}_{  -0.087}$ & $^{   +0.15}_{   -0.18}$ & $^{  +0.079}_{  -0.083}$ & $^{   +0.15}_{   -0.17}$ & $^{  +0.079}_{  -0.083}$ & $^{   +0.15}_{   -0.17}$  \\
\cuB &     $-0.021$ & $^{  +0.043}_{  -0.046}$ & $^{  +0.082}_{  -0.094}$ &     $-0.021$ & $^{  +0.043}_{  -0.046}$ & $^{  +0.082}_{  -0.094}$ & $^{  +0.042}_{  -0.045}$ & $^{  +0.080}_{  -0.092}$ & $^{  +0.042}_{  -0.045}$ & $^{  +0.080}_{  -0.092}$  \\
\cuG &      ~~\,$0.030$ & $^{  +0.078}_{  -0.078}$ & $^{   +0.16}_{   -0.14}$ &      ~~\,$0.030$ & $^{  +0.077}_{  -0.078}$ & $^{   +0.16}_{   -0.15}$ & $^{  +0.079}_{  -0.074}$ & $^{   +0.16}_{   -0.14}$ & $^{  +0.078}_{  -0.075}$ & $^{   +0.16}_{   -0.14}$  \\
\cuH &      $-0.29$ & $^{    +1.4}_{    -1.5}$ & $^{    +2.7}_{    -3.0}$ &      $-0.30$ & $^{    +1.4}_{    -1.6}$ & $^{    +2.6}_{    -3.3}$ & $^{    +1.4}_{    -1.5}$ & $^{    +2.7}_{    -3.0}$ & $^{    +1.4}_{    -1.5}$ & $^{    +2.5}_{    -3.3}$  \\
\cdH &       ~~\,$0.63$ & $^{    +1.4}_{    -1.3}$ & $^{    +2.9}_{    -2.5}$ &       ~~\,$0.61$ & $^{    +1.3}_{    -1.3}$ & $^{    +2.5}_{    -2.7}$ & $^{    +1.4}_{    -1.3}$ & $^{    +2.8}_{    -2.4}$ & $^{    +1.3}_{    -1.4}$ & $^{    +2.5}_{    -2.7}$  \\
\ceH &        ~~\,$5.8$ & $^{     +13}_{     -12}$ &                      ~~{--}  &        ~~\,$1.9$ & $^{    +5.6}_{    -5.7}$ & $^{    +8.9}_{    -8.9}$ & $^{     +13}_{     -12}$ &                      ~~{--}  & $^{    +9.1}_{    -5.2}$ & $^{     +12}_{    -8.4}$  \\
\midrule
\cfn{Hq}{3} &     $-0.027$ & $^{  +0.091}_{  -0.081}$ & $^{   +0.19}_{   -0.15}$ &     $-0.037$ & $^{  +0.096}_{   -0.21}$ & $^{   +0.17}_{   -0.34}$ & $^{   +0.10}_{  -0.089}$ & $^{   +0.20}_{   -0.17}$ & $^{  +0.085}_{   -0.12}$ & $^{   +0.16}_{   -0.29}$  \\
\cfn{Hq}{1} &        ~~\,$1.9$ & $^{    +1.7}_{    -2.0}$ & $^{    +3.1}_{    -4.1}$ &      ~~\,$0.029$ & $^{   +0.20}_{   -0.22}$ & $^{   +0.35}_{   -0.37}$ & $^{    +2.0}_{    -2.3}$ & $^{    +3.6}_{    -4.8}$ & $^{   +0.30}_{   -0.28}$ & $^{   +0.44}_{   -0.41}$  \\
\cfn{Hl}{3} &      $-0.15$ & $^{   +0.28}_{   -0.28}$ & $^{   +0.52}_{   -0.58}$ &      $-0.15$ & $^{   +0.28}_{   -0.28}$ & $^{   +0.52}_{   -0.58}$ & $^{   +0.26}_{   -0.28}$ & $^{   +0.50}_{   -0.57}$ & $^{   +0.26}_{   -0.28}$ & $^{   +0.50}_{   -0.57}$  \\
\cfn{Hl}{1} &         ~~~{--} &                      ~~{--}  &                      ~~{--}  &        ~~\,$4.4$ & $^{    +6.8}_{    -6.9}$ & $^{     +12}_{     -12}$ & $^{     +13}_{     -15}$ &                      ~~{--}  & $^{     +16}_{    -7.8}$ &                      ~~{--}   \\
\cf{Hu} &      $-0.97$ & $^{   +0.79}_{   -0.67}$ & $^{    +1.7}_{    -1.2}$ &      $-0.14$ & $^{   +0.30}_{   -0.24}$ & $^{   +0.51}_{   -0.41}$ & $^{   +0.96}_{   -0.82}$ & $^{    +2.0}_{    -1.5}$ & $^{   +0.32}_{   -0.43}$ & $^{   +0.49}_{   -0.61}$  \\
\cf{Hd} &        ~~\,$3.4$ & $^{    +2.2}_{    -2.6}$ & $^{    +4.0}_{    -5.4}$ &      ~~\,$0.070$ & $^{   +0.33}_{   -0.36}$ & $^{   +0.55}_{   -0.60}$ & $^{    +2.7}_{    -3.1}$ & $^{    +4.9}_{    -6.5}$ & $^{   +0.51}_{   -0.44}$ & $^{   +0.73}_{   -0.67}$  \\
\cf{He} &         -- &                      --  &                      --  &        7.3 & $^{     +10}_{     -11}$ &                      --  &                      --  &                      --  &                      --  &                      --   \\
\midrule
\cHbox &       ~~\,$0.68$ & $^{    +1.5}_{    -1.4}$ & $^{    +3.1}_{    -2.7}$ &       ~~\,$0.63$ & $^{    +1.4}_{    -1.4}$ & $^{    +2.7}_{    -2.7}$ & $^{    +1.5}_{    -1.4}$ & $^{    +3.0}_{    -2.6}$ & $^{    +1.4}_{    -1.4}$ & $^{    +2.8}_{    -2.9}$  \\
\cHD &      $-0.21$ & $^{   +0.42}_{   -0.44}$ & $^{   +0.79}_{   -0.91}$ &      $-0.21$ & $^{   +0.42}_{   -0.45}$ & $^{   +0.79}_{   -0.93}$ & $^{   +0.41}_{   -0.43}$ & $^{   +0.77}_{   -0.88}$ & $^{   +0.40}_{   -0.43}$ & $^{   +0.76}_{   -0.89}$  \\
\midrule
\cfn{qq}{3} &       ~~\,$0.72$ & $^{    +3.4}_{    -2.8}$ & $^{    +7.3}_{    -5.0}$ &      $-0.20$ & $^{   +0.55}_{   -0.18}$ & $^{   +0.69}_{   -0.32}$ & $^{    +3.2}_{    -2.6}$ & $^{    +6.8}_{    -4.7}$ & $^{   +0.29}_{   -0.31}$ & $^{   +0.43}_{   -0.46}$  \\
\cfn{qq}{3}[\prime] &      ~~\,$0.042$ & $^{   +0.37}_{   -0.28}$ & $^{   +0.83}_{   -0.50}$ &      $-0.30$ & $^{   +0.52}_{   -0.19}$ & $^{   +0.67}_{   -0.34}$ & $^{   +0.38}_{   -0.29}$ & $^{   +0.84}_{   -0.49}$ & $^{   +0.21}_{   -0.44}$ & $^{   +0.36}_{   -0.60}$  \\
\cfn{qq}{1} &        ~~\,$2.0$ & $^{     +14}_{     -11}$ &                      ~~{--}  &      $-0.20$ & $^{   +0.69}_{   -0.30}$ & $^{   +0.90}_{   -0.52}$ & $^{     +14}_{     -11}$ &                      ~~{--}  & $^{   +0.44}_{   -0.45}$ & $^{   +0.67}_{   -0.68}$  \\
\cfn{qq}{1}[\prime] &      ~~\,$0.097$ & $^{   +0.79}_{   -0.60}$ & $^{    +1.7}_{    -1.0}$ &      $-0.50$ & $^{   +0.92}_{   -0.31}$ & $^{    +1.2}_{   -0.57}$ & $^{   +0.79}_{   -0.61}$ & $^{    +1.8}_{    -1.0}$ & $^{   +0.39}_{   -0.73}$ & $^{   +0.66}_{    -1.0}$  \\
\cf{ll}[\prime] &       ~~\,$0.30$ & $^{   +0.53}_{   -0.56}$ & $^{    +1.1}_{    -1.0}$ &       ~~\,$0.30$ & $^{   +0.52}_{   -0.56}$ & $^{    +1.1}_{    -1.0}$ & $^{   +0.55}_{   -0.51}$ & $^{    +1.1}_{   -0.98}$ & $^{   +0.54}_{   -0.51}$ & $^{    +1.1}_{   -0.99}$  \\
\cf{uu} &        ~~\,$1.4$ & $^{     +13}_{    -9.9}$ &                      ~~{--}  &      $-0.25$ & $^{   +0.85}_{   -0.37}$ & $^{    +1.1}_{   -0.64}$ & $^{     +13}_{    -9.9}$ &                      ~~{--}  & $^{   +0.53}_{   -0.56}$ & $^{   +0.81}_{   -0.84}$  \\
\cf{uu}[\prime] &      ~~\,$0.098$ & $^{   +0.80}_{   -0.61}$ & $^{    +1.8}_{    -1.1}$ &      $-0.50$ & $^{   +0.92}_{   -0.31}$ & $^{    +1.2}_{   -0.57}$ & $^{   +0.81}_{   -0.61}$ & $^{    +1.8}_{    -1.1}$ & $^{   +0.39}_{   -0.72}$ & $^{   +0.66}_{    -1.0}$  \\
\cfn{qu}{1} &         ~~~{--} &                      ~~{--}  &                      ~~{--}  &      $-0.30$ & $^{    +1.1}_{   -0.48}$ & $^{    +1.4}_{   -0.81}$ &                      ~~{--}  &                      ~~{--}  & $^{   +0.68}_{   -0.70}$ & $^{    +1.0}_{    -1.0}$  \\
\cfn{qu}{8} &       ~~\,$0.15$ & $^{    +1.3}_{   -0.97}$ & $^{    +2.8}_{    -1.7}$ &       $-1.8$ & $^{    +2.6}_{   -0.82}$ & $^{    +3.3}_{    -1.5}$ & $^{    +1.3}_{   -0.97}$ & $^{    +2.8}_{    -1.7}$ & $^{   +0.84}_{    -2.3}$ & $^{    +1.5}_{    -3.0}$  \\
\cfn{qd}{1} &         ~~~{--} &                      ~~{--}  &                      ~~{--}  &       ~~\,$0.75$ & $^{   +0.94}_{    -2.4}$ & $^{    +1.7}_{    -3.2}$ &                      ~~{--}  &                      ~~{--}  & $^{    +1.5}_{    -1.5}$ & $^{    +2.3}_{    -2.3}$  \\
\cfn{qd}{8} &       ~~\,$0.53$ & $^{    +5.5}_{    -4.3}$ & $^{     +12}_{    -7.5}$ &       $-2.3$ & $^{    +4.8}_{    -2.3}$ & $^{    +6.4}_{    -3.9}$ & $^{    +5.6}_{    -4.3}$ & $^{     +12}_{    -7.5}$ & $^{    +2.4}_{    -4.1}$ & $^{    +4.0}_{    -5.8}$  \\
\cfn{ud}{1} &         ~~~{--} &                      ~~{--}  &                      ~~{--}  &       ~~\,$0.75$ & $^{   +0.93}_{    -2.4}$ & $^{    +1.7}_{    -3.1}$ &                      ~~{--}  &                      ~~{--}  & $^{    +1.5}_{    -1.5}$ & $^{    +2.3}_{    -2.3}$  \\
\cfn{ud}{8} &       ~~\,$0.53$ & $^{    +5.5}_{    -4.3}$ & $^{     +12}_{    -7.5}$ &       $-2.5$ & $^{    +5.1}_{    -2.0}$ & $^{    +6.7}_{    -3.6}$ & $^{    +5.6}_{    -4.4}$ & $^{     +12}_{    -7.6}$ & $^{    +2.4}_{    -4.1}$ & $^{    +4.0}_{    -5.7}$  \\
\bottomrule
\end{tabular}}
\end{table}


\pagebreak
\FloatBarrier

\subsection{Simultaneous measurement of SMEFT parameters}
This appendix presents the complete results of the simultaneous measurement of SMEFT parameters described in Section~\ref{sec:results:smeft:evn}, shown in part in Table~\ref{tab:eft:results} and illustrated in Figure~\ref{fig:eft:results}.
The measurement parameters \EVn\ are shown in Table~\ref{tab:eft:EVs}.
Confidence intervals at 68\% and 95\% CL for the \EVn\ parameters defined in Table~\ref{tab:eft:EVs} are computed both in observed data and in an Asimov data set generated under the SM hypothesis. Results for the linear SMEFT parameterization are shown in Table~\ref{tab:eft:results_linear} and for the linear+quadratic parameterization in Table~\ref{tab:eft:results_bsm}.

\begin{table}
\caption{Measurement directions corresponding to the 12 largest eigenvalues of the Fisher information matrix of the SMEFT interpretation of the STXS measurement, shown as a decomposition in terms of Wilson coefficients in the Warsaw basis. The information matrix is obtained from the covariance matrix $C^{-1}_{\text{STXS}}$ of the STXS measurement, computed using an Asimov data set generated in the SM hypothesis, propagated to the SMEFT measurement using the linear parameterization. Each linear combination is normalized to unit Euclidean norm. Only Wilson coefficients with a coefficient larger than 0.01 are shown.}
\renewcommand*{\arraystretch}{1.3}
\centering
\resizebox{\linewidth}{!}{
\begin{tabular}{p{0.1\textwidth} p{0.9\textwidth}}
\toprule
Eigenvalue & Eigenvector\\
\midrule
350000 & $-0.53\cHG -0.02\cuG +0.23\cHW +0.71\cHB -0.4\cHWB +0.02\cW +0.02\cuW +0.04\cuB$ \\
34000 & $-0.85\cHG -0.02\cuG -0.14\cHW -0.44\cHB +0.25\cHWB -0.01\cW -0.01\cuW -0.02\cuB +0.01\cfn{Hq}{3}$ \\
110 & $-0.01\cHG +0.05\cuG -0.17\cHW +0.03\cHB -0.04\cHWB -0.01\cfn{Hl}{3} -0.98\cfn{Hq}{3} -0.07\cf{Hu} +0.02\cf{Hd} +0.03\cfn{Hq}{1} +0.01\cfn{qq}{1}[\prime] +0.03\cfn{qq}{3}[\prime] +0.01\cf{uu}[\prime]$ \\
20 & $-0.01\cHG +0.68\cuG -0.06\cuH -0.08\cHW +0.01\cHB -0.04\cHWB +0.13\cfn{Hl}{3} -0.07\cf{ll}[\prime] -0.01\cHD +0.08\cfn{Hq}{3} +0.14\cG +0.01\cfn{qq}{1} +0.27\cfn{qq}{1}[\prime] +0.06\cfn{qq}{3} +0.56\cfn{qq}{3}[\prime] +0.02\cf{uu} +0.26\cf{uu}[\prime] +0.04\cfn{ud}{8} +0.17\cfn{qu}{8} +0.04\cfn{qd}{8}$ \\
2.9 & $-0.02\cHG +0.64\cuG -0.09\cuH -0.24\cHW +0.04\cHB -0.06\cHWB +0.15\cfn{Hl}{3} -0.09\cf{ll}[\prime] -0.01\cHD +0.05\cfn{Hq}{3} +0.02\cf{Hu} -0.02\cfn{Hq}{1} -0.19\cG -0.02\cfn{qq}{1} -0.28\cfn{qq}{1}[\prime] -0.04\cfn{qq}{3} -0.52\cfn{qq}{3}[\prime] -0.01\cf{uu} -0.27\cf{uu}[\prime] -0.03\cfn{ud}{8} -0.16\cfn{qu}{8} -0.03\cfn{qd}{8}$ \\
1.8 & $-0.24\cuG +0.01\cuH -0.9\cHW +0.21\cHB -0.14\cHWB +0.01\cuB -0.11\cfn{Hl}{3} +0.01\cHD +0.15\cfn{Hq}{3} +0.1\cf{Hu} -0.03\cf{Hd} -0.08\cfn{Hq}{1} +0.03\cG +0.05\cfn{qq}{1}[\prime] +0.08\cfn{qq}{3}[\prime] +0.05\cf{uu}[\prime] +0.03\cfn{qu}{8}$ \\
0.89 & $+0.03\cuG +0.03\cuH +0.09\cHW +0.15\cHB +0.32\cHWB +0.02\cfn{Hl}{3} +0.01\cf{ll}[\prime] +0.05\cHD -0.1\cfn{Hq}{3} +0.83\cf{Hu} -0.25\cf{Hd} -0.31\cfn{Hq}{1} -0.04\cf{He} -0.05\cfn{Hl}{1}$ \\
0.075 & $+0.27\cuG +0.38\cuH +0.02\cHW +0.06\cHB +0.1\cHWB +0.02\cuW -0.78\cfn{Hl}{3} +0.37\cf{ll}[\prime] +0.07\cHD +0.01\cfn{Hq}{3} -0.04\cf{Hu} +0.09\cfn{Hq}{1} +0.09\cG -0.03\cfn{qq}{1}[\prime] -0.06\cfn{qq}{3} -0.04\cfn{qq}{3}[\prime] -0.03\cf{uu}[\prime] -0.02\cfn{qu}{8}$ \\
0.038 & $+0.01\cuG +0.03\cuH +0.09\cHW -0.38\cHB -0.65\cHWB -0.08\cuW -0.17\cfn{Hl}{3} +0.03\cf{ll}[\prime] -0.08\cHD -0.02\cfn{Hq}{3} +0.13\cf{Hu} +0.04\cf{Hd} -0.56\cfn{Hq}{1} +0.09\cf{He} +0.12\cfn{Hl}{1} +0.02\cG +0.18\cfn{qq}{3} -0.02\cfn{qq}{3}[\prime]$ \\
0.027 & $+0.06\cuH +0.02\cHW -0.09\cHB -0.13\cHWB +0.37\cuW +0.05\cfn{Hl}{3} -0.02\cf{ll}[\prime] +0.02\cf{Hd} -0.14\cfn{Hq}{1} +0.02\cf{He} +0.03\cfn{Hl}{1} -0.05\cG +0.04\cfn{qq}{1}[\prime] -0.89\cfn{qq}{3} +0.06\cfn{qq}{3}[\prime] +0.03\cf{uu}[\prime] +0.02\cfn{qu}{8}$ \\
0.011 & $+0.04\cuH -0.03\cHB -0.05\cHWB -0.02\cuW -0.1\cfn{Hl}{3} +0.03\cf{ll}[\prime] +0.06\cf{Hu} -0.05\cf{Hd} +0.11\cfn{Hq}{1} +0.01\cf{He} +0.02\cfn{Hl}{1} -0.95\cG +0.15\cfn{qq}{1}[\prime] +0.05\cfn{qq}{3} +0.11\cfn{qq}{3}[\prime] +0.13\cf{uu}[\prime] -0.01\cfn{qu}{1} +0.09\cfn{qu}{8}$ \\
0.0067 & $-0.01\cuG -0.15\cuH +0.01\cHW -0.2\cHB -0.36\cHWB +0.02\cuW -0.13\cfn{Hl}{3} -0.16\cf{ll}[\prime] -0.06\cHD +0.37\cf{Hu} -0.3\cf{Hd} +0.69\cfn{Hq}{1} +0.1\cf{He} +0.14\cfn{Hl}{1} +0.14\cG -0.02\cfn{qq}{1}[\prime] -0.05\cfn{qq}{3} -0.01\cfn{qq}{3}[\prime] -0.02\cf{uu}[\prime] -0.01\cfn{qu}{8}$ \\
\bottomrule
\end{tabular}
}
\label{tab:eft:EVs}
\end{table}


\begin{table}[ht]
\caption{Measured values of the \EVn\ parameters in data (observed) and in an Asimov data set generated under the SM hypothesis (expected). The linear SMEFT parameterization is used. Numbers in bold script indicate that the uncertainty band is truncated at the value for which the model pdf becomes negative.}
\label{tab:eft:results_linear}
\centering
\renewcommand{\arraystretch}{1.5}
\adjustbox{max width=0.67\textwidth}{
\begin{tabular}{llllll}
\toprule
Model parameter &  \multicolumn{3}{c}{Observed} &  \multicolumn{2}{c}{Expected} \\
\cmidrule{2-6}
& \multirow{2}{*}{Value} & \multicolumn{2}{c}{Uncertainty}  & \multicolumn{2}{c}{Uncertainty} \\
\cmidrule{3-6}
& & 68\% CL  &  95\% CL & 68\% CL  &  95\% CL\\
\midrule
EV1 &   $-0.0008$ & $^{       +0.0017}_{-0.0018}$ & $^{       +0.0032}_{       -0.0037}$ & $^{+0.0016}_{-0.0018}$ & $^{+0.0031}_{-0.0036}$ \\
EV2 &     ~~\,$0.000$  & $                  \pm 0.006$ & $^{        +0.012}_{        -0.010}$ & $^{+0.006}_{-0.005}$ & $^{ +0.011}_{ -0.010}$ \\
EV3 &      ~~\,$0.04$ & $                   \pm 0.10$ & $^{         +0.18}_{         -0.21}$ & $^{ +0.09}_{  -0.10}$ & $^{  +0.18}_{  -0.20}$ \\
EV4 &     $-0.04$ & $^{         +0.25}_{  -0.22}$ & $^{         +0.5}_{         -0.4}$ & $^{  +0.24}_{  -0.21}$ & $^{  +0.5}_{  -0.4}$ \\
EV5 &      $-0.2$ & $                  \pm 0.6$ & $^{          +1.2}_{          -1.3}$ & $           \pm 0.6$ & $^{   +1.1}_{   -1.3}$ \\
EV6 &       ~~\,$0.2$ & $                \pm 0.8$ & $^{          +1.7}_{          -1.6}$ & $^{  +0.8}_{  -0.7}$ & $               \pm 1.5$ \\
EV7 &       $-1.7$ & $               \pm 1.0$ & $^{          +2.0}_{ \mathbf{-1.3}}$ & $^{   +1.1}_{   -1.0}$ & $^{   +2.2}_{   -2.1}$ \\
EV8 &      $-0.7$ & $^{          +3.5}_{   -3.2}$ & $^{          +7}_{          -6}$ & $^{   +3.9}_{   -3.4}$ & $^{   +8}_{   -7}$ \\
EV9 &        ~~\,$7.5$ & $^{ \mathbf{+2.5}}_{   -5.2}$ & $^{ \mathbf{+2.5}}_{           -11}$ & $^{   +5}_{   -5}$ & $^{   +10}_{    -11}$ \\
EV10 &       ~~\,$0$ & $^{          +7}_{   -9}$ & $^{ \mathbf{+8}}_{           -19}$ & $^{   +5}_{   -7}$ & $^{   +9}_{    -16}$ \\
EV11 &       $-6$ & $^{          +9}_{   -10}$ & $^{           +18}_{           -19}$ & $          \pm 10$ & $           \pm 19$ \\
EV12 &        ~~\,$3$ & $^{  \mathbf{+12}}_{    -13}$ & $^{  \mathbf{+12}}_{           -25}$ & $       \pm 12$ & $          \pm 24$ \\
\bottomrule
\end{tabular}
}
\end{table}


\begin{table}[hb]
\caption{Measured values of the \EVn\ parameters in data (observed) and in an Asimov data set generated under the SM hypothesis (expected). The linear+quadratic SMEFT parameterization is used.}
\label{tab:eft:results_bsm}
\centering
\renewcommand{\arraystretch}{1.5}
\adjustbox{max width=0.67\textwidth}{
\begin{tabular}{llllll}
\toprule
Model parameter &  \multicolumn{3}{c}{Observed} &  \multicolumn{2}{c}{Expected} \\
\cmidrule{2-6}
& \multirow{2}{*}{Value} & \multicolumn{2}{c}{Uncertainty}  & \multicolumn{2}{c}{Uncertainty} \\
\cmidrule{3-6}
& & 68\% CL  &  95\% CL & 68\% CL  &  95\% CL\\
\midrule
EV1 &  ~~\,$0.004$ & $^{+0.007}_{-0.010}$ & $^{+0.014}_{-0.049}$ & $^{  +0.14}_{-0.01}$ & $^{ +0.20}_{-0.03}$ \\
EV2 & $-0.006$  & $^{+0.008}_{-0.009}$ & $^{+0.017}_{-0.030}$ & $^{+0.006}_{-0.095}$ & $^{+0.014}_{ -0.16}$ \\
EV3 &   ~~\,$0.04$ & $^{  +0.11}_{ -0.08}$ & $^{ +0.37}_{ -0.21}$ & $^{  +0.14}_{ -0.11}$ & $^{ +0.40}_{ -0.27}$ \\
EV4 &  $-0.08$ & $^{  +0.29}_{  -0.35}$ & $^{ +0.5}_{  -1.2}$ & $^{  +0.23}_{ -0.48}$ & $^{ +0.5}_{  -1.1}$ \\
EV5 &    ~~\,$0.29$ & $^{  +0.30}_{  -0.69}$ & $^{ +0.7}_{  -2.2}$ & $^{  +0.5}_{ -0.6}$ & $^{ +0.9}_{  -1.7}$ \\
EV6 &   ~~\,$0.0$ & $^{  +0.8}_{  -0.5}$ & $^{  +1.7}_{ -1.0}$ & $^{  +0.9}_{ -0.7}$ & $^{  +2.2}_{  -1.2}$ \\
EV7 &   $-0.9$ & $^{   +1.2}_{  -0.5}$ & $^{  +1.4}_{  -1.5}$ & $^{  +0.7}_{  -1.7}$ & $^{  +1.2}_{  -3.3}$ \\
EV8 &    $-1.2$ & $^{   +2.5}_{   -1.0}$ & $^{   +10}_{  -1.8}$ & $^{   +3.1}_{  -1.7}$ & $^{  +9.0}_{  -2.4}$ \\
EV9 &     ~~\,$1.7$ & $^{   +1.4}_{   -1.6}$ & $^{  +4.8}_{  -3.3}$ & $^{   +3.9}_{  -1.8}$ & $^{  +8.4}_{  -2.9}$ \\
EV10 &    ~~\,$0.4$ & $^{  +0.5}_{  -0.6}$ & $              \pm 1.2$ & $^{  +0.9}_{ -0.5}$ & $^{  +1.8}_{ -0.8}$ \\
EV11 &    ~~\,$0.05$ & $^{  +0.47}_{  -0.21}$ & $^{  +1.6}_{ -0.5}$ & $^{  +0.7}_{ -0.4}$ & $^{  +1.8}_{ -0.6}$ \\
EV12 &     ~~\,$1.2$ & $^{  +0.8}_{   -1.0}$ & $^{  +2.3}_{  -2.1}$ & $^{   +2.4}_{  -1.1}$ & $^{  +4.8}_{  -1.8}$ \\
\bottomrule
\end{tabular}
}
\end{table}


\pagebreak
\FloatBarrier
 
\subsection{Results including SMEFT propagator corrections}
\label{app:results:LP}
 
This appendix presents results similar to those in Section~\ref{sec:results:smeft}, but with SMEFT corrections applied to the mass and width parameters of off-shell SM particles, as implemented in the \SMEFTsim\ generator~\cite{SMEFTsim3}. These corrections are applied to the propagators of the $W$ and $Z$ boson, the Higgs boson and the top quark in each process, at first order in the SMEFT. These corrections are only available for the linear SMEFT parameterization.
 
Table~\ref{tab:eft:results_LP} shows the observed results with the propagator corrections included, for comparison with the ones in Table~\ref{tab:eft:results_linear}. The \EVn\ parameters are defined in the same way as for the baseline linear parameterization.
Differences from the baseline results are visible in the measurement of EV3, due to the impact of $W$ and $Z$ propagator corrections on the \qqtoVH\ processes. Small changes in the principal components of the measurements due to the propagator corrections also lead to changes in the uncertainties in other parameters, in particular EV1. This also leads to generally larger correlations between the measurements than in the baseline linear parameterization, as shown in Figure~\ref{fig:eft:correlations_LP}.
 
\begin{table}[ht]
\caption{Observed values of the \EVn\ parameters in data together with their 68\% CL and 95\% CL intervals in data for the linear SMEFT parameterization including corrections to the $W$, $Z$, Higgs boson and top quark propagators as described in the text. Numbers in bold script indicate than the uncertainty band is truncated at the value for which the probability distribution function of the fit becomes negative.}
\centering
\renewcommand{\arraystretch}{1.5}
\adjustbox{max width=\textwidth}{
\begin{tabular}{llll}
\toprule
\multirow{2}{*}{Model parameter} & \multirow{2}{*}{Value} & \multicolumn{2}{c}{Uncertainty} \\
\cmidrule{3-4}
& & 68\% CL  &  95\% CL \\
\midrule
EV1 &    ~~\,$0.0001$ & $^{       +0.0031}_{       -0.0038}$ & $^{       +0.006}_{       -0.007}$ \\
EV2 &    ~~\,$0.000$ & $                           \pm 0.006$ & $^{        +0.012}_{        -0.010}$ \\
EV3 &     ~~\,$0.05$ & $                           \pm 0.12$ & $^{         +0.22}_{         -0.26}$ \\
EV4 &   $-0.03$   & $^{            +0.25}_{         -0.23}$  & $^{          +0.5}_{          -0.4}$ \\
EV5 &      $-0.2$ & $                           \pm 0.6$ & $^{          +1.2}_{          -1.3}$ \\
EV6 &       ~~\,$0.2$ & $                           \pm 0.8$ & $^{          +1.7}_{          -1.6}$ \\
EV7 &       $-2.0$ & $^{          +1.1}_{\mathbf{-0.9}}$ & $^{          +2.2}_{\mathbf{-0.9}}$ \\
EV8 &      $-0.5$ & $^{          +3.4}_{          -3.2}$ & $^{          +7}_{          -6}$ \\
EV9 &        ~~\,$8.2$ & $^{ \mathbf{+2.5}}_{          -5.8}$ & $^{ \mathbf{+2.5}}_{           -12}$ \\
EV10 &       ~~\,$1$ & $^{          +7}_{          -9}$ & $^{ \mathbf{+8}}_{           -19}$ \\
EV11 &       $-5$ & $^{          +9}_{          -10}$ & $^{           +18}_{           -19}$ \\
EV12 &        ~~\,$4$ & $^{           +12}_{           -13}$ & $^{  \mathbf{+13}}_{           -25}$ \\
 
\hline
\bottomrule
\end{tabular}
}
\label{tab:eft:results_LP}
\end{table}
\begin{figure}[!htbp]
\centering
\includegraphics[width=0.60\textwidth,page=1]{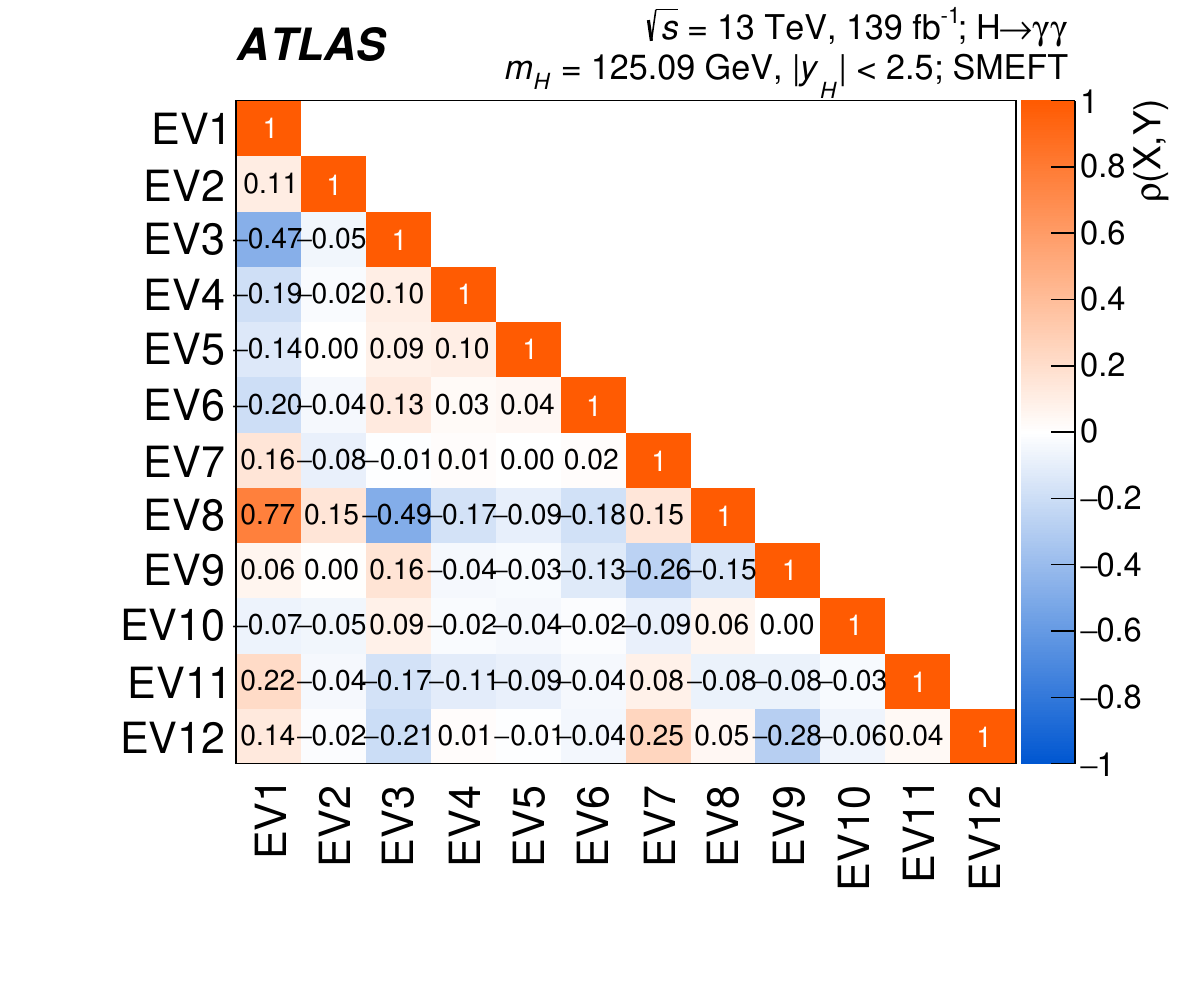}
\caption{Observed linear correlation coefficients of the EV$n$ parameters in the linear SMEFT parameterization including corrections to $W$, $Z$, Higgs boson and top quark propagators.}
\label{fig:eft:correlations_LP}
\end{figure}

\clearpage

\printbibliography
 
\clearpage
 
\begin{flushleft}
\hypersetup{urlcolor=black}
{\Large The ATLAS Collaboration}

\bigskip

\AtlasOrcid[0000-0002-6665-4934]{G.~Aad}$^\textrm{\scriptsize 101}$,
\AtlasOrcid[0000-0002-5888-2734]{B.~Abbott}$^\textrm{\scriptsize 119}$,
\AtlasOrcid[0000-0002-7248-3203]{D.C.~Abbott}$^\textrm{\scriptsize 102}$,
\AtlasOrcid[0000-0002-1002-1652]{K.~Abeling}$^\textrm{\scriptsize 55}$,
\AtlasOrcid[0000-0002-8496-9294]{S.H.~Abidi}$^\textrm{\scriptsize 29}$,
\AtlasOrcid[0000-0002-9987-2292]{A.~Aboulhorma}$^\textrm{\scriptsize 35e}$,
\AtlasOrcid[0000-0001-5329-6640]{H.~Abramowicz}$^\textrm{\scriptsize 150}$,
\AtlasOrcid[0000-0002-1599-2896]{H.~Abreu}$^\textrm{\scriptsize 149}$,
\AtlasOrcid[0000-0003-0403-3697]{Y.~Abulaiti}$^\textrm{\scriptsize 116}$,
\AtlasOrcid[0000-0003-0762-7204]{A.C.~Abusleme~Hoffman}$^\textrm{\scriptsize 136a}$
% ...
\bigskip
\\

% ...
$^{29}$Physics Department, Brookhaven National Laboratory, Upton NY; United States of America.\\
% ...
$^{35}$$^{(a)}$Facult\'e des Sciences Ain Chock, R\'eseau Universitaire de Physique des Hautes Energies - Universit\'e Hassan II, Casablanca;$^{(b)}$Facult\'{e} des Sciences, Universit\'{e} Ibn-Tofail, K\'{e}nitra;$^{(c)}$Facult\'e des Sciences Semlalia, Universit\'e Cadi Ayyad, LPHEA-Marrakech;$^{(d)}$LPMR, Facult\'e des Sciences, Universit\'e Mohamed Premier, Oujda;$^{(e)}$Facult\'e des sciences, Universit\'e Mohammed V, Rabat;$^{(f)}$Institute of Applied Physics, Mohammed VI Polytechnic University, Ben Guerir; Morocco.\\
% ...
$^{55}$II. Physikalisches Institut, Georg-August-Universit\"{a}t G\"ottingen, G\"ottingen; Germany.\\
% ...
$^{101}$CPPM, Aix-Marseille Universit\'e, CNRS/IN2P3, Marseille; France.\\
$^{102}$Department of Physics, University of Massachusetts, Amherst MA; United States of America.\\
% ...
$^{116}$Department of Physics, New York University, New York NY; United States of America.\\
% ...
$^{119}$Homer L. Dodge Department of Physics and Astronomy, University of Oklahoma, Norman OK; United States of America.\\
% ...
$^{136}$$^{(a)}$Departamento de F\'isica, Pontificia Universidad Cat\'olica de Chile, Santiago;$^{(b)}$Millennium Institute for Subatomic physics at high energy frontier (SAPHIR), Santiago;$^{(c)}$Instituto de Investigaci\'on Multidisciplinario en Ciencia y Tecnolog\'ia, y Departamento de F\'isica, Universidad de La Serena;$^{(d)}$Universidad Andres Bello, Department of Physics, Santiago;$^{(e)}$Instituto de Alta Investigaci\'on, Universidad de Tarapac\'a, Arica;$^{(f)}$Departamento de F\'isica, Universidad T\'ecnica Federico Santa Mar\'ia, Valpara\'iso; Chile.\\
% ...
$^{149}$Department of Physics, Technion, Israel Institute of Technology, Haifa; Israel.\\
$^{150}$Raymond and Beverly Sackler School of Physics and Astronomy, Tel Aviv University, Tel Aviv; Israel.\\
% ...
$^{a}$ Also Affiliated with an institute covered by a cooperation agreement with CERN.\\
$^{b}$ Also at An-Najah National University, Nablus; Palestine.\\
$^{c}$ Also at Borough of Manhattan Community College, City University of New York, New York NY; United States of America.\\
$^{d}$ Also at Bruno Kessler Foundation, Trento; Italy.\\
$^{e}$ Also at Center for High Energy Physics, Peking University; China.\\
$^{f}$ Also at Centro Studi e Ricerche Enrico Fermi; Italy.\\
% ...

\end{flushleft}


\FloatBarrier

\end{document}